\documentclass[11pt]{article}
\usepackage[cp1250]{inputenc}
\usepackage{amsmath} 
\usepackage{amssymb} 
\usepackage{a4wide} 
\usepackage[left=2cm,right=2cm]{geometry} 
\usepackage{braket}
\usepackage{slashed}
\usepackage{bbm}
\usepackage{yfonts}
\usepackage{hyperref}
\usepackage{subcaption}
\usepackage[boxsize=0.7em,centertableaux]{ytableau}
\usepackage{tikz}
\usetikzlibrary{matrix,arrows,automata,positioning}
\usetikzlibrary{decorations.markings, decorations.pathreplacing}

\tikzset{
    state/.style={
           rectangle,
           rounded corners,
           draw=black, thick,
           minimum height=2em,
           inner sep=2pt,
           text centered,
           },
}

\allowdisplaybreaks

\numberwithin{equation}{section}

\DeclareMathOperator{\res}{res}

\DeclareMathOperator{\lcm}{lcm}
\DeclareMathOperator{\Refl}{Ref}
\DeclareMathOperator{\Rot}{Rot}

\title{On W-algebras and ODE/IM correspondence}
\author{Mat?j Kudrna, Tomáš Procházka}

\begin{document}

\bibliographystyle{hieeetr}

\vskip 2.1cm

\centerline{\large \bf On $\mathcal{W}$-algebras and ODE/IM correspondence}
\vspace*{8.0ex}

\centerline{\large \rm Mat\v{e}j Kudrna\footnote{Email: {\tt kudrnam@fzu.cz}}, Tom\'{a}\v{s} Proch\'{a}zka\footnote{Email: {\tt prochazkat@fzu.cz}}}

\vspace*{8.0ex}

\centerline{\large \it Institute of Physics AS CR}
\centerline{\large \it Na Slovance 2, Prague 8, Czech Republic}
\vspace*{2.0ex}

\vspace*{6.0ex}

\centerline{\bf Abstract}
\bigskip

We study the ODE/IM correspondence for two-dimensional conformal field theories with Virasoro and $\mathcal{W}_N$ symmetry. Building on earlier work establishing the correspondence, we develop a systematic algorithm for calculating the eigenvalues of local integrals of motion in terms of the Bethe roots using formal WKB expansions of wave functions associated to the differential operators. The method is demonstrated explicitly for Virasoro, $\mathcal{W}_3$, and $\mathcal{W}_4$ algebras, yielding closed expressions for the eigenvalues of the first few local quantum KdV Hamiltonians. A key geometric structure emerging from our analysis is the mirror curve, a three-punctured sphere that is naturally covered by the WKB curve. We show how the algebraic properties of the $\mathcal{W}$-symmetry algebras are reflected in the geometry of these curves, and how period integrals on these curves reproduce the spectral data of the integrable system. Applications to Argyres–Douglas minimal models allow us to test the prescription both analytically and numerically and we find complete agreement between the calculations in different triality frames. Finally, we examine large rank limits of ground state eigenvalues and show that they match the genus expansion of the topological string partition function on $\mathbbm{C}^3$.

\vfill \eject

\tableofcontents

\setcounter{footnote}{0}

\newpage

\section{Introduction}
Two-dimensional conformal field theories are very rich yet to large extent solvable models of quantum field theory. The usual approach to solvability is largely based on representation theory of the underlying worldsheet symmetry algebra, the Virasoro algebra, combined with the bootstrap approach \cite{Belavin:1984vu,DiFrancesco:1997nk}. More recently, many new insights about Virasoro algebra and its higher spin extensions \cite{Bouwknegt:1992wg,Gaberdiel:2010pz,Gaberdiel:2011wb,Candu:2012tr} have been obtained by applying tools from other areas, such as the quantum integrability \cite{Bazhanov:1994ft,Bazhanov:1996dr,Bazhanov:1996aq,Bazhanov:1998dq} and the related ODE/IM correspondence \cite{Dorey:1998pt,Dorey:1999uk,Dorey:2001uw,Dorey:2007zx,Dorey:2019ngq}, higher dimensional gauge theories \cite{Nekrasov:2002qd,Alday:2009aq,Alba:2010qc,Beem:2013sza}, topological strings \cite{Aganagic:2003db,Aganagic:2003qj,Okounkov:2003sp,Iqbal:2003ds} and (geometric) representation theory \cite{feigin:2011a,feigin:2011b,feigin2012quantum,Maulik:2012wi,schifvas}.

One of the central questions of quantum integrability is the construction of commuting operators and studying their spectra. In the context of 2d CFT, the ODE/IM correspondence provides a remarkable bridge between the integrable structures underlying 2d CFT and spectral theory of ordinary differential operators. In the simplest setting, simultaneous eigenstates of commuting operators on CFT side are mapped to Schr\"odinger-type ordinary differential operators and the the eigenvalues are encoded in spectral and monodromy data associated to these ODEs.

Despite the fact that the basic ODE/IM correspondence for Virasoro algebra and its higher spin extensions has been established long ago \cite{Bazhanov:2003ni,Dorey:2006an,Feigin:2007mr,Masoero:2018rel,Dorey:2019ngq}, the explicit expressions for eigenvalues of local integrals of motion in terms of solutions of Bethe equations, especially in the context of $\mathcal{W}_N$ algebras, have not been worked out. In this work, we provide a systematic algorithm based on formal WKB expansion of the wave functions associated to ODEs for extracting the eigenvalues of local integrals of motion from Bethe roots characterizing a given eigenstate. We demonstrate this concretely in the case of Virasoro and $\mathcal{W}_3$ and $\mathcal{W}_4$ algebras.

A key geometric insight that emerges from our analysis is the central role played by the mirror curve associated to the system. This curve, a three-punctured sphere, arises naturally in uniformization of the WKB curve associated to the differential operator. In the context of the topological string, it is the mirror curve to $\mathbbm{C}^3$ which in turn is very closely related to the symmetry algebra itself. The algebraic properties of the symmetry algebra are nicely reflected in the geometric properties of the three-punctured sphere.

\subsection{Organization of the paper}

The remainder of this paper is organized as follows: Section \ref{secvirasoro} is to large extent a review of the simplest instance of ODE/IM, the case of Virasoro algebra. We introduce infinitely many commuting local charges labeled by odd integers. These are usually called the quantum KdV or BLZ charges \cite{Bazhanov:1994ft,Bazhanov:1996dr,Bazhanov:1998dq} and can be thought of as a quantum version of the Hamiltonians of the classical KdV hierarchy \cite{Zabrodin:2018uwz}. Next we discuss the problem of their diagonalization. Since $L_0$ (the Virasoro zero Fourier mode) is a part of the commuting family and since it has finite dimensional eigenspaces, the diagonalization problem reduces to a problem of linear algebra. A useful description of simultaneous eigenvectors was found by Bazhanov, Lukyanov and Zamolodchikov \cite{Bazhanov:2003ni} in the context of ODE/IM correspondence \cite{Dorey:1998pt,Dorey:1999uk,Dorey:2001uw,Dorey:2007zx,Dorey:2019ngq}. According to this correspondence, to every simultaneous eigenvector of commuting Hamiltonians on the IM (Virasoro) side, there corresponds an ordinary differential operator. The spectral or Stokes data associated to this differential operator encode the simultaneous eigenvalues of our family of Hamiltonians. We first focus on the highest weight states (the primaries). The differential operators that we find in this case are Schr\"odinger-like differential operators with singular points at zero and infinity encoding the central charge and the primary field we are interested in \cite{Bazhanov:1998wj}. Following \cite{Dorey:2019ngq} we explain how the formal WKB wave function and its periods exactly agree with the eigenvalues of quantum KdV charges. The basic geometric object that arises from the WKB analysis is the WKB curve $\mathcal{W}$, in this case a hyperelliptic curve. The differential operator on ODE side compactly encodes an infinite collection of $1$-forms living on $\mathcal{W}$. We calculate period integrals of these $1$-forms and find a nice agreement with explicit calculations of eigenvalues using the commutation relations of Virasoro algebra.

Next we include the descendants (excited states) following \cite{Bazhanov:2003ni}. We dress the differential operator by a finite collection of regular singular points, their positions being analogous to rapidities parametrizing off-shell Bethe states. To get a finite number of states corresponding to finite number of states on the IM side, we impose quantization conditions (Bethe equations): we requite that the local monodromy of solutions of ODE around the added regular singular points is trivial. For generic values of parameters this gives exactly the right number of states \cite{Conti:2020zft}. We conclude the section by repeating the WKB analysis, this time including the contribution of descendants. We find that the WKB curve is not modified and but the information about descendants deforms the infinite tower of $1$-forms living on $\mathcal{W}$. Evaluating their periods, we find an exact agreement with explicit calculations on the CFT side.

In the next section, we review some basic properties of $\mathcal{W}_{\infty}$ that are needed in the rest of this article \cite{Prochazka:2014gqa,Prochazka:2015deb,Prochazka:2024xyd}. Virasoro algebra is the first member of a family of algebras called $\mathcal{W}_N$. The parameter $N$ is rank-like parameter and Virasoro algebra corresponds to $N=2$. In general, each $\mathcal{W}_N$ is generated by local fields of dimension $2, 3, \ldots, N$. All of these algebras can be understood as quotients of two-parametric family of algebras $\mathcal{W}_\infty$. This larger family enjoys symmetries and properties that are not manifest at $\mathcal{W}_N$ level. The prime example of this being the triality symmetry \cite{Gaberdiel:2012ku}.

Even though from the perspective of $\mathcal{W}_N$ algebras and their Bethe equations it would be most natural to consider the parameters to be integer $N$ and generic central charge $c$, the geometric considerations simplify if we specialize to discrete family of $\mathcal{W}_N$ minimal models which have special values of the central charge $c$. We call these Argyres-Douglas minimal models, because via 4d SCFT/VOA correspondence of \cite{Beem:2013sza,Jeong:2019pzg} they correspond to a class of 4d $\mathcal{N}=2$ superconformal field theories, the Argyres-Douglas theories \cite{Shapere:1999xr,Aharony:2007dj,Shapere:2008zf,Cecotti:2010fi,Gaiotto:2012sf,Xie:2012hs,Tachikawa:2013kta,Beem:2014zpa,Buican:2015ina,Cordova:2015nma,Agarwal:2017roi,Fredrickson:2017yka,Ito:2017ypt,Song:2017oew,Bah:2021mzw}. It becomes clear from the analysis that most of the calculations can be done for more general rational values of the central charge, but the specialization that we choose makes the discussion simpler. On the level of the Bethe ansatz equations and spectra of the commuting quantities, this specialization also leads to no loss of generality, because all the expressions that we find depend algebraically on the parameters -- once we have the expressions for Argyres-Douglas minimal models, the same expressions are true also in general.

We continue the section by defining the quantum commuting Hamiltonians in $\mathcal{W}_N$ or $\mathcal{W}_\infty$ setting and finding explicit expressions for the first few of these. We conclude by discussing different parametrizations of the highest weight vectors in the context of $\mathcal{W}_N$ or $\mathcal{W}_{\infty}$, generalizing the conformal weight that is familiar in the Virasoro setting. For Argyres-Douglas models, there is a particularly nice class of representations labeled by Young diagrams. These minimal models are not unitary, so the translationally-invariant vacuum representation differs from the representation of minimal energy (whose primary corresponds to the ground state of the system on cylinder). Since the minimal energy representation plays a special role in the following, we determine the correct labels parametrizing this state.

In Section \ref{secgeometry}, we discuss the geometry of algebraic curves $\mathcal{W}$ that will later appear as WKB curves associated to Argyres-Douglas minimal models. These curves are labeled by a coprime pair of integers $(K,N)$. In the 4d setup the corresponding Argyres-Douglas theories SCFTs are usually labeled as $(A_{K-1},A_{N-1})$. We calculate the genus of these curves and their symmetry groups. Quotienting out by these automorphisms, the WKB curves naturally project to a three-punctured sphere $\mathcal{M}$. We identify this curve with the mirror curve appearing in the mirror description of the topological vertex \cite{Okounkov:2003sp}. The three punctures naturally correspond to three parameters of $\mathcal{W}_\infty$ algebra.

Since a part of the prescription for calculating the eigenvalues of quantum Hamiltonians is to calculate the contour integral of $1$-forms along various cycles, it is useful to have a good understanding of the homology of $\mathcal{W}$. It turns out that the usual description using symplectic basis of cycles is not the most convenient one for our purposes, because it does not respect the symmetries that the WKB curve naturally has. We instead use a collection of $1$-cycles on $\mathcal{W}$ that are lifts of a fundamental Pochhammer cycle on $\mathcal{M}$. We find that the intersection form of these cycles naturally reflects the basic structure function of $\mathcal{W}_\infty$ (which in turn is an algebraic manifestation of the fundamental box in the associated box counting problems). Even though the calculation of the periods only requires understanding of the homology of $\mathcal{W}$, since $\mathcal{W}$ are compact Riemann surfaces, we can do better and find canonical representatives of the homology classes. In order to do that, we put a hyperbolic metric on $\mathcal{W}$. The related problem of uniformization nicely connects our curves to the Schwarzian triangle maps and hyperbolic tilings studied in 19th century \cite{caratheodory1954theory,siegel1969topics,gray2008linear}. Finally we discuss how to calculate the period integrals. In general period integrals on $\mathcal{W}$ could be complicated transcendental functions because the Riemann surfaces $\mathcal{W}$ are hyperbolic, but the high symmetry of $\mathcal{W}$ allows translating all the calculations down to the mirror curve $\mathcal{M}$. The period integrals then reduce to Euler beta functions so everything we need can be calculated in closed form.

In Section \ref{secgennk} we generalize the ODE/IM prescription to general $N$. The general prescription is already known from \cite{Dorey:2006an,Feigin:2007mr,Masoero:2018rel}, so after fixing the conventions we focus on the explicit algorithm for calculation of the WKB periods. This in principle allows one to find expressions for any local integral of motion in our family in terms of the parameters of the associated differential operator. We compare the period integrals with the explicit eigenvalues evaluated on the CFT side and in this way we find a precise identification of the parameters. With suitable choice of normalization, everything becomes manifestly invariant under the triality symmetry of the algebra. In the following Section \ref{secw3} we discuss the explicit form of the Bethe equations for $\mathcal{W}_3$ and $\mathcal{W}_4$. The $\mathcal{W}_3$ case was already considered in \cite{Masoero:2019wqf}. We see that the Bethe equations are getting more and more complicated as we increase the rank $N$ and writing a system that is more uniform in $N$ requires ideas that are outside of the scope of the present paper.

In the following Section \ref{secexamples}, we test the validity of our calculations by explicitly making analytic and numerical tests for simpler modes, in particular for the Lee-Yang $(A_1,A_2)$ model and for the non-trivial $(A_2,A_3)$ minimal model which is the simplest Argyres-Douglas model that does not reduce to Virasoro algebra. We use the fact that the Argyres-Douglas models have two dual descriptions, as $\mathcal{W}$-algebras of two different ranks. The ODE/IM map and the associated Bethe equations pick one of these ranks as a preferred one so the differential operator and Bethe equations are not manifestly symmetric under the exchange of the two ranks. The fact that WKB periods nevertheless agree is a non-trivial test of our calculations. The explicit calculations also illustrate many general properties of solutions of the Bethe equations. For generic values of parameters corresponding to Verma modules (generic central charge and highest weights) all the solutions of Bethe equations are regular and their number agrees with the number of states. As soon as we tune the parameters to special values (such as those of $\mathcal{W}_N$ minimal models), some of the physical solutions become singular and conversely there are solutions that are regular but do not correspond to states in the irreducible modules. We illustrate these properties on examples, but we do not aim at finding a prescription that would identify exactly which solutions correspond to physical states in $\mathcal{W}$-algebra irreducible modules.

In the final Section \ref{seclargelimit} we have a look at few examples of large $N$ and $K$ limits of the eigenvalues of quantum Hamiltonians. The representation theory of $\mathcal{W}_\infty$ minimal models in particular implies that the characters of Argyres-Douglas minimal models should reduce as $N, K \to \infty$ to partition functions calculated by the topological vertex \cite{Aganagic:2003db,Okounkov:2003sp,Prochazka:2015deb,Prochazka:2023zdb}. This is because the topological vertex corresponds to characters of $\mathcal{W}_{1+\infty}$ at generic parameters and the $N, K \to \infty$ regime is exactly the limit where the null states go to infinity so the characters in the limit agree with the generic ones. With this motivation, we calculate the first few eigenvalues of quantum Hamiltonian for Argyres-Douglas ground state and we find that the leading order behavior agrees with the asymptotic expansion of the MacMahon function which is the relevant topological $A$-string partition function with $\mathbbm{C}^3$ target. The mirror symmetric B model calculation essentially calculates the quantum periods of the three-punctured sphere which is our curve $\mathcal{M}$, so it is not surprising that we get mathematically the same result. What is surprising is the interpretation: we are not calculating a partition function, but a ground state expectation value of quantum Hamiltonians. The result that we find is that for a specific linear combination of the Hamiltonians, the result agrees with the partition function. Finally, we discuss a large $K$ limit at fixed $N$, where the quantum curve reduces to Schr\"odinger equation with Liouville potential an its generalization to higher order kinetic terms. We again find that the WKB calculation applied to this quantum curve captures the large central charge behavior of the ground state Hamiltonian eigenvalues.

In Appendix \ref{apporbitsums} we list some useful formulas for summation over orbits under $\mathbbm{Z}_N$ group of rotations. In Appendix \ref{appuniform} we give more details about the uniformization of $\mathcal{W}$ and $\mathcal{M}$, i.e. the the procedure of putting a hyperbolic metric on these spaces. In Appendix \ref{appiom} we list expressions for the first few quantum Hamiltonians as well as formulas for their eigenvalues in terms of solutions of Bethe equations in the case of Virasoro and $\mathcal{W}_3$ algebra. The final Appendix \ref{appw3comparison} discusses the change of coordinates that maps our coordinate system and choice of differential operator to the one of Masoero and Raimondo \cite{Masoero:2019wqf}.

\section{Virasoro algebra and quantum KdV}
\label{secvirasoro}
In this section we review the commuting local integrals of motion in Virasoro algebra (the quantum KdV hierarchy) and discuss the problem of describing their spectra. We use the ODE/IM map to translate this problem to the realm of ordinary differential equations and explain how to use the WKB techniques to find explicit expressions for the eigenvalues in terms of solutions of Bethe equations. Since the Virasoro example is the most important and also the simplest one, we discuss the calculations in detail.

\subsection{Local integrals of motion}

\subsubsection{Virasoro algebra}
In this section we will look for the local integrals of motion in Virasoro algebra. The Virasoro algebra
\begin{equation}
\label{vircomrel}
\left[ L_m, L_n \right] = (m-n) L_{m+n} + \frac{c}{12} (m-1)m(m+1) \delta_{m+n,0}
\end{equation}
is the algebra of Fourier modes\footnote{This Fourier expansion corresponds to $z$ being a coordinate on the \emph{complex plane}, rather than on the \emph{cylinder} which will be the preferred geometry used in the following, see \cite{Prochazka:2019dvu} for discussion.} of the stress-energy current $T(z)$,
\begin{equation}
\label{fourierTplane}
T(z) = \sum_{m \in \mathbbm{Z}} z^{-m-2} L_m.
\end{equation}
Equivalently, the same algebraic structure is captured by the operator product expansion of $T(z)$ with itself,
\begin{equation}
\label{virasoroope}
T(z) T(w) \sim \frac{c/2}{(z-w)^4} + \frac{2T(w)}{(z-w)^2} + \frac{\partial T(w)}{z-w} + reg.
\end{equation}
where it is the singular terms in the OPE that encode the commutation relations \eqref{vircomrel}. The quantity $c$, usually called the central charge of the algebra, is the most important characteristic of the physical models with two-dimensional conformal symmetry \cite{Belavin:1984vu,DiFrancesco:1997nk}.

\subsubsection{Local fields from $T(z)$}
Starting from the stress-energy tensor $T(z)$, we can construct other local fields using two basic operations: given a local field $\phi(z)$ we can define its derivative $\partial \phi(z)$ and given two fields $\phi(z)$ and $\chi(z)$ we can define the normal ordered product $(\phi \chi)(z)$ \cite{DiFrancesco:1997nk,Prochazka:2014gqa}. The resulting vector space of local fields is graded by the engineering dimension of the fields, with $T(z)$ being of dimension $2$. The dimension of the product of two local fields is additive under taking the normal ordered product while the derivative increases the dimension by one. The product itself is neither commutative nor associative, but it satisfies generalizations of these properties \cite{DiFrancesco:1997nk,Prochazka:2014gqa,Bouwknegt:1992wg}. The derivative satisfies the Leibniz property with respect to the product. Table \ref{tabvirfields} shows a list of fields of lower dimension that can be constructed out of $T(z)$.
\begin{table}
\centering
\begin{tabular}{|c|c|}
\hline
dim & fields \\
\hline
$0$ & $\mathbbm{1}$ \\
$1$ & - \\
$2$ & $T$ \\
$3$ & $\partial T$ \\
$4$ & $(TT), \partial^2 T$ \\
$5$ & $(\partial T T), \partial^3 T$ \\
$6$ & $(T(TT)), (\partial T \partial T), (\partial^2 T T), \partial^4 T$ \\
$7$ & $(\partial T(TT)), (\partial^2 T \partial T), (\partial^3 T T), \partial^5 T$ \\
$8$ & $(T(T(TT))), (\partial T(\partial T T)), (\partial^2 T(TT)),$ \\
& $(\partial^2 T \partial^2 T), (\partial^3 T \partial T), (\partial^4 T T), \partial^6 T$ \\
\hline
\end{tabular}
\caption{List of linearly independent local fields of lower dimension that can be constructed out of stress-energy tensor $T(z)$.}
\label{tabvirfields}
\end{table}
Only linearly independent fields are given in the table, i.e. we use the relations such as
\begin{equation}
(T \partial T) = (\partial T T) + \frac{1}{6} \partial^3 T
\end{equation}
or
\begin{equation}
\partial (TT) = (\partial T T) + (T \partial T) = 2(\partial T T) + \frac{1}{6} \partial^3 T
\end{equation}
to solve for linearly dependent fields. The algebra of local fields is also filtered by the number of derivatives and therefore in order to count the number of fields of a given dimension we can forget about the non-commutativity and non-associativity of the normal ordering and count the (commutative) polynomials in variables $\partial^k T$. The corresponding generating function
\begin{equation}
\label{virvacchar}
\prod_{n=2}^\infty \frac{1}{1-q^n} = 1 + q^2 + q^3 + 2q^4 + 2q^5 + 4q^6 + 4q^7 + 7q^8 + 8q^9 + 12q^{10} + \ldots
\end{equation}
is easily seen to indeed count the number of fields constructed out of $T(z)$. Up to an overall prefactor, it agrees with the character of vacuum Verma module of Virasoro algebra. For special values of the central charge (for Virasoro minimal models) there are additional relations between the local fields, but we are focusing on the generic values of $c$ where \eqref{virvacchar} is the correct counting function.

\subsubsection{Fourier modes on cylinder}
We want to study a family of mutually commuting local integrals of motion (or Hamiltonians), i.e. family of Fourier zero modes of local fields that all mutually commute. For this it is convenient to switch from the coordinate $z$ used in \eqref{fourierTplane} which we often for simplicity call the coordinate on the complex plane, to the coordinate $w = \log z$ parametrizing the cylinder. The corresponding Fourier expansion is now
\begin{equation}
\label{fourierTcylinder}
T^{(pl)}(w) = \sum_{m \in \mathbbm{Z}} e^{-mw} T_m
\end{equation}
We will mostly use the cylinder frame as the corresponding Fourier modes $T_m$ are more natural from the point of view of canonical formalism. As operators acting on the Hilbert space they do not agree with the Fourier modes $L_m$ in \eqref{fourierTplane}, but they almost do so:
\begin{equation}
\label{cylplanevirasoro}
T_m = L_m - \frac{c}{24} \delta_{m,0}.
\end{equation}
The shift of the zero mode is important and physically corresponds to the Casimir energy on the cylinder. More generally, the shift in the labeling of Fourier modes as in \eqref{fourierTplane} is chosen in such a way that for the primary fields the Fourier modes in both plane and cylinder frame agree, but this is not true in general. For general local fields the relation between the Fourier modes is governed by the transformation property of a given field under the change of local coordinates (conformal transformation). For the stress-energy tensor $T(z)$, we have the well-known transformation rule
\begin{equation}
T(z) \to \tilde{T}(\tilde{z}) = \left(\frac{d\tilde{z}}{dz}\right)^{-2} T(z) - \frac{c}{12} \left(\frac{d\tilde{z}}{dz}\right)^{-2} \left[ \left(\frac{d^3 \tilde{z}}{dz^3}\right) \left( \frac{d\tilde{z}}{dz} \right)^{-1} - \frac{3}{2} \left( \frac{d^2 \tilde{z}}{dz^2} \right)^2 \left( \frac{d\tilde{z}}{dz} \right)^{-2} \right]
\end{equation}
and it is easy to check that the second term involving the Schwarzian derivative and proportional to central charge $c$ is responsible for the shift in \eqref{cylplanevirasoro}. For more details about the map between the complex plane and the cylinder and its relation to Fourier expansions and normal ordered product see Appendix A of \cite{Prochazka:2023zdb}.

\subsubsection{Commuting local integrals of motion}
\label{virasoroimconstr}
Now we are ready to look for local fields whose zero Fourier modes mutually commute. First of all, notice that the zero mode of a total derivative (in the cylinder frame, this would not be the case in the complex plane!) always vanishes,
\begin{equation}
(\partial \phi)_0 = \int \partial \phi = 0.
\end{equation}
Therefore, we have a unique independent zero mode of dimension $2$ field $T(z)$ (which is $L_0$ or $T_0$), no such field in dimension $3$ and a unique zero Fourier mode in dimension $4$, $(TT)_0$. We should check if $T_0$ and $(TT)_0$ commute. In order to do that, we use the formula \cite{Prochazka:2023zdb}
\begin{equation}
\label{zeromodeaction}
\left[ \phi_0, \chi(w) \right] = \res_{z \to w} \phi(z) \chi(w) \equiv \{\phi,\chi\}_{-1},
\end{equation}
i.e. the commutator of $\phi_0$ with a local field $\chi(w)$ is given by the local field that appears as a first order pole of the OPE of $\phi(z)$ with $\chi(w)$ when expanded at $z = w$. In our case, we need the OPE of $T(z)$ and $(TT)(w)$ which is\footnote{All these OPEs follow from the basic OPE \eqref{virasoroope} using the generalization of associativity conditions, see for example Appendix 6.B of \cite{DiFrancesco:1997nk}, but the calculations soon get very involved. Fortunately one can use the Mathematica package \texttt{OPEdefs} by Kris Thielmans \cite{Thielemans:1991uw} which is an invaluable tool for all the OPE calculations.}
\begin{align}
\nonumber
T(z) (TT)(w) & = \frac{3c}{(z-w)^6} + \frac{(c+8)T(w)}{(z-w)^4} + \frac{3\partial T(w)}{(z-w)^3} \\
& + \frac{4(TT)(w)}{(z-w)^2} + \frac{\partial(T T)(w)}{z-w} + reg.
\end{align}
The commutator of $T_0$ with $(TT)_0$ is simply the zero mode of the local field appearing as the coefficient of the first order pole, i.e. $\partial(TT)_0$. But this is a zero mode of a total derivative so we find
\begin{equation}
\left[ T_0, (TT)_0 \right] = (\partial(TT))_0 = 0.
\end{equation}
The commutativity with $T_0$ is actually automatic as long as $T(z)$ is the stress-energy tensor because in this case for any local field $\phi(w)$ the first order pole with $T(z)$ is simply $\partial \phi(w)$. It is easy to see the local fields of dimension $5$ are all local derivatives (this actually happens in all odd dimensions) so the next possible commuting quantity can appear in dimension $6$. Here we have two independent local fields up to total derivatives, $(T(TT))$ and $(\partial T \partial T)$. To see the commutativity with $(TT)_0$ we need to look at the OPEs
\begin{align}
(TT)(z) (T(TT))(w) & \sim \cdots + \frac{\frac{c+2}{2} (\partial T(\partial T \partial T)) + \partial(\ldots)}{z-w} + reg. \\
(TT)(z) (\partial T \partial T))(w) & \sim \cdots + \frac{6 (\partial T(\partial T \partial T))  + \partial(\ldots)}{z-w} + reg.
\end{align}
Therefore, although neither $(T(TT))_0$ nor $(\partial T \partial T)_0$ commute with $(TT)_0$, we see that the combination
\begin{equation}
I_5 \equiv (T(TT))_0 - \frac{c+2}{12} (\partial T \partial T)_0
\end{equation}
does indeed commute with both $I_1$ and $I_3$. In a similar fashion, we can look for commuting quantities of higher dimensions. Due to exponential growth of the number of fields, the commutativity with the previous Hamiltonians gives an overdetermined system of equations which would generically not have any solution. But explicit calculation in lower dimension shows that at least up to dimension $20$ there is up to normalization a unique local conserved quantity $I_{2n-1}$ that commutes with all the ones constructed previously. The first few of these are \cite{Sasaki:1987mm,Eguchi:1989hs,Bazhanov:1994ft}:
\begin{align}
\label{virasoroqham}
\nonumber
I_1 & = T_0 \\
\nonumber
I_3 & = (TT)_0 \\
\nonumber
I_5 & = (T(TT))_0 - \frac{c+2}{12} (\partial T \partial T)_0 \\
\nonumber
I_7 & = (T(T(TT)))_0 + \frac{c^2+14 c-21}{15} (\partial T(\partial T T))_0 \\
& + \frac{c^2+19 c+19}{30} (\partial^2 T (TT))_0 \\
\nonumber
I_9 & = (T(T(T(TT))))_0 + \frac{c^2+24 c+59}{6} (\partial T(\partial T(TT)))_0 \\
\nonumber
& + \frac{c^2+29 c+129}{18} (\partial^2 T(T(TT)))_0 \\
\nonumber
& + \frac{10c^3+287c^2+1448c-3842}{504} (\partial^2 T(\partial T\partial T))_0
\end{align}
Note that even though for each even dimension the conserved quantity $I_{2n-1}$ is uniquely determined up to normalization, the corresponding local field is only determined up to total derivatives, so the explicit expressions \eqref{virasoroqham} depend on some choices. The total derivative terms can be uniquely fixed if we require the local fields to be quasi-primary with respect to $T$. This choice can be convenient when studying properties of $I_{2n-1}$ under modular transformations \cite{Maloney:2018hdg}. The number of fields on the right hand side grows rapidly with increasing dimension. The expressions for the first $10$ non-trivial expressions up to $I_{19}$ are given in the attached Mathematica notebook.

\subsubsection{Action of Hamiltonians on representation spaces}
Having found the expressions for a collection of commuting quantities $I_{2n-1}$ as in \eqref{virasoroqham}, we would like to find their simultaneous eigenvalues in the representation spaces of the Virasoro algebra. In order to do this explicitly, it is convenient to express $I_{2n-1}$ in terms of the Virasoro Fourier modes. All that we need are formulas expressing the Fourier modes of a derivative and a normal ordered product (in cylinder coordinates) \cite{Prochazka:2023zdb}. For derivative, we have simply
\begin{equation}
\label{derivativemodes}
(\partial \phi)_m = -m \phi_m
\end{equation}
while for the normal ordered product the corresponding formula is
\begin{equation}
\label{normordmodes}
(\phi \chi)_m = \sum_{k > 0} \phi_{-k} \chi_{m+k} + \sum_{k \geq 0} \chi_{m-k} \phi_k - \sum_{k>0} \frac{B^-_k}{k!} \left(\{ \phi, \chi \}_{-k}\right)_m
\end{equation}
Here the curly bracket $\{\phi,\chi\}_{-k}$ denotes the operator that appears in the OPE of $\phi(z)$ and $\chi(w)$ as a coefficient of $k$-th order pole,
\begin{equation}
\phi(z) \chi(w) = \sum_{k>0} \frac{\{ \phi, \chi \}_{-k}(w)}{(z-w)^k} + reg.
\end{equation}
The coefficient $B^-_k$ is the $k$-th Bernoulli number with the minus convention such that $B_1^- = -\frac{1}{2}$ (this is the convention used in Mathematica). The expression \eqref{normordmodes} is quite different from the corresponding expression in terms of plane Fourier modes, in which case there is no correction due to singular part of $\phi$ and $\chi$ OPE, but the splitting between creation and annihilation operators in the first two terms on the right-hand side depends on the conformal dimension of $\phi$.

Finally, in order to work with the Fourier modes, we also need a formula for the commutator of two Fourier modes. This is given by \cite{Prochazka:2023zdb}
\begin{equation}
\left[ \phi_m, \chi_n \right] = \sum_{k>0} \frac{m^{k-1}}{(k-1)!} \left( \{\phi,\chi\}_{-k} \right)_{m+n}.
\end{equation}
In other words, the right-hand side is the weighted sum over all $m+n$-th Fourier modes of local operators that appear in the singular part of the OPE of $\phi$ and $\chi$. We already used a special case of this formula with $m=0$ in equation \eqref{zeromodeaction}.

Using the expressions discussed so far, it is straightforward to write
\begin{align}
\label{blzintegrals}
\nonumber
I_1 & = T_0 = L_0 - \frac{c}{24} \\
I_3 & = L_0^2 + 2\sum_{m>0} L_{-m} L_m - \frac{c+2}{12} L_0 + \frac{c(5c+22)}{2880} \\
\nonumber
I_5 & = \sum_{m_1+m_2+m_3=0} : L_{m_1} L_{m_2} L_{m_3} : + \sum_{m>0} \left( \frac{c+11}{6} m^2 - 1 - \frac{c}{4} \right) L_{-m} L_m \\
\nonumber
& + \frac{3}{2} \sum_{m>0} L_{1-2m} L_{2m-1} - \frac{c+4}{8} L_0^2 + \frac{(c+2)(3c+20)}{576} L_0 - \frac{c(3c+14)(7c+68)}{290304}.
\end{align}
in agreement with the expressions given in \cite{Bazhanov:1994ft}. Here $: L_{m_1} L_{m_2} L_{m_3} :$ used by \cite{Bazhanov:1994ft} is the analogue of free field creation-annihilation normal ordering, where the mode operators with larger value of the mode index are placed on the right. The conformal normal ordering $(AB)$ that we use here is more practical than the non-commutative analogue of the free field creation-annihilation ordering $:AB:$ especially when one considers higher (and therefore more complicated) $I_{2n-1}$ or their analogues in more general $\mathcal{W}_N$ algebras. For calculations in Mathematica it is sufficient to recurrently express $I_{2n-1}$ in terms of $L_m$ using the relations \eqref{derivativemodes} and \eqref{normordmodes}.

We can now have a look at the action of $I_{2n-1}$ on the low lying states. We have
{\small
\begin{align}
\nonumber
I_3 \ket{\Delta} & = \left( \Delta^2 - \frac{c+2}{12} \Delta + \frac{c(5c+22)}{2880} \right) \ket{\Delta} \equiv I_{3,hw} \ket{\Delta} \\
(I_3-I_{3,hw}) L_{-1} \ket{\Delta} & = \left( 6\Delta - \frac{c+2}{12} + 1 \right) L_{-1} \ket{\Delta} \\
\nonumber
(I_3-I_{3,hw}) \begin{pmatrix} L_{-2} \ket{\Delta} \\ L_{-1}^2 \ket{\Delta} \end{pmatrix} & = \begin{pmatrix} 12\Delta+\frac{5c+22}{6} & 12\Delta \\ 6 & 12\Delta-\frac{c+2}{6}+8 \end{pmatrix} \begin{pmatrix} L_{-2} \ket{\Delta} \\ L_{-1}^2 \ket{\Delta} \end{pmatrix} \\
\nonumber
(I_3-I_{3,hw}) \begin{pmatrix} L_{-3} \ket{\Delta} \\ L_{-2} L_{-1} \ket{\Delta} \\ L_{-1}^3 \ket{\Delta} \end{pmatrix} & = \begin{pmatrix} 18 \Delta + \frac{3(5c+22)}{4} & 24\Delta  & 48\Delta \\ 18 & 18\Delta + \frac{3(c+22)}{4} & 36\Delta + 12 \\ 0 & 6 & 18\Delta - \frac{c+2}{4} + 21 \end{pmatrix} \begin{pmatrix} L_{-3} \ket{\Delta} \\ L_{-2} L_{-1} \ket{\Delta} \\ L_{-1}^3 \ket{\Delta} \end{pmatrix}
\end{align}
}for the action of $I_3$,
{\small
\begin{align}
\nonumber
I_5 \ket{\Delta} & = \left( \Delta^3 -\frac{c+4}{8}\Delta^2 + \frac{(c+2)(3c+20)}{576}\Delta -\frac{c(3c+14)(7c+68)}{290304} \right) \ket{\Delta} \\
\nonumber
& \equiv I_{5,hw} \ket{\Delta} \\
(I_5-I_{5,hw}) L_{-1} \ket{\Delta} & = \left[ 15 \Delta^2 - \frac{5(c-16)}{12}\Delta + \frac{3c^2-46c+328}{576} \right] L_{-1} \ket{\Delta} \\
\nonumber
(I_5 -I_{5,hw}) \begin{pmatrix} L_{-2} \ket{\Delta} \\ L_{-1}^2 \ket{\Delta} \end{pmatrix} & = {\tiny \begin{pmatrix} 30\Delta^2 + \frac{5(5c+64)}{6} \Delta +\frac{(7c+68)(9c+26)}{288} & 60\Delta^2 + \frac{5(c+20)}{2}\Delta \\ 30\Delta + \frac{5(c+20)}{4} & 30\Delta^2 - \frac{5(c-88)}{6}\Delta + \frac{3c^2-166c+6568}{288} \end{pmatrix}} \begin{pmatrix} L_{-2} \ket{\Delta} \\ L_{-1}^2 \ket{\Delta} \end{pmatrix}
\end{align}
}for the action of $I_5$ and finally
{\small
\begin{align}
\nonumber
I_7 \ket{\Delta} & = \Big[ \Delta^4 - \frac{1}{6} (c+6) \Delta^3 + \frac{(15c^2+194c+568)}{1440} \Delta^2 - \frac{(c+2)(c+10)(3c+28)}{10368} \Delta \\
& + \frac{c(3c+46)(25c^2+426c+1400)}{24883200} \Big] \ket{\Delta} \equiv I_{7,hw} \ket{\Delta} \\
\nonumber
(I_7-I_{7,hw}) L_{-1} \ket{\Delta} & = \Big[ 28 \Delta^3 -\frac{7(c-22)}{6} \Delta^2 +\frac{7(3c^2-62c+584)}{720} \Delta -\frac{15c^3-220c^2+3636c-17648}{51840} \Big] L_{-1} \ket{\Delta}
\end{align}
}for the action of $I_7$. These illustrate some properties of $I_{2n-1}$. On each Virasoro level ($L_0$ eigenspace) they act as finite dimensional matrices with coefficients that are polynomials in $c$ and $\Delta$ with rational coefficients. The degree of these polynomials is bounded by the spin of $I_{2n-1}$. The matrix representatives of $I_{2n-1}$ also commute among themselves as can be easily checked. The spectrum of $I_{2n-1}$ however is given by rather complicated algebraic expressions in $c$ and $\Delta$.

In order to understand the spectrum of $I_{2n-1}$, we have to diagonalize these finite matrices for every $n$ and every Virasoro level independently and there are no obvious relations between these matrices or their spectra that we could use to simplify the calculation. In fact, there is even no known explicit formula for the quantities $I_{2k-1,hw}$, which are polynomials in $c$ and $\Delta$, and are generalizations of the Casimir energy on the cylinder to higher spin currents (the Casimir energy corresponding to $I_{1,hw}$). There is an alternative description of spectra of $I_{2n-1}$ in the spirit of quantum integrability, where this problem essentially factorizes into two different problems: first we parametrize the simultaneous eigenvectors of $I_{2n-1}$ in terms of collections of commuting complex numbers, the Bethe roots (there are as many Bethe roots as is the Virasoro level). This parametrization must be found for all states at each Virasoro level, but works uniformly for all $I_{2n-1}$. In the second step, we need to find the eigenvalues of $I_{2n-1}$. These are given by symmetric polynomials of the Bethe roots and work uniformly for states at all Virasoro levels. In the case of Casimir quantities $I_{2k-1,hw}$, there are no Bethe roots (because we are at level $0$), but we will see that all of $I_{2k-1,hw}$ are encoded via ODE/IM in a single second order differential operator.

\subsection{ODEs and Bethe equations}
The first step of the procedure that we described, i.e. finding a way of labeling eigenstates of $I_{2n-1}$ in terms of $M$-tuples of complex numbers, was achieved in \cite{Bazhanov:2003ni}, see also \cite{Fioravanti:2004cz}, in the context of ODE/IM correspondence \cite{Dorey:1998pt,Dorey:2001uw,Bazhanov:1998wj,Dorey:2007zx}. The authors observed that certain quantities in 2d CFT constructed in \cite{Bazhanov:1994ft,Bazhanov:1996dr,Bazhanov:1998dq} such as the eigenvalues of the $Q$-operator satisfy the same functional equations as quantities associated to an ordinary differential equation in the complex domain. As originally formulated, it applied to the ground state of a given model, but it was generalized in \cite{Bazhanov:1998wj} to arbitrary primaries and finally in \cite{Bazhanov:2003ni} to all the states (descendants) in Virasoro lowest weight representations.

\subsubsection{Primary states}

Under the ODE/IM correspondence, the simultaneous eigenstates of Hamiltonians $I_{2n-1}$ are identified with certain second order ordinary differential operators in the complex plane. Let us first of all parametrize the central charge $c$ via
\begin{equation}
c = -\frac{(K-1)(3K+2)}{K+2}.
\end{equation}
If $K$ are odd integers, these central charges correspond to a family of non-unitary minimal models as will be explained later in Section \ref{secwinf}. For instance, $K=3$ corresponds to $c = -\frac{22}{5}$ which is the Lee-Yang minimal model (corresponding to the simplest interacting Argyres-Douglas theory $(A_1,A_2)$). This parametrization of the central charge is symmetric under
\begin{equation}
K \leftrightarrow -\frac{2K}{K+2}
\end{equation}
which is the Feigin-Fuchs duality of the Virasoro algebra.

We start by associating to a Virasoro primary of conformal dimension
\begin{equation}
\Delta \equiv \frac{(2\ell+1)^2-K^2}{8(K+2)}
\end{equation}
a differential operator
\begin{equation}
\label{virasoroprimaryoper}
\partial_x^2 + x^K - u - \frac{\ell(\ell+1)}{x^2}
\end{equation}
(this follows the conventions of \cite{Bazhanov:2003ni} with $K=2\alpha$ up to signs, but these can be changed by rescaling $x$ and $u$). For generic values of parameters $K$ and $\ell$, this differential operator has irregular singular points at $x=0$ and $x=\infty$ and the local behavior around these points encodes both the central charge $c$ and the conformal dimension $\Delta$. For $\ell$ an integer and $K$ a non-negative integer, the singularity at the origin is regular.

The recipe for calculating the higher integrals of motion $I_{2n-1}$ as spelled out in \cite{Dorey:2019ngq} is as follows: we first find a formal WKB solution to \eqref{virasoroprimaryoper}, which gives us a WKB curve associated to the minimal model we are studying, together with an infinite collection of $1$-forms $Y_n$ living on the curve. Up to normalization, the integrals of motion $I_{2n-1}$ are given by the periods of these $1$-forms, and their formal generating functions are usually called the quantum periods \cite{marino2021advanced}.

\subsubsection{WKB analysis}
Let us do this calculation concretely. First of all, the WKB expansion corresponds to an expansion in large values of the spectral parameter $u$ (which in \eqref{virasoroprimaryoper} enter as quantum mechanical energy). We can rescale the coordinate as $x \to u^{\frac{1}{K}} x$ and bring \eqref{virasoroprimaryoper} to the form
\begin{equation}
\label{virasoroprimaryoperhbar}
\hbar^2\partial_x^2 + x^K - 1 - \frac{\hbar^2\ell(\ell+1)}{x^2}
\end{equation}
with
\begin{equation}
\label{virasorouhbar}
\hbar = u^{-\frac{K+2}{2K}}.
\end{equation}
We see that we can think of $u \to \infty$ expansion as a semiclassical expansion $\hbar \to 0$. The WKB ansatz for the wave function
\begin{equation}
\psi(x) = \exp \left( \hbar^{-1} \int^x Y(x^\prime) dx^\prime \right)
\end{equation}
transforms the Schr\"odinger equation \eqref{virasoroprimaryoperhbar} to Riccati form
\begin{equation}
Y(x)^2 + \hbar Y^\prime(x) + x^K - 1 - \frac{\hbar^2\ell(\ell+1)}{x^2} = 0.
\end{equation}
Up to this point everything was exact. Now we look for a formal power series solution of the form
\begin{equation}
Y(x) = \sum_{k=0}^\infty \hbar^k Y_k(x).
\end{equation}
At the leading order we find the equation
\begin{equation}
\label{virasorowkbcurveY0}
Y_0(x)^2 + x^K = 1
\end{equation}
while at higher orders we find the recurrence relations for quantum corrections
\begin{equation}
\label{virriccati}
\sum_{j=0}^n Y_j Y_{n-j} + \partial Y_{n-1} = \delta_{n,2} \frac{\ell(\ell+1)}{x^2}, \qquad n \geq 1.
\end{equation}
These equations determine an infinite collection of one-forms $Y_n(x)dx$ which are differential polynomials of $Y_0$ and of the potential.

The leading order classical equation \eqref{virasorowkbcurveY0} can be though of as determining a plane algebraic curve $\mathcal{W} \subset \mathbbm{C}^2$ with coordinates $(x,y)$ given by a set of points satisfying
\begin{equation}
\label{virasorowkbcurve}
x^K + y^2 = 1.
\end{equation}
In the following, we call this curve $\mathcal{W}$ the \emph{WKB curve} associated to our problem. The other quantities $Y_n(x)dx$ can be interpreted as meromorphic $1$-forms living on the WKB curve. Explicitly, we find for the first few of these
\begin{align}
\label{virprimoneforms}
\nonumber
Y_0(x) dx & = y dx \\
\nonumber
Y_1(x) dx & = \frac{Kx^K}{4y^2} \frac{dx}{x} \\
\nonumber
Y_2(x) dx & = \Big[ -\frac{5K^2 x^{2K}}{32 y^4} - \frac{K(K-1) x^K}{8y^2} + \frac{\ell(\ell+1)}{2} \Big] \frac{dx}{x^2 y} \\
\nonumber
Y_3(x) dx & = \Big[ \frac{15K^3}{64} \frac{x^{3K}}{y^6} + \frac{9K^2(K-1)}{32} \frac{x^{2K}}{y^4} \\
& + \frac{K(K^2-3K-4\ell^2-4\ell+2)}{16} \frac{x^K}{y^2} + \frac{\ell(\ell+1)}{2} \Big] \frac{dx}{x^3 y^2} \\
\nonumber
Y_4(x) dx & = \Big[ -\frac{1105K^4}{2048} \frac{x^{4K}}{y^8} -\frac{221K^3(K-1)}{256} \frac{x^{3K}}{y^6} \\
\nonumber
& -\frac{K^2(47K^2-122K+75-50\ell(\ell+1))}{128} \frac{x^{2K}}{y^4} \\
\nonumber
& -\frac{K(K^3-6K^2+11K-6-6K\ell(\ell+1)+26\ell(\ell+1))}{32} \frac{x^K}{y^2} \\
\nonumber
& -\frac{(\ell-2)\ell(\ell+1)(\ell+3)}{8} \Big] \frac{dx}{x^4 y^3}
\end{align}
The recursion relation \eqref{virriccati} provides quite efficient for calculating the coefficients of $Y_n(x)dx$. Writing
\begin{equation}
Y_n(x) dx = \sum_{l} Y_{n,l} \frac{x^{Kl}}{y^{2l}} \frac{dx}{x^n y^{n-1}}
\end{equation}
and restricting for simplicity to the case of $\ell=0$, we see that the only non-zero $Y_{n,l}$ are those that have $1 \leq l \leq n$ and for those we have the following recurrence relation
\begin{equation}
Y_{n,k} = -\frac{1}{2} \sum_{j=1}^{n-1} \sum_l Y_{j,l} Y_{n-j,k-l} - \frac{Kk-n+1}{2} Y_{n-1,k} - \frac{K(n+2k-4)}{4} Y_{n-1,k-1}
\end{equation}
with the initial condition $Y_{1,1}=K/4$.

\subsubsection{Period integrals}
Finally, we need to calculate the period integrals of these differentials around a suitable contour\footnote{The period integrals along the cycles of a Seiberg-Witten curve are known to encode a lot of interesting information \cite{Lerche:1996xu,Mironov:2009uv,Mironov:2009dv,Tachikawa:2013kta,Kimura:2016ebq,Bourgine:2017jan,Grassi:2019coc,Ito:2025pfo}.}. The geometry of the WKB curves \eqref{virasorowkbcurve} is discussed more generally and in much more detail in Section \ref{secgeometry} and in Appendix \ref{appuniform}. For the purposes of the present discussion, let us state the following facts: for $K$ an odd integer, we can add a finite number of points at infinity to \eqref{virasorowkbcurve} and after this we obtain a smooth algebraic curve (a compact Riemann surface) of genus
\begin{equation}
g = \frac{K-1}{2}.
\end{equation}
This means that the homology group $H_1(\mathcal{W},\mathbbm{Z})$ is given by a $(K-1)$-dimensional lattice $\mathbbm{Z}^{K-1}$ and we can choose $K-1$ independent cycles generating the homology (for $K$ odd this number is even as it must be for a compact Riemann surface). Conventionally, one would use a symplectic basis of $H_1(\mathcal{W},\mathbbm{Z})$, but in our case we will choose a different basis for the following reason: there is a $\mathbbm{Z}_K \times \mathbbm{Z}_2$ group of automorphisms of $\mathcal{W}$, generated by the rotation symmetries
\begin{equation}
x \mapsto e^{\frac{2\pi i}{K}} x
\end{equation}
and by
\begin{equation}
\quad y \mapsto -y.
\end{equation}
which is simply the hyperelliptic involution of $\mathcal{W}$, exchanging the two sheets in the covering map from $\mathcal{W}$ to the complex $x$-plane. The usual symplectic basis does not transform simply under this group of automorphisms of the WKB curve. In Section \ref{secgeometry} it is shown that there is another convenient basis: we can choose a generating $1$-cycle $\mathcal{C}$ in $\mathcal{W}$ such that the whole basis of $H_1(\mathcal{W},\mathbbm{Z})$ is generated by its rotations by $\mathbbm{Z}_K \times \mathbbm{Z}_2$ automorphims. As a consequence of this, the periods along all other cycles differ only by an unessential overall phase.

Looking at the form of the meromorphic $1$-forms $Y_n(x)dx$, we see that we need to evaluate integrals of the type
\begin{equation}
\int_{\mathcal{C}} x^j y^{k-1} dx.
\end{equation}
The evaluation of these can be done in the complex plane of parameter $t = x^K = 1 - y^2$ (that we will later call the \emph{mirror curve}). This can be done easily using Euler's beta integral (see Section \ref{pochhammerintegral} for details of this calculation) and the result is
\begin{equation}
\int_{\mathcal{C}} x^j y^{k-1} dx = \frac{1}{K} \left( 1 - e^{2\pi i \frac{j+1}{K}} \right) \left( 1 - e^{2\pi i \frac{k+1}{2}} \right) B \left( \frac{j+1}{K}, \frac{k+1}{2} \right).
\end{equation}
The periods of $Y_n(x)dx$ calculated in this way are transcendental due to Euler beta functions, but it turns out that we can choose a normalization of the period integrals such that the results are rational functions of all the parameters. The integrals that we are going to calculate are therefore
\begin{equation}
\mathcal{I}_n \equiv \mathcal{N}_n^{-1} \int_{\mathcal{C}} Y_n(x) dx
\end{equation}
with normalization factor
\begin{equation}
\mathcal{N}_n = (-1)^n (K+2)^{n/2-1} \left(1-e^{-2\pi i\frac{n-1}{K}}\right) \left(1-e^{-2\pi i\frac{n-1}{2}}\right) B \left(-\frac{n-1}{K},-\frac{n-1}{2}\right).
\end{equation}

Using these expressions to evaluate the period integrals of \eqref{virprimoneforms}, we finally find
\begin{align}
\nonumber
\mathcal{I}_0 & = 1 \\
\nonumber
\mathcal{I}_2 & = -\frac{K-1}{24(K+2)} + \frac{\ell(\ell+1)}{2(K+2)} \\
\mathcal{I}_4 & = \frac{(K-3)(K-1)(2K+3)}{1920(K+2)^2} + \frac{(K-3)\ell(\ell+1)}{16(K+2)^2} - \frac{(\ell-1)\ell(\ell+1)(\ell+2)}{8(K+2)^2} \\
\nonumber
\mathcal{I}_6 & = -\frac{(K-5)(K-1)(24K^3+22K^2-117K-139)}{193536(K+2)^3} -\frac{(K-5)(2K^2-9K+13)\ell(\ell+1)}{768(K+2)^3} \\
\nonumber
& -\frac{5(K-5)(\ell-1)\ell(\ell+1)(\ell+2)}{64(K+2)^3} +\frac{(\ell-2)(\ell-1)\ell(\ell+1)(\ell+2)(\ell+3)}{16(K+2)^3}.
\end{align}
Comparing these with the highest weight state eigenvalues of the quantities $I_{2n-1}$ evaluated using commutation relations, we find the identification
\begin{align}
\nonumber
\mathcal{I}_0 & = I_{-1} \\
\nonumber
\mathcal{I}_2 & = I_{1} \\
\mathcal{I}_4 & = -\frac{1}{2} I_{3} \\
\nonumber
\mathcal{I}_6 & = \frac{1}{2} I_{5} \\
\nonumber
\mathcal{I}_8 & = -\frac{5}{8} I_{7}
\end{align}
and in general
\begin{equation}
\label{virasoroperiodiomnorm}
\mathcal{I}_{2n} = \frac{(-1)^{n+1} (2n-3)!!}{n!} I_{2n-1}.
\end{equation}
This completely fixes the relative normalization between the periods of $Y_n(x)dx$ that we normalized in rather ad hoc way and the standard normalization of higher quantum KdV charges which are normalized such that the non-derivative term of the corresponding current is $(T(T\cdots(TT)))$ ($n$ stress-energy tensors). We will see later how these normalization factors simplify in the context of $\mathcal{W}_\infty$.

Let us summarize where we are. We introduced higher charges of Virasoro algebra and calculated their eigenvalues when acting on the highest weight states. Although there is no general closed form formula for those, we found that all of these can be encoded in a single second order differential operator \eqref{virasoroprimaryoper} as its quantum periods up to a normalization factor given in \eqref{virasoroperiodiomnorm}. Next we will to turn to the descendants.

\subsubsection{Descendants}
The original ODE/IM correspondence of \cite{Dorey:1998pt,Dorey:2001uw,Bazhanov:1998wj,Dorey:2007zx} addressed the correspondence between the Virasoro primary and a differential operator of the form \eqref{virasoroprimaryoper}. Later, \cite{Bazhanov:2003ni} extended the correspondence to arbitrary descendant states. Recall that the differential operator \eqref{virasoroprimaryoper} is regular everywhere except for $x = 0$ and $x=\infty$ where there are generically irregular singular points, encoding the central charge $c$ and the conformal dimension of the primary $\Delta$. The idea of \cite{Bazhanov:2003ni} is to modify the potential by adding additional regular singular points in the complex plane
\begin{equation}
\label{viroperdesc}
\partial_x^2 + x^K - u - \frac{\ell(\ell+1)}{x^2} \longrightarrow \partial_x^2 + x^K - u - \frac{\ell(\ell+1)}{x^2} + \sum_j \left( \frac{-2}{(x-x_j)^2} + \frac{\beta_j}{x(x-x_j)} \right).
\end{equation}
Here $x_j \in \mathbbm{C}$ are new parameters that parametrize the positions of the regular singular points in the complex plane and $\beta_j \in \mathbbm{C}$ are additional parameters associated to these singular points. The coefficient of the quadratic pole at $x = x_j$ is fixed in such a way that the Frobenius indices of solutions around $x = x_j$ are $\sigma = -1$ and $\sigma = 2$\footnote{This is an important part of the proposal -- if we choose the Frobenius indices to have larger difference, we will end up at more complicated Bethe equations of higher degree. Our choice of writing the simple pole is convenient as it does not change the leading asymptotic behavior as $x \to \infty$.}. In order to determine the positions of singularities $x_j$ and the coefficients $\beta_j$, \cite{Bazhanov:2003ni} require that there are \emph{no logarithmic terms} in solutions around the singular points (i.e. the local solutions around added singularities have trivial monodromy). In more detail, since the Frobenius indices associated to $x = x_j$ differ by an integer, which is sometimes called the resonant case, the generic local solutions around $x = x_j$ involve $\log(x-x_j)$: let us parametrize the local behavior of the differential operator \eqref{viroperdesc} around $x = x_j$ as
\begin{equation}
\partial_x^2 - \frac{2}{(x-x_j)^2} + \sum_{m=-1}^\infty \nu^{(j)}_m (x-x_j)^m.
\end{equation}
The Frobenius solution with leading order behavior $(x-x_j)^2$ can be written as a power series
\begin{equation}
\chi^{(j)}(x) = (x-x_j)^2 + \sum_{m=3}^\infty \chi^{(j)}_m (x-x_j)^m
\end{equation}
with
\begin{align}
\nonumber
\chi^{(j)}_3 & = -\frac{\nu^{(j)}_{-1}}{4} \\
\chi^{(j)}_4 & = \frac{\nu^{(j)2}_{-1}-4\nu^{(j)}_0}{40} \\
\nonumber
\chi^{(j)}_5 & = -\frac{\nu^{(j)3}_{-1}-14\nu^{(j)}_0\nu^{(j)}_{-1}+40\nu^{(j)}_1}{720}
\end{align}
etc. uniquely determined by the differential equation around $x = x_j$. In particular, there is never any obstruction in determining the coefficients $\chi^{(j)}_m$ of the subleading solution and the resulting power series converges with the radius of convergence being determined by the distance to the singularity nearest to $x=x_j$. On the other hand, if the Frobenius indices satisfy the resonance condition and the values of $\nu^{(j)}_m$ are generic, the dominant local solution is not of the form of a generalized power series, but instead takes the form
\begin{equation}
\psi^{(j)}(x) = (x-x_j)^{-1} + \sum_{m=0}^\infty \psi^{(j)}_m (x-x_j)^m + \rho^{(j)} \log (x-x_j) \chi^{(j)}(x).
\end{equation}
With this ansatz, the coefficients $\psi^{(j)}_m$ as well as $\rho$ are uniquely determined in terms of $\nu^{(j)}_m$ except for the fact that $\psi^{(j)}_2$ can be chosen to be arbitrary (it corresponds to addition of $\chi^{(j)}(x)$) and for concreteness we choose it to be zero. The explicit expressions for the first few coefficients are
\begin{align}
\nonumber
\psi^{(j)}_0 & = \frac{\nu^{(j)}_{-1}}{2} \\
\nonumber
\psi^{(j)}_1 & = \frac{\nu^{(j)2}_{-1}+2\nu^{(j)}_0}{4} \\
\psi^{(j)}_2 & = 0 \qquad \text{(by choice)} \\
\nonumber
\psi^{(j)}_3 & = \frac{-5 \nu^{(j)4}_{-1}-32\nu^{(j)}_0\nu^{(j)2}_{-1}-44\nu^{(j)}_1\nu^{(j)}_{-1}-24\nu^{(j)2}_0-48\nu^{(j)}_2}{192}
\end{align}
and most importantly
\begin{equation}
\rho^{(j)} = -\frac{\nu^{(j)3}_{-1}+4\nu^{(j)}_{-1}\nu^{(j)}_0+4\nu^{(j)}_1}{12}.
\end{equation}
We see that indeed for generic values of $\nu^{(j)}_m$, we have $\rho^{(j)} \neq 0$ so the logarithmic terms are necessary in order to find two linearly independent solutions of \eqref{viroperdesc}. The quantization condition of \cite{Bazhanov:2003ni} determining the positions of $x_j$ as well as the values of $\beta_j$ comes from putting $\rho^{(j)}$ equal to zero \emph{for all values of the spectral parameter $u$}.

For our specific potential \eqref{viroperdesc} we have
\begin{align}
\nonumber
\nu^{(j)}_{-1} & = \frac{\beta_j}{x_j} \\
\nu^{(j)}_0 & = x_j^K -u - \frac{\ell(\ell+1)}{x_j^2} - \frac{\beta_j}{x_j^2} + \sum_{k \neq j} \left[ - \frac{2}{(x_j-x_k)^2} + \frac{\beta_k}{x_j(x_j-x_k)} \right] \\
\nonumber
\nu^{(j)}_1 & = Kx_j^{K-1} + \frac{2\ell(\ell+1)}{x_j^3} + \frac{\beta_j}{x_j^3} + \sum_{k \neq j} \left[ \frac{4}{(x_j-x_k)^3} - \frac{(2x_j-x_k)\beta_k}{x_j^2(x_j-x_k)^2} \right].
\end{align}
Since $\rho^{(j)}$ should vanish for every $u$ which enters only linearly in $\nu^{(j)}_0$, we see that we need to have
\begin{equation}
\nu^{(j)}_{-1} = 0 \qquad \forall j,
\end{equation}
which implies $\beta_j = 0$ for all $j$ and this significantly simplifies the equations. The remaining equations are $\nu^{(j)}_1 = 0$ or
\begin{equation}
\label{virbetheequations}
0 = Kx_j^{K-1} + \frac{2\ell(\ell+1)}{x_j^3} + \sum_{k \neq j} \frac{4}{(x_j-x_k)^3} \qquad \forall j.
\end{equation}
These can be interpreted as Bethe equations determining the positions of singularities $x_j$. We can think of them as equations for equilibrium positions of particles interacting via classical potential
\begin{equation}
\label{virbethepotential}
0 = \sum_j x_j^{K} - \sum_j \frac{\ell(\ell+1)}{x_j^2} - \sum_{j < k} \frac{2}{(x_j-x_k)^2}.
\end{equation}
which describes (a complexification of) a system of particles interacting pairwise by Calogero inverse square potential in an external field with potential $x^K$ and additional $\frac{\ell(\ell+1)}{2}$ of Calogero-like particles fixed at the origin.

\subsubsection{Orbifold}
Due to the form of the potential \eqref{virbethepotential} or equivalently the original differential operator \eqref{viroperdesc}, there is a $\mathbbm{Z}_{K+2}$ rotation symmetry (sometimes called Symanzik rotations) whose generator acts on $x_j$ as
\begin{equation}
\label{virrootaction}
x_j \mapsto e^{\frac{2\pi i}{K+2}} x_j.
\end{equation}
This symmetry acts transitively on the Stokes sectors around the irregular singular point at $x = \infty$. The solutions of \eqref{virbetheequations} corresponding to Virasoro descendants are invariant under the action of this group, or in other words the physical solutions always have full orbits of Bethe roots under \eqref{virrootaction}. The number of Bethe roots therefore has to be a multiple of $K+2$ and it is in fact equal to $M(K+2)$ where $M$ is the Virasoro level. In order to solve the equations \eqref{virbetheequations} (either analytically at lower levels or numerically), it is convenient to reduce the number of equations and unknowns from $M(K+2)$ to $M$ by rewriting these equations in terms of the invariants of Bethe roots under $\mathbbm{Z}_{K+2}$ rotations,
\begin{equation}
X_j \equiv x_j^{K+2}.
\end{equation}
One can do this either on the level of the differential operator by changing the coordinate to $X = x^{K+2}$ and rederiving the conditions for the absence of monodromy (see discussion in Section \ref{virfeiginfrenkel}), or directly on the level of Bethe equations by using the identities \cite{Fioravanti:2004cz} that are summarized in Appendix \ref{apporbitsums}. The resulting Bethe equations are
\begin{align}
\label{virbetheequationsorb}
\nonumber
0 = \, & KX_j + 2\ell(\ell+1) - \frac{K^2-1}{2} \\
& + 2(K+2) X_j \sum_{\substack{k=1 \\ k \neq j}}^M \frac{2X_j^2 + (K+1)(K+6) X_j X_k + K(K+1) X_k^2}{(X_j-X_k)^3}.
\end{align}
This is a form analogous to the one discussed in \cite{Litvinov:2013zda}.

The differential operator \eqref{viroperdesc} with $\beta_j = 0$ can be written more compactly using a $\tau$-function which in this case is simply a polynomial whose zeros are the positions of singularities,
\begin{equation}
\tau(x) = \prod_j (x-x_j) = \prod_{j=1}^M (x^{K+2}-X_j).
\end{equation}
We have
\begin{equation}
\partial_x^2 + x^K - u - \frac{\ell(\ell+1)}{x^2} + 2 \partial_x^2 \log \tau(x).
\end{equation}
This is close to the original form proposed in \cite{Bazhanov:2003ni}. The dependence of the differential operator on the highest weight is parametrized by the coefficient of the quadratic pole at the origin (where we can think of having non-dynamical Calogero particles, see \eqref{virbetheequations}), and these can be also absorbed into $\tau$ function by replacing
\begin{equation}
\prod_j (x-x_j) \to x^{\frac{\ell(\ell+1)}{2}} \prod_j (x-x_j).
\end{equation}
For integer values of $\ell$ this $\tau$ function is still a polynomial, but has a zero of higher multiplicity at the origin than the zeros corresponding to descendants away from the origin.

\subsubsection{WKB analysis including descendants}
At this point we have parametrization of states in Virasoro representation spaces in terms of $M$-tuples of complex numbers $X_j$ solving the Bethe equations \eqref{virbetheequationsorb} and it remains to find the spectrum of higher conserved quantities $I_{2n-1}$ in terms of these Bethe roots. Fortunately, we can use the same WKB analysis as we did for the primary state and the WKB curve turns out to be the same as before. The reason is that the scaling transformation which trades the spectral parameter $u$ for the Planck constant $\hbar$ brings the differential operator \eqref{viroperdesc} to the form
\begin{equation}
\label{viroperdeschbar}
\hbar^2 \partial_x^2 + x^K - 1 - \frac{\hbar^2\ell(\ell+1)}{x^2} + \hbar^2\sum_j \left[ \frac{-2}{\left(x-\hbar^{\frac{2}{K+2}}x_j\right)^2} + \frac{\beta_j}{x\left(x-\hbar^{\frac{2}{K+2}}x_j\right)} \right].
\end{equation}
The leading order classical curve does not get any corrections from $\ell$ or from the descendants, only the higher differentials living on $\mathcal{W}$ are corrected. Furthermore, expanding the contribution from the descendants at small $\hbar$, we see that the corrections to higher $1$-forms $Y_n(x)dx$ involve symmetric functions of $x_j$ of bounded degree. In fact, expanding the differential operator \eqref{viroperdeschbar} in $\hbar$ (and putting $\beta_j = 0$ to simplify the equations), we find
\begin{multline}
\hbar^2 \partial_x^2 + x^K - 1 - \frac{\hbar^2\ell(\ell+1)}{x^2} -2\hbar^2\sum_j \frac{1}{\left(x-\hbar^{\frac{2}{K+2}}x_j\right)^2} = \\
= \hbar^2 \partial_x^2 + x^K - 1 - \frac{\hbar^2\ell(\ell+1)}{x^2} -\frac{2\hbar^2}{x^2} \sum_{m=0}^\infty \frac{(m+1)\hbar^{\frac{2m}{K+2}} p_m(x_j)}{x^m}
\end{multline}
where
\begin{equation}
p_m(x_j) = \sum_j x_j^m
\end{equation}
are the Newton power sums and we also use the same formula at $m=0$ by putting
\begin{equation}
p_0(x_j) = (K+2)M.
\end{equation}
Now because for physical solutions of Bethe equations the roots $x_j$ appear in $\mathbbm{Z}_{K+2}$ orbits of $K+2$ elements, only the terms where the summation index $m$ is a multiple of $K+2$ are non-vanishing. We can therefore rewrite the sum as
\begin{equation}
\hbar^2 \partial_x^2 + x^K - 1 - \frac{\hbar^2\ell(\ell+1)}{x^2} -2 (K+2) \sum_{m=0}^\infty \frac{((K+2)m+1)\hbar^{2m+2} p_m(X_j)}{x^{(K+2)m+2}}.
\end{equation}
(and we put $p_0(X_j)=M$). In this form we are ready to calculate the WKB differentials $Y_n(x)dx$. The leading curve is unmodified, i.e. we still have \eqref{virasorowkbcurveY0}. This is also true for $Y_1(x)dx$ because the leading corrections appear at order $\hbar^2$. The Riccati equation has now sources at all even orders of $\hbar$ so we have recurrence relations
\begin{equation}
\sum_{j=0}^k Y_j Y_{k-j} + \partial Y_{k-1} = S_k
\end{equation}
where the sources are
\begin{equation}
S_{2k} = \delta_{k,1} \frac{\ell(\ell+1)}{x^2} + \frac{2(K+2)((K+2)(k-1)+1)p_{k-1}(X_j)}{x^{(K+2)(k-1)+2}}, \qquad k \geq 1.
\end{equation}
The first two $1$-forms $Y_n(x)dx$ that are not total derivatives are
\begin{align}
Y_2(x)dx & = \left[ -\frac{5K^2}{32} \frac{x^{2K}}{y^4} - \frac{K(K-1)}{8} \frac{x^K}{y^2} + \frac{\ell(\ell+1)}{2} + (K+2)M \right] \frac{dx}{x^2 y} \\
\nonumber
Y_4(x) dx & = \Bigg[ -\frac{1105K^4}{2048} \frac{x^{4K}}{y^8} -\frac{221K^3(K-1)}{256} \frac{x^{3K}}{y^6} \\
\nonumber
& -\frac{K^2(47K^2-122K+75-50\ell(\ell+1))}{128} \frac{x^{2K}}{y^4} \\
\nonumber
& -\frac{K(K^3-6K^2+11K-6-6K\ell(\ell+1)+26\ell(\ell+1))}{32} \frac{x^K}{y^2} \\
& -\frac{(\ell-2)\ell(\ell+1)(\ell+3)}{8} + \frac{25K^2(K+2)}{32} M \frac{x^{2K}}{y^4} \\
\nonumber
& + \frac{K(K+2)(3K-13)}{8} M \frac{x^K}{y^2} - \frac{(K+2)\ell(\ell+1)}{2} M \\
\nonumber
& + \frac{3(K+2)}{2} M - \frac{(K+2)^2}{2} M^2 + (K+2)(K+3) p_1(X) \frac{x^{-K}}{y^{-2}} \Bigg] \frac{dx}{x^4 y^3}.
\end{align}
We see that these expressions get quickly rather complicated. The expressions for higher $Y_{2k}(x)dx$ up to $Y_{16}(x)dx$ are given in the attached Mathematica notebook. We are more interested in the associated period integrals. These are easy to evaluate using \eqref{masterintegral} and the resulting expressions are listed in appendix \ref{appvirasoroperiods} and up to $I_{11}$ in the attached Mathematica notebook.

\subsubsection{Feigin-Frenkel duality}
\label{virfeiginfrenkel}
On the level of $\mathcal{W}_\infty$, we have triality symmetry permuting the three $\lambda$-parameters. Choosing one of these to be $N=2$ so that we are working with Virasoro algebra, the triality symmetry is reduced to a simple duality exchanging the other two $\lambda$-parameters. This is the well-known Feigin-Frenkel duality. On the level of our (orbifolded) Bethe equations \eqref{virbetheequationsorb}, this symmetry is the involutive transformation
\begin{equation}
\label{feiginfrenkel}
K \leftrightarrow -\frac{2K}{K+2} \equiv K^\prime.
\end{equation}
The original Bethe equations that were calculated in the frame of \eqref{virasoroprimaryoper} involved $M(K+2)$ Bethe roots distributed in orbits of $\mathbbm{Z}_{K+2}$ where $M$ is the Virasoro level. Since the duality \eqref{feiginfrenkel} maps integer $K$ to a non-integer $K$, this frame is not suitable if $K$ is not an integer. We can instead use the orbifold frame, i.e. we can first transform the differential operator \eqref{virasoroprimaryoper} to coordinate
\begin{equation}
\label{virorbifoldingmap}
X \equiv x^{K+2}.
\end{equation}
Under this change of coordinates, the differential operator becomes
\begin{equation}
\label{viroperlframe}
\partial_X^2 - \frac{u}{(K+2)^2} X^{-\frac{2(K+1)}{K+2}} + \frac{1}{(K+2)^2 X} + \frac{(K+1)(K+3)-4\ell(\ell+1)}{4(K+2)^2 X^2}
\end{equation}
for the highest weight vector and with additional contribution
\begin{equation}
- \sum_{j=1}^M \frac{2}{(X-X_j)^2} + \frac{2(K+1)}{K+2} \sum_{j=1}^M \frac{1}{X(X-X_j)}.
\end{equation}
representing the descendants. The factors of $(K+2)$ could be eliminated by rescaling of $X$ but we wanted to keep the map \eqref{virorbifoldingmap} simple without additional prefactors. We want to apply the transformation
\begin{equation}
X \mapsto X^\prime \equiv -\frac{4u}{(K+2)^2} X^{\frac{2}{K+2}}
\end{equation}
to get the operator of the same form \eqref{virasoroprimaryoper} in terms of primed variables. The prefactor and the exponent are chosen to preserve the form of the operator \eqref{viroperlframe} with parameters mapped as
\begin{align}
\nonumber
K^\prime & = -\frac{2K}{K+2} \\
\ell^\prime & = -\frac{1}{2} \pm \frac{2\ell+1}{K+2} \\
\nonumber
u^\prime & = i^K \left(\frac{K+2}{2}\right)^K u^{-1-\frac{K}{2}}.
\end{align}
We see in particular that under this transformation the large $u$ and small $u$ regimes are exchanged under this transformation \cite{masoero2023q}. When calculating the local charges, the natural variable is $\hbar$ instead of $u$ with these two being related by \eqref{virasorouhbar}. The Feigin-Frenkel duality maps $\hbar$ to
\begin{equation}
\hbar^\prime = (-1)^{1/2} \frac{K+2}{2} \hbar,
\end{equation}
i.e. the WKB expansion agrees order by order in $\hbar$ in both frames related by Feigin-Frenkel duality. The transformation of descendants is much more non-trivial. In particular, since the transformation of coordinate $X$ involves the spectral parameter $u$ and fractional powers of $X$, we do not know a simple relation between the Bethe roots in the original frame with parameter $K$ and in the new frame with rank $K^\prime$.

\section{$\mathcal{W}_\infty$ and Argyres-Douglas minimal models}
\label{secwinf}

\subsection{$\mathcal{W}_N$, $\mathcal{W}_\infty$ and triality}

The Virasoro algebra is a member of a two-parameter family of algebras $\mathcal{W}_N$. Conventionally these two parameters are chosen to be the rank parameter $N$ and the central charge $c$. The Virasoro algebra corresponds to $N=2$. For $N = 3$ we obtain the algebra $\mathcal{W}_3$ generated by Virasoro stress-energy tensor $T$ and spin $3$ primary $W$ constructed originally in \cite{Zamolodchikov:1985wn}. The rank $N$ is equal to the rank of $\mathfrak{sl}(N)$ family of simple Lie algebras which are the starting point of various constructions of $\mathcal{W}_N$ algebras \cite{Bouwknegt:1992wg} and $\mathcal{W}_N$ at generic $c$ is generated by stress-energy tensor and additional primary fields of dimension $3, 4, \ldots, N$.

There is a universal two-parametric family of algebras called $\mathcal{W}_{\infty}$ that interpolates between all the algebras of $\mathcal{W}_N$ family \cite{Hornfeck:1994is,Gaberdiel:2012ku,Prochazka:2014gqa,Linshaw:2017tvv}. It is parametrized by rank-like parameter $\lambda$ and the central charge $c$ and generated by stress-energy tensor and a single primary of every spin $\geq 3$. If we specialize $\lambda$ to be a positive integer $N \geq 2$, the primaries of spin $N+1, N+2, \ldots$ generate an infinite dimensional ideal and quotienting out by this ideal reduces the algebra to $\mathcal{W}_N$. The algebra $\mathcal{W}_\infty$ itself is meaningful for any generic complex values of $\lambda$ and $c$.

The classical versions of $\mathcal{W}_\infty$ have the usual duality symmetry relating $N$ to $-N$, but in \cite{Gaberdiel:2012ku} it was found that at quantum level this symmetry is modified to a triality symmetry $\mathcal{S}_3$. To make this symmetry manifest, we can trade $\lambda$ and $c$ for three parameters $\lambda_1, \lambda_2$ and $\lambda_3$ satisfying \cite{Prochazka:2014gqa}
\begin{equation}
\label{winfcentralcharge}
c = (\lambda_1-1)(\lambda_2-1)(\lambda_3-1), \qquad \frac{1}{\lambda_1} + \frac{1}{\lambda_2} + \frac{1}{\lambda_3} = 0, \qquad \lambda_3 = \lambda.
\end{equation}
The structure constants of the algebra in an appropriate basis are manifestly invariant under permutations of $\lambda_j$. This triality symmetry manifests also in the counting of local fields in the algebra, i.e. in the calculation of vacuum character \cite{Gaberdiel:2012ku}. For this to be more symmetric, it is useful to introduce $\mathcal{W}_{1+\infty}$ as a product of $\mathcal{W}_\infty$ with an independent decoupled free boson $\widehat{\mathfrak{gl}(1)}$ (i.e. additional spin $1$ generator corresponding to the center of Lie algebra $\mathfrak{gl}(N)$). The vacuum character of $\mathcal{W}_{1+\infty}$ for generic values of $c$ and $\lambda$ is easily seen to be
\begin{equation}
\label{macmahon}
\prod_{n=1}^\infty \frac{1}{(1-q^n)^n} \simeq 1 + q + 3q^2 + 6q^3 + 13q^4 + 24q^5 + 48q^6 + \ldots.
\end{equation}
We recognize the famous MacMahon function which counts the plane partitions, i.e. the three-dimensional generalization of Young diagrams. The representation theory of $\mathcal{W}_{1+\infty}$ is closely related to combinatorics of these \cite{feigin2012quantum,Prochazka:2015deb}. The triality symmetry acts as a symmetry permuting the coordinate axes of the space where the plane partitions live.

\subsubsection{Truncations}

For $c$ and $\lambda$ non-generic, the vacuum Verma module can become reducible. More concretely, whenever the parameters $\lambda_j$ satisfy a relation of the form \cite{Prochazka:2014gqa,Prochazka:2017qum}
\begin{equation}
\label{winftrunccurve}
\frac{N_1}{\lambda_1} + \frac{N_2}{\lambda_2} + \frac{N_3}{\lambda_3} = 1
\end{equation}
with $N_j$ non-negative integers, there is a singular vector at level
\begin{equation}
(N_1+1)(N_2+1)(N_3+1)
\end{equation}
and the character of the vacuum Verma module differs starting at order
\begin{equation}
q^{(N_1+1)(N_2+1)(N_3+1)}
\end{equation}
from the character of the irreducible vacuum representation. Combinatorially, if we impose the condition \eqref{winftrunccurve} but keep the remaining parameter generic, the vectors in the irreducible vacuum representation correspond to plane partitions with the box at coordinates $(N_1+1,N_2+1,N_3+1)$ not allowed \cite{Bershtein:2018pcf,Prochazka:2017qum}. In particular, for $(N_1,N_2,N_3) = (0,0,N)$ which is the reduction to $\mathcal{W}_N$, the plane partitions are allowed to have only up to $N$ layers in the 3rd direction (and similarly for the other two coordinate axes).

The other choices of $(N_1,N_2,N_3)$ correspond to more general truncations of $\mathcal{W}_{1+\infty}$ which were coined $Y$-algebras in \cite{Gaiotto:2017euk}, see also \cite{Prochazka:2017qum,Prochazka:2018tlo}. For example the choice $(N_1,N_2,N_3)=(2,1,0)$ corresponds to so called parafermion algebras. These algebras are generated by spins $1,2,\ldots,(N_1+1)(N_2+1)(N_3+1)-1$ but not freely \cite{Prochazka:2018tlo}. Imposing two independent conditions of the form \eqref{winftrunccurve} fixes the parameters of the algebra completely resulting in so-called minimal models. The most familiar family of these are the Virasoro unitary minimal models which are obtained by imposing \eqref{winftrunccurve} with $(N_1,N_2,N_3)=(0,0,2)$ and with $(N_1,N_2,N_3)=(k+1,k,0)$. The central charge that follows from \eqref{winfcentralcharge} is
\begin{equation}
\frac{k(k+5)}{(k+2)(k+3)}
\end{equation}
which for $k=1,2,\ldots$ gives $c = \frac{1}{2}, \frac{7}{10}, \frac{4}{5}, \ldots$. Imposing two conditions \eqref{winftrunccurve} modifies the combinatorics of plane partitions, effectively making the plane partitions periodic \cite{Foda:2015bsa,Prochazka:2023zdb}. The counting problem for such periodic plane partitions reproduces the vacuum character of $\mathcal{W}_N$ minimal models (these are usually written in terms of theta functions, see \cite{Bouwknegt:1992wg}).

Apart from the vacuum representation, we can consider other highest weight representations of the algebra. There are various possible choices of these but in the context of the minimal models the most interesting ones correspond to maximally degenerate primaries, i.e. those that have the maximal number of null vectors. In $\mathcal{W}_{1+\infty}$ a large class of these corresponds combinatorially to counting plane partitions with non-trivial Young diagram asymptotics along the three coordinate axes. For generic values of parameters the corresponding characters can be identified with the topological vertex \cite{Aganagic:2003db,Prochazka:2015deb}.

The choice of primaries can be combined with the specializations of parameters of $\mathcal{W}_{\infty}$ leading to $\mathcal{W}_N$ algebras and their minimal models. For $\mathcal{W}_N$ with generic central charge the maximally degenerate primaries can be labeled by two Young diagrams with at most $N$ rows. This restriction comes from the restriction on plane partitions to have at most $N$ layers in one of the directions. This prohibits one of the non-trivial asymptotics as well as restricting the size of the other two asymptotics \cite{Prochazka:2015deb}.

\subsubsection{Argyres-Douglas VOAs}
\label{secadintro}

For most of this article we will restrict our attention to the special class of minimal models that can be viewed simultaneously as $\mathcal{W}_N$ and $\mathcal{W}_K$ minimal models, i.e. we specialize to
\begin{equation}
\label{parameterchoice}
\lambda_3 = N, \qquad \lambda_1 = K, \qquad \lambda_2 = -\frac{KN}{K+N}, \qquad c = -\frac{(K-1)(N-1)(KN+K+N)}{K+N}
\end{equation}
with $K$ and $N$ positive coprime integers. We call these the Argyres-Douglas minimal models because they agree with the vertex operator algebra associated to four-dimensional $\mathcal{N}=2$ superconformal field theory of Argyres-Douglas type $(A_{K-1},A_{N-1})$ \cite{Cecotti:2010fi,Beem:2013sza,Cordova:2015nma,Ito:2017ypt}. From the two-dimensional CFT perspective these theories are non-unitary, but are arguably the simplest minimal models of $\mathcal{W}$-algebras. The first representative of this class is the $(2,3)$ model which is the Lee-Yang minimal model of Virasoro and $\mathcal{W}_3$ algebra with $c = -\frac{22}{5}$.

Every maximally degenerate primary of these models is labeled by a single Young diagram that fits a $N \times K$ rectangle (this is due to $(0,0,N)$ and $(K,0,0)$ restrictions on plane partitions). There are
\begin{equation}
{N + K \choose N} = {N + K \choose K}
\end{equation}
such Young diagrams (we just need to count the number of north-east paths from the lower left corner to the upper right corner of the $K \times N$ rectangle). Not all of these give distinct $\mathcal{W}_\infty$ primary though. In fact, for $N$ and $K$ coprime there are exactly $K+N$ equivalent Young diagrams for each primary\footnote{The reason for these identifications is that adding a full row or a full column to the asymptotic Young diagram corresponds to a translation in the spectral parameter space, i.e. change of the $\widehat{\mathfrak{gl}(1)}$ charge, and this does not change the representation of $\mathcal{W}_{\infty}$.} and therefore the number of maximally degenerate primaries of these minimal models is
\begin{equation}
\frac{1}{K+N} \frac{(K+N)!}{K!N!} = \frac{(K+N-1)!}{K!N!}.
\end{equation}
For the Lee-Yang model this equals $2$, i.e. we have primaries with conformal dimensions
\begin{equation}
\Delta = 0 \qquad \text{and} \qquad \Delta=-\frac{1}{5}
\end{equation}
which are exactly the two primaries we expect. Let us mention that these conformal dimensions are encoded in the combinatorics of the plane partitions as well, see for instance \cite{Prochazka:2015deb} or Appendix D of \cite{Prochazka:2017qum} and Section \ref{secwinfhw}.

\subsection{Integrable structures}
We want to diagonalize quantities that commute with the $L_0$ generator of the Virasoro algebra\footnote{$L_0$ is the holomorphic half of the standard Hamiltonian generating time evolution on the cylinder or the generator of scaling transformations in the complex plane. We always focus on one copy of Virasoro or $\mathcal{W}_N$ algebra associated to say left-movers and so as an abuse of notation, we use the words energy and spin interchangeably.}. There are different inequivalent choices of such commuting quantities. The quantities studied in \cite{Sasaki:1987mm,Eguchi:1989hs,Bazhanov:1994ft} in the context of the Virasoro algebra (see Section \ref{secvirasoro}) correspond to the zero modes of local operators of dimension $4, 6, 8, \ldots$. As we will see in the following, a natural generalization of such local commuting conserved quantities exists also in $\mathcal{W}_\infty$ and these will be our main focus in this article.

There is another interesting choice of quantities that commute with $L_0$. These are obtained by applying the standard procedure of algebraic Bethe ansatz (quantum inverse scattering method) to instanton $\mathcal{R}$-matrix of Maulik and Okounkov \cite{Maulik:2012wi,Litvinov:2013zda,Zhu:2015nha,Prochazka:2019dvu,Prochazka:2023zdb}. First of all, the Maulik-Okounkov $\mathcal{R}$-matrix is defined as an operator that intertwines between two ways of representing $\widehat{\mathfrak{gl}(1)} \, \times$ Virasoro algebra in terms of a pair of free bosons. This object is an operator acting on a pair of bosonic Fock spaces and is closely related to Zamolodchikov's Liouville reflection operator \cite{zamolodchikov2007lectures}. The $\mathcal{R}$-matrix satisfies the Yang-Baxter equation so we can consider a spin chain with $N$ sites (that will eventually lead to $\mathcal{W}_N$ algebra) and associate a monodromy matrix to it. The last step in the construction of commuting quantities in QISM is to take the trace of the monodromy matrix over the auxiliary space to get the transfer matrix, the generating function of commuting conserved quantities. The simplest choice of the auxiliary space is the infinite dimensional bosonic Fock space, but due to its infinite dimension, the trace over it is not well defined. We can regularize the trace by inserting a \emph{twist} $q^{L_0^{(A)}}$ (the Virasoro $L_0$ acting in the auxiliary space), effectively putting the auxiliary boson on a torus with complex structure controlled by $q$. This does not spoil the commutativity of the conserved quantities, but introduces a very specific non-locality controlled by $q$. The corresponding commuting Hamiltonians are related to quantum intermediate long wave hierarchy \cite{Litvinov:2013zda}. In the limit $q \to 0$ (or $q \to \infty$) the auxiliary boson is effectively at low temperature and the corresponding conserved quantities reduce to non-local Yangian or Benjamin-Ono Hamiltonians (whose spectrum can be described explicitly in terms of combinatorics of Young diagrams \cite{Prochazka:2015deb,Prochazka:2023zdb}). On the other hand, the limit $q \to 1$ is the high temperature limit of the auxiliary boson and it is the limit where we are effectively removing the regulator. For this reason, the $q \to 1$ limit is singular, but one can still extract conserved quantities which turn out to be local \cite{Prochazka:2023zdb} and generalize the Virasoro conserved quantities of \cite{Bazhanov:1994ft}.

To summarize, the procedure of QISM constructs a one-parametric family of inequivalent integrable structures labeled by the twist parameter $q$. The corresponding Hamiltonians are in general non-local, i.e. the conserved quantities are not zero Fourier modes of local currents. They interpolate between Yangian (or Benjamin-Ono) and quantum KP (or Bazhanov-Lukyanov-Zamolodchikov) integrable structures. The Hamiltonians of the latter are local and are essentially uniquely determined in the following sense: the algebra $\mathcal{W}_{\infty}$ has a unique stress-energy tensor $T(z)$ as well as spin $3$ primary field $W_3(z)$. Their zero modes (on cylinder) commute automatically ($L_0$ in fact commutes with any zero mode on the cylinder). An explicit calculation shows that starting from spin $4$, for every spin there is a unique local field whose zero mode commutes with zero mode of $W_3$ (up to total derivatives and overall normalization). We will explicitly construct first few of these local fields in the following.

There were few implicit choices that we made in this construction: first of all, apart from Virasoro $L_0$, we based the construction on the choice of the zero Fourier mode of spin $3$ field. We could have instead started from a zero mode of a $4$ field, but in this case there are choices to make, because in $\mathcal{W}_\infty$ there are $4$ spin $4$ fields and only two of them are total derivatives, i.e. there is one-parametric freedom of integrable structures in this case. The second implicit choice we made is related to the choice of working with $\mathcal{W}_\infty$ instead of $\mathcal{W}_{1+\infty}$. Having additional spin $1$ generator makes the ambiguities even worse, in particular it leads to $6$ fields of dimension $3$, only $2$ of which are total derivatives. Therefore when looking for local conserved quantities, there is much more freedom in $\mathcal{W}_{1+\infty}$ than in $\mathcal{W}_\infty$. One possible justification for the choice of $\mathcal{W}_3$ zero mode in $\mathcal{W}_\infty$ comes from a careful study of $q \to 1$ limit of ILW conserved quantities. It turns out \cite{Prochazka:2023zdb} that the finite local quantities that one extracts in the $q \to 1$ limit correspond to decoupled conserved quantities of $\mathcal{W}_\infty$ algebra together with those of $\widehat{\mathfrak{gl}(1)}$. The mechanism of this is very interesting and is related to the fact that the Bethe roots in $q \to 1$ limit naturally split into two groups, those that are associated to $\widehat{\mathfrak{gl}(1)}$ and those that are associated to $\mathcal{W}_\infty$. The Bethe roots of $\mathcal{W}_\infty$ stay finite as $q \to 1$ while those of $\widehat{\mathfrak{gl}(1)}$ go to infinity in a very specific way that encodes the Young diagram labeling the corresponding state in the free boson Fock space. This decoupling of $\widehat{\mathfrak{gl}(1)}$ from $\mathcal{W}_\infty$ only happens at $q \to 1$, for other values of $q$ the ILW quantities mix the $\widehat{\mathfrak{gl}(1)}$ and $\mathcal{W}_\infty$ subalgebras of $\mathcal{W}_{1+\infty}$. For more details, see the discussion in \cite{Prochazka:2023zdb}.

\paragraph{Explicit calculation}
Having discussed the choice of the integrable structure, let us calculate the first few of the local Hamiltonians explicitly. Since these quantities are written in terms of $\mathcal{W}_\infty$ generators, it is practically convenient to first decouple the $\widehat{\mathfrak{gl}(1)}$ from $\mathcal{W}_{\infty}$. Recall that in $\mathcal{W}_{1+\infty}$ there exists a basis of local fields $U_j(z)$, $j=1,2,\ldots$ in terms of which the OPEs have only \emph{quadratic} non-linearities (and can in fact be written in a bi-local way in a closed form) \cite{Prochazka:2014gqa}. These fields $U_j(z)$ are neither primary nor quasi-primary and mix the $\widehat{\mathfrak{gl}(1)}$ and $\mathcal{W}_{\infty}$ subalgebras.

There exists a similar set of fields $V_j(z)$, $j=2,3,\ldots$ which commute with $U_1(z)$, generate $\mathcal{W}_\infty$ subalgebra and still have quadratic (and bi-local) OPEs. They are related to fields $U_j$ by a transformation that is linear in $U_j(z)$, $j \geq 2$ but non-linear in $U_1(z)$ (i.e. $U_j$ fields are dressed by fields corresponding to $\widehat{\mathfrak{gl}(1)}$). A compact way of writing this transformation is using the Miura operator \cite{Prochazka:2014gqa}
\begin{multline}
\sum_{k=0}^N U_k(z) (\alpha_0 \partial)^{N-k} = \\
: \exp\left[-\frac{1}{\alpha_0 N} \int^z U_1(z) dz\right]: \left[ (\alpha_0 \partial)^N + \sum_{k=2}^N V_k(z) (\alpha_0 \partial)^{N-k} \right] : \exp\left[\frac{1}{\alpha_0 N} \int^z U_1(z) dz\right]:.
\end{multline}
In other words we are conjugating the differential operator (or equivalently making a multiplicative redefinition of the wave function) such that the coefficient of the subleading derivative vanishes (this is the differential operator analogue of bringing the polynomial to its depressed form). The object that we are conjugating by is a free-field normal ordered vertex operator. For the first few of these quantities we find
\begin{align}
\label{utov}
\nonumber
V_2 & = U_2 - \frac{N-1}{2N} (U_1 U_1) - \frac{(N-1)\alpha_0}{2} U_1^\prime \\
\nonumber
V_3 & = U_3 -\frac{N-2}{N} (U_1 U_2) + \frac{(N-1)(N-2)}{3N^2} (U_1(U_1 U_1)) -\frac{(N-1)(N-2)\alpha_0^2}{6} U_1^{\prime\prime} \\
\nonumber
V_4 & = U_4 -\frac{N-3}{N} (U_1 U_3) +\frac{(N-2)(N-3)}{2N^2} (U_1 (U_1 U_2)) \\
& -\frac{(N-1)(N-2)(N-3)}{8N^3} (U_1 (U_1 (U_1 U_1))) -\frac{(N-2)(N-3)\alpha_0}{2N} (U_1^\prime U_2) \\
\nonumber
& +\frac{(N-1)(N-2)(N-3)\alpha_0}{4N^2} (U_1^\prime(U_1 U_1)) +\frac{(N-1)(N-2)(N-3) \alpha_0^2}{8N} (U_1^\prime U_1^\prime) \\
\nonumber
& -\frac{(N-1)(N-2)(N-3)\alpha_0^3}{24} U_1^{(3)}
\end{align}
Notice that the coefficients of the various local fields on the right-hand side are simple polynomial functions of $N$ and in fact one can write a Fa\`a di Bruno-style formula for these local fields, but we will not need this in the following.

As already mentioned, the OPE of $V_j(z)$ close quadratically just as was the case with $U_j(z)$ and all the derivatives can be resummed and one can write a bi-local expansion of product of two such fields, in complete parallel with the situation in $\mathcal{W}_{1+\infty}$ \cite{Prochazka:2014gqa}. The advantage of working with $V_j(z)$ is that there are significantly fewer local fields of a given dimension that we can write in $\mathcal{W}_\infty$ than in $\mathcal{W}_{1+\infty}$ so for practical purposes it is easier to write an ansatz in terms of $V_j(z)$ and if needed to express the result in terms of $U_j(z)$ afterwards.

The procedure of finding local commuting quantities follows the same steps as the analogous procedure in the Virasoro case in Section \ref{virasoroimconstr}. We start with $\mathcal{W}_{\infty}$ stress-energy tensor
\begin{equation}
T_\infty = -U_2 + \frac{N-1}{2N} (U_1 U_1) + \frac{(N-1)\alpha_0}{2} U_1^\prime = -V_2
\end{equation}
and the unique (up to normalization) spin $3$ primary in $\mathcal{W}_\infty$
\begin{align}
\nonumber
W_3 & = -U_3 + \frac{N-2}{N} (U_1 U_2) - \frac{(N-1)(N-2)}{3N^2} (U_1 (U_1 U_1)) \\
\nonumber
& - \frac{(N-1)(N-2)\alpha_0}{2N} (U_1^\prime U_1) + \frac{(N-2)\alpha_0}{2} U_2^\prime - \frac{(N-1)(N-2)\alpha_0^2}{12} U_1^{\prime\prime} \\
& = -V_3 + \frac{(N-2)\alpha_0}{2} V_2^\prime.
\end{align}
For future reference, note that this field is invariant under the triality symmetry only up to an overall normalization. In fact, the leading part of its OPE is
\begin{equation}
W_3(z) W_3(w) \sim \frac{(\lambda_1-1)(\lambda_2-1)(\lambda_3-1)(\lambda_1-2)(\lambda_2-2)(\lambda_3-2)}{6N(z-w)^6} + \ldots
\end{equation}
so the triality-invariant normalization is
\begin{equation}
N^{1/2} W_3(z).
\end{equation}
An analogous issue with normalization happens already at the level of spin $1$ field -- the field $U_1$ has OPE
\begin{equation}
U_1(z) U_1(w) \sim \frac{N}{(z-w)^2} + reg.
\end{equation}
so the triality invariant normalization is $N^{-1/2} U_1(z)$.

Starting from dimension $4$ there is in every dimension a unique field $\mathcal{J}_n$ (up to normalization and total derivatives) whose zero mode commutes with $(W_3)_0$, the zero Fourier mode of $W_3(z)$. Since we work in the cylinder frame, the commutator of zero mode of the field $W_3(z)$ with $\mathcal{J}_n(w)$ is given simply by the residue of their OPE, i.e. the coefficient of their simple pole
\begin{equation}
\left[ (W_3)_0, \mathcal{J}_n(w) \right] = \oint_w \frac{dz}{2\pi i} W_3(z) \mathcal{J}_n(w).
\end{equation}
Furthermore, in the cylinder frame the zero Fourier modes of total derivatives vanish\footnote{See Appendix A in \cite{Prochazka:2023zdb} where the cylinder and plane frames are compared in more detail.}. We can fix the total derivatives in $\mathcal{J}_n(z)$ by requiring that it is quasi-primary. The number of conditions imposed by this requirement is given by the number of plane partitions of $n-1$ boxes which is exactly the number of total derivatives, so generically we expect to find a unique field up to an overall normalization. The overall normalization can be fixed in triality-invariant manner by imposing that the coefficient of $U_n$ in $\mathcal{J}_n$ is $-N^{\frac{n-2}{2}}$. In this way, we find the following list of currents
\begin{align}
\nonumber
\mathcal{J}_1 & = -N^{-1/2} U_1 \\
\nonumber
\mathcal{J}_2 & = -V_2 \\
\nonumber
\mathcal{J}_3 & = N^{1/2} \left( -V_3 + \frac{(N-2)\alpha_0}{2} V_2^\prime \right) \\
\mathcal{J}_4 & = N \left( - V_4 + \frac{N-3}{2N} (V_2 V_2) + \frac{(N-3)\alpha_0}{2} V_3^\prime - \frac{(N-3)(2\alpha_0^2 N^2 - 4\alpha_0^2 N - 3)}{20N} V_2^{\prime\prime} \right) \\
\nonumber
\mathcal{J}_5 & = N^{3/2} \Bigg( -V_5 + \frac{(N-4)}{N} (V_2 V_3) + \frac{(N-4) \alpha_0}{2} V_4^\prime - \frac{(N-2)(N-4)\alpha_0}{2N} (V_2^\prime V_2) \\
\nonumber
& - \frac{3(N-4) \left(\alpha_0^2 N^2-3\alpha_0^2 N-2\right)}{28N} V_3^{\prime\prime} + \frac{(N-2)(N-4)\alpha_0 \left(\alpha_0^2 N^2-3\alpha_0^2 N-2\right)}{84N} V_2^{(3)} \Bigg)
\end{align}
and
{\small
\begin{align}
\nonumber
\mathcal{J}_6 & = N^2 \Bigg( - V_6 + \frac{N-5}{N} (V_2 V_4) + \frac{N-5}{2N} (V_3 V_3) - \frac{(N-5)(2N-5)}{6N^2} (V_2(V_2 V_2)) \\
\nonumber
& + \frac{(N-5)\alpha_0}{2} V_5^\prime - \frac{(N-2)(N-5)\alpha_0}{2N} (V_2^\prime V_3) - \frac{(N-3)(N-5)\alpha_0}{2N} (V_2 V_3^\prime) \\
\nonumber
& + \frac{(N-5) \left(12\alpha_0^2 N^3 - 64\alpha_0^2 N^2 + 82\alpha_0^2 N - 48N + 103\right)}{108N^2} (V_2^{\prime\prime} V_2) \\
& + \frac{(N-5) \left(24\alpha_0^2 N^3 - 83\alpha_0^2 N^2 + 65\alpha_0^2 N - 96N + 215\right)}{216N^2} (V_2^\prime V_2^\prime) \\
\nonumber
& - \frac{(N-5)\left(\alpha_0^2 N^2-4\alpha_0^2 N-4\right)}{9N} V_4^{\prime\prime} \\
\nonumber
& + \frac{(N-5)\alpha_0 \left(\alpha_0^2 N^3 - 7\alpha_0^2 N^2  +12\alpha_0^2 N - 7N + 30\right)}{72N} V_3^{(3)} \\
\nonumber
& - \frac{(N-5) \left(9\alpha_0^4 N^5 - 81\alpha_0^4 N^4 + 234\alpha_0^4 N^3 + 345\alpha_0^2 N^3 - 216\alpha_0^4 N^2 - 1033\alpha_0^2 N^2 + 691\alpha_0^2 N - 12N + 55\right)}{9072N^2} V_2^{(4)} \Bigg)
\end{align}
}The non-derivative terms in these currents follow a simple structure (which we checked up to dimension $12$): we have
\begin{equation}
\mathcal{J}_n = N^{\frac{n}{2}-1} \sum_{2k_2+3k_3+\ldots=n} \frac{(-1)^{\sum_j k_j} \left( \frac{1-n}{N} + 1\right)_{-1+\sum_j k_j}}{k_2! k_3! \cdots} (V_2^{k_2} V_3^{k_3} \cdots) + derivatives
\end{equation}
where $(x)_k$ is the raising factorial $x(x+1)\cdots(x+k-1)$. We do not need to care about the order of the local fields $V_j$ or the nesting of the parentheses as we only write the expression up to terms involving derivatives of $V_j$ fields. Returning back to $\mathcal{W}_{1+\infty}$, we can express the currents $\mathcal{J}_n$ in terms of $\mathcal{W}_{1+\infty}$ generating fields $U_j$ using \eqref{utov}. The non-derivative part of the expressions for the local currents takes the same form as in the case of $\mathcal{W}_{\infty}$, namely
\begin{equation}
\mathcal{J}_n = N^{\frac{n}{2}-1} \sum_{k_1+2k_2+3k_3+\ldots=n} \frac{(-1)^{\sum_j k_j} \left( \frac{1-n}{N} + 1\right)_{-1+\sum_j k_j}}{k_1! k_2! k_3! \cdots} (U_1^{k_1} U_2^{k_2} U_3^{k_3} \cdots) + derivatives,
\end{equation}
the only difference being the fact that we allow for fields containing powers of $U_1$.

\subsection{Highest weights}
\label{secwinfhw}
In this section we will discuss various ways of parametrizing the highest weights of a highest weight representation. The highest weight vector $\ket{\mu}$ satisfies
\begin{equation}
\left(U_j^{(cyl)}\right)_0 \ket{\mu} = u^{(cyl)}_j \ket{\mu}, \qquad \left(U_j^{(cyl)}\right)_m \ket{\mu} = 0, \quad m \geq 1.
\end{equation}
Here we put the subscript $cyl$ to emphasize that we use Fourier modes of $U_j(z)$ in the cylinder frame. Since the fields $U_j(z)$ are not primary, their Fourier modes in the plane and cylinder frame differ by a triangular transformation. The analogous highest weight condition in the plane frame is
\begin{equation}
\left(U_j^{(pl)}\right)_0 \ket{\mu} = u^{(pl)}_j \ket{\mu}, \qquad \left(U_j^{(pl)}\right)_m \ket{\mu} = 0, \quad m \geq 1
\end{equation}
and we first want to find the transformation relating $u^{(pl)}_j$ to $u^{(cyl)}_j$. All $u_k^{(pl)}$ can be conveniently collected into a generating function \cite{Prochazka:2015deb}
\begin{equation}
\label{hwgeneratingfn}
\mathcal{U}_{\mu}(u) = 1 + \sum_{k > 0} \frac{u_k^{(pl)}}{(-u)(-u+\alpha_0)\cdots(-u+(k-1)\alpha_0)}.
\end{equation}
For $\mathcal{W}_N$ this sum truncates at $k=N$ (since for $k$ the in the denominator vanish) so $\mathcal{U}_{\mu}(u)$ is a rational function of $N$-th degree. If we represent $\mathcal{W}_N$ in terms of $N$ free bosons $J_j(z)$, the zero modes $u^{(pl)}_j$ can be parametrized in terms of their zero modes $a_j^{(pl)}$ as \cite{Prochazka:2014gqa}
\begin{align}
\nonumber
u_1^{(pl)} & = \sum_{j=1}^N a_j^{(pl)} \\
u_2^{(pl)} & = \sum_{j<k} a_j^{(pl)} a_k^{(pl)} - \alpha_0 \sum_{j=1}^N (j-1) a_j^{(pl)} \\
\nonumber
u_3^{(pl)} & = \sum_{j<k<l} a_j^{(pl)} a_k^{(pl)} a_l^{(pl)} - \alpha_0 \sum_{j<k} (j+k-3) a_j^{(pl)} a_k^{(pl)} + \alpha_0^2 \sum_{j=1}^N (j-1) (j-2) a_j^{(pl)}
\end{align}
etc. In terms of the generating function \eqref{hwgeneratingfn}, we have \cite{Prochazka:2015deb}
\begin{equation}
\mathcal{U}_{\mu}(u) = \prod_{j=1}^N \frac{u-a_j^{(pl)}-\alpha_0 j+\alpha_0}{u-\alpha_0 j+\alpha_0}.
\end{equation}
Since neither $J_j(z)$ nor $U_j(z)$ are primary fields, the transformation between the cylinder frame and the complex plane is non-trivial. In particular, due to anomalous term in the transformation of $J_j(z)$ \cite{Prochazka:2019dvu}
\begin{equation}
\tilde{J}_j(\tilde{z}) = \left(\frac{d\tilde{z}}{dz}\right)^{-1} \left[ J_j(z) + \frac{\alpha_0(N+1-2j)}{2} \left(\frac{d\tilde{z}}{dz}\right)^{-1} \left(\frac{d^2\tilde{z}}{dz^2}\right) \right],
\end{equation}
the plane and cylinder zero modes are related by a $j$-dependent shift
\begin{equation}
a_j^{(pl)} = a_j^{(cyl)} + \frac{\alpha_0}{2}(N+1-2j).
\end{equation}
The generating function \eqref{hwgeneratingfn} can therefore be written as
\begin{equation}
\mathcal{U}_{\mu}(u) = \prod_{j=1}^N \frac{u-a_j^{(cyl)}-\frac{\alpha_0}{2}(N-1)}{u-\alpha_0 j+\alpha_0},
\end{equation}
i.e. as a \emph{symmetric} function of the zero modes $a_j^{(cyl)}$ (this was not the case when written in terms of $a_j^{(pl)}$). In the cylinder frame, the charges $u_j^{(cyl)}$ are simply the elementary symmetric polynomials of $a_j^{(cyl)}$ \cite{Prochazka:2014gqa},
\begin{align}
\nonumber
u_1^{(cyl)} & = \sum_{j=1}^N a_j^{(cyl)} \\
u_2^{(cyl)} & = \sum_{j<k} a_j^{(cyl)} a_k^{(cyl)} \\
\nonumber
u_3^{(cyl)} & = \sum_{j<k<l} a_j^{(cyl)} a_k^{(cyl)} a_l^{(cyl)}
\end{align}
etc. so the generating function $\mathcal{U}_{\mu}(u)$ can be written as
\begin{equation}
\mathcal{U}_{\mu}(u) = \frac{\sum_{j=0}^N (-1)^j \left(u-\frac{\alpha_0}{2}(N-1)\right)^{N-j} u_j^{(cyl)}}{\prod_{j=1}^N \left( u-\alpha_0 j+\alpha_0 \right)}.
\end{equation}
Comparing this to \eqref{hwgeneratingfn}, we find a linear triangular system of relations
\begin{align}
\nonumber
u_1^{(cyl)} & = u_1^{(pl)} \\
u_2^{(cyl)} & = u_2^{(pl)} + \frac{(N-1)\alpha_0}{2} u_1^{(pl)} - \frac{(N-1)N(N+1)\alpha_0^2}{24} \\
\nonumber
u_3^{(cyl)} & = u_3^{(pl)} + (N-2)\alpha_0 u_2^{(pl)} - \frac{(N-1)(N-2)(N-3)\alpha_0^2}{24} u_1^{(pl)}
\end{align}
which are the same relations as we would get from the transformation \cite{Prochazka:2019dvu}
\begin{equation}
\sum_{k=0}^N U_k(z) (\alpha_0 \partial_z)^{N-k} = \left( \frac{d\tilde{z}}{dz} \right)^{\frac{N+1}{2}} \sum_{k=0}^N \tilde{U}_k(\tilde{z}) (\alpha_0 \partial_{\tilde{z}})^{N-k} \left( \frac{d\tilde{z}}{dz} \right)^{\frac{N-1}{2}}
\end{equation}
applied to $\tilde{z} = e^z$, i.e.
\begin{align}
\nonumber
\sum_{k=0}^N u_k^{(cyl)} \left(\alpha_0 \tilde{z} \partial_{\tilde{z}} \right)^{N-k} & = \tilde{z}^{\frac{N+1}{2}} \left[ \sum_{k=0}^N u_k^{(pl)} \tilde{z}^{-k} \left(\alpha_0 \partial_{\tilde{z}}\right)^{N-k} \right] \tilde{z}^{\frac{N-1}{2}} \\
& = \tilde{z}^{-\frac{N-1}{2}} \left[ \sum_{k=0}^N \alpha_0^{N-k} u_k^{(pl)} \tilde{z}^{N-k} \partial_{\tilde{z}}^{N-k} \right] \tilde{z}^{\frac{N-1}{2}} \\
\nonumber
& = \sum_{k=0}^N \alpha_0^{N-k} u_k^{(pl)} \left(\tilde{z} \partial_{\tilde{z}}+\frac{2k+1-N}{2}\right) \cdots \left(\tilde{z} \partial_{\tilde{z}}+\frac{N-3}{2}\right) \left(\tilde{z} \partial_{\tilde{z}}+\frac{N-1}{2}\right).
\end{align}

\paragraph{Maximally degenerate representations} The previous discussion summarizes the parametrization of the highest weights in terms of the zero modes of $U_j(z)$ fields as well as those of free bosons $J_j(z)$ in both the cylinder frame and on the complex plane. There is one more useful parametrization in terms of discrete data, namely in terms of the asymptotic Young diagram along one of the coordinate axes in the space of plane partitions. Identifying parameters as in \eqref{parameterchoice}, the only non-trivial allowed asymptotic is along 2nd coordinate axis and is labeled by a Young diagram that has at most $N$ rows (3rd direction) and $K$ columns (1st direction). These Young diagrams parametrize the maximally degenerate primaries, i.e. $\mathcal{W}_N$ generalization of the primaries listed in the Kac table in the case of Virasoro algebra. Let us label the row lengths by $\mu_j, j=1,\ldots,N$ with $\mu_1 \geq \mu_2 \geq \cdots \mu_N$ and $0 \leq \mu_j \leq K$. The generating function of Yangian charges $\psi(u)$ corresponding to such primary is given by
\cite{Prochazka:2015deb}
\begin{equation}
\psi_\mu(u) = \frac{u+\psi_0 \epsilon_1 \epsilon_2 \epsilon_3}{u} \prod_{\ell=1}^\infty \prod_{j=1}^N \prod_{k=1}^{\mu_j} \varphi(u-k\epsilon_1-\ell\epsilon_2-j\epsilon_3),
\end{equation}
i.e. as an infinite product over all boxes of the minimal configuration with the asymptotics specified by $\mu$. Here $\epsilon_j$ are the Nekrasov-like parameters appearing in the Yangian description of the algebra and are related to $\lambda_j$ by
\begin{equation}
\lambda_j = -\frac{\psi_0 \epsilon_1 \epsilon_2 \epsilon_3}{\epsilon_j}, \qquad \psi_0 = N, \qquad \epsilon_1 \epsilon_2 = -1.
\end{equation}
or\footnote{The appearance of square roots is a consequence of the fixing of the freedom of rescaling $\psi_0$ by putting it equal to $N$ which are the conventions used in \cite{Prochazka:2015deb}.}
\begin{equation}
\epsilon_1 = \sqrt{\frac{N}{K+N}}, \qquad \epsilon_2 = -\sqrt{\frac{K+N}{N}}, \qquad \epsilon_3 = \frac{K}{\sqrt{N(K+N)}}, \qquad \psi_0 = N.
\end{equation}
The function $\varphi(u)$,
\begin{equation}
\label{yangianstructurefunction}
\varphi(u) = \frac{(u+\epsilon_1)(u+\epsilon_2)(u+\epsilon_3)}{(u-\epsilon_1)(u-\epsilon_2)(u-\epsilon_3)}
\end{equation}
is the structure function of the algebra and controls its representation theory. It encodes the properties of the fundamental box entering the box counting calculations, see Figure \ref{figbasiccube}. Almost all the infinitely many factors in the product over $\ell$ cancel in pairs and we are left with a finite product over all boxes of the asymptotic Young diagram,
\begin{equation}
\psi_\mu(u) = \frac{u+\psi_0 \epsilon_1 \epsilon_2 \epsilon_3}{u} \prod_{j=1}^N \prod_{k=1}^{\mu_j} \frac{(u-k\epsilon_1-j\epsilon_3)(u-k\epsilon_1-j\epsilon_3+\epsilon_1+\epsilon_3)}{(u-k\epsilon_1-j\epsilon_3+\epsilon_1)(u-k\epsilon_1-j\epsilon_3+\epsilon_3)}.
\end{equation}
Finally, this product has cancellations along every row so we can reduce it to product over the rows
\begin{equation}
\psi_\mu(u) = \prod_{j=1}^N \frac{u-\mu_j\epsilon_1-j\epsilon_3}{u-\mu_j\epsilon_1-(j-1)\epsilon_3}.
\end{equation}
Apart from the overall prefactor coming from the vacuum representation, the expression is symmetric in $\epsilon_1$ and $\epsilon_3$ if we simultaneously transpose the Young diagram, i.e. we also have
\begin{equation}
\psi_\mu(u) = \frac{u-N\epsilon_3}{u-K\epsilon_1} \prod_{k=1}^K \frac{u-\mu^T_k\epsilon_3-k\epsilon_1}{u-\mu^T_k\epsilon_3-(k-1)\epsilon_1}.
\end{equation}
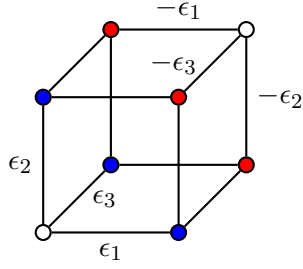
\begin{figure}
\centering
\begin{tikzpicture}[scale=0.9, vertex/.style={circle, draw, fill, inner sep=2pt}, thick]
\node[vertex, fill=white] (A) at (0,0) {};
\node[vertex, fill=blue] (B) at (2,0) {};
\node[vertex, fill=red] (C) at (2,2) {};
\node[vertex, fill=blue] (D) at (0,2) {};
\node[vertex, fill=blue] (E) at (1,1) {};
\node[vertex, fill=red] (F) at (3,1) {};
\node[vertex, fill=white] (G) at (3,3) {};
\node[vertex, fill=red] (H) at (1,3) {};
\draw[black,thick] (A) -- (B) node[midway, below] {$\epsilon_1$};
\draw[black,thick] (B) -- (C);
\draw[black,thick] (C) -- (D);
\draw[black,thick] (A) -- (D) node[midway, left] {$\epsilon_2$};
\draw[black,thick] (E) -- (F);
\draw[black,thick] (G) -- (F) node[midway, right] {$-\epsilon_2$};
\draw[black,thick] (G) -- (H) node[midway, above] {$-\epsilon_1$};
\draw[black,thick] (H) -- (E);
\draw[black,thick] (A) -- (E) node[midway, right] {$\,\epsilon_3$};
\draw[black,thick] (B) -- (F);
\draw[black,thick] (C) -- (G) node[midway, left] {$-\epsilon_3\,$};
\draw[black,thick] (D) -- (H);
\end{tikzpicture}
\caption{The basic combinatorial object underlying the representation theory of $\mathcal{W}_\infty$ is the cube which we can interpret as a fundamental excitation governed by the algebra. There are three Nekrasov-like parameters $\epsilon_1, \epsilon_2$ and $\epsilon_3$ associated to the coordinate axes satisfying the constraint $\epsilon_1+\epsilon_2+\epsilon_3=0$. The vertices separated along one of the axes by one step in positive or negative direction correspond to poles and zeros of the basic structure function $\varphi(u)$ \eqref{yangianstructurefunction}.}
\label{figbasiccube}
\end{figure}
To relate the generating functions $\psi_\mu(u)$ and $\mathcal{U}_\mu(u)$, we can use the formula (5.65) of \cite{Prochazka:2015deb},
\begin{equation}
\psi_\mu(u) = \frac{u+\psi_0\epsilon_1\epsilon_2\epsilon_3}{u} \frac{\mathcal{U}_\mu(u-\epsilon_3)}{\mathcal{U}_\mu(u)}.
\end{equation}
Comparing the two expressions, we find
\begin{equation}
\prod_{j=1}^N \frac{u-\mu_j\epsilon_1-j\epsilon_3}{u-\mu_j\epsilon_1-(j-1)\epsilon_3} = \prod_{j=1}^N \frac{u-a_j^{(pl)}-j\epsilon_3}{u-a_j^{(pl)}-(j-1)\epsilon_3 j}
\end{equation}
and we arrive at the relation between bosonic zero modes $a_j^{(pl)}$ and the Young diagram parameters $\mu_j$ which is simply
\begin{equation}
a_j^{(pl)} = \epsilon_1 \, \mu_j
\end{equation}
in agreement with equation (5.43) of \cite{Prochazka:2015deb}. This identification let us calculate all the highest weight charges $u_j^{(cyl)}$ or $u_j^{(pl)}$ in terms of the asymptotic Young diagram with rows of length $\mu_j$.


\paragraph{Higher spin charges of primaries}
Calculating the first few higher spin charges of the primary state explicitly, we find
\begin{align}
\label{primarytricharges}
\nonumber
\mathcal{I}_1 & = -\frac{u_1^{(cyl)}}{\sqrt{N}} = -\frac{1}{\sqrt{K+N}} \sum_j \mu_j \\
\nonumber
\mathcal{I}_2 & = -\frac{N-1}{24} - u_2^{(cyl)} + \frac{N-1}{2N} u_1^{(cyl)2} \\
\nonumber
& = - u_2^{(pl)} + \frac{N-1}{2N} u_1^{(pl)2} - \frac{(N-1)\alpha_0}{2} u_1^{(pl)} - \frac{(N-1)\left(1-N(N+1)\alpha_0^2\right)}{24} \\
\mathcal{I}_3 & = -\sqrt{N} u_3^{(cyl)} + \frac{N-2}{\sqrt{N}} u_1^{(cyl)} u_2^{(cyl)} - \frac{(N-1)(N-2)}{3N^{3/2}} u_1^{(cyl)3} \\
\nonumber
\mathcal{I}_4 & = - N u_4^{(cyl)} + (N-3) u_1^{(cyl)} u_3^{(cyl)} + \frac{N-3}{2} u_2^{(cyl)2} - \frac{(N-3)(2N-3)}{2N} u_1^{(cyl)2} u_2^{(cyl)} \\
\nonumber
& + \frac{(N-3)(N-1)(2N-3)}{8N^2} u_1^{(cyl)4} + \frac{N-3}{8} u_2^{(cyl)} - \frac{(N-1)(N-3)}{16N} u_1^{(cyl)2} \\
\nonumber
& - \frac{(N-3)(N-1)(2N\alpha_0^2-9)}{1920}.
\end{align}

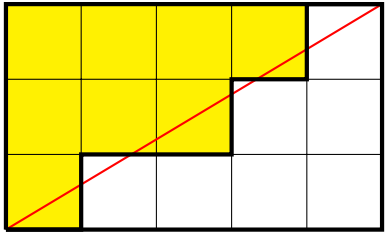
\begin{figure}
\centering
\begin{tikzpicture}[scale=1]
  \fill[yellow] (0,0) rectangle (1,1);   
  \fill[yellow] (0,1) rectangle (1,2);   
  \fill[yellow] (1,1) rectangle (2,2);   
  \fill[yellow] (2,1) rectangle (3,2);   
  \fill[yellow] (0,2) rectangle (1,3);   
  \fill[yellow] (1,2) rectangle (2,3);   
  \fill[yellow] (2,2) rectangle (3,3);   
  \fill[yellow] (3,2) rectangle (4,3);   
  \draw[step=1cm,black] (0,0) grid (5,3);
  
  \draw[red, thick] (0,0) -- (5,3);
	\draw[black, ultra thick] (0,0) -- (1,0) -- (1,1) -- (3,1) -- (3,2) -- (4,2) -- (4,3) -- (5,3);
	\draw[black, ultra thick] (0,0) -- (5,0) -- (5,3) -- (0,3) -- (0,0);
\end{tikzpicture}
\caption{Asymptotic Young diagram corresponding to ground state of $(N,K)=(3,5)$ Argyres-Douglas VOA. Yellow squares are the squares whose area is at least 50\% above the diagonal red line. The square in the center is cut exactly into two halves so we can consider it either as a yellow square or as a white square and both choices lead to the same $\mathcal{W}_{\infty}$ primary.}
\label{figgroundstateyd}
\end{figure}

\paragraph{Ground state} The Argyres-Douglas minimal models are not unitary (see in particular the negative central charge \eqref{parameterchoice}). The translationally invariant vacuum state $\ket{0}$ with $\Delta=0$ is not the lowest energy primary on the cylinder. In fact, the conformal dimension of the primary field labeled by partition $\mu$ is given by
\begin{align}
\label{confdimyt}
\nonumber
\Delta(\mu) & = -u_2^{(pl)} + \frac{N-1}{2N} u_1^{(pl)2} - \frac{(N-1)\alpha_0}{2} u_1^{(pl)} \\
& = -\sum_{j<k} a_j^{(pl)} a_k^{(pl)} + \frac{N-1}{2N} \sum_j a_j^{(pl)} \sum_k a_k^{(pl)} - \frac{\alpha_0}{2} \sum_j (N+1-2j)a_j^{(pl)} \\
\nonumber
& = -\frac{N}{K+N} \sum_{j<k} \mu_j \mu_k + \frac{N-1}{2(K+N)} \sum_j \mu_j \sum_k \mu_k - \frac{1}{2} \frac{K}{K+N} \sum_j (N+1-2j) \mu_j \\
\nonumber
& = \frac{N}{K+N} \sum_j \frac{\mu_j(\mu_j+1)}{2} + \frac{K}{K+N} \sum_j j \mu_j - \frac{1}{2(K+N)} \sum_{j,k} \mu_j \mu_k - \frac{KN+N+K}{2(K+N)} \sum_j \mu_j
\end{align}
where we used the explicit expression for $\mathcal{W}_{\infty}$ stress-energy tensor
\begin{equation}
T_{\infty} = -U_2 + \frac{N-1}{2N} (U_1 U_1) + \frac{(N-1)\alpha_0}{2} U_1^\prime.
\end{equation}
The last expression in \eqref{confdimyt} is manifestly invariant under transposition of Young diagram $\mu$ together with exchange $K \leftrightarrow N$. It also does not change if we add a row of length $K$ or a column of length $N$. These transformations correspond to a constant shift of the spectral parameter in $\mathcal{W}_{1+\infty}$ or in other words shifting the zero mode of $U_1$ current without touching the $\mathcal{W}_{\infty}$ subalgebra. Since the stress-energy tensor $T_{\infty}$ lies in $\mathcal{W}_{\infty}$ subalgebra, this transformation must not change the eigenvalues of $L_0$. Repeating the operations of addition of column or removal of a row results in $\mathbbm{Z}_{K+N}$ group of identifications of primaries, i.e. every primary is represented by $K+N$ different Young diagrams \cite{Prochazka:2023zdb} (here it is important that $K$ and $N$ are assumed to be coprime).

In order to find the value of $\mu_j$ for the lowest energy state, let us for a moment forget that $\mu_j$ are integers and let us minimize the energy \eqref{confdimyt} with respect to $\mu_j$. The equations for minimum are
\begin{equation}
0 = (K+N) \frac{\partial \Delta}{\partial \mu_j} = N \mu_j + \frac{N \mu_j}{2} + K j - \sum_k \mu_k - \frac{KN+K+N}{2}
\end{equation}
which can be written as
\begin{equation}
\label{grounddiag}
\mu_j - \frac{1}{N} \sum_k \mu_k = - \frac{K j}{N} - \frac{1}{2} + \frac{KN+K+N}{2N}.
\end{equation}
This form is consistent with the fact that the minimum of \eqref{confdimyt} is determined only up to simultaneous shift of all $\mu_j$ by a constant and up to this shift the solution is unique. Since we always assume that $K$ and $N$ are coprime, this solution minimizing $\Delta(\mu)$ does not have integer values of $\mu_j$, but as can be checked explicitly for concrete values of $K$ and $N$, the minimum of $\Delta(\mu)$ as $\mu$ runs over all Young diagrams (that fit in $N \times K$ rectangle) is the nearest Young diagram to \eqref{grounddiag}, i.e.
\begin{equation}
\label{groundmu}
\mu_j = \left\lfloor K-\frac{K}{N} \left(j-\frac{1}{2}\right) \right\rceil.
\end{equation}
Here $\lfloor x \rceil$ denotes the nearest integer to $x$. In the case that $x$ happens to be a half-integer, we can round either up or down and the resulting $\mu$ will be related by the $\mathbbm{Z}_{K+N}$ group of identifications. This happens for example in the case of $(N,K)=(3,5)$ as illustrated in Figure \ref{figgroundstateyd} (more generally when both $N$ and $K$ are odd).

Plugging \eqref{groundmu} in \eqref{confdimyt}, we find the formula for the scaling dimension of the lowest energy primary ($L_0$ eigenvalue),
\begin{equation}
\Delta_{gr} = -\frac{(K^2-1)(N^2-1)}{24(K+N)}.
\end{equation}
This expression is only symmetric under exchange of $K$ and $N$ and not under the full triality symmetry of the algebra because the lowest energy state has non-trivial asymptotic along one of the coordinate directions and therefore breaks the triality symmetry $S_3 \to S_2$. The corresponding eigenvalue of the zero mode of $T(z)$ on the cylinder (holomorphic half of the ground state energy) is
\begin{equation}
\Delta_{gr} - \frac{c}{24} = \Delta_{gr} + \frac{(K-1)(N-1)(KN+K+N)}{24(K+N)} = -\frac{(K-1)(N-1)}{24(K+N)}.
\end{equation}

\section{Geometry of the WKB curve}
\label{secgeometry}

In this section we will describe the geometry of plane algebraic curves
\begin{equation}
\label{spectralalgcurve}
x^K + y^N = 1
\end{equation}
where $K$ and $N$ are coprime integers greater than or equal to $2$. Sometimes these are called \emph{Catalan curves} \cite{silverman1987quantitative,hazama1997hodge,arul2019torsion,goodson2023sato}. The curves themselves make sense even when $K$ and $N$ are not coprime, but in that case the discussion would be slightly more complicated so we focus on the coprime case. The extreme non-coprime case is the case of $K = N$ in which case the curves are often called \emph{Fermat curves}.

\begin{figure}
\centering
\includegraphics[scale=0.80]{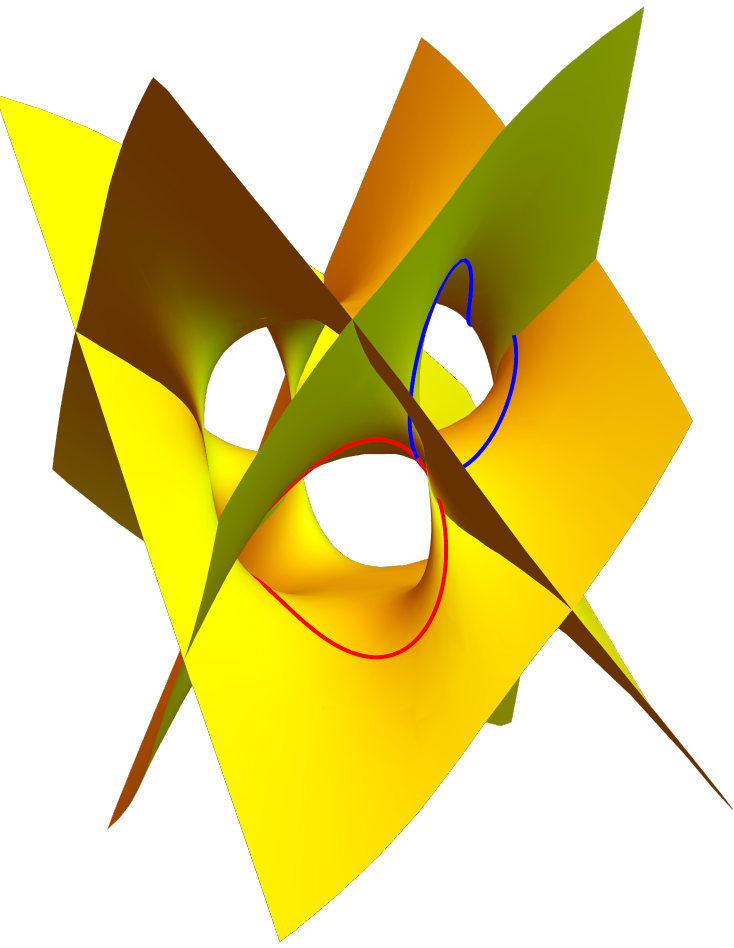}
\caption{Three-dimensional projection of genus $3$ $(A_2,A_3)$ Argyres-Douglas curve. As a plane algebraic curve, it is naturally a two-dimensional surface in four-dimensional Euclidean space $(\text{Re}\,x,\text{Im}\,x,\text{Re}\,y,\text{Im}\,y)$, and we can visualize it by projecting it to three-dimensional subspace ignoring one of the four real coordinates. The apparent self-intersections are an artifact of this projection. Nevertheless, we can see that the genus is indeed $3$. The red and blue curves are two intersecting geodesics representing two homology cycles. The attached Mathematica notebook has a version of this figure that can be rotated.}
\label{fig3dprojection}
\end{figure}

Plane algebraic curves such as \eqref{spectralalgcurve} are two-dimensional surfaces in four-dimensional space. We can visualize these quite concretely by projecting to three dimensions as illustrated in Figure \ref{fig3dprojection}, but in the following we will mostly resort to more indirect ways of investigating these curves.

\subsection{Projections to plane, ramification points, genus}

The first property of the curve \eqref{spectralalgcurve} that we want to understand is the genus of the associated compact Riemann surface $\mathcal{W}$. For that we will use the Riemann-Hurwitz formula applied to covering of a sphere by $\mathcal{W}$. To count the ramification points correctly, we also need to understand how many points at infinity we need to add to \eqref{spectralalgcurve} to get the compact surface $\mathcal{W}$ \cite{miranda1995algebraic}.

\begin{figure}
\centering
\begin{tikzpicture}[xscale=0.8, yscale=0.8]
\draw[very thick] (0,0) ellipse (3.5 and 2);
\draw[very thick, black, bend right=40] (-2.5,0.4) to (-0.7,0.4);
\draw[very thick, black, bend left=40] (-2.35,0.3) to (-0.85,0.3);
\draw[very thick, black, bend right=40] (0.7,0.4) to (2.5,0.4);
\draw[very thick, black, bend left=40] (0.85,0.3) to (2.35,0.3);
\draw[very thick, black, bend right=40] (-0.9,-0.9) to (0.9,-0.9);
\draw[very thick, black, bend left=40] (-0.75,-1.0) to (0.75,-1.0);
\node at (2,-0.7) {$(x,y)$};
\draw[very thick] (-4,-5) circle (1.5);
\node at (-4+0.5,-5+0.7) {$x$};
\draw[very thick] (4,-5) circle (1.5);
\node at (4+0.7,-4.3) {$y$};
\draw[very thick] (0,-8.2) circle (1.5);
\node at (0.8,-7.7) {$t$};
\draw[thick, ->] (-2,-2) -- (-3.3,-3.3) node[midway, above left] {$\pi_x$};
\draw[thick, ->] (2,-2) -- (3.3,-3.3) node[midway, above right] {$\pi_y$};
\draw[thick, ->] (-3.1,-6.5) -- (-1.8,-7.8) node[midway, below left] {$\tau_x$};
\draw[thick, ->] (3.1,-6.5) -- (1.8,-7.8) node[midway, below right] {$\tau_y$};
\draw[thick, ->] (0,-2.5) -- (0,-6.2) node[midway, right] {$\pi$};
\fill[red] (-2.1+0.4,-0.5+0) circle (2pt);
\fill[red] (-2.1-0.2,-0.5+0.35) circle (2pt);
\fill[red] (-2.1-0.2,-0.5-0.35) circle (2pt);
\fill[blue] (-0.1+0.4,0.9) circle (2pt);
\fill[blue] (-0.1-0.4,0.9) circle (2pt);
\fill[blue] (-0.1,0.9+0.4) circle (2pt);
\fill[blue] (-0.1,0.9-0.4) circle (2pt);
\fill[black] (1.5,1.2) circle (2pt);
\fill[blue] (-4+0.8,-5) circle (2pt);
\fill[blue] (-4-0.8,-5) circle (2pt);
\fill[blue] (-4,-5+0.8) circle (2pt);
\fill[blue] (-4,-5-0.8) circle (2pt);
\fill[red] (-4,-5) circle (2pt);
\fill[black] (-3.2,-5.7) circle (2pt);
\fill[red] (4+0.8,-5+0) circle (2pt);
\fill[red] (4-0.4,-5+0.7) circle (2pt);
\fill[red] (4-0.4,-5-0.7) circle (2pt);
\fill[blue] (4,-5) circle (2pt);
\fill[black] (4.8,-5.7) circle (2pt);
\fill[red] (0-0.25-0.25,-8.2-0.43) circle (2pt);
\fill[blue] (0+1-0.25-0.25,-8.2-0.43) circle (2pt);
\fill[black] (0+0.5-0.25-0.25,-8.2+0.87-0.43) circle (2pt);
\end{tikzpicture}
\caption{The WKB curve $x^K + y^N = 1$ (in this case $K=4, N=3$) together with its projections to $x$-plane, $y$-plane and to $t$-plane (the mirror curve). The $x$-plane is a $K$-sheeted branched covering of $\mathcal{M}$ with respect to $\tau_x$ with a ramification point at $x=0$ (red) and $x=\infty$. Analogously, $y$-plane is a $N$-sheeted branched covering of $\mathcal{M}$ with respect to $\tau_y$ with a ramification point at $y=0$ (blue) and $y=\infty$. The corresponding branch points in $\mathcal{M}$ are points $t=0$ (red), $t=1$ (blue) and $t=\infty$ (black). From the point of view of projections $\pi_x$, we see $\mathcal{W}$ as a $N$-sheeted branched covering of $x$-plane ramified at $K$ ramification points $\left(e^{\frac{2\pi ij}{K}},0\right)$ (blue) and an additional ramification point at infinity. Analogously, with respect to $\pi_y$, $\mathcal{W}$ is a $K$-sheeted branched covering of the $y$-plane with $N$ ramification points $\left(0,e^{\frac{2\pi ij}{N}}\right)$ (red). There is again an additional ramification point at infinity.}
\label{figcoverings}
\end{figure}
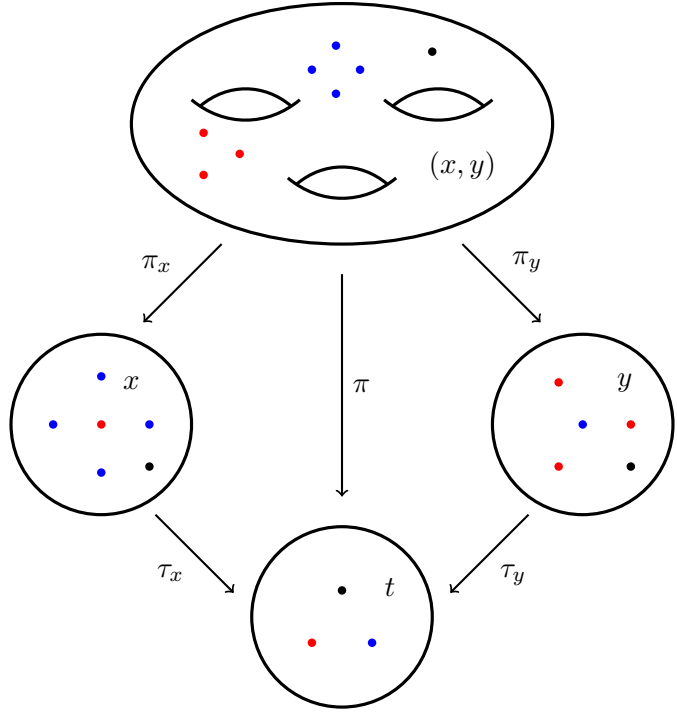

\paragraph{Covering maps}
In the following, we will often use the projections of \eqref{spectralalgcurve} to the $x$-plane $\mathbbm{C}P^1_x$,
\begin{equation}
\pi_x: (x,y) \mapsto x
\end{equation}
as well as the projection to the $y$-plane $\mathbbm{C}P^1_y$,
\begin{equation}
\pi_y: (x,y) \mapsto y.
\end{equation}
$\pi_x$ is $N$-sheeted branched covering of the Riemann sphere $\mathbbm{C}P^1_x$, since for general value of $x$ there are $N$ choices of $y$ solving \eqref{spectralalgcurve}. Analogously, $\pi_y$ is $K$-sheeted branched covering of the Riemann sphere $\mathbbm{C}P^1_y$. It is also convenient to introduce one more $\mathbbm{C}P^1$ with coordinate
\begin{equation}
t = x^K = 1-y^N
\end{equation}
which we will call the \emph{mirror curve} $\mathcal{M}$. The covering maps from $\mathbbm{C}P^1_x$ or $\mathbbm{C}P^1_y$ to $\mathcal{M}$ will be denoted by $\tau_x$ and $\tau_y$, i.e.
\begin{equation}
\tau_x: x \mapsto x^K, \qquad \tau_y: y \mapsto 1 - y^N.
\end{equation}
The map $\tau_x$ is a $K$-sheeted branched covering map of $\mathcal{M}$ by $\mathbbm{C}P^1$ and analogously for $\tau_y$. The compositions
\begin{equation}
\pi \equiv \tau_x \circ \pi_x = \tau_y \circ \pi_y
\end{equation}
represent $\mathcal{W}$ as a $NK$-sheeted branched covering map over $\mathcal{M}$ with three branch points at $t=0, 1$ and $\infty$. This is illustrated in Figure \ref{figcoverings}.

\paragraph{Points at infinity}
The set of points in $\mathbbm{C}^2$ satisfying \eqref{spectralalgcurve} defines an open subset of a compact Riemann surface $\mathcal{W}$. The algebraic curve defined by \eqref{spectralalgcurve} is smooth, because there are no points $(x,y) \in \mathbbm{C}^2$ that would satisfy both \eqref{spectralalgcurve} and be zeros of its derivative with respect to both $x$ and $y$, i.e. $x^{K-1} = 0 = y^{N-1}$. So all the points satisfying \eqref{spectralalgcurve} are smooth. In order to get the associated compact Riemann surface it is therefore enough to understand how many points to add to \eqref{spectralalgcurve} at infinity to make it a compact surface.

Recall \cite{miranda1995algebraic} that even in the hyperelliptic case $N=2$ the number of points at infinity depends on the value of $K$. For $K$ odd we need to add one point while for $K$ even we need to add two points. This can be seen by representing the curve as two-sheeted covering over $x$-plane. For $K$ even there are $K$ branch points in the finite part of the $x$-plane and the covering map is not branched at infinity, so there is one point over infinity on each of the two sheets. For $K$ odd there are $K$ finite branch points in the $x$ plane but furthermore there is one branch point at infinity of the $x$-plane. At this point the two sheets meet so we need to add only one point at infinity in order to get a compact Riemann surface.

Returning to our more general case, we will use the monodromy representation of the covering map $\pi: \mathcal{W} \to \mathcal{M}$ in the neighborhood of infinity to see how many points are there to add\footnote{We would like to thank to Davide Masoero for the explanation.} \cite{miranda1995algebraic}. Over general points of $\mathcal{M}$, we have $NK$ separate sheets of the covering map $\pi: \mathcal{W} \to \mathcal{M}$. As we go once around the point at infinity in $t$-plane $\mathcal{M}$, the coordinates $(x,y)$ change their phase by
\begin{equation}
(x,y) \mapsto \left( e^{\frac{2\pi i}{K}} x, e^{\frac{2\pi i}{N}} y \right).
\end{equation}
We see that for $K$ and $N$ coprime, all the preimages of a point in the neighborhood of the point at infinity are in a single orbit under the group generated by this monodromy transformation and therefore there is only one point above infinity of ramification index $KN$ that we need to add to get a compact Riemann surface from \eqref{spectralalgcurve}. If $K$ and $N$ were not coprime, there would be $\gcd(K,N)$ different orbits each of order $\lcm(K,N)$ and there would be $\gcd(K,N)$ points at infinity that we need to add.

As a consistency check, we can see that it agrees with the well-known situation of the hyperelliptic curve. For $N=2$ and $K$ odd, the greatest common divisor is $1$ and we are indeed adding a single ramification point at infinity, while for $N=2$ and $K$ even we need to add $\gcd(K,N)=2$ points and the projection is not ramified at these two points.

\paragraph{Branch points of the covering maps and genus}
In order to apply the Riemann-Hurwitz theorem to our covering maps, we need to find the ramification and branch points and calculate the ramification indices. Let us first consider $\pi_x: \mathcal{W} \to \mathbbm{C}P^1_x$. This map is an $N$-sheeted branched covering map to the Riemann sphere. The branch points in the finite part of the plane are points where $y$-coordinate cannot be expressed locally uniquely in terms of $x$ coordinate, i.e. the points where the $y$-derivative of \eqref{spectralalgcurve} vanishes. These are exactly the points with $y^{N-1} = 0$ and $x^K = 1$. There are $K$ such ramification points (the blue points in Figure \ref{figcoverings}) and the ramification index at each such point is $N$. The point at infinity is also a ramification point with the same value of the ramification index $N$. Applying now the Riemann-Hurwitz formula, we have
\begin{equation}
2g(\mathcal{W})-2 = N \cdot (2g(\mathbbm{C}P^1_x)-2) + (K+1)\cdot(N-1)
\end{equation}
or
\begin{equation}
\label{spectralcurvegenus}
g(\mathcal{W}) = \frac{(K-1)(N-1)}{2}.
\end{equation}
This formula is symmetric in $K$ and $N$ so we would get exactly the same result if we considered the covering $\pi_y$ instead, but this time we would have $N$ ramification points with $x=0$ and ramification index $K$ (the red points in Figure \ref{figcoverings}) plus another ramification point at infinity with index $K$ as well.

We can verify the Riemann-Hurwitz formula for maps $\tau_x$ and $\tau_y$ as well. Consider $\tau_x$ for concreteness. This time $\tau$ gives a $K$-sheeted branched covering of $\mathcal{M}$ by $\mathbbm{C}P^1_x$, both of which have genus $0$. The branch points are the points $t=0$ and $t=\infty$. The corresponding ramification points are points $x=0$ and $x=\infty$ (the blue point and the black point in Figure \ref{figcoverings}) and the ramification index is $K$ in both cases. The Riemann-Hurwitz formula gives us
\begin{equation}
2g(\mathbbm{C}P^1_x)-2 = K \cdot (2g(\mathcal{M})-2) + 2\cdot(K-1)
\end{equation}
which indeed holds. The situation of $\tau_y$ is entirely analogous.

Let us finally look at the covering map $\pi: \mathcal{W} \to \mathcal{M}$ from the WKB curve to the mirror curve. This map is a branched $NK$-sheeted covering of a Riemann sphere. There are three branch points, $t=0$, $t=1$ and $t=\infty$. There are $N$ ramification points over $t=0$ (red), each of which has ramification index $K$. There are $K$ ramification points over $t=1$ (blue), each with ramification index $N$. Finally, there is only one point on $\mathcal{W}$ mapped to $t=\infty$ and this point has ramification index $NK$. The Riemann-Hurwitz formula in this case reads
\begin{equation}
2g(\mathcal{W})-2 = NK \cdot (2g(\mathcal{M})-2) + K \cdot (N-1) + N \cdot (K-1) + 1 \cdot (NK-1)
\end{equation}
which again leads to correct genus of $\mathcal{W}$. A holomorphic map from the WKB curve $\mathcal{W}$ to $\mathbbm{C}P^1$ ramified at three points is a \emph{Belyi function} and such maps play a prominent role in Grothendieck's dessins d'enfant.

\subsection{Uniformization, covering map and fundamental polygon}
We determined the genus of Catalan curves \eqref{spectralalgcurve} to be given by \eqref{spectralcurvegenus}. For $(K,N)$ coprime and greater than or equal to $2$ the genus is always greater than one (except for $(2,3)$ Lee-Yang theory where the genus is $1$ and we should treat this case separately). From the general theory of compact Riemann surfaces we know that the universal covering space of $\mathcal{W}$ is a hyperbolic disk. Therefore there exists a holomorphic map from the hyperbolic disk (where we choose the coordinate to be $z$) to $\mathcal{W}$. Composing this map with the projection $\mathcal{W} \to \mathcal{M}$ we get a holomorphic covering map of three-punctured sphere by a hyperbolic disk. In Appendix \ref{appuniform} we construct a multi-valued inverse function to this map. It is given by (analytic continuation of) Schwarz triangle map $s_{\alpha,\beta,\gamma}(w)$, see \eqref{unifmap}.

Let us specialize the parameters to our situation. We have a special point $t = \infty$ around which the map $\mathcal{W} \to \mathcal{M}$ has ramification index $KN$ and we want to map this point to the center of the hyperbolic disk. Let us therefore choose the parameters of the Schwarz triangle map to be
\begin{equation}
\label{alphabetagammank}
\alpha = \frac{1}{KN}, \qquad \beta = \frac{1}{N}, \qquad \gamma = \frac{1}{K}, \qquad w = t^{-1}.
\end{equation}
We therefore define the map around $t = \infty$
\begin{equation}
\label{uniformizationmap}
z \equiv s_{\frac{1}{KN},\frac{1}{N},\frac{1}{K}}\left(\frac{1}{t}\right) = t^{-\frac{1}{KN}} \left[ 1 + \frac{(N^2-1)(K^2+1)}{2KN(K^2N^2-1)} t^{-1} + \mathcal{O}(t^{-2}) \right]
\end{equation}
and its inverse map
\begin{align}
\nonumber
\label{automorphict}
t & = z^{-KN} \Bigg[ 1 + \frac{(K^2+1)(N^2-1)}{2(K^2N^2-1)} z^{KN} + \frac{(K^2-1)(N^2-1)(5K^2N^2+3K^2+3N^2+1)}{16 (K^2N^2-1)(4K^2N^2-1)} z^{2KN} \\
& + \frac{(K^2-1)(N^2-1)(K^2-N^2)(2K^2N^2+K^2+N^2)}{6(K^2N^2-1)^2(9K^2N^2-1)} z^{3KN} + \mathcal{O}(z^{4KN}) \Bigg].
\end{align}
This defines a meromorphic function in the hyperbolic disk mapping the hyperbolic disk to the three-punctured sphere $\mathcal{M}$\footnote{Note that since we want to keep $t=\infty$ at the center of $\mathcal{D}$ to make symmetry between $N$ and $K$ manifest, we choose $w = t^{-1}$ and a result of this, the image of the upper half $t$-plane is a hyperbolic triangle in the \emph{lower} half-disk. This is related to the fact that if we run around a large circle in $t$-plane counter-clockwise, in local coordinate $t^{-1}$ around $t^{-1}=0$ this is mapped to a small circle around which we run in the clockwise direction.}.

Taking the $K$-th root of $t$, we get the projection from the disk to $x$-plane
\begin{equation}
\label{automorphicx}
x = z^{-N} \left[ 1 + \frac{(K^2+1)(N^2-1)}{2K(K^2N^2-1)} z^{KN} + \mathcal{O}(z^{2KN}) \right].
\end{equation}
and analogously taking a $N$-th root of $1-t$ we get the $y$-projection
\begin{equation}
\label{automorphicy}
y = e^{\frac{\pi i}{N}} z^{-K} \left[ 1 - \frac{(K^2-1)(N^2+1)}{2N(K^2N^2-1)} z^{KN} + \mathcal{O}(z^{2KN}) \right].
\end{equation}
We see that $x(z)$ has $N$-th order pole at the origin while $y(z)$ has $K$-th order pole at the origin. We could have chosen different phases of $x(z)$ and $y(z)$, the choice we make here is such that $x$ is positive for $z$ positive while $y$ has phase $e^{\frac{\pi i}{N}}$ along the positive $z$-axis. One can easily check that for any choice of these global phases, the transformations \eqref{xrotationtoz} and \eqref{yrotationtoz} correspond basic rotations in $x$-plane and $y$-plane.

\begin{figure}
\centering
\includegraphics[scale=0.60]{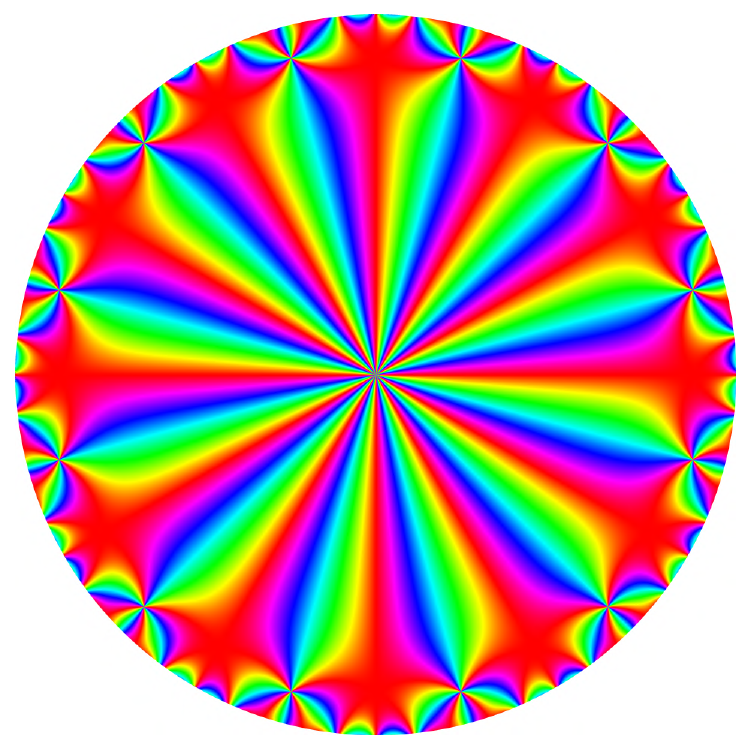}
\caption{Complex plot of the $t(z)$ in the case of $(A_2,A_3)$ model, i.e. a plot where the color corresponds to the phase of $t$. We see that the function respects the symmetries of the tiling, i.e. is indeed an automorphic function.}
\label{figautomorphict}
\end{figure}

By construction, all of these functions are meromorphic function on the hyperbolic disk and are periodic with respect to all of the symmetries of the tiling. Such functions are called \emph{automorphic functions} and are generalizations of both elliptic and modular functions. The specialization to $(N,K)=(2,3)$ which is the elliptic case is discussed in Appendix \ref{leeyangelliptic}. For $(A_2,A_3)$ model the complex plot of $t$ as a function of the disk coordinate $z$ is shown in Figure \ref{figautomorphict}.

\subsection{Homology and cohomology of $\mathcal{W}$}

\paragraph{Holomorphic differentials}
We can determine the genus of $\mathcal{W}$ in another way. Recall that for compact Riemann surface of genus $g$ there are $g$ linearly independent holomorphic $1$-forms (and $g$ linearly independent anti-holomorphic $1$-forms which together give a basis over $\mathbbm{C}$ of the first cohomology group $H^1(\mathcal{W},\mathbbm{C})$). The holomorphic $1$-forms on the plane algebraic curves of the form $f(x,y)=0$ such as \eqref{spectralalgcurve} are of the form \cite{miranda1995algebraic}
\begin{equation}
\frac{g(x,y)dx}{\frac{\partial f(x,y)}{\partial y}} = -\frac{g(x,y)dy}{\frac{\partial f(x,y)}{\partial x}}
\end{equation}
where $g(x,y)$ is a polynomial. In our case this is
\begin{equation}
\frac{g(x,y)dx}{N y^{N-1}} = -\frac{g(x,y)dy}{K x^{K-1}}.
\end{equation}
From the left-hand side we see that the only possible singularities are at infinity and at points with $y=0$, but the right-hand side shows that the points $y=0$ are actually regular. So the only possible singularity is at infinity. Requiring no pole at infinity gives us restriction on the degree of the polynomial $g(x,y)$. In order to find it, we need to introduce a holomorphic coordinate on $\mathcal{W}$ around the point at infinity. For large $x$ and $y$, \eqref{spectralalgcurve} is not far from the singular curve
\begin{equation}
x^K + y^N = 0
\end{equation}
which has a simple parametrization
\begin{equation}
x = z^{-N}, \qquad y = e^{\frac{\pi i}{N}} z^{-K}.
\end{equation}
The neighborhood of infinity corresponds to a neighborhood of $z=0$. If the right-hand side of \eqref{spectralalgcurve} does not vanish, we can instead use the automorphic functions \eqref{automorphict}, \eqref{automorphicx} and \eqref{automorphicy}
\begin{equation}
t = z^{-KN} \left(1+\mathcal{O}(z^{KN})\right), \qquad x = z^{-N} \left(1+\mathcal{O}(z^{KN})\right), \qquad y = e^{\frac{\pi i}{K}} z^{-K} \left(1+\mathcal{O}(z^{KN})\right)
\end{equation}
where the functions in parentheses have convergent power series expansion in powers of $z^{KN}$ in the neighborhood of $z=0$ (i.e. the neighborhood of infinity).

\begin{figure}
\centering
\begin{tikzpicture}[scale=0.7]
  \draw[thin,gray] (0,0) grid (6,4);
  \foreach \x in {0,...,6} {
    \foreach \y in {0,...,4} {
      \fill[black] (\x,\y) circle (2pt);
    }
  }
  \coordinate (A) at (22/5,0);
  \coordinate (B) at (0,22/7);
  \fill[red,opacity=0.2] (0,0) -- (A) -- (B) -- cycle;
  \draw[thick,red] (0,0) -- (A);
  \draw[thick,red] (0,0) -- (B);
  \draw[thick,->] (A) -- (6.5,0) node[below] {$j$};
  \draw[thick,->] (B) -- (0,4.5) node[left] {$k$};
  \draw[thick,red] (A) -- (B) ;
	\foreach \x/\y in {0/0, 1/0, 2/0, 3/0, 4/0, 0/1, 1/1, 2/1, 3/1, 0/2, 1/2, 0/3} {\fill[red] (\x,\y) circle (2pt);}
\end{tikzpicture}
\caption{Illustration of counting of holomorphic differentials for $N=5, K=7$ model. There are $12$ red points corresponding to $12$ holomorphic differentials that we find on the WKB curve which in this case has genus $12$.}
\label{differentialcounting}
\end{figure}
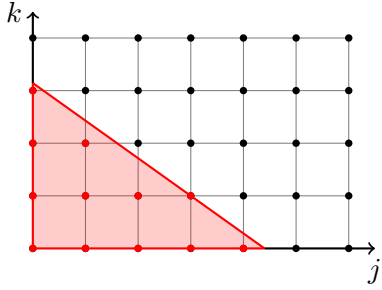

Consider now the monomial $g(x,y) = x^j y^k$, i.e. the meromorphic $1$-form
\begin{equation}
\frac{x^j y^k dx}{N y^{N-1}} \sim z^{-jN-kK-N-1+K(N-1)} dz \qquad \text{as} \qquad z \sim 0
\end{equation}
This is regular at $z = 0$ if
\begin{equation}
\label{holdifineq}
jN + kK \leq KN-K-N-1 = (K-1)(N-1)-2.
\end{equation}
For any concrete values of $K$ and $N$, it is quite easy to check that there are exactly $\frac{(K-1)(N-1)}{2}$ pairs $(j,k) \geq (0,0)$ satisfying this inequality and this agrees with the previously found genus of $\mathcal{W}$. This can be visualized by drawing all possible monomials $x^j y^k$ as a lattice $\mathbbm{Z}^2$ and marking the allowed ones as follows: the holomorphicity at $x=0$ and $y=$ requires that $j \geq 0$ and $k \geq 0$, i.e. the allowed values must lie in the first quadrant of $\mathbbm{Z}^2$ (with the axes included). The regularity at $z=0$  \eqref{holdifineq} implies that the allowed values must lie below or on the line
\begin{equation}
\label{ehrhartline}
jN + kK = (K-1)(N-1)-2,
\end{equation}
see Figure \ref{differentialcounting}. We cannot directly use Pick's theorem to count the number of points, because the intersections of \eqref{ehrhartline} with positive semi-axes have rational but non-integer coordinates. But in the case of $K$ and $N$ coprime, the rescaled versions of the triangle have integer intersections with the positive semi-axes and we can use Ehrhart polynomial do determine the number of points in the triangle. Rescaling the triangle by factor of $d$, the number of points is given by
\begin{equation}
\frac{KN}{2} d^2 - \frac{K+N+1}{2}d + 1.
\end{equation}
The first term is the volume of the triangle in the continuum limit, the last term is its Euler characteristic. The second term reflects the perimeter correction to the number of points at large $d$. Evaluating this polynomial at $d=1$ gives the correct number of points in the original triangle, $\frac{(K-1)(N-1)}{2}$ which is the genus of our WKB curve. We have therefore found all holomorphic $1$-forms on $\mathcal{W}$ and they are given as explicit rational functions in terms of $(x,y)$ coordinates.

\begin{figure}
\centering
\begin{subfigure}{0.40\textwidth}
\includegraphics[width=\linewidth]{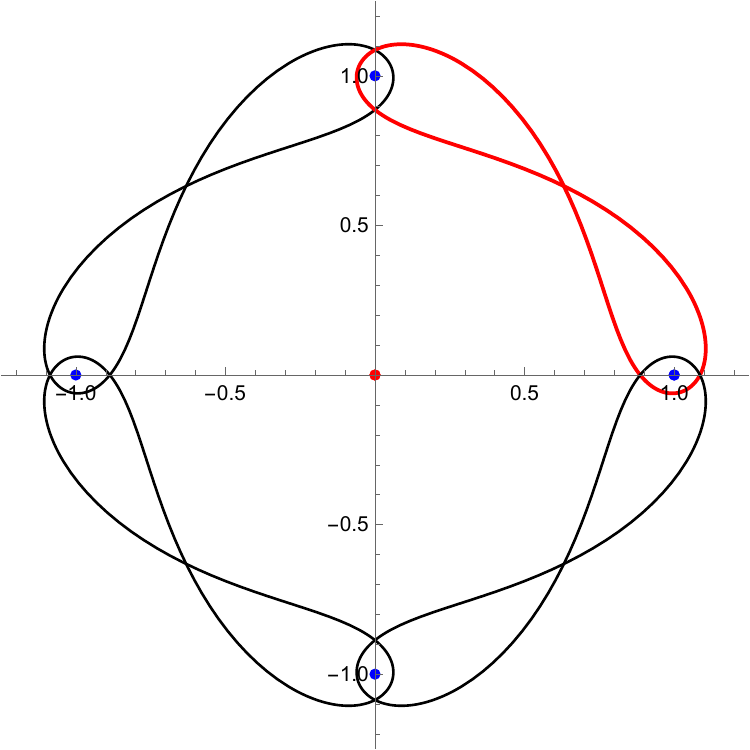}
\caption{$x$-plane projection}
\label{fighomologycyclesx}
\end{subfigure}
\hspace{3em}
\begin{subfigure}{0.40\textwidth}
\includegraphics[width=\linewidth]{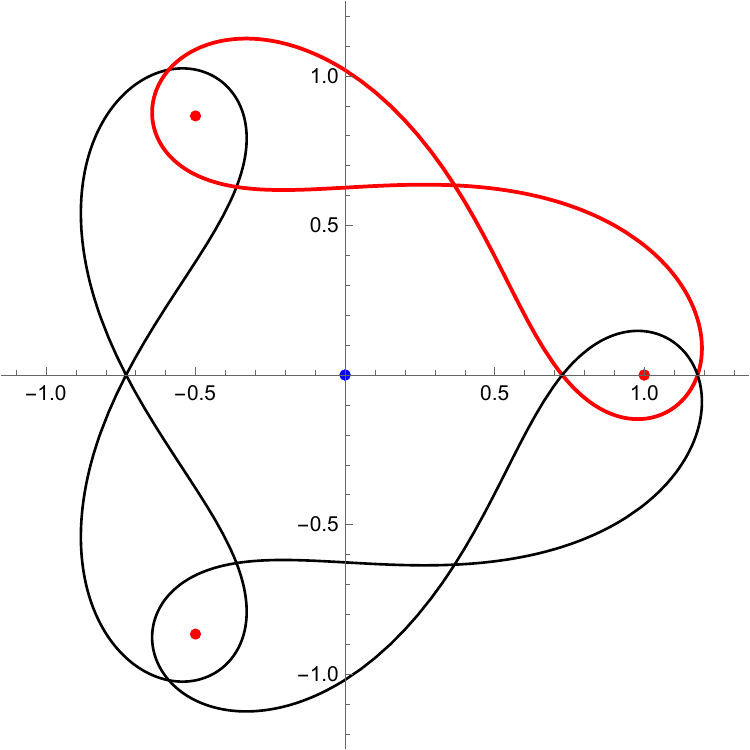}
\caption{$y$-plane projection}
\label{fighomologycyclesy}
\end{subfigure}
\begin{subfigure}{0.40\textwidth}
\includegraphics[width=\linewidth]{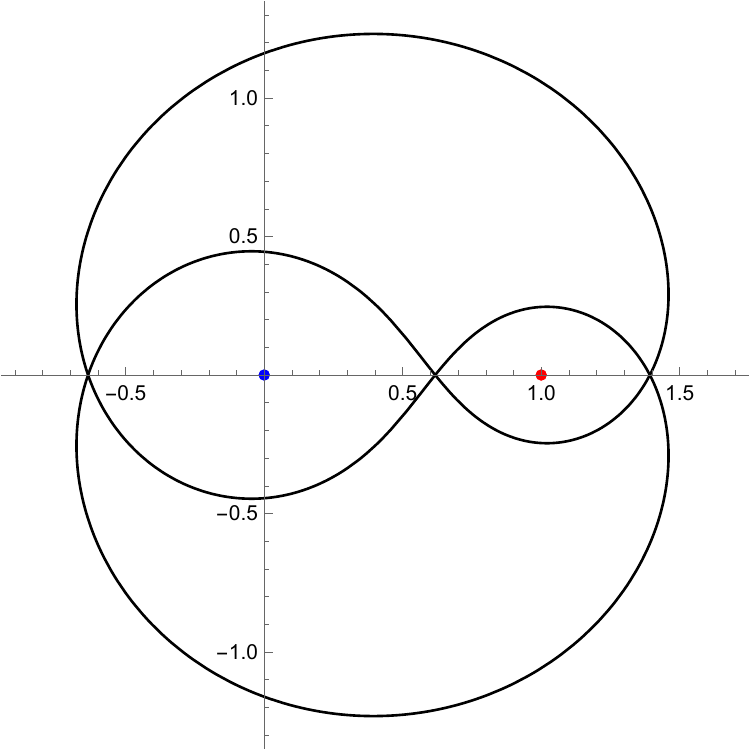}
\caption{$t$-plane projection}
\label{fighomologycyclest}
\end{subfigure}
\caption{Projections of the basic homology cycles on $\mathcal{W}$ to the $x$-plane, $y$-plane and the mirror curve $\mathcal{M}$. We choose the parameters so that $N=3, K=4$, i.e. we consider the WKB curve $y^3+x^4=1$. The upper left figure shows the $x$-plane with $K=4$ ramification points $x^4=1, y=0$ (blue) and projections of $\mathcal{C}_{0,0}$ (red) and $\mathcal{C}_{j,0}, j=1,2,3$ (black). The upper right figure shows the $y$-plane with $N=3$ ramification points $x=0, y^3=1$ (red) and the projections of $\mathcal{C}_{0,0}$ (red) and $\mathcal{C}_{0,j}, j=1,2$ (black). Finally the lower picture shows the projection of all $\mathcal{C}_{j,k}$ curves down to the mirror curve $\mathcal{M}$. The image is the Pochhammer contour encircling the three branch points $t = 0, 1$ and $\infty$. The curves are the actual shortest distance curves of the corresponding homology class (geodesics).}
\label{fighomologycycles}
\end{figure}

\paragraph{Homology}
On the Riemann surface $\mathcal{W}$ of genus $g = \frac{(K-1)(N-1)}{2}$ there exists a (non-canonical) symplectic basis of homology given by $g$ $A$-cycles and $g$ $B$-cycles with the intersection form
\begin{equation}
A_j \cap A_k = 0, \qquad B_j \cap B_k = 0, \qquad A_j \cap B_k = \delta_{jk}, \qquad j,k=1,\ldots,g.
\end{equation}
These $1$-cycles form a basis of the first homology group $H_1(\mathcal{W},\mathbbm{Z})$ over $\mathbbm{Z}$ and such a basis is unique up to symplectic transformations $Sp(2g,\mathbbm{Z})$. In our situation such a basis is not the most convenient one, because it does not respect the group of automorphisms of $\mathcal{W}$, $\mathbbm{Z}_K \times \mathbbm{Z}_N$.

Consider a Pochhammer contour in the $t$-plane around the three branch points $t = 0, 1$ and $\infty$ as in Figure \ref{fighomologycyclest}. For concreteness we choose the orientation such that it encircles $t=0$ counter-clockwise, then we encircle $t=1$ clockwise, then $t=0$ clockwise and finally $t=1$ counter-clockwise. Since we encircle all the points $t=0, 1$ and $\infty$ once clockwise and once counter-clockwise, the WKB $1$-forms that are multi-valued around these branch points will have well-defined integrals around this Pochhammer contour, i.e. the lifts of the Pochhammer contour to $\mathcal{W}$ will close.

There are $NK$ lifts of the Pochhammer contour to $\mathcal{W}$. We fix one of these lifts and we will call it $\mathcal{C}$ (the precise lift will only be needed in Section \ref{pochhammerintegral} when evaluating the contour integrals). All the other $NK$ lifts can be obtained from $\mathcal{C}$ by the action of $\mathbbm{Z}_K \times \mathbbm{Z}_N$. We will call $\mathcal{C}_{j,k}$ the contour that is obtained by rotating $\mathcal{C}$ by $e^{\frac{2\pi i j}{K}}$ in $x$-plane and by $e^{\frac{2\pi i k}{N}}$ in the $y$-plane, see the Figure \ref{fighomologycyclesx} and \ref{fighomologycyclesy}. For convenience it is useful to consider the labels $j$ and $k$ to be periodic, i.e. $\mathcal{C}_{j+K,k} \equiv \mathcal{C}_{j,k}$ and $\mathcal{C}_{j,k+N} \equiv \mathcal{C}_{jk}$. The homology classes on $\mathcal{W}$ corresponding to curves $C_{j,k}$ are not linearly independent: we have two relations,
\begin{equation}
\label{homologyconstraints}
\sum_{j=0}^{K-1} \mathcal{C}_{j,k} = 0, \qquad \sum_{k=0}^{N-1} \mathcal{C}_{j,k} = 0.
\end{equation}
\begin{figure}
\centering
\includegraphics[scale=0.80]{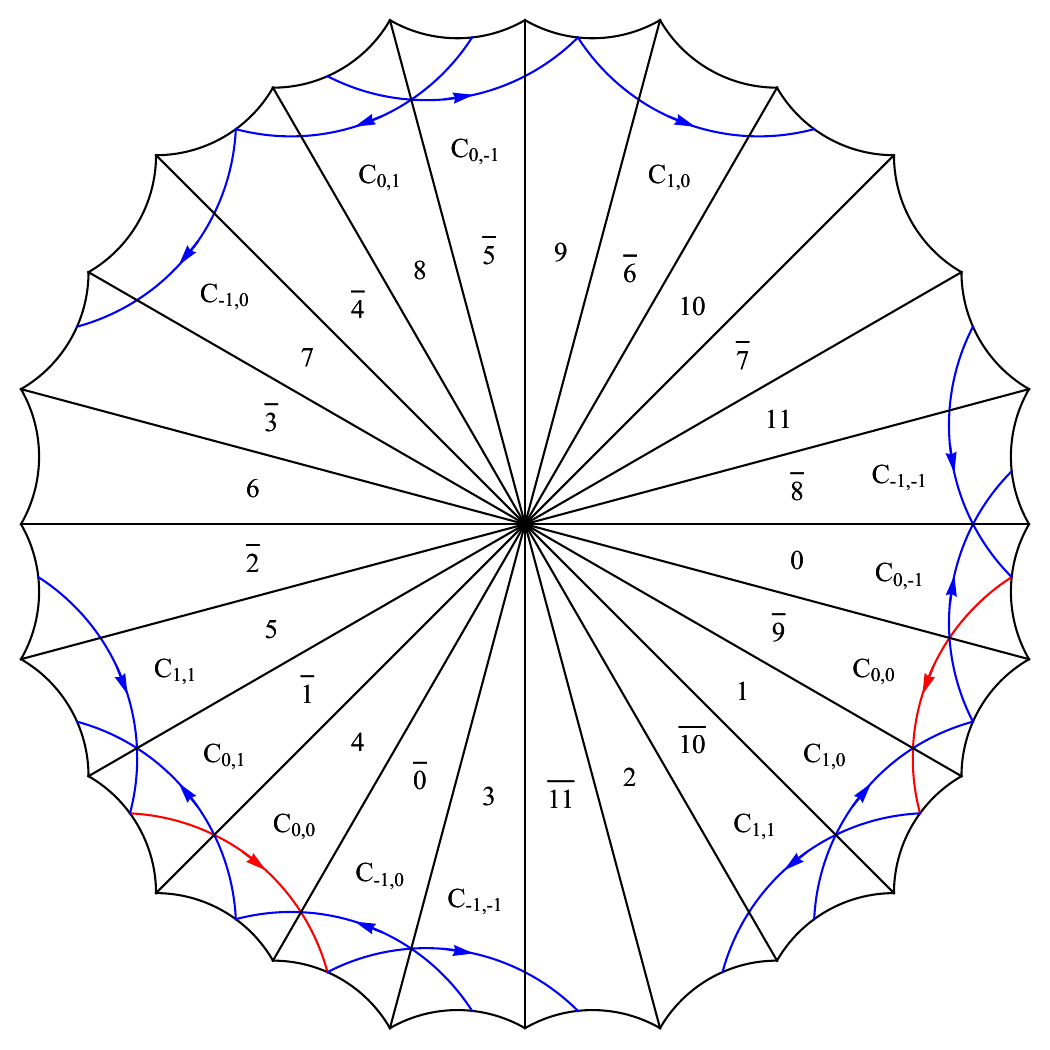}
\caption{Some of the geodesics representing the homology classes in $H_1(\mathcal{W},\mathbbm{Z})$ in the fundamental polygon of $(A_2,A_3)$ theory. Every geodesic $\mathcal{C}_{j,k}$ is represented by two circular arcs. The geodesic $\mathcal{C}_{0,0}$ is shown in red. $\mathcal{C}_{0,0}$ intersects six other geodesics that are shown in blue. The intersection pairing can be seen to agree with \eqref{homologypairings}. The inner labels label the white triangles $T_j$ and black triangles $\bar{T}_j$. The while and black triangles are exchanged relative to Figure \ref{figbwtiling} due to $t^{-1}$ in the argument of Schwarz triangle function.}
\label{fighypergeo}
\end{figure}
In order to calculate the intersection pairing between $C_{j,k}$, we can have a look at their projections to $x$- and $y$-planes. The image of $\mathcal{C}_{j,k}$ in $x$-plane is a figure eight curve encircling a neighboring pair of ramification points $x = e^{\frac{2\pi i j}{K}}$ and $x = e^{\frac{2\pi i (j+1)}{K}}$, one of them clockwise and the other one counter-clockwise, see Figure \ref{fighomologycyclesx}. The projection to $y$-plane is analogous, we find a figure eight curve encircling the points $y = e^{\frac{2\pi i k}{N}}$ and $y = e^{\frac{2\pi i (k+1)}{N}}$, see Figure \ref{fighomologycyclesy}. By rotation symmetry, the intersection pairing $\mathcal{C}_{j,k} \cap \mathcal{C}_{j^\prime,k^\prime}$ depends only on the differences $j^\prime-j$ and $k^\prime-k$ and furthermore can be non-zero only if $|j^\prime-j| \leq 1$ and $|k^\prime-k| \leq 1$ because if this condition is not satisfied, the $x$ or $y$ projections of $\mathcal{C}_{j,k}$ and $\mathcal{C}_{j^\prime k^\prime}$ do not intersect. By inspecting the remaining cases, we find that all the non-trivial intersections follow from the basic relations
\begin{align}
\label{homologypairings}
\nonumber
\mathcal{C}_{j,k} \cap \mathcal{C}_{j+1,k} & = +1 \\
\mathcal{C}_{j,k} \cap \mathcal{C}_{j,k+1} & = +1 \\
\nonumber
\mathcal{C}_{j,k} \cap \mathcal{C}_{j+1,k+1} & = -1
\end{align}
by antisymmetry of the intersection pairing. In particular, for each fixed curve $\mathcal{C}_{j,k}$ there are exactly $6$ neighboring curves that have non-zero intersection with it. This structure of the intersection pairing reflects the basic structure constant of the algebra or the elementary box in Figure \ref{figbasiccube}, where the index $j$ of $\mathcal{C}_j$ is associated to $\epsilon_1$ direction and the index $k$ is associated to $\epsilon_3$ direction. We did not use the three-index labeling of the fundamental cycles because the second direction is treated asymmetrically in Argyres-Douglas models.

Figure \ref{fighypergeo} shows the lifts of $\mathcal{C}_{0,0}$ as well as its neighbors to the fundamental $2KN$-gon in the hyperbolic disk covering $\mathcal{W}$. Each curve $\mathcal{C}_{j,k}$ is represented by two circular arcs, one arc connecting two nearest white triangles and the other arc connecting two nearest black triangles. The curve $\mathcal{C}_{0,0}$ meets the four curves $\mathcal{C}_{\pm 1,0}$ and $\mathcal{C}_{0,\pm 1}$ inside of the polygon while the curves $\mathcal{C}_{1,1}$ and $\mathcal{C}_{-1,-1}$ intersect with $\mathcal{C}_{0,0}$ on its boundary. Examining the orientations of the curves at the intersection point, we see again that \eqref{homologypairings} is satisfied.

\paragraph{Inverse matrix to the intersection form}
For completeness, we can find the inverse matrix to the intersection pairing. Such an object enters for example Riemann bilinear relations. The main difficulty in inverting \eqref{homologypairings} is that we have to work modulo the constraints \eqref{homologyconstraints}. Due to rotation symmetry, the inverse intersection pairing is of the form
\begin{equation}
\Pi = \sum_{j,l=0}^{K-1} \sum_{k,m=0}^{N-1} \beta_{j-l,k-m} \mathcal{C}_{j,k} \times \mathcal{C}_{l,m}.
\end{equation}
We furthermore impose the restrictions
\begin{equation}
\label{betavanishingrowcolsums}
\sum_{j=0}^{K-1} \beta_{j,k} = 0, \qquad \sum_{k=0}^{N-1} \beta_{j,k} = 0.
\end{equation}
These conditions on vanishing of row and column sums of $\beta$ fix the redundancy in the choice of cycles \eqref{homologyconstraints}. The requirement that $\Pi$ is the inverse of the intersection form $\mathsf{I}$ is of the form
\begin{equation}
\label{periodintersectiondelta}
\Pi \cdot \mathsf{I} = \Delta
\end{equation}
where $\Delta$ is the discrete analogue of delta function centered at zero but in the subspace restricted by \eqref{betavanishingrowcolsums}. Since the matrix representing $\mathsf{I}$ in the basis of $\mathcal{C}_{j,k}$ is degenerate due to \eqref{homologyconstraints}, it does not have a naive inverse. In components, the equation \eqref{periodintersectiondelta} reads
\begin{equation}
\label{betaequations}
\beta_{j-1,k} + \beta_{j,k-1} + \beta_{j+1,k+1} - \beta_{j+1,k} - \beta_{j,k+1} - \beta_{j-1,k-1} = \mathfrak{d}_{j,k}
\end{equation}
with the delta function
\begin{equation}
\mathfrak{d}_{j,k} = \begin{cases} \frac{(K-1)(N-1)}{KN} & j = 0 = k \\ -\frac{K-1}{KN} & j=0, k \neq 0 \\ -\frac{N-1}{KN} & j \neq 0, k=0 \\ \frac{1}{KN} & j \neq 0 \neq k \end{cases}
\end{equation}
This function is the projection of the matrix which has $1$ at position $(0,0)$ and zero everywhere else, to the subspace of matrices satisfying \eqref{betavanishingrowcolsums}. If we did not apply this projection, it would not be possible to satisfy the equation \eqref{periodintersectiondelta}.

Due to symmetries of $\mathcal{W}$, the equations \eqref{betaequations} can be solved by a discrete Fourier transform. We can write
\begin{equation}
\label{betafourier}
\beta_{jk} = \sum_{a=0}^{K-1} \sum_{b=0}^{N-1} \widehat{\beta_{ab}} e^{\frac{2\pi i a j}{K}} e^{\frac{2\pi i b k}{N}}.
\end{equation}
The Fourier transform of the delta function source is
\begin{equation}
\widehat{\mathfrak{d}_{ab}} = \frac{1}{KN} (1-\delta_{a,0})(1-\delta_{b,0}),
\end{equation}
i.e. it is a constant except for the zero Fourier modes which have to be modified due to the constraints \eqref{betavanishingrowcolsums}. The Fourier transform of \eqref{betaequations}
is
\begin{equation}
\left( e^{-\frac{2\pi i a}{N}} + e^{-\frac{2\pi i b}{K}} + e^{\frac{2\pi i a}{N} + \frac{2\pi i b}{K}} - e^{\frac{2\pi i a}{N}} - e^{\frac{2\pi i b}{K}} - e^{-\frac{2\pi i a}{N} - \frac{2\pi i b}{K}} \right) \widehat{\beta_{ab}} = \frac{1}{KN} (1-\delta_{a,0})(1-\delta_{b,0})
\end{equation}
with solution
\begin{equation}
\widehat{\beta_{ab}} = \frac{i(1-\delta_{a,0})(1-\delta_{b,0})}{8KN \sin\left(\frac{\pi a}{K}\right) \sin\left(\frac{\pi b}{N}\right) \sin\left(\frac{\pi a}{K}+\frac{\pi b}{N}\right)}.
\end{equation}
Plugging this back into \eqref{betafourier} gives an explicit expression for the inverse intersection pairing as a discrete Fourier transform. Even though it involves trigonometric functions, it is a fun exercise to check that $\widehat{\beta_{ab}}$ take rational values as they should.

\subsection{Period integrals along Pochhammer contour}
\label{pochhammerintegral}

\begin{figure}
\centering
\begin{tikzpicture}
\draw[thick] ({1*cos(0)}, {1*sin(0)}) coordinate (A1) arc [start angle=0, end angle=340, radius=1] coordinate (B1);
\draw[thick] ({1.5*cos(15)}, {1.5*sin(15)}) coordinate (C1) arc [start angle=15, end angle=295, radius=1.5] coordinate (D1);
\draw[thick] ({8+1*cos(181)}, {1*sin(181)}) coordinate (A2) arc [start angle=181, end angle=515, radius=1] coordinate (B2);
\draw[thick] ({8+1.5*cos(193)}, {1.5*sin(193)}) coordinate (D2) arc [start angle=193, end angle=470, radius=1.5] coordinate (C2);
\draw[thick, decoration={markings, mark=at position 0.15 with {\arrow[scale=1.5]{>}}}, postaction={decorate}] (D2) -- (B1);
\draw[thick, decoration={markings, mark=at position 0.20 with {\arrow[scale=1.5,draw=red]{>}}}, postaction={decorate}] (A1) -- (A2);
\draw[thick, decoration={markings, mark=at position 0.85 with {\arrow[scale=1.5]{>}}}, postaction={decorate}] (B2) -- (C1);
\draw[thick, decoration={markings, mark=at position 0.80 with {\arrow[scale=1.5]{>}}}, postaction={decorate}] (D1) -- (C2);
\node[fill=black, circle, inner sep=1.5pt] (pt0) at (0,0) {};
\node[anchor=south, xshift=7, yshift=-12] at (pt0) {$0$};
\node[fill=black, circle, inner sep=1.5pt] (pt1) at (8,0) {};
\node[anchor=south, xshift=7, yshift=-12] at (pt1) {$1$};
\end{tikzpicture}
\caption{Pochhammer contour in $t$-plane. The red arrow corresponds to the point where we start the description of the contour.}
\label{usualpochhammer}
\end{figure}
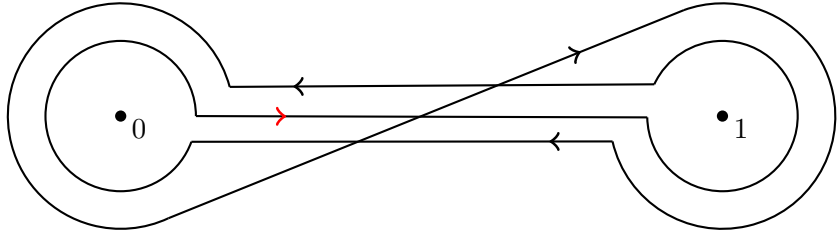

In the following, when evaluating the WKB integrals, we will need to integrate differential forms of the form
\begin{equation}
\int_{\mathcal{C}} \frac{x^j y^k dt}{KN x^{K-1} y^{N-1}} = \int_{\mathcal{C}} \frac{x^j y^k dx}{N y^{N-1}} = - \int_{\mathcal{C}} \frac{x^j y^k dy}{K x^{K-1}}
\end{equation}
along the lift of Pochhammer contour from $\mathcal{M}$ to $\mathcal{W}$. It is easiest to perform the integration on the mirror curve $\mathcal{M}$ in $t$ coordinate. Concretely, we perform the calculation as follows (see Figure \ref{usualpochhammer}):
\begin{enumerate}
\item We start at the point $t = \epsilon$ where $\epsilon$ is a small positive number. For this value of $t$, we choose the $x$ coordinate of the initial point on $\mathcal{W}$ to be $x = \sqrt[K]{\epsilon} \in \mathbbm{R}$ and the $y$ coordinate to be $y = \sqrt[N]{1-\epsilon^K} \in \mathbbm{R}$, i.e. both in the interval $(0,1)$ with $x$ near $x=0$ and $y$ near $y=1$. This specifies the initial point of a lift of the Pochhammer contour from the $t$-plane $\mathcal{M}$ to the WKB curve $\mathcal{W}$ and we will call this specific lift $\mathcal{C}$ in the following. With this choice of the lift we get expressions that are symmetric in $K$ and $N$. Other lifts would lead to integrals that differ by $KN$-th roots of unity. The integrand is
\begin{equation}
\frac{1}{KN} t^{\frac{j+1}{K}-1} (1-t)^{\frac{k+1}{N}-1} dt.
\end{equation}
\item We integrate from $t = \epsilon$ to $t = 1-\epsilon$ along the real axis in the positive direction. This is the standard Euler beta integral and we get the first contribution
\begin{equation}
\frac{1}{KN} B \left( \frac{j+1}{K}, \frac{k+1}{N} \right).
\end{equation}
\item Next we analytically continue the integrand around a small circle centered at $t = 1$ in a counter-clockwise (positive) direction. The integrand becomes
\begin{equation}
\frac{1}{KN} e^{2\pi i \frac{k+1}{N}} t^{\frac{j+1}{K}-1} (1-t)^{\frac{k+1}{N}-1} dt.
\end{equation}
\item We integrate from $t=1-\epsilon$ to $t=\epsilon$ along the real axis (in the negative direction) and get the second contribution
\begin{equation}
-\frac{1}{KN} e^{2\pi i \frac{k+1}{N}} B \left( \frac{j+1}{K}, \frac{k+1}{N} \right).
\end{equation}
\item Now we continue the integrand around $t=0$ counter-clockwise from $t=\epsilon$ to the same point. The integrand changes to
\begin{equation}
\frac{1}{KN} e^{2\pi i \frac{j+1}{K}} e^{2\pi i \frac{k+1}{N}} t^{\frac{j+1}{K}-1} (1-t)^{\frac{k+1}{N}-1} dt.
\end{equation}
\item Next we integrate along the positive real axis from $t=\epsilon$ to $t = 1-\epsilon$ in the positive direction. This gives the third contribution
\begin{equation}
\frac{1}{KN} e^{2\pi i \frac{j+1}{K}} e^{2\pi i \frac{k+1}{N}} B \left( \frac{j+1}{K}, \frac{k+1}{N} \right).
\end{equation}
\item As a next step, we analytically continue the integrand along a circle around $t = 1$ in a clockwise direction, starting and ending at $t = 1-\epsilon$. The integrand after this continuation is
\begin{equation}
\frac{1}{KN} e^{2\pi i \frac{j+1}{K}} t^{\frac{j+1}{K}-1} (1-t)^{\frac{k+1}{N}-1} dt.
\end{equation}
\item The final contribution comes from the integration from $t = 1-\epsilon$ to $t = \epsilon$ along the positive real axis in the negative direction. This integration gives
\begin{equation}
-\frac{1}{KN} e^{2\pi i \frac{j+1}{K}} B \left( \frac{j+1}{K}, \frac{k+1}{N} \right).
\end{equation}
\item In order to close the contour, we can make the last analytic continuation along a circle with center at $t = 0$ in the clockwise direction and we can verify that we are left with the same choice of branches as we started, i.e. the contour is closed.
\end{enumerate}
Adding all four contributions, the final result for the contour integral is
\begin{equation}
\label{masterintegral}
\boxed{
\int_{\mathcal{C}} \frac{x^j y^k dx}{N y^{N-1}} = \frac{1}{NK} \left( 1 - e^{2\pi i \frac{j+1}{K}} \right) \left( 1 - e^{2\pi i \frac{k+1}{N}} \right) B \left( \frac{j+1}{K}, \frac{k+1}{N} \right).}
\end{equation}
This is the main formula from this section that we will use in the following to evaluate the period integrals.

\section{ODE/IM for general $N$ and $K$}
\label{secgennk}

In this section we want to first identify the differential operators corresponding to primary states in $\mathcal{W}_N$ algebras and later generalize them to operators associated to the excited states.

\subsection{Primaries and their local charges}

Generalizing the Virasoro expression \eqref{virasoroprimaryoperhbar} as well as those discussed for $\mathcal{W}_3$ in \cite{Bazhanov:2001xm,Masoero:2019wqf,Ashok:2024zmw,Ashok:2024ygp} assuming scaling symmetry with respect to $x$ variable, we start with a differential operator \cite{Dorey:2006an,Feigin:2007mr}
\begin{equation}
\label{hwoper}
\partial_x^N + x^K - u + \sum_{k=1}^N \frac{b_k}{x^k} \partial_x^{N-k}.
\end{equation}
We want to apply the WKB expansion so we first rescale the coordinate $x$ by
\begin{equation}
\label{operscaling}
x \to \hbar^{-\frac{N}{K+N}} x
\end{equation}
and up to an overall rescaling we obtain
\begin{equation}
\hbar^N \partial_x^N + x^K - \hbar^{\frac{KN}{K+N}} u + \hbar^N \sum_{k=1}^N \frac{b_k}{x^k} \partial_x^{N-k}.
\end{equation}
Choosing
\begin{equation}
\hbar = u^{-\frac{K+N}{KN}} \qquad \text{or} \qquad u = \hbar^{-\frac{KN}{K+N}}
\end{equation}
we arrive at $\hbar$-dependent differential operator
\begin{equation}
\mathcal{O} \equiv \hbar^N \partial_x^N + x^K - 1 + \hbar^N \sum_{k=1}^N \frac{b_k}{x^k} \partial_x^{N-k}.
\end{equation}
The WKB ansatz for the wave function is just like before
\begin{equation}
\label{psiwkb}
\psi(x) = \exp \left( \sum_{n=0}^\infty \hbar^{n-1} \int^x Y_n(x^\prime) dx^\prime \right).
\end{equation}
The sum over $n$ is a formal power series, we do not assume its convergence. At the leading order in $\hbar$, the eigenfunction equation
\begin{equation}
\label{schroedingerkernel}
\mathcal{O} \psi = 0
\end{equation}
is just the classical WKB curve
\begin{equation}
\label{wkbcurve}
Y_0^N + x^K = 1
\end{equation}
which (after the identification $y \leftrightarrow Y_0$) is the Catalan curve \eqref{spectralalgcurve} that we studied previously. The $1$-form appearing in the wave function is simply the tautological (Liouville) $1$-form
\begin{equation}
Y_0(x) \, dx = y dx.
\end{equation}
At the next order $\mathcal{O}(\hbar)$, we have
\begin{equation}
\frac{N(N-1)}{2} Y_0^\prime + N Y_0 Y_1 + \frac{b_1}{x} Y_0 = 0
\end{equation}
which can be easily solved for $Y_1$ and we find
\begin{equation}
Y_1 = - \frac{N-1}{2} \frac{Y_0^\prime}{Y_0} - \frac{b_1}{N x}.
\end{equation}
In order to write this as a differential on the WKB curve, we use the derivative of \eqref{spectralalgcurve} to write
\begin{equation}
\frac{dy}{dx} = -\frac{K}{N} \frac{x^{K-1}}{y^{N-1}}.
\end{equation}
The differential $Y_1(x) dx$ becomes
\begin{equation}
Y_1 \, dx = \left( \frac{(N-1)K}{2N} \frac{x^K}{y^N} - \frac{b_1}{N} \right) \frac{dx}{x}.
\end{equation}
We can proceed analogously at order $\mathcal{O}(\hbar^2)$. We have
\begin{align}
Y_2 & = -\frac{b_2}{N Y_0 x^2} - \frac{(N-1) b_1 Y_1}{N Y_0 x} - \frac{(N-1) Y_1^2}{2Y_0} - \frac{(N-1)(N-2) b_1 Y_0^\prime}{2N Y_0^2 x} \\
\nonumber
& - \frac{(N-1)(N-2) Y_0^\prime Y_1}{2 Y_0^2} - \frac{(N-1)(N-2)(N-3) Y_0^{\prime 2}}{8 Y_0^3} - \frac{(N-1) Y_1^\prime}{2Y_0} - \frac{(N-1)(N-2) Y_0^{\prime\prime}}{6Y_0^2}
\end{align}
and using the expression that we already know for $Y_1$ as well as the derivatives \eqref{spectralalgcurve} we find
\begin{align}
\label{y2hw1form}
\nonumber
Y_2 \, dx & = -\Bigg( \frac{K^2(N^2-1)(2N+1)}{24N^2} \frac{x^{2K}}{y^{2N}} + \frac{K(K-1)(N^2-1)}{12N} \frac{x^K}{y^N} \\
& + \frac{2Nb_2+(N-1)(N-b_1)b_1}{2N^2} \Bigg) \frac{dx}{x^2 y}.
\end{align}
At order $\mathcal{O}(\hbar^3)$ we have analogously
\begin{align}
\nonumber
Y_3 \, dx & = \Bigg( \frac{K^3(N^2-1)(N+1)(2N+1)}{24N^3} \frac{x^{3N}}{y^{3K}} + \frac{K^2(K-1)(N^2-1)(N+1)}{8N^2} \frac{x^{2N}}{y^{2K}} \\
& + \frac{K(K-1)(K-2)(N^2-1)}{24N} \frac{x^N}{y^K} + \frac{K\left(2Nb_2+b_1(N-1)(N-b_1)\right)}{2N^3} \frac{x^N}{y^K} \\
\nonumber
& - \frac{b_3}{N} + \frac{N-2}{N^2} b_1 b_2- \frac{N-1}{N} b_2 - \frac{(N-1)(N-2)}{3N^3} b_1^3 + \frac{(N-1)^2}{2N^2} b_1^2 - \frac{N^2-1}{6N} b_1 \Bigg).
\end{align}
These $1$-forms on WKB curve can be integrated along the contour $\mathcal{C}$ using \eqref{masterintegral}.

\paragraph{Normalization of the period integrals}
In order to match with the the triality invariant quantities calculated from the $\mathcal{W}_\infty$ point of view, we normalize the periods as
\begin{equation}
\mathcal{N}_n^{-1} \mathcal{P}_n = \mathcal{N}_n^{-1} \int_{\mathcal{C}} Y_n(x) dx
\end{equation}
with
\begin{align}
\label{Ynormalization}
\mathcal{N}_n & = (-1)^n (K+N)^{n/2-1} \left(1-e^{-2\pi i\frac{n-1}{K}}\right) \left(1-e^{-2\pi i\frac{n-1}{N}}\right) B \left(-\frac{n-1}{K},-\frac{n-1}{N}\right) \\
\nonumber
& = \underbrace{(K+N)^{(n-1)/2} e^{-i\pi\left(\frac{1}{K}+\frac{1}{N}\right)(n-1)}}_{\text{can be absorbed into $\hbar^{n-1}$}} \times \underbrace{\frac{KN}{\sqrt{K+N}} \frac{4\pi^2 (-1)^{n-1}}{(n-1)^2 \Gamma\left(\frac{n-1}{K}\right)\Gamma\left(\frac{n-1}{N}\right)\Gamma\left(-\frac{n-1}{K}-\frac{n-1}{N}\right)}}_{\text{triality invariant}}.
\end{align}
This choice of normalization has the advantage of removing all the transcendental factors from the periods. More concretely, the $1$-forms take the form
\begin{equation}
Y_n(x) dx = \sum_{\ell} Y_{n,\ell} \frac{x^{K\ell}}{y^{N\ell}} \frac{dx}{x^n y^{n-1}}
\end{equation}
where the coefficients $Y_{n,\ell}$ are certain rational functions of $N, K$ and $b_j$. The powers $\ell$ that appear here are integers (in general both positive and negative!). The normalized period integrals are
\begin{equation}
\mathcal{N}_n^{-1} \int_{\mathcal{C}} Y_n(x) dx = \sum_{\ell} \frac{N}{K(K+N)^{n/2}} (-1)^{n+\ell+1} \frac{\left(1-\frac{n-1}{K}\right)_{\ell-1}}{\left(1+\frac{n-1}{N}\right)_{\ell-1}} Y_{n,\ell}.
\end{equation}
For the first few periods normalized in this way we find
\begin{align}
\nonumber
\mathcal{N}_0^{-1} \mathcal{P}_0 & = 1 \\
\nonumber
\mathcal{N}_1^{-1} \mathcal{P}_1 & = \frac{b_1}{N\sqrt{K+N}} - \frac{N-1}{2\sqrt{K+N}} \\
\nonumber
\mathcal{N}_2^{-1} \mathcal{P}_2 & = -\frac{(K-1)(N-1)}{24(K+N)} - \frac{b_2}{(K+N)N} + \frac{(N-1) b_1^2}{2(K+N)N^2} - \frac{(N-1) b_1}{2(K+N)N} \\
\nonumber
\mathcal{N}_3^{-1}\mathcal{P}_3 & = \frac{(N-1)(N-2)b_1^3}{3N^3(K+N)^{3/2}} - \frac{(N-1)(N-2) b_1^2}{2N^2 (K+N)^{3/2}} - \frac{b_1 b_2 (N-2)}{N^2 (K+N)^{3/2}} \\
\nonumber
& + \frac{(N-2) b_2}{N (K+N)^{3/2}} + \frac{(N-1)(N-2) b_1}{6N (K+N)^{3/2}} + \frac{b_3}{N (K+N)^{3/2}} \\
\nonumber
\mathcal{N}_4^{-1} \mathcal{P}_4 & = \frac{(N-1)(N-3)(2N-3) b_1^4}{8N^4(K+N)^2} - \frac{(N-1)(N-3)(2N-3)b_1^3}{4N^3(K+N)^2} \\
& - \frac{(N-3)(2N-3) b_1^2 b_2}{2N^3(K+N)^2} + \frac{(N-1)(N-3)(4N-K-7) b_1^2}{16N^2(K+N)^2} \\
\nonumber
& + \frac{(N-3)(3N-5) b_1 b_2}{2N^2(K+N)^2} + \frac{(N-3) b_1 b_3}{N^2(K+N)^2} \\
\nonumber
& + \frac{(N-3) b_2^2}{2N^2(K+N)^2} + \frac{(N-1)(N-3)(K+1)b_1}{16N(K+N)^2} \\
\nonumber
& - \frac{b_4}{N(K+N)^2} - \frac{3(N-3) b_3}{2N(K+N)^2} - \frac{(N-3)(4N-K-9) b_2}{8N(K+N)^2} \\
\nonumber
& - \frac{(K-1)(K-3)(N-1)(N-3)(2K+2N-1)}{1920(K+N)^2}
\end{align}
The case of $\mathcal{P}_1$ is a slightly subtle because the normalization factor is singular at $n=1$. We can regularize it by taking the limit $n \to 1$ of the normalized integral. The associated charge anyway does not belong to $\mathcal{W}_\infty$ subalgebra of $\mathcal{W}_{1+\infty}$ so in the following we will usually focus on periods of $Y_n$ with $n \geq 2$.

In order to compare these periods with the eigenvalues of integrals $\mathcal{I}_n$ acting on the primary field, it is useful to express the coefficients $b_j$ in terms of the Frobenius indices $\sigma_j$ of local solutions around $x=0$. Plugging the ansatz
\begin{equation}
\psi(x) \sim x^\sigma (1 + \ldots)
\end{equation}
into \eqref{schroedingerkernel} and looking at the leading behavior around $x = 0$ we see that
\begin{equation}
\sigma(\sigma-1)\cdots(\sigma-N+1) + \sum_{k=1}^N b_k \sigma(\sigma-1)\cdots(\sigma-N+k+1) = 0
\end{equation}
which is the equation whose roots are precisely the Frobenius indices $\sigma_j, j=1,\ldots,N$ characterizing the regular singular point at $x=0$. For the first few $b_j$ we find in this way
\begin{align}
\nonumber
b_1 & = -e_1(\sigma_j) + \frac{N(N-1)}{2} \\
\nonumber
b_2 & = e_2(\sigma_j) - \frac{(N-1)(N-2)}{2} e_1(\sigma_j) + \frac{N(N-1)(N-2)(3N-5)}{24} \\
\nonumber
b_3 & = -e_3(\sigma_j) +\frac{(N-2)(N-3)}{2} e_2(\sigma_j) - \frac{(N-1)(N-2)(N-3)(3N-8)}{24} e_1(\sigma_j) \\
& + \frac{N(N-1)(N-2)^2(N-3)^2}{48} \\
\nonumber
b_4 & = e_4(\sigma_j) - \frac{(N-3)(N-4)}{2} e_3(\sigma_j) + \frac{(N-2)(N-3)(N-4)(3N-11)}{24} e_2(\sigma_j) \\
\nonumber
& - \frac{(N-1)(N-2)(N-3)^2(N-4)^2}{48} e_1(\sigma_j) \\
\nonumber
& + \frac{N(N-1)(N-2)(N-3)(N-4) \left(15N^3-150N^2+485N-502\right)}{5760}
\end{align}
where $e_j$ are the elementary symmetric polynomials. In terms of the Frobenius indices, the first few normalized periods are
\begin{align}
\nonumber
\mathcal{N}_1^{-1} \mathcal{P}_1 & = -\frac{e_1}{N \sqrt{K+N}} \\
\nonumber
\mathcal{N}_2^{-1} \mathcal{P}_2 & = -\frac{e_2}{N(K+N)} + \frac{(N-1)e_1^2}{2N^2(K+N)} - \frac{N-1}{24} \\
\nonumber
\mathcal{N}_3^{-1}\mathcal{P}_3 & = -\frac{e_3}{N(K+N)^{3/2}} + \frac{(N-2) e_1 e_2}{N^2(K+N)^{3/2}} - \frac{(N-1)(N-2) e_1^3}{3N^3(K+N)^{3/2}} \\
\mathcal{N}_4^{-1}\mathcal{P}_4 & = \frac{(N-1)(N-3)(2N-3) e_1^4}{8N^4(K+N)^2} - \frac{(N-3)(2N-3) e_1^2 e_2}{2N^3(K+N)^2} \\
\nonumber
& -\frac{(N-1)(N-3) e_1^2}{16N^2(K+N)} + \frac{(N-3) e_1 e_3}{N^2(K+N)^2} + \frac{(N-3) e_2^2}{2N^2(K+N)^2} \\
\nonumber
& +\frac{(N-3) e_2}{8N(K+N)} - \frac{e_4}{N(K+N)^2} + \frac{(N-1)(N-3)\left(-2K^2+9K+9N\right)}{1920(K+N)}.
\end{align}
This exactly agrees with the primary charges calculated in \eqref{primarytricharges} assuming that we identify
\begin{equation}
e_k(\sigma_j) \leftrightarrow N^{k/2} (K+N)^{k/2} u_k^{(cyl)}.
\end{equation}
In other words, the Frobenius indices agree with free boson zero modes on the cylinder up to an overall rescaling,
\begin{equation}
\sigma_j \leftrightarrow \sqrt{N(K+N)} a_j^{(cyl)} = N\mu_j + Kj - \frac{K(N+1)}{2}, \qquad j=1,\ldots,N.
\end{equation}
With this identification we have
\begin{equation}
\label{periodstointegrals}
\mathcal{P}_n = \mathcal{N}_n \mathcal{I}_n,
\end{equation}
i.e. we have a precise identification between the quantum periods and local integrals of motion in $\mathcal{W}_\infty$. Inspecting \eqref{Ynormalization}, we see that the normalization factor $\mathcal{N}_n$ is triality invariant up to an overall $(n-1)$st power of a constant depending only on $K$ and $N$. Since the exponent in the WKB wave function $\psi(x)$ involves $\hbar^{n-1}$, all the triality non-invariant part of the periods can be absorbed into $\hbar$, in particular the combination
\begin{equation}
\sqrt{K+N} e^{-i\pi\left(\frac{1}{K}+\frac{1}{N}\right)} \, \hbar
\end{equation}
should be triality invariant.

\paragraph{Translation invariant vacuum and triality}
We can test the triality invariance of \eqref{periodstointegrals}. In general the triality invariance is broken by the non-trivial asymptotic $\mu \neq \emptyset$, but for the translation invariant vacuum $\mu = \emptyset$ we expect the full triality symmetry. Specializing to $\mu_j = 0$, the Frobenius indices are equal to
\begin{equation}
\rho_j = - \frac{K(N+1-2j)}{2}
\end{equation}
and with this choice, we have
\begin{align}
\nonumber
e_1(\rho) & = 0 \\
\nonumber
e_2(\rho) & = -\frac{K^2(N-1)N(N+1)}{24} \\
e_3(\rho) & = 0 \\
\nonumber
e_4(\rho) & = \frac{K^4(N-3)(N-2)(N-1)N(N+1)(5N+7)}{5760}.
\end{align}
The eigenvalues of $\mathcal{I}_n$ in translationally invariant vacuum $\ket{0}$ are then
\begin{align}
\nonumber
\mathcal{I}_1 & = 0 \\
\nonumber
\mathcal{I}_2 & = \frac{(K-1)(N-1)(KN+K+N)}{24(K+N)} \\
\mathcal{I}_3 & = 0 \\
\nonumber
\mathcal{I}_4 & = \frac{(K-3)(K-1)(N-3)(N-1)(KN+K+N)(KN+3K+3N)}{1920(K+N)^3}
\end{align}
etc. and it is easy to see that these are invariant under the triality symmetry of the algebra.

\paragraph{Lowest energy state}
Since the VOAs associated to Argyres-Douglas theories are non-unitary, the translationally invariant vacuum $\mu = \emptyset$ does not agree with the primary of the lowest conformal dimension. Such a primary field is instead characterized by having all $b_k$ vanishing, i.e. the Frobenius indices are
\begin{equation}
\left\{ \sigma_j \right\} = \left\{ 0, 1, 2, \ldots, N-1 \right\}.
\end{equation}
In other words, the point $x=0$ is in this case an ordinary regular point of the differential operator $\mathcal{O}$. We can choose
\begin{equation}
a_j^{(cyl)} = \frac{\sigma_j}{\sqrt{N(K+N)}}
\end{equation}
and the associated Young diagram has shape determined by
\begin{equation}
\mu_j = \frac{1}{N} \left( \sigma_j - Kj + \frac{K(N+1)}{2} \right)
\end{equation}
We want these to be the row lengths of the Young diagram labeling the lowest energy primary determined in \eqref{groundmu}. This is indeed the case up to two details: first of all, we possibly need to shift all $\mu_j$ simultaneously by the same constant factor (this corresponds to spectral shift in the Heisenberg subalgebra and does not affect the $\mathcal{W}_\infty$ charges). Second, not all of $N!$ permutations of $\sigma_j$ produce a valid Young diagram with $\mu_j \geq \mu_{j+1}$. Running however over all permutations, we find all the Young diagrams that are in the identification orbit of the lowest energy primary.

\subsection{Including the descendants}
\label{descwkbtheory}
In order to include the descendants, we allow for additional regular singular points in the $x$-plane just as in the case of Virasoro algebra. A general such operator generalizing \eqref{hwoper} is of the form
\begin{equation}
\label{generaloper}
\partial_x^N + x^K - u + \sum_{k=1}^N \frac{b_k}{x^k} \partial_x^{N-k} + \sum_{k=1}^N \sum_{l=0}^{k-1} \sum_j \frac{c^j_{kl}}{x^l (x-x_j)^{k-l}} \partial_x^{N-k}
\end{equation}
For example, for $N=2$ we have
\begin{equation}
\label{opern2}
\partial_x^2 + x^K - u + \left( \frac{b_1}{x} + \sum_j \frac{c_{10}^j}{x-x_j} \right) \partial_x + \left( \frac{b_2}{x^2} + \sum_j \frac{c_{20}^j}{(x-x_j)^2} + \frac{c_{21}^j}{x(x-x_j)} \right)
\end{equation}
which generalizes \eqref{viroperdesc} and for $N=3$ we have
\begin{multline}
\label{opern3}
\partial_x^3 + x^K - u + \left( \frac{b_1}{x} + \sum_j \frac{c_{10}^j}{x-x_j} \right) \partial_x^2 + \left( \frac{b_2}{x^2} + \sum_j \frac{c_{20}^j}{(x-x_j)^2} + \frac{c_{21}^j}{x(x-x_j)} \right) \partial_x + \\
+ \left( \frac{b_3}{x^3} + \sum_j \frac{c_{30}^j}{(x-x_j)^3} + \frac{c_{31}^j}{x(x-x_j)^2} + \frac{c_{32}^j}{x^2(x-x_j)} \right).
\end{multline}
The values of $x_j$ and $c^j_{kl}$ should be determined by requiring trivial monodromy around all $x_j$, except for $c_{10}^j$ which can be freely chosen by a gauge choice, i.e. conjugation of the operator (in the following we will usually choose $b_1$ and $c_{10}^j$ to be zero) and $c_{k0}^j,\, k > 0$ which we fix following \cite{Feigin:2007mr,Masoero:2018rel}: the coefficients $c_{k0}^j$ determine the collection of Frobenius indices which characterize the leading order local behavior of solutions around the singular points $x=x_j$. The prescription of \cite{Feigin:2007mr} (in particular the formula (5.21) and the surrounding discussion) is to choose $c_{k0}^j$ such that the Frobenius indices are
\begin{equation}
\sigma \in \left\{ -1, 1, 2, \ldots, N-3, N-2, N \right\}
\end{equation}
i.e. all integers in the range from $-1$ to $N$ without $0$ and $N-1$. For example, for $N=5$ the relevant part of the differential operator around $x=x_j$ is
\begin{equation}
\partial_x^5 - \frac{5}{(x-x_j)^2} \partial_x^3 + \frac{15}{(x-x_j)^3} \partial_x^2 - \frac{30}{(x-x_j)^4} \partial_x + \frac{30}{(x-x_j)^5}
\end{equation}
and the corresponding indical equation
\begin{equation}
\sigma(\sigma-1)(\sigma-2)(\sigma-3)(\sigma-4) - 5\sigma(\sigma-1)(\sigma-2) + 15\sigma(\sigma-1) - 30\sigma + 30 = 0
\end{equation}
has roots $\sigma \in \left\{ -1, 1, 2, 3, 5 \right\}$.

Since we do not know how to write the resulting Bethe equations explicitly for every $N$, we will write them explicitly for $\mathcal{W}_3$ and $\mathcal{W}_4$ in the following section. In this section instead, we will calculate the quantum periods associated to these differential operators and therefore find the expressions for $\mathcal{I}_n$ in terms of the parameters of the differential operator. In order to proceed with WKB analysis, we apply the scaling \eqref{operscaling} to \eqref{generaloper} and find
\begin{equation}
\label{generaloperrescaled}
\hbar^N \partial_x^N + x^K - 1 + \hbar^N \sum_{k=1}^N \frac{b_k}{x^k} \partial_x^{N-k} + \hbar^N \sum_{k=1}^N \sum_{l=0}^{k-1} \sum_j \frac{c^j_{kl}}{x^l \left(x-\hbar^{\frac{N}{K+N}}x_j\right)^{k-l}} \partial_x^{N-k}.
\end{equation}
Note that after this rescaling, the positions of the singularities move as we vary $\hbar$ and in the semiclassical limit $\hbar \to 0$ around which we are doing the WKB expansion these singularities all merge at $x=0$.

Expanding the descendant contribution around $\hbar=0$, we have
\begin{equation}
\hbar^N \sum_{k=1}^N \sum_{l=0}^{k-1} \sum_{m=0}^\infty \sum_{j=1}^{(K+N)M} \frac{c^j_{kl}}{x^{k+m}} {k-l+m-1 \choose m} \hbar^{\frac{Nm}{K+N}} x_j^m \partial_x^{N-k}.
\end{equation}
Since all Bethe roots corresponding to physical solutions appear in $\mathbbm{Z}_{K+N}$ orbits corresponding to rotations generated by $x \to e^{\frac{2\pi i}{K+N}} x$, only the terms where $m$ is a multiple of $K+N$ give a non-vanishing contribution (by symmetry all $c^j_{kl}$ corresponding to the same $\mathbbm{Z}_{K+N}$ orbit are equal). We can therefore write the descendant contribution to \eqref{generaloperrescaled} as
\begin{equation}
\sum_{m=0}^\infty \hbar^{N(m+1)} \sum_{k=1}^N \left[ \sum_{l=0}^{k-1} \sum_j c^j_{kl} {k-l+(K+N)m-1 \choose (K+N)m} x_j^{(K+N)m} \right] x^{-k-(K+N)m} \partial_x^{N-k}
\end{equation}
and the differential operator \eqref{generaloperrescaled} can be written as
\begin{equation}
\label{operwithdescexp}
\hbar^N \partial_x^N + x^K - 1 + \sum_{m=0}^\infty \sum_{k=1}^N \hbar^{Nm+k} \frac{\rho_{km}}{x^{k+(K+N)m}} \hbar^{N-k} \partial_x^{N-k}
\end{equation}
with
\begin{equation}
\rho_{km} = \delta_{m,0} b_k + \sum_{l=0}^{k-1} \sum_{j=1}^{(K+N)M} c^j_{kl} {k-l+(K+N)m-1 \choose (K+N)m} x_j^{(K+N)m}.
\end{equation}
The next step is to use the WKB ansatz for the wave function \eqref{psiwkb} and require that it is a formal solution of our differential equation. It is useful to have a convenient formula for terms such as
\begin{equation}
\psi^{-1} \hbar^n \partial_x^n \psi
\end{equation}
with
\begin{equation}
\psi = \exp \left[ \hbar^{-1} \int^x Y(x^\prime) dx^\prime \right].
\end{equation}
Applying Fa\`a di Bruno formula to a derivative of an exponential function, we find
\begin{equation}
\sum_{\sum_{i m_i = n}} \frac{\hbar^n n!}{m_1! 1^{m_1} \cdots m_n! n!^{m_n}} \prod_{j \geq 1} \left[ \hbar^{-1} \partial_x^{j-1} Y(x) \right]^{m_j}.
\end{equation}
We are summing over all partitions $\mu$ of $n$ with $m_i$ parts of length $i$. Next, we would like to rewrite the formula in such a way that it makes sense for $n$ that is not necessarily an integer. In order to do that, we factor out the contribution of undifferentiated $Y(x)$, i.e. $j=1$ terms (so that in the remaining factors the order of derivative $n$ appears rationally) and we introduce a Young diagram $\nu$ obtained from $\mu$ by removing the first column. The result can be written as a sum over \emph{all} Young diagrams $\nu$ with $n$-dependence being a rational function of $n$ times $Y^n$,
\begin{equation}
\sum_{\nu} \hbar^{|\nu|} \left[n\right]_{|\nu|+\ell(\nu)} Y^{n-|\nu|-\ell(\nu)} \frac{\prod_j \left(\partial_x^j Y\right)^{n_j}}{n_1! (1+1)!^{n_1} \cdots n_j! (j+1)!^{n_j} \cdots}.
\end{equation}
Here $|\nu|$ is the total number of boxes in Young diagram $\nu$ while $\ell(\nu)$ is the number of rows. Finally, $n_j$ is the number of parts of $\nu$ of length $j$. If we specialize $n$ to be an integer, only those Young diagrams $\nu$ that have at most $n$ boxes will contribute.

Using this form of Fa\`a di Bruno formula, we can now evaluate the quantum WKB periods associated to \eqref{operwithdescexp}. The zeroth order terms in $\hbar$ still determine the same WKB curve \eqref{wkbcurve} as before (which is independent of the primary fields or the descendants). The order $\mathcal{O}(\hbar^1)$ terms in WKB expansion give
\begin{equation}
Y_1 \, dx = -\frac{\rho_{10}}{N} \frac{dx}{x} - \frac{(N-1)}{2} \frac{y^\prime dx}{y} = \left( -\frac{\rho_{10}}{N} + \frac{K(N-1)}{2N} \frac{x^K}{y^N} \right) \frac{dx}{x}
\end{equation}
which is exactly what we found before with the replacement $b_1 \to \rho_{10}$. For the next $1$-form we find analogously
\begin{align}
\nonumber
Y_2 \, dx & = \frac{(N-1)\rho_{10}^2-N(N-1)\rho_{10}-2N\rho_{20}}{2N^2} \frac{dx}{x^2 y} \\
& -\frac{K(K-1)(N^2-1)}{12N} \frac{x^K}{y^N} \frac{dx}{x^2 y} -\frac{K^2 (N^2-1)(2N+1)}{24N^2} \frac{x^{2K}}{y^{2N}} \frac{dx}{x^2 y}
\end{align}
and this is in agreement with \eqref{y2hw1form} found previously.

At higher orders we encounter potential instabilities in dependence of $Y_n$ on $N$, i.e. terms in $Y_n$ that appear only for certain values of $N$ and are absent generically. The origin of these terms comes from the terms in \eqref{operwithdescexp} that have $m \neq 0$. The order in $\hbar$ at which such terms appear is proportional to $\hbar^{Nm}$ and so as we increase $N$, they contribute to higher and higher $Y_n$. This first happens for $\mathcal{I}_3$ which is for $N \geq 3$ given by
\begin{align}
\nonumber
\mathcal{I}_3 & = \frac{(N-2)(N-1) \rho_{10}^3}{3N^3(K+N)^{3/2}} - \frac{(N-2)(N-1) \rho_{10}^2}{2N^2(K+N)^{3/2}} - \frac{(N-2) \rho_{10} \rho_{20}}{N^2 (K+N)^{3/2}} + \frac{(N-2)(N-1)\rho_{10}}{6N(K+N)^{3/2}} \\
&  + \frac{(N-2) \rho_{20}}{N(K+N)^{3/2}} + \frac{\rho_{30}}{N(K+N)^{3/2}}.
\end{align}
For $N=2$ the same formula holds if we put the term involving $\rho_{30}$ to zero. The situation is more interesting for $\mathcal{I}_4$. For $N \geq 4$ we have
\begin{align}
\nonumber
\mathcal{I}_4 & = \frac{(N-3)(N-1)(2N-3)\rho_{10}^4}{8N^4(K+N)^2} - \frac{(N-3)(N-1)(2N-3)\rho_{10}^3}{4N^3(K+N)^2} - \frac{(N-3)(2N-3)\rho_{20}\rho_{10}^2}{2N^3(K+N)^2} \\
& - \frac{(N-3)(N-1)\rho_{10}^2(K-4N+7)}{16N^2(K+N)^2} + \frac{(N-3)(3N-5)\rho_{20}\rho_{10}}{2N^2(K+N)^2} + \frac{(N-3)\rho_{30}\rho_{10}}{N^2(K+N)^2} \\
\nonumber
& + \frac{(N-3)\rho_{20}^2}{2N^2(K+N)^2} + \frac{(K+1)(N-3)(N-1)\rho_{10}}{16N(K+N)^2} + \frac{(N-3)\rho_{20}(K-4N+9)}{8N(K+N)^2} - \frac{\rho_{40}}{N(K+N)^2} \\
\nonumber
& -\frac{3(N-3)\rho_{30}}{2N(K+N)^2}-\frac{(K-3)(K-1)(N-3)(N-1)(2K+2N-1)}{1920(K+N)^2}.
\end{align}
Finally the formulas for $\mathcal{I}_5$ and $\mathcal{I}_6$ for $N \geq 5$ and $N \geq 6$ respectively are
\begin{align}
\nonumber
\mathcal{I}_5 & = -\frac{(N-4)(N-2)\rho_{20}^2}{N^2(K+N)^{5/2}} - \frac{(N-4)\rho_{30}\rho_{20}}{N^2(K+N)^{5/2}} - \frac{(K+2)(N-4)(N-2)\rho_{20}}{3N(K+N)^{5/2}} + \frac{2(N-4)\rho_{40}}{N(K+N)^{5/2}} \\
& + \frac{\rho_{50}}{N(K+N)^{5/2}} - \frac{(N-4)\rho_{30}(K-3N+11)}{3N(K+N)^{5/2}}
\end{align}
and
{\small
\begin{align}
\nonumber
\mathcal{I}_6 & = \frac{(N-5) D_1 \rho_{20}}{1152N(K+N)^3} - \frac{(N-5)(2N-5)\rho_{20}^3}{6N^3(K+N)^3} - \frac{(N-5)\rho_{20}^2\left(15KN-35K-40N^2+205N-257\right)}{48N^2(K+N)^3} \\
& + \frac{(N-5)(5N-13)\rho_{30}\rho_{20}}{2N^2(K+N)^3} +\frac{(N-5)\rho_{40}\rho_{20}}{N^2(K+N)^3} + \frac{(N-5)\rho_{30}^2}{2N^2(K+N)^3} + \frac{15(K+3)(N-5)(N-3)\rho_{30}}{16N(K+N)^3} \\
\nonumber
& + \frac{5(N-5)\rho_{40}(3K-8N+41)}{24N(K+N)^3} - \frac{\rho_{60}}{N(K+N)^3} - \frac{5(N-5)\rho_{50}}{2N(K+N)^3} - \frac{(K-5)(K-1)(N-5)(N-1) D_2}{580608(K+N)^3}
\end{align}}
with
\begin{align}
\nonumber
D_1 & = 10K^3N-22K^3+10K^2N^2-113K^2N+205K^2+288KN^2-1750KN+2410K+192N^3 \\
\nonumber
& -730N^2-235N+1895 \\
D_2 & = 8K^3N^2+48K^3N-152K^3+8K^2N^3+96K^2N^2-474K^2N+478K^2+48KN^3 \\
\nonumber
& -474KN^2+969KN-309K-152N^3+478N^2-309N+61.
\end{align}
Here we simplified the expressions by putting $\rho_{10} = 0$ which is a choice that is always possible to achieve by a suitable gauge transformation.

The expressions for $\mathcal{I}_n$ charges get modified for $n>N$. First of all, the coefficients $\rho_{km}$ vanish if $k>N$ and furthermore as follows from \eqref{operwithdescexp} there appear contributions from $\rho_{km}$ with $m>0$. For example, $\mathcal{I}_4$ gets an additional contribution
\begin{equation}
-\frac{K \rho_{21}}{4(K+2)^2(K+3)} \delta_{N,2}
\end{equation}
and for $\mathcal{I}_5$ a contribution
\begin{equation}
\frac{K \rho_{21}}{(K+3)^{5/2} (9K+36)} \delta_{N,3}
\end{equation}
(both assuming for simplicity $\rho_{10}=0$). Finally, the corrections to $\mathcal{I}_6$ are
\begin{multline}
- \frac{K\left((2K+5)\rho_{21}\left(12\rho_{20}+2K^2+13K+15\right)+12K\rho_{22}\right)}{32(K+2)^3(K+5)(2K+5)} \delta_{N,2} \\
- \frac{K((K+5)\rho_{21}+2\rho_{31})}{9(K+3)^3(K+5)} \delta_{N,3} -\frac{K\rho_{21}}{16(K+4)^3(K+5)} \delta_{N,4}
\end{multline}

\subsection{Duality symmetries}
The algebra $\mathcal{W}_\infty$ has various symmetries such as the triality symmetry or charge conjugation symmetry. Let us see how these are realized on the level of the associated differential operators.

\subsubsection{Feigin-Frenkel duality}
For every fixed $\lambda_3 = N$, we can understand the duality exchanging $\lambda_1 = K$ with $\lambda_2 = -\frac{NK}{N+K}$ acting on the differential operator \eqref{hwoper} that represents the primary field as a transformation
\begin{equation}
x \to x^\prime \equiv \left(\frac{N e^{-\frac{2\pi i}{N}} u^{\frac{1}{N}}}{K+N}\right)^{\frac{K+N}{N}} x^{\frac{K+N}{N}}.
\end{equation}
Under this transformation, the operator \eqref{hwoper} is mapped to an operator of the same form if we map the energy parameter as follows:
\begin{equation}
u \to u^\prime \equiv e^{\frac{\pi i K}{N}} \left(\frac{K+N}{N}\right)^K u^{-\frac{K+N}{N}}.
\end{equation}
In terms of the parameter $\hbar$ which is more natural from the point of view of WKB expansion and which is related to $u$ via
\begin{equation}
u = \hbar^{-\frac{KN}{K+N}}
\end{equation}
we have
\begin{equation}
\hbar \to \hbar^\prime \equiv e^{\frac{\pi i}{N}} \frac{K+N}{N} \hbar,
\end{equation}
i.e. the WKB expansion agrees order by order in both frames. Therefore from the point of view of the local charges, the WKB calculations in both frames give access order by order to the same local charges. The parameters $b_j$ in the two frames are related by a triangular transformation,
\begin{align}
b_1^\prime & = \frac{N}{K+N} b_1 \\
b_2^\prime & = \frac{N^2}{(K+N)^2}b_2 - \frac{K(N-1)N}{2(K+N)^2}b_1 + \frac{KN(N-1)(N+1)(K+2N)}{24(K+N)^2}
\end{align}
etc. The transformation of descendants on the level of differential operators is not so clear, i.e. the operation of dressing the highest weight operator by apparent singularities does not commute with the Feigin-Frenkel transformation of the coordinates.

\subsubsection{Fourier transform}
The other duality exchanging $K$ and $N$ is slightly more involved. Let us again focus our attention to the case of primaries, i.e. no descendants. For the ground state differential operator
\begin{equation}
\partial_x^N + x^K - u
\end{equation}
the duality exchanging $N$ and $K$ corresponds simply to a Fourier transform
\begin{equation}
\partial_x \leftrightarrow y, \qquad x \leftrightarrow -\partial_y
\end{equation}
that brings the operator to the form
\begin{equation}
(-1)^K \partial_y^K + y^N - u
\end{equation}
followed by a trivial phase change of $y$ and $u$ that removes the $(-1)^K$ sign. If we formally apply the same procedure to the operator corresponding to other primary fields \eqref{hwoper}, we find
\begin{equation}
y^N + (-1)^K \partial_y^K - u + \sum_{k=1}^N b_k (-1)^k \partial_y^{-k} y^{N-k},
\end{equation}
i.e. we end up with a pseudo-differential operator involving negative powers of $\partial_y$. Fortunately, there is a procedure which formally brings the operator to the form of a differential operator. Choosing for simplicity $b_1 = 0$ (which can be always achieved by an automorphism of the algebra or in other words by conjugating the operator by a function), we need to conjugate our differential operator by a transformation of the form
\begin{equation}
\chi = 1 + \chi_1(y) \partial_y^{-1} + \chi_2(y) \partial_y^{-2} + \ldots.
\end{equation}
Let us illustrate it on the example of $(\mathcal{W}_3,\mathcal{W}_4)$ minimal model. In this case we start with
\begin{equation}
\mathcal{O} = \partial_x^3 + \frac{b_2}{x^2} \partial_x + \frac{b_3}{x^3} + x^4 - u.
\end{equation}
The formal Fourier transform gives
\begin{equation}
\tilde{\mathcal{O}} = \partial_y^4 + y^3 + b_2 \partial_y^{-2} y - b_3 \partial_y^{-3} - u.
\end{equation}
Using the algebra of pseudo-differential operators we can move the derivatives to the right (following the usual Leibniz rule), e.g.
\begin{equation}
\partial_y^{-2} y = y \partial_y^{-2} - 2 \partial_y^{-3}
\end{equation}
and arrive at
\begin{equation}
\tilde{\mathcal{O}} = \partial_y^4 + y^3 -u + b_2 y \partial_y^{-2} - 2b_2 \partial_y^{-3} - b_3 \partial_y^{-3}.
\end{equation}
Now we want to conjugate this by $\chi$, i.e. consider an operator
\begin{equation}
\chi^{-1} \tilde{\mathcal{O}} \chi.
\end{equation}
First of all, the pseudo-differential operator $\chi^{-1}$ is uniquely determined by requiring $\chi^{-1} \chi = 1$. The coefficients of powers of $\partial_y^{-k}$ are differential polynomials in $\chi_j(y)$, i.e.
\begin{equation}
\chi^{-1} = 1 - \chi_1 \partial_y^{-1} - \left( \chi_2 - \chi_1^2 \right) \partial_y^{-2} - \left(\chi_3-2\chi_1 \chi_2+\chi_1^3+\chi_1 \chi_1^\prime\right) \partial_y^{-3} + \ldots.
\end{equation}
We want to determine the coefficients $\chi_j(y)$ such that the resulting operator is the usual differential operator of form \eqref{hwoper}, i.e. in our case
\begin{equation}
\chi^{-1} \tilde{\mathcal{O}} \chi \stackrel{!}{=} \partial_y^4 + \frac{\tilde{b}_2}{y^2} \partial_y^2 + \frac{\tilde{b}_3}{y^3} \partial_y + \frac{\tilde{b}_4}{y^4} + y^3 - \tilde{u}.
\end{equation}
This in turn determines the functions $\chi_j(y)$ in terms of the parameters:
\begin{align}
\nonumber
\chi_1 & = -\frac{\tilde{b}_2}{4y} \\
\nonumber
\chi_2 & = \frac{\tilde{b}_2^2}{32 y^2}-\frac{3 \tilde{b}_2}{8 y^2}-\frac{\tilde{b}_3}{8 y^2} \\
\nonumber
\chi_3 & = -\frac{\tilde{b}_2^3}{384 y^3}+\frac{\tilde{b}_2^2}{6 y^3}+\frac{\tilde{b}_2 \tilde{b}_3}{32 y^3}-\frac{5 \tilde{b}_2}{8 y^3}-\frac{3 \tilde{b}_3}{8 y^3}-\frac{\tilde{b}_4}{12 y^3} \\
\nonumber
\chi_4 & = \frac{\tilde{b}_2^4}{6144 y^4}-\frac{23 \tilde{b}_2^3}{768 y^4}-\frac{\tilde{b}_2^2 \tilde{b}_3}{256 y^4}+\frac{101 \tilde{b}_2^2}{128 y^4}+\frac{17 \tilde{b}_2 \tilde{b}_3}{64 y^4} \\
& +\frac{\tilde{b}_2 \tilde{b}_4}{48 y^4}-\frac{15 \tilde{b}_2}{16 y^4}+\frac{\tilde{b}_3^2}{128 y^4}-\frac{15 \tilde{b}_3}{16 y^4}-\frac{3 \tilde{b}_4}{8 y^4} \\
\nonumber
\chi_5 & = \frac{\#}{y^5} + \left( \frac{3 \tilde{b}_2}{32}-\frac{b_2}{8} \right) y^2 \\
\nonumber
\chi_6 & = \frac{\#}{y^6} + \left( \frac{b_2 \tilde{b}_2}{32}+\frac{7 b_2}{8}+\frac{b_3}{4}-\frac{3 \tilde{b}_2^2}{128}-\frac{3 \tilde{b}_2}{32}+\frac{3 \tilde{b}_3}{16} \right) y \\
\nonumber
\chi_7 & = \frac{\#}{y^7} + \left( \frac{3 \tilde{b}_2^2}{128}-\frac{3 \tilde{b}_2}{32}-\frac{9 \tilde{b}_3}{32}-\frac{3 \tilde{b}_4}{16} \right) \log y.
\end{align}
The symbol $\#$ denotes polynomials in $b_j$ and $\tilde{b}_j$ that we do not spell out explicitly. We will require that the functions $\chi_j(y)$ are homogeneous, i.e. that it is of the form
\begin{equation}
\chi_j(y) = \frac{c_j}{y^j}.
\end{equation}
This gives us three equations for three unknowns $\tilde{b}_2, \tilde{b}_3$ and $\tilde{b}_4$:
\begin{align}
\nonumber
\tilde{b}_2 & = \frac{4b_2}{3} \\
\tilde{b}_3 & = -\frac{4}{3}\left(b_3+3b_2\right) \\
\nonumber
\tilde{b}_4 & = \frac{2}{9} \left(b_2^2+24b_2+9b_3\right).
\end{align}
Continuing the calculation and using these expressions for $\tilde{b}_j$ in terms of $b_k$, the requirement of homogeneity of $\chi_8$ requires
\begin{equation}
(b_2+6)(b_2+b_3) = 0.
\end{equation}
Similarly, $\chi_9$ is homogeneous in $y$ of degree $-9$ only if
\begin{equation}
10b_2^3 + 12b_2^2b_3 - 99b_2^2 - 54b_2b_3 - 954b_2 + 9b_3^2 - 864b_3 = 0.
\end{equation}
These equations have only $5$ solutions that correspond to five primaries that the irreducible $(\mathcal{W}_3,\mathcal{W}_4)$ minimal model has, see Table \ref{w3w4primaries}. It is rather interesting that the formal conjugation by $\chi$ takes rather simple form: for all five primaries it is a formal series that can be compactly written as a pseudo-differential operator
\begin{align}
\nonumber
\chi = 1 + \sum_{n=1}^\infty \frac{(n-1)!}{y^n} \partial_y^{-n} = 1 + \partial_y^{-1} y^{-1} & \qquad \text{for} \quad b_2=-3, \quad b_3=3 \\
\nonumber
\chi = 1 + 2\sum_{n=1}^\infty \frac{(n-1)!}{y^n} \partial_y^{-n} = 1 + 2\partial_y^{-1} y^{-1} & \qquad \text{for} \quad b_2=-6, \quad b_3=0 \\
\chi = 1 + 2\sum_{n=1}^\infty \frac{n!}{y^n} \partial_y^{-n} = 1 + 2y\partial_y^{-1} y^{-2} & \qquad \text{for} \quad b_2=-6, \quad b_3=12 \\
\nonumber
\chi = 1 + \frac{5}{2}\sum_{n=1}^\infty \frac{(n+1)!}{y^n} \partial_y^{-n} = 1 + 5y^2 \partial_y^{-1} y^{-3} & \qquad \text{for} \quad b_2=-15, \quad b_3=15
\end{align}
(for the ground state there is no need for the conjugation as we saw at the beginning of this section).

Let us consider a slightly more complicated example, say $(\mathcal{W}_5,\mathcal{W}_4)$. Consider an operator
\begin{equation}
\mathcal{O} = \partial_x^5 + \frac{b_2}{x^2} \partial_x^3 + \frac{b_3}{x^3} \partial_x^2 + \frac{b_4}{x^4} \partial_x + \frac{b_5}{x^5} + x^4 - u.
\end{equation}
representing a primary of $\mathcal{W}_5$ algebra. Its formal Fourier transform
\begin{equation}
y^5 + b_2 \partial_y^{-2} y^3 - b_3 \partial_y^{-3} y^2 + b_4 \partial_y^{-4} y - b_5 \partial_y^{-5} + \partial_y^4 - u
\end{equation}
is equivalent to
\begin{equation}
\partial_y^4 + \frac{\tilde{b}_2}{y^2} \partial_y^2 + \frac{\tilde{b}_3}{y^3} \partial_y + \frac{\tilde{b}_4}{y^4} + y^5 - u
\end{equation}
by a formal conjugation by $\chi$ that is homogeneous in $y$ and $\partial_y$, i.e.
\begin{equation}
\chi = 1 + \frac{\#}{y} \partial_y^{-1} + \frac{\#}{y^2} \partial_y^{-2} + \ldots
\end{equation}
only for $14$ choices of $(b_2,b_3,b_4,b_5)$ as given in Table \ref{w4w5fourier}. These exactly correspond to
\begin{equation}
\frac{1}{4+5} {4+5 \choose 4} = 14
\end{equation}
primaries of this minimal model.

\begin{table}
\centering
\begin{tabular}{|c|c|c|c|c|c|c|c|c|c|}
\hline
$\Delta$ & $b_2$ & $b_3$ & $b_4$ & $b_5$ & $\tilde{b}_2$ & $\tilde{b}_3$ & $\tilde{b}_4$ & $\chi$ \\
\hline
$0$ & $-75$ & $225$ & $495$ & $-1440$ & $-60$ & $120$ & $216$ & $1+\frac{12}{5}\partial_y^{-1}y^{-1}-\frac{21}{2}y\partial_y^{-1}y^{-2}+\frac{231}{10}y^5\partial_y^{-1}y^{-6}$ \\
$-2/3$ & $-45$ & $75$ & $270$ & $-270$ & $-36$ & $120$ & $0$ & $1-12y\partial_y^{-1}y^{-2}+21y^2\partial_y^{-1}y^{-3}$ \\
$-2/3$ & $-45$ & $195$ & $-90$ & $-630$ & $-36$ & $24$ & $144$ & $1-\frac{3}{2}\partial_y^{-1}y^{-1}+\frac{21}{2}y^4\partial_y^{-1}y^{-5}$ \\
$-8/9$ & $-35$ & $105$ & $-105$ & $0$ & $-28$ & $56$ & $-56$ & $1+7y^3\partial_y^{-1}y^{-4}$ \\
$-1$ & $-30$ & $60$ & $0$ & $0$ & $-24$ & $72$ & $-72$ & $1+6y^2\partial_y^{-1}y^{-3}$ \\
$-1$ & $-30$ & $120$ & $-180$ & $0$ & $-24$ & $24$ & $0$ & $1+6y^3\partial_y^{-1}y^{-4}$ \\
$-11/9$ & $-20$ & $60$ & $0$ & $-120$ & $-16$ & $32$ & $40$ & $1-2\partial_y^{-1}y^{-1}+6y\partial_y^{-1}y^{-2}$ \\
$-4/3$ & $-15$ & $15$ & $0$ & $0$ & $-12$ & $48$ & $-72$ & $1+3\partial_y^{-1}y^{-1}$ \\
$-4/3$ & $-15$ & $45$ & $-45$ & $0$ & $-12$ & $24$ & $0$ & $1+3y\partial_y^{-1} y^{-2}$ \\
$-4/3$ & $-15$ & $75$ & $-180$ & $180$ & $-12$ & $0$ & $0$ & $1+3y^2\partial_y^{-1} y^{-3}$ \\
$-13/9$ & $-10$ & $20$ & $-20$ & $0$ & $-8$ & $24$ & $-24$ & $1+2\partial_y^{-1} y^{-1}$ \\
$-13/9$ & $-10$ & $40$ & $-80$ & $80$ & $-8$ & $8$ & $0$ & $1+2y\partial_y^{-1} y^{-2}$ \\
$-14/9$ & $-5$ & $15$ & $-30$ & $30$ & $-4$ & $8$ & $-8$ & $1+\partial_y^{-1} y^{-1}$ \\
$-5/3$ & $0$ & $0$ & $0$ & $0$ & $0$ & $0$ & $0$ & $1$ \\
\hline
\end{tabular}
\caption{List of primaries in $(\mathcal{W}_4,\mathcal{W}_5)$ minimal models as well as the corresponding coefficients of the differential operators and the pseudo-differential operator $\chi$ used to conjugate the pseudo-differential Fourier transform to a differential operator.}
\label{w4w5fourier}
\end{table}

To summarize, there is a systematic procedure implementing $K \leftrightarrow N$ duality for the primary states of $(\mathcal{W}_K,\mathcal{W}_N)$ minimal models generalizing the Fourier transformation that we have for the ground state primary. It involves a formal Fourier transformation together with a conjugation by a pseudo-differential operator $\chi$. Requiring homogeneity, this procedure works precisely for the finite number of primaries that are compatible with the symmetry algebra. This is actually what we should have expected, because the Verma modules of both algebras have different number of parameters, so in order to find a one-to-one identification with allowed primaries, we need to restrict to those that are compatible with the vacuum null states, i.e. the irreducible representations. Note that the transformation of $b_j$ is rather non-trivial, in terms of Frobenius indices that are encoded by $b_j$ the operation involves the transposition of the relevant Young diagram (which is manifest in the box counting description of primaries as discussed in Section \ref{secwinfhw}).

\subsubsection{Charge conjugation symmetry}
Another symmetry that $\mathcal{W}_\infty$ has is the charge conjugation symmetry which flips sign of the odd spin primary generators of the algebra while leaving the even spin generators intact. On the level of differential operator \eqref{hwoper} we simply write the transpose operator with $\partial \to -\partial$, i.e.
\begin{align}
\partial_x^N + x^K - u + \sum_{k=1}^N \frac{b_k}{x^k} \partial_x^{N-k} & \to (-1)^N \partial_x^N + x^K - u + \sum_{k=1}^N (-1)^{N-k} \partial_x^{N-k} \frac{b_k}{x^k}.
\end{align}
(for $N$ odd we might need to make a phase transformation of $x$ and $u$ to map it to the form \eqref{hwoper}). For the first few $b_j$ we find
\begin{align}
\nonumber
b_1 & \mapsto -b_1 \\
\nonumber
b_2 & \mapsto b_2 + (N-1)b_1 \\
b_3 & \mapsto -b_3 - 2(N-2)b_2 -(N-1)(N-2)b_1 \\
\nonumber
b_4 & \mapsto b_4 + 3(N-3)b_3 + 3(N-2)(N-3)b_2 + (N-1)(N-2)(N-3)b_1
\end{align}
from where we can actually immediately see a general formula in terms of Pochhammer symbols,
\begin{equation}
b_n \mapsto (-1)^n \sum_{j=0}^{n-1} {n-1 \choose j} (N-n+1)_j b_{n-j}.
\end{equation}
For $(\mathcal{W}_3,\mathcal{W}_4)$ example see the Table \ref{w3w4primaries}.

\subsection{Rational $K$ and minimal models}
In most of the discussion we focused on the case of both $N$ and $K$ positive coprime integers (i.e. the VOAs related to Argyres-Douglas 4d SCFTs \cite{Beem:2013sza}). Having $N$ an integer is important in order to have a positive degree differential operator on the ODE side of the correspondence. On the other hand, having integer values of $K$ is not necessary at all from ODE/IM perspective. The differential operators are meaningful even for non-integer values of $K$ in the potential, even though one has to specify the choice of the branches of the logarithms. Restricting to $K$ integer and coprime to $N$ in particular considerably simplified the discussion of the geometry of the associated WKB curve. On the other hand, we saw that the period calculations naturally happen on the mirror curve, the three-punctured sphere. This geometry most naturally encodes the symmetries of $\mathcal{W}_\infty$ such as the triality symmetry. After a suitable normalization, the integrals of motion are rational functions of the Bethe roots so no transcendental functions related to higher genus WKB curves really enter. Due to high symmetry of WKB curves there was essentially a unique integration cycle along which we integrated the periods, the cycle coming from the Pochhammer contour on the mirror curve. We could have therefore completely ignored the geometry of the WKB curves and simply focused on the mirror curve. The differential $1$-forms that were single-valued meromorphic functions on WKB curves can be expressed as pull-backs of $1$-forms living on the mirror curve, but which have non-trivial monodromy around the three punctures (which in fact encodes the $\lambda_j$ parameters of the model). The natural cohomology in this setting is the twisted cohomology \cite{kita1994intersection,aomoto2011theory,Casali:2019ihm}.

It is nevertheless interesting to have a look what would the associated WKB curves making the $1$-forms $Y_n$ single-valued look like. Let us specialize to $\mathcal{W}_N$ minimal models with parameters $p,p^\prime$ such that
\begin{equation}
p \geq N, \qquad p^\prime \geq N, \qquad \text{and} \qquad \gcd(p,p^\prime)=1.
\end{equation}
Let us find a minimal branched covering surface $\mathcal{W}$ over the mirror curve $\mathcal{M}$ such that the $1$-forms $Y_n$ which on $\mathcal{M}$ have abelian monodromies
\begin{equation}
\exp \left(\frac{2\pi i}{\lambda_j} \right)
\end{equation}
around the punctures become single-valued. The corresponding $\lambda$-parameters are
\begin{equation}
\lambda_1 = \frac{N\left(p^\prime-p\right)}{p}, \qquad \lambda_2 = -\frac{N\left(p^\prime-p\right)}{p^\prime}, \qquad \lambda_3 = N.
\end{equation}
These parameters satisfy the relations
\begin{equation}
\frac{1}{\lambda_1} + \frac{1}{\lambda_2} + \frac{1}{\lambda_3} = 0, \qquad \frac{N}{\lambda_3} = 1, \qquad \frac{p^\prime-N}{\lambda_1} + \frac{p-N}{\lambda_2} = 1
\end{equation}
and these relations generate all the relations of the form
\begin{equation}
\frac{N_1}{\lambda_1} + \frac{N_2}{\lambda_2} + \frac{N_3}{\lambda_3} = 1
\end{equation}
with integer $N_j$. From here we see that the degree of the covering is
\begin{equation}
d = \left| \det \begin{pmatrix} 1 & 1 & 1 \\ 0 & 0 & N \\ p^\prime & p & N \end{pmatrix} \right| = N \left| p^\prime - p \right|.
\end{equation}
For example for the Argyres-Douglas models where $p=N$ and $p^\prime=K+N$ we recover the degree $d = KN$.

We want to calculate the genus of the covering surface using Riemann-Roch formula. To use that, we need to know the form of local monodromy around the punctures. As we go once around the singularity associated to $\lambda_1$, the phase of $Y_n(t) \, dt$ changes by the phase
\begin{equation}
e^{\frac{2\pi i}{\lambda_1}} = e^{2\pi i \frac{p}{d}}.
\end{equation}
The denominator of $\frac{p}{d}$ is
\begin{equation}
\frac{d}{\gcd(p,d)}
\end{equation}
and we therefore find that going around the first puncture generates a cyclic group
\begin{equation}
\mathbbm{Z}\Bigg/\left(\frac{d}{\gcd(p,d)}\mathbbm{Z}\right)
\end{equation}
and analogously for the second puncture. We have $\gcd(p,d)$ points on $\mathcal{W}$ covering the first puncture on $\mathcal{M}$. Going around third puncture generates simply $\mathbbm{Z}/N\mathbbm{Z}$. We are ready to use the Riemann-Roch formula. The genus of the covering surface satisfies
\begin{equation}
2 - 2g = \chi(\mathcal{W}) = d \chi(\mathcal{M}) - \left(d - \gcd(p,d)\right) - \left(d - \gcd(p^\prime,d)\right) - \left(d - \frac{d}{N}\right)
\end{equation}
which means that
\begin{equation}
g = \frac{1}{2} \left(d - \frac{d}{N} - \gcd(p,d) - \gcd(p^\prime,d)\right)
\end{equation}
which is our final formula for the genus. Let us list genera of $\mathcal{W}$ of the first few minimal models:
\begin{center}
\begin{tabular}{|c|c|c|c|c|c|c|}
\hline
minimal model & $N$ & $c$ & $p^\prime$ & $p$ & degree & genus \\
\hline
Ising & $2$ & $1/2$ & $3$ & $4$ & $2$ & $0$ \\
Lee-Yang $(A_1,A_2)$ model & $2$ & $-22/5$ & $2$ & $5$ & $6$ & $1$ \\
tricritical Ising & $2$ & $7/10$ & $4$ & $5$ & $2$ & $0$ \\
$3$-state Potts (Virasoro) & $2$ & $4/5$ & $5$ & $6$ & $2$ & $0$ \\
$3$-state Potts ($\mathcal{W}_3$) & $3$ & $4/5$ & $4$ & $5$ & $3$ & $1$ \\
hard hexagon & $2$ & $-3/5$ & $3$ & $5$ & $4$ & $1$ \\
tricritical $3$-state Potts & $2$ & $6/7$ & $6$ & $7$ & $2$ & $0$ \\
$(A_2,A_3)$ model & $3$ & $-114/7$ & $3$ & $7$ & $12$ & $3$ \\
\hline
\end{tabular}
\end{center}
We see that many of these minimal models have $\mathcal{W}$ of low genus, unlike the situation with $(\mathcal{W}_K,\mathcal{W}_N)$ Argyres-Douglas models where the genus is one only for Lee-Yang and all other models have genus at least two. This is actually true in general: all Virasoro unitary minimal models with central charge
\begin{equation}
c = \frac{k (k+5)}{(k+2) (k+3)}, \qquad k \geq 1
\end{equation}
have $\mathcal{W}$ of genus $0$ (i.e. $\mathcal{W}$ is rational). For unitary $\mathcal{W}_3$ minimal models with
\begin{equation}
c = \frac{2 k (k+7)}{(k+3) (k+4)}, \qquad k \geq 1
\end{equation}
we have genus zero for $k = 3j$ and $k = 3j+2$ and genus one for $k = 3j+1$ (where $j$ is an integer). For $3$-state Potts model we see that the genus of $\mathcal{W}$ depends on whether we use the Virasoro (with non-diagonal modular invariant) or $\mathcal{W}_3$ description. This is not surprising: in Virasoro description, the spin $3$ field generator is additional primary field independent from the chiral algebra so it is not included in the quantum KdV integrals of motion, while in $\mathcal{W}_3$ description this is one of the generating fields of the algebra and as such enters the conserved quantities we are diagonalizing.

\section{Bethe equations for $\mathcal{W}_3$ and $\mathcal{W}_4$}
\label{secw3}

In this section we will spell out explicitly the trivial monodromy conditions which we interpret as Bethe equations for $\mathcal{W}_N$ algebras. Since the equations get quite complicated very quickly, we will focus on $\mathcal{W}_3$ and $\mathcal{W}_4$ algebras.

\subsection{Bethe equations for $\mathcal{W}_3$}
Starting with the differential operator \eqref{opern3}, the indical equation around $x = x_j$ is
\begin{equation}
\sigma(\sigma-1)(\sigma-2) + c_{10}^j \sigma(\sigma-1) + c_{20}^j \sigma + c_{30}^j = 0.
\end{equation}
Following \cite{Masoero:2019wqf,Ashok:2024ygp}, we choose
\begin{equation}
c_{10}^j = 0, \qquad c_{20}^j = -3, \qquad c_{30}^j = 3
\end{equation}
so that the corresponding indices are
\begin{equation}
\sigma \in \left\{ -1, 1, 3 \right\}.
\end{equation}
We write the operator in the form
\begin{equation}
\label{ode3simple}
\partial_x^3 + v_2(x) \partial_x + v_3(x) + u.
\end{equation}
The expansion of $v_2$ and $v_3$ around $x=x_j$ is of the form
\begin{align}
v_2(x) & = -\frac{3}{(x-x_j)^2} + \sum_{m=-1}^\infty v_{2,m} (x-x_j)^m \\
v_3(x) & = \frac{3}{(x-x_j)^3} + \sum_{m=-2}^\infty v_{3,m} (x-x_j)^m.
\end{align}
The three linearly independent local solutions around $x=x_j$ are
\begin{align}
\label{ode3ansatz}
\nonumber
\psi_1(x) & = (x-x_j)^3 + \chi_{1,4} (x-x_j)^4 + \chi_{1,5} (x-x_j)^5 + \ldots \\
\nonumber
\psi_2(x) & = (x-x_j)^1 + \chi_{2,2} (x-x_j)^2 + \chi_{2,3} (x-x_j)^3 + \ldots + \rho \log(x-x_j) \psi_1(x) \\
\psi_3(x) & = (x-x_j)^{-1} + \chi_{3,0} + \chi_{3,1} (x-x_j) + \chi_{3,2} (x-x_j)^2 + \chi_{3,3} (x-x_j)^3 + \ldots \\
\nonumber
& + \rho_2 \log(x-x_j) \psi_2(x) + \rho_3 \log^2(x-x_j) \psi_1(x) + \rho_4 \log(x-x_j) \psi_1(x).
\end{align}
where all the coefficients except for $\chi_{2,3}, \chi_{3,1}$ and $\chi_{3,3}$ are determined by plugging this ansatz into the differential equation. We are free to choose these coefficients the way we like by taking a linear combination of solutions, i.e. $\chi_{2,3}$ can be chosen arbitrarily by adding a multiple of $\psi_2$ to $\psi_1$ etc. For simplicity we choose these to be zero. The differential equation will have trivial monodromy around $x = x_j$ precisely when all the coefficients of the logarithmic terms $\rho_k$ vanish. Plugging the ansatz \eqref{ode3ansatz} into \eqref{ode3simple} we find
\begin{align}
\nonumber
\rho_1 & = \frac{1}{24} \left[ -2v_{2,-1}^2 - 3v_{2,0} - 3v_{2,-1} v_{3,-2} - v_{3,-2}^2 - 3 v_{3,-1} \right] \\
\nonumber
\rho_2 & = \frac{1}{12} \left[ -3v_{2,0} + v_{2,-1} v_{3,-2} - v_{3,-2}^2 + 3 v_{3,-1} \right] \\
\rho_3 & = \frac{1}{48} \rho_2  \left[ 2v_{2,-1}^2 + 3v_{2,0} + 3v_{2,-1} v_{3,-2} + v_{3,-2}^2 + 3v_{3,-1} \right] \\
\nonumber
\rho_4\Big|_{\rho_1=0=\rho_2} & = \frac{1}{216} \Big[ 2v_{2,-1}^4+18v_{2,-1} v_{2,1}+27v_{2,2}+3v_{2,-1}^3 v_{3,-2}+9v_{2,1} v_{3,-2} \\
\nonumber
& -3v_{2,-1}^2 v_{3,-2}^2-2v_{2,-1} v_{3,-2}^3-27v_{2,-1} v_{3,0}-27v_{3,1}\Big] - \frac{u}{8} v_{2,-1}
\end{align}
Requiring that all $\rho_k$ vanish for all values of $u$ ($u$ appears only linearly in $\rho_4$), the conditions reduce to
\begin{align}
\label{w3monofree}
\nonumber
0 & = v_{2,-1} \\
\nonumber
0 & = v_{3,-1} \\
0 & = 3v_{2,0} + v_{3,-2}^2 \\
\nonumber
0 & = 3v_{2,2} - 3v_{3,1} + v_{3,-2} v_{2,1}.
\end{align}
It remains to evaluate these Laurent coefficients of the potentials $v_2(x)$ and $v_3(x)$. We find
\begin{equation}
v_{2,-1} = \frac{c_{21}^j}{x_j}
\end{equation}
and
\begin{equation}
v_{3,-1} = \frac{c_{32}^j-c_{31}^j}{x_j^2}.
\end{equation}
Vanishing of these requires
\begin{equation}
c_{21}^j = 0 \qquad \text{and} \qquad c_{31}^j = c_{32}^j \equiv \beta_j.
\end{equation}
Using these equations, the remaining information carried by singularity at $x=x_j$ is its position $x_j$ and the coefficient $\beta_j$. In terms of these two parameters, the remaining Laurent coefficients of $v_2(x)$ and $v_3(x)$ that we need are
\begin{align}
\nonumber
v_{2,0} & = \frac{b_2}{x_j^2} - \sum_{k \neq j} \frac{3}{(x_j-x_k)^2} \\
\nonumber
v_{2,1} & = -\frac{2b_2}{x_j^3} + \sum_{k \neq j} \frac{6}{(x_j-x_k)^3} \\
v_{2,2} & = \frac{3b_2}{x_j^4} - \sum_{k \neq j} \frac{9}{(x_j-x_k)^4} \\
\nonumber
v_{3,-2} & = \frac{\beta_j}{x_j} \\
\nonumber
v_{3,1} & = K x_j^{K-1} - \frac{3b_3}{x_j^4} + \frac{2\beta_j}{x_j^4} - \sum_{k \neq j} \left( \frac{9}{(x_j-x_k)^4} + \frac{(3x_j-2x_k)\beta_k}{x_j^3(x_j-x_k)^2} + \frac{(3x_j-x_k)\beta_k}{x_j^2(x_j-x_k)^3} \right).
\end{align}
Plugging these values into the remaining two equations in \eqref{w3monofree}, we finally find
\begin{equation}
\label{w3bethe1}
\frac{\beta_j^2}{x_j^2} + \frac{3b_2}{x_j^2} = \sum_{k \neq j} \frac{9}{(x_j-x_k)^2}
\end{equation}
and
\begin{multline}
\label{w3bethe2}
K x_j^{K-1} + \frac{2\beta_j}{x_j^4} - \frac{3(b_2+b_3)}{x_j^4} + \frac{2b_2 \beta_j}{3x_j^4} = \\
= \frac{2\beta_j}{x_j} \sum_{k \neq j} \frac{1}{(x_j-x_k)^3} + \sum_{k \neq j} 2\beta_k \left( \frac{1}{x_j(x_j-x_k)^3} + \frac{1}{x_j^2(x_j-x_k)^2} + \frac{1}{x_j^3(x_j-x_k)} \right)
\end{multline}
which are the Bethe equations for $\mathcal{W}_3$ that we were looking for.

\paragraph{Orbifold}
The equations that we found have somewhat simpler structure than those derived in \cite{Masoero:2019wqf} and used in \cite{Ashok:2024zmw,Ashok:2024ygp} (for comparison see Appendix \ref{appw3comparison}), but for numerical studies it is convenient to sum over orbits under $\mathbbm{Z}_{K+N}$ as we already did in the Virasoro case. Just like before, we use the identities of Appendix \ref{apporbitsums}. Since the Bethe equations \eqref{w3bethe1} and \eqref{w3bethe2} are homogeneous with respect to $\mathbbm{Z}_{K+N}$ rotations (assuming that $x_j$ rotate while $\beta_j$ are invariant), we need to introduce the new variables
\begin{equation}
X_j \equiv x_j^{K+N} = x_j^{K+3}
\end{equation}
while we use the same value of $\beta_j$. Our notation is such that we pick one representative $(X_j,\beta_j)$ from each orbit of $(K+N)$ pairs $(x_j,\beta_j)$. The orbifolded Bethe equations are
\begin{equation}
\label{w3betheorb1}
\beta_j^2 + 3b_2 = -\frac{3(K^2-4)}{4} + 9(K+3)\sum_{k \neq j}^M \frac{X_j^2 + (K+2)X_j X_k}{(X_j-X_k)^2}
\end{equation}
and
\begin{multline}
\label{w3betheorb2}
K X_j + 2\beta_j - 3(b_2+b_3) + \frac{2}{3} b_2 \beta_j = -\frac{2}{3} (K^2-4) \beta_j \\
+ (K+3) \sum_{k \neq j}^M \beta_j \frac{X_j(2X_j^2+(K+2)(K+7)X_j X_k+(K+1)(K+2)X_k^2)}{(X_j-X_k)^3} \\
+ (K+3) \sum_{k \neq j}^M \beta_k \frac{X_j(6X_j^2+(K^2+11K+12)X_j X_k+K(K+1)X_k^2)}{(X_j-X_k)^3}.
\end{multline}
Following the procedure of Section \ref{descwkbtheory}, we find the expressions for $\mathcal{I}_n$ in terms of Bethe roots $X_j$. The first few of these are given in Appendix \ref{appw3im} and more are given in the attached Mathematica notebook.

\subsection{Bethe equations for $\mathcal{W}_4$}
The differential operator is (after making a gauge choice that makes $c_{10}^j = 0$ and $b_1 = 0$)
\begin{multline}
\partial_x^4 + x^K - u + \left( \frac{b_2}{x^2} + \sum_j \frac{-4}{(x-x_j)^2} + \frac{c_{21}^j}{x(x-x_j)} \right) \partial_x^2 + \\
+ \left( \frac{b_3}{x^3} + \sum_j \frac{8}{(x-x_j)^3} + \frac{c_{31}^j}{x(x-x_j)^2} + \frac{c_{32}^j}{x^2(x-x_j)} \right) \partial_x \\
+ \left( \frac{b_4}{x^3} + \sum_j \frac{-8}{(x-x_j)^4} + \frac{c_{41}^j}{x(x-x_j)^3} + \frac{c_{42}^j}{x^2(x-x_j)^2} + \frac{c_{43}^j}{x^3(x-x_j)} \right).
\end{multline}
Requiring trivial monodromy around $x = x_j$ for all $u$ determines all the coefficients in terms of $c_{31}^j \equiv \beta_j$, namely
\begin{align}
\nonumber
c_{21}^j & = 0 \\
\nonumber
c_{31}^j & \equiv \beta_j \\
\nonumber
c_{32}^j & = \beta_j \\
c_{41}^j & = -\beta_j \\
\nonumber
c_{42}^j & = -2b_2 - \beta_j - \frac{\beta_j^2}{4} + \sum_{k \neq j} \frac{8x_j^2}{(x_j-x_k)^2} \\
\nonumber
c_{43}^j & = -2b_2 - \beta_j - \frac{\beta_j^2}{2} - \sum_{k \neq j} \frac{8x_j^3}{(x_j-x_k)^3} + \sum_{k \neq j} \frac{16x_j^2}{(x_j-x_k)^2}
\end{align}
and on top of that imposes two non-linear constraints
\begin{equation}
0 = 8b_2 + 4b_3 - 4\beta_j - 2b_2 \beta_j - \frac{\beta_j^3}{4} + \sum_{k \neq j} \frac{8\beta_j x_j^2+4\beta_k x_j^2}{(x_j-x_k)^2} + \sum_{k \neq j} \frac{4\beta_k x_j}{x_j-x_k}
\end{equation}
and
\begin{align}
\nonumber
0 & = -4c_{42}^j + 6c_{43}^j + 2\beta_j - \frac{\beta_j^2}{2} - \frac{b_2 \beta_j^2}{8} +\frac{3b_2 \beta_j}{2} - 8b_2 + \frac{3b_3 \beta_j}{4} -6 b_3 - 4b_4 + Kx_j^{K+4} \\
\nonumber
& + \sum_{k \neq j} \Bigg( -\frac{4 c_{42}^k x_j^3}{(x_j-x_k)^3} + \frac{2c_{42}^k x_j^2 x_k}{(x_j-x_k)^3} - \frac{4c_{43}^k x_j^2}{(x_j-x_k)^2} + \frac{3c_{43}^k x_j x_k}{(x_j-x_k)^2} + \frac{16x_j^5}{(x_j-x_k)^5} \\
& - \frac{2x_j^4 \beta_k}{(x_j-x_k)^4} + \frac{3x_j^3 x_k \beta_k}{(x_j-x_k)^4} - \frac{x_j^2 x_k^2 \beta_k}{(x_j-x_k)^4} + \frac{\frac{1}{2}x_j^3 \beta_j^2}{(x_j-x_k)^3} - \frac{6x_j^3 \beta_k}{(x_j-x_k)^3} + \frac{8x_j^2 x_k \beta_k}{(x_j-x_k)^3} \\
\nonumber
& - \frac{3x_j x_k^2 \beta_k}{(x_j-x_k)^3} + \frac{\frac{3}{4} x_j^3 \beta_j \beta_k}{(x_j-x_k)^3} - \frac{\frac{1}{4} x_j^2 x_k \beta_j \beta_k}{(x_j-x_k)^3} + \frac{\frac{3}{4}x_j^2 \beta_j \beta_k}{(x_j-x_k)^2} - \frac{\frac{1}{2} x_j x_k \beta_j \beta_k}{(x_j-x_k)^2} \Bigg).
\end{align}
We interpret these as Bethe equations for $\mathcal{W}_4$. For each Bethe root, they depend on a pair of parameters $(x_j,\beta_j)$ just like in the case of $\mathcal{W}_3$. We therefore have a system of $2M(K+4)$ equations for $2M(K+4)$ unknowns. As always, the equations are invariant under rotations
\begin{equation}
x_j \to e^{\frac{2\pi i}{K+4}} x_j, \qquad \beta_j \to \beta_j
\end{equation}
and the physically interesting solutions are those solutions that are invariant under this symmetry. Note that the only term in the equations that is not invariant under arbitrary $GL(1)$ rotations of $x_j$ is the potential term $Kx_j^{K+4}$.

We could continue and write the equations also for $\mathcal{W}_5$ etc. but the equations get complicated very quickly. They share the same features as what we saw for $\mathcal{W}_3$ and $\mathcal{W}_4$, namely there are always two parameters $(x_j,\beta_j)$ per singularity that satisfy non-linear equations and all other coefficients can be determined in terms of these. It would be very useful to have a closed form formulation of the Bethe equations that would be uniform in the rank $N$.


\section{Examples}
\label{secexamples}

\subsection{Lee-Yang model}
In this section we will compare the results of calculating the $\mathcal{I}_n$ eigenvalues on the lowest energy levels of Lee-Yang model -- $(A_1,A_2)$ Argyres-Douglas VOA. We can do it in three different ways, by directly acting with $\mathcal{I}_n$ or by solving Bethe equations for Virasoro or for $\mathcal{W}_3$ algebra. Lee-Yang VOA has two primaries of dimension $\Delta=0$ and $\Delta=-\frac{1}{5}$.

\paragraph{Characters}
In order to know the expected number of states (and therefore the solutions of Bethe equations), let us write the formula for Lee-Yang characters. All Argyres-Douglas VOAs have simple product formulas for their characters expressed in terms of the asymptotic Young diagram in the plane partition picture. This is explained for instance in Section 7 of \cite{Prochazka:2023zdb}. For Lee-Yang, the two characters are
\begin{align}
\chi_{\Delta=0} & = \frac{q^{11/60}}{(q^2;q^5)_\infty (q^3;q^5)_\infty} \simeq q^{11/60}(1+q^2+q^3+q^4+q^5+2q^6+\ldots) \\
\chi_{\Delta=-\frac{1}{5}} & = \frac{q^{-1/60}}{(q;q^5)_\infty (q^4;q^5)_\infty} \simeq q^{-1/60} (1+q+q^2+q^3+2q^4+2q^5+3q^6+\ldots).
\end{align}

\subsubsection{$\Delta=-\frac{1}{5}$, Virasoro Bethe equations}
We consider for concreteness equations \eqref{virbetheequationsorb} with $K=3$ $\left(c=-\frac{22}{5}\right)$ and $\ell=0$ $\left(\Delta=-\frac{1}{5}\right)$.

\paragraph{Level 0} This is mainly to check the normalization. The eigenvalues of $\mathcal{I}_n$ charges evaluated on the highest weight state are
\begin{equation}
\label{leeyangtestlevel0}
(\mathcal{I}_2,\mathcal{I}_4,\mathcal{I}_6,\mathcal{I}_8) = \left(-\frac{1}{60},0,\frac{89}{1512000},\frac{211}{12960000} \right)
\end{equation}

\paragraph{Level $1$} At level $1$ we have just one linear equation for $X_1$ with the solution
\begin{equation}
X_1 = \frac{4}{3}.
\end{equation}
The eigenvalues of integrals of motion calculated by direct action with Fourier modes of Virasoro algebra are
\begin{equation}
\label{leeyangtestlevel1}
(\mathcal{I}_2,\mathcal{I}_4,\mathcal{I}_6,\mathcal{I}_8) = \left(\frac{59}{60},0,-\frac{59131}{1512000},\frac{122371}{12960000} \right)
\end{equation}
and these agree with formulas given in Appendix \ref{appiom} if we plug in $M=1$ and the value of the Bethe root $X_1$.

\paragraph{Level $2$} At the next level, the Bethe equations are two polynomial equations in $X_1$ and $X_2$. Eliminating $X_2$, we find that the Bethe roots need to satisfy the equation
\begin{equation}
\label{leeyanggslevel1eq1}
X^2-6X-6 = 0
\end{equation}
so the solution to BAE is given by a pair of Bethe roots
\begin{equation}
\left\{X_1,X_2\right\} = \left\{3-\sqrt{15},3+\sqrt{15}\right\}.
\end{equation}
This agrees with the calculation using Virasoro mode expansions and in both ways we find the following eigenvalues of first few integrals of motion acting on level $2$ state:
\begin{equation}
\label{leeyangtestlevel2}
(\mathcal{I}_2,\mathcal{I}_4,\mathcal{I}_6,\mathcal{I}_8) = \left(\frac{119}{60},0,-\frac{1630351}{1512000},\frac{14154931}{12960000} \right)
\end{equation}
Virasoro Verma module has two states at level $2$ and apart from the state surviving in the irreducible quotient we find also another (singular) solution
\begin{equation}
\left\{X_1,X_2\right\} = \left\{0,0\right\}
\end{equation}
with charges
\begin{equation}
(\mathcal{I}_2,\mathcal{I}_4,\mathcal{I}_6,\mathcal{I}_8) = \left( \frac{119}{60}, -\frac{9}{5}, \frac{4493249}{1512000}, -\frac{72800189}{12960000} \right)
\end{equation}
representing a null state. This state is not present in the irreducible $(\mathcal{W}_2,\mathcal{W}_3)$ representation because such states should have vanishing $\mathcal{I}_4$. Nevertheless, by continuity, the values of $\mathcal{I}_j$ for this state calculated using BAE and using commutation relations in the corresponding Verma module agree.

\paragraph{Level $3$} At the level $3$ we still have a single physical state so we expect rational values of the $\mathcal{I}_n$. The three roots $X_1, X_2$ and $X_3$ corresponding to the state at level $3$ satisfy the algebraic equation
\begin{equation}
729X^3 - 10206X^2 + 34398X - 66248 = 0
\end{equation}
and using these, we find the eigenvalues of $\mathcal{I}_n$
\begin{equation}
\label{leeyangtestlevel3}
(\mathcal{I}_2,\mathcal{I}_4,\mathcal{I}_6,\mathcal{I}_8) = \left(\frac{179}{60}, 0, -\frac{12273571}{1512000}, \frac{238657891}{12960000}\right).
\end{equation}

\paragraph{Level $4$} Finally, let us have a look at level $4$ where there are two states (so that the values of $\mathcal{I}_n$ can take irrational values which is the typical situation at higher levels). The Bethe roots corresponding to two physical states are given by roots of the irreducible polynomial
\begin{multline}
\label{leeyangbethepoly}
19683X^8 - 997272X^7 + 19058976X^6 - 212644656X^5 + 861149160X^4 \\
- 13185260064X^3 + 28988614336X^2 + 90419780352X - 33449456832.
\end{multline}
The first solution has the roots
\begin{equation}
\label{leeyangbaelevel4sol1}
\left\{X_1,X_2,X_3,X_4\right\} = \left\{ -1.8630,\, 4.0501,\, 11.5731-14.1017i,\, 11.5731-14.1017i \right\}
\end{equation}
while the second one
\begin{equation}
\label{leeyangbaelevel4sol2}
\left\{X_1,X_2,X_3,X_4\right\} = \left\{ -2.1508-7.9735i,\, -2.1508+7.9735i,\, 0.3387,\, 29.2962 \right\}.
\end{equation}
The eigenvalues of $\mathcal{I}_n$ for the first one are given by
\begin{equation}
\label{leeyanglevel4crtest1}
(\mathcal{I}_2,\mathcal{I}_4,\mathcal{I}_6,\mathcal{I}_8) = \left( \frac{239}{60}, 0, -\frac{29871991+302400\sqrt{5149}}{1512000}, \frac{956072851+11491200\sqrt{5149}}{12960000} \right)
\end{equation}
while for the second one they are
\begin{equation}
\label{leeyanglevel4crtest2}
(\mathcal{I}_2,\mathcal{I}_4,\mathcal{I}_6,\mathcal{I}_8) = \left( \frac{239}{60}, 0, -\frac{29871991-302400\sqrt{5149}}{1512000}, \frac{956072851-11491200 \sqrt{5149}}{12960000} \right).
\end{equation}
We can match this with calculation using Virasoro modes. First of all, the $c=-\frac{22}{5}, \Delta=-\frac{1}{5}$ primary has a singular vectors at levels $2$ and $3$. In the Verma module we would have $5$ states at level $4$,
\begin{equation}
L_{-4} \ket{-\frac{1}{5}}, \qquad L_{-3} L_{-1} \ket{-\frac{1}{5}}, \qquad L_{-2}^2 \ket{-\frac{1}{5}}, \qquad L_{-2} L_{-1}^2 \ket{-\frac{1}{5}}, \qquad L_{-1}^4 \ket{-\frac{1}{5}}
\end{equation}
but due to presence of null states, in the simple module (i.e. when we put all the null states to zero) we have the following relations:
\begin{equation}
L_{-2} \ket{-\frac{1}{5}} \sim \frac{5}{2} L_{-1}^2 \ket{-\frac{1}{5}}, \qquad L_{-3} \ket{-\frac{1}{5}} \sim \frac{25}{12} L_{-1}^3 \ket{-\frac{1}{5}}.
\end{equation}
Therefore, the two linearly independent states at level $4$ can be chosen as
\begin{equation}
L_{-4} \ket{-\frac{1}{5}} \qquad \text{and} \qquad L_{-1}^4 \ket{-\frac{1}{5}}
\end{equation}
and the remaining states can be eliminated using
\begin{align}
\nonumber
L_{-3} L_{-1} \ket{-\frac{1}{5}} & \sim -2 L_{-4} \ket{-\frac{1}{5}} + \frac{25}{12} L_{-1}^4 \ket{-\frac{1}{5}} \\
L_{-2}^2 \ket{-\frac{1}{5}} & \sim 5 L_{-4} \ket{-\frac{1}{5}} - \frac{25}{6} L_{-1}^4 \ket{-\frac{1}{5}} \\
\nonumber
L_{-2} L_{-1}^2 \ket{-\frac{1}{5}} & \sim 2 L_{-4} \ket{-\frac{1}{5}} - \frac{5}{3} L_{-1}^4 \ket{-\frac{1}{5}}.
\end{align}
Acting in this basis with $\mathcal{I}_n$, we find the matrix for $\mathcal{I}_6$ to be
\begin{equation}
\begin{pmatrix} -57995191/1512000 & -168/5 \\ 25/6 & -1748791/1512000 \end{pmatrix}
\end{equation}
and for $\mathcal{I}_8$
\begin{equation}
\begin{pmatrix} 2024754451/12960000 & 3724/25 \\ -665/36 & -112608749/12960000 \end{pmatrix}
\end{equation}
and it is easy to check that their eigenvalues agree with \eqref{leeyanglevel4crtest1} and \eqref{leeyanglevel4crtest2}.

This example illustrates some typical features of the solutions of BLZ Bethe ansatz equations. In particular, Bethe roots are always algebraic numbers and Bethe roots corresponding to different eigenstates are roots of the same polynomial as it happened in \eqref{leeyangbethepoly}. Therefore they cannot be simply separated from each other and different eigenstates of $\mathcal{I}_n$ at a given Virasoro level are related by Galois symmetries of \eqref{leeyangbethepoly}. We see this also from the explicit expressions for $\mathcal{I}_n$ where the map $\sqrt{5149} \to -\sqrt{5149}$ exchanges the eigenvalues of $\mathcal{I}_n$ corresponding to different states. This feature was also observed when diagonalizing the quantum ILW Hamiltonians in \cite{Prochazka:2023zdb}.

From the numerical point of view, reducing the diagonalization of $\mathcal{I}_n$ to solution of Bethe equations might not simplify the calculation itself, but there is an important factorization property of the Bethe ansatz description of the spectrum: to a given quantum eigenstate of $\mathcal{I}_n$ there corresponds a single collection of Bethe roots as in \eqref{leeyangbaelevel4sol1}. From this solution we can in principle calculate arbitrary $\mathcal{I}_n$ which are given by universal formulas in terms of Bethe roots $X_j$,
\begin{equation}
\mathcal{I}_n(X_j).
\end{equation}
Therefore we need to find the $X_j$ for each quantum state as well as the universal functions $\mathcal{I}_n(X_j)$, but both of these calculations are independent. In the usual calculation using Fourier modes of generators of $\mathcal{W}_N$ acting on the states the calculation has to be done independently for every $\mathcal{I}_n$ at every Virasoro level and there are no obvious relations connecting the calculations of a fixed $\mathcal{I}_n$ at different levels or different $\mathcal{I}_n$ at a fixed Virasoro level.

\subsubsection{$\Delta=-\frac{1}{5}$, $\mathcal{W}_3$ Bethe equations}
Let us compare the calculation of Lee-Yang eigenvalues of $\mathcal{I}_n$ using Virasoro Bethe equations to calculation using $\mathcal{W}_3$ Bethe equations of Section \ref{secw3}.

\paragraph{Level $1$} In order to reproduce the correct eigenvalues of $\mathcal{I}_n$ at level $1$ as in \eqref{leeyangtestlevel1}, we need to choose a singular solution
\begin{equation}
X_1 = 0, \qquad \beta_1 = 0.
\end{equation}

\paragraph{Level $2$}
The solution of orbifolded Bethe equations for $N=3$ and $K=2$ \eqref{w3betheorb1} and \eqref{w3betheorb2} at level $M=2$ is
\begin{equation}
\left\{X_1,X_2\right\} = \left\{ \frac{3i\sqrt{15}}{8}, -\frac{3i\sqrt{15}}{8} \right\}, \qquad \left\{ \beta_1, \beta_2 \right\} = \left\{ \frac{3i\sqrt{15}}{2}, -\frac{3i\sqrt{15}}{2} \right\}
\end{equation}
and the corresponding eigenvalues of $\mathcal{I}_n$ reproduce \eqref{leeyangtestlevel2}. Apart from this solution, we find other solutions corresponding to states in the Verma module that are quotiented out when we pass to the irreducible module. These states are still captured by our Bethe equations: we have a pair of conjugate solutions
\begin{equation}
\left\{X_1,X_2\right\} = \left\{ \pm 12 \sqrt{5}, 0 \right\}, \qquad \left\{ \beta_1, \beta_2 \right\} = \left\{ 3\sqrt{5}, 0 \right\}
\end{equation}
with charges
\begin{equation}
\left( \mathcal{I}_2, \mathcal{I}_3, \mathcal{I}_4, \mathcal{I}_5, \mathcal{I}_6, \mathcal{I}_7, \mathcal{I}_8 \right) = \left( \frac{119}{60}, \pm 2, 0, \mp \frac{76}{15}, -\frac{12970351}{1512000}, 0, \frac{339234931}{12960000} \right)
\end{equation}
which can be found analytically using the Gr\"obner basis to eliminate $\beta_1, \beta_2, X_2$ resulting in equation
\begin{equation}
X_1^3(X_1^2-720)(64X_1^2+135)
\end{equation}
for $X_1$. We are missing other two solutions for states in Verma module which turn out to be singular. They correspond to the limit
\begin{equation}
X_1 \to 0, \quad X_2 \to 0, \quad  \beta_1 \to \infty, \quad \beta_2 \to \infty
\end{equation}
such that
\begin{equation}
\beta_1 + \beta_2 \to \pm 3.
\end{equation}
The corresponding charges are
\begin{equation}
\left( \mathcal{I}_2, \mathcal{I}_3, \mathcal{I}_4, \mathcal{I}_5, \mathcal{I}_6, \mathcal{I}_7, \mathcal{I}_8 \right) = \left( \frac{119}{60}, \pm \frac{2}{\sqrt{5}}, 0, \mp \frac{52}{15 \sqrt{5}}, -\frac{3293551}{1512000}, 0, \frac{48930931}{12960000} \right)
\end{equation}
and these agree with calculation using $\mathcal{W}_3$ commutation relations.

\paragraph{Level $3$}
Level $M=3$ solution of \eqref{w3betheorb1} and \eqref{w3betheorb2} is
\begin{equation}
\left\{ X_1, X_2, X_3 \right\} = \left\{ \frac{63\sqrt{5}}{8}, -\frac{63\sqrt{5}}{8}, 0 \right\}, \qquad \left\{ \beta_1, \beta_2, \beta_3 \right\} = \left\{ \frac{3\sqrt{5}}{2}, -\frac{3\sqrt{5}}{2}, 0 \right\}
\end{equation}
which again correctly reproduces the results calculated using commutation relations or using Virasoro Bethe equations.

\paragraph{Level $4$}
At level $4$ we have two eigenstates of $\mathcal{I}_n$. The Bethe roots $X_j$ of both solutions are roots of irreducible polynomial
\begin{equation}
65536X^8 - 427253760X^6 + 564707116800X^4 + 3896840340000X^2 + 473314640625
\end{equation}
while the variables $\beta_j$ are roots of
\begin{equation}
4752400\beta^8 - 182745040\beta^6 - 16496762084\beta^4 + 350508659040\beta^2 + 17431428211425.
\end{equation}
The eigenvalues corresponding to state \eqref{leeyangbaelevel4sol1} are reproduced by the solution
\begin{align}
\nonumber
\left\{X_1,X_2,X_3,X_4\right\} & \approx \left\{68.36, -68.36, 0.35i, -0.35i\right\} \\
\left\{\beta_1,\beta_2,\beta_3,\beta_4\right\} & \approx \left\{ 7.50, -7.50, 5.81i, -5.81i \right\}
\end{align}
while the eigenvalues corresponding to \eqref{leeyangbaelevel4sol2} by the solution
\begin{align}
\nonumber
\left\{X_1,X_2,X_3,X_4\right\} & \approx \left\{43.05, -43.05, 2.60i, -2.60i\right\} \\
\left\{\beta_1,\beta_2,\beta_3,\beta_4\right\} & \approx \left\{ -7.26, 7.26, -6.06i, 6.06i \right\}.
\end{align}

\subsubsection{$\Delta = -\frac{1}{5}$ Virasoro Bethe equations with $K = -\frac{6}{5}$}
Even though we mostly focus on Argyres-Douglas parametrization of the minimal models where both $N$ and $K$ are integers, on the level of orbifolded Bethe equation the parameter $K$ parametrizing the minimal model is meaningful for generic $K \in \mathbbm{C}$ and so are the expressions for charges in Appendix \eqref{appvirasoroperiods}. There is therefore a description of Lee-Yang model with $K=-\frac{6}{5}$ which is related to $K=3$ description by the usual Feigin-Frenkel duality (which is a part of triality symmetry that preserves $N=2$). Let us test the solutions of the corresponding Bethe equations for this dual value of $K$.

\paragraph{Level $0$}
To get the correct charges for the highest weight state, we have to choose
\begin{equation}
\ell = -\frac{3}{10} \qquad \text{or} \qquad \ell = -\frac{7}{10}.
\end{equation}
With this choice, we reproduce \eqref{leeyangtestlevel0}.

\paragraph{Level $1$}
The solution of Bethe equations at level $1$ is just
\begin{equation}
X_1 = -\frac{8}{15}
\end{equation}
which reproduces the charges \eqref{leeyangtestlevel1}.

\paragraph{Level $2$}
There is one physical state on level $2$ corresponding to solution
\begin{equation}
\left\{X_1, X_2 \right\} = \left\{\frac{1}{5} \left(-6-2 i \sqrt{6}\right),\frac{1}{5} \left(-6+2 i \sqrt{6}\right)\right\}
\end{equation}
and this solution reproduces \eqref{leeyangtestlevel2}.

\paragraph{Level $3$}
Finally, for level $3$ there is up to permutations a unique solution of Bethe equations
\begin{equation}
101920 + 68796 X + 20412 X^2 + 3645 X^3 = 0
\end{equation}
which is
\begin{equation}
\left\{X_1, X_2, X_3 \right\} \sim \left\{ -2.5157, -1.5422 - 2.9558i, -1.5422 + 2.9558i \right\}
\end{equation}
and reproduces the charges given in \eqref{leeyangtestlevel3}. Unfortunately, we do not know an explicit map that would map solutions of Bethe equations in the Feigin-Frenkel dual frames. This is the simplest example of the triality symmetry as the rank $N$ is being fixed so the form of the differential operator or the Bethe equations does not change.

\subsection{$(\mathcal{W}_3,\mathcal{W}_4)$ model}
As another example, let us consider the $(\mathcal{W}_3,\mathcal{W}_4)$ minimal model. Unlike the Lee-Yang model, it is not a minimal model of the Virasoro algebra, so there exist primaries that are not self-conjugate. We can therefore test also the conserved charges of odd degree. There are five different primaries with conformal dimensions
\begin{equation}
\Delta = \left\{ -\frac{5}{7}, -\frac{4}{7}, -\frac{3}{7}, -\frac{3}{7}, 0 \right\}.
\end{equation}
The ground state primary has $\Delta = -\frac{5}{7}$. The central charge of this model is
\begin{equation}
c = -\frac{114}{7}
\end{equation}
while the effective central charge is
\begin{equation}
c_{eff} = c - 24h_{min} = \frac{6}{7}.
\end{equation}
The characters of the irreducible representations are \cite{Prochazka:2023zdb}
\begin{align}
\label{w3w4chars}
\nonumber
\chi_{\Delta=0} & = \frac{q^{\frac{19}{28}}}{(q^2;q^7)_\infty (q^3;q^7)_\infty^2 (q^4;q^7)_\infty^2 (q^5;q^7)_\infty} = q^{\frac{19}{28}} \left( 1 + q^2 + 2q^3 + 3q^4 + \ldots \right) \\
\nonumber
\chi_{\Delta=-\frac{3}{7}} & = \frac{q^{\frac{7}{28}}}{(q;q^7)_\infty (q^2;q^7)_\infty (q^3;q^7)_\infty (q^4;q^7)_\infty (q^5;q^7)_\infty (q^6;q^7)_\infty} \\
& = q^{\frac{7}{28}} \left(1 + q + 2q^2 + 3q^3 + 5q^4 + \ldots \right) \\
\nonumber
\chi_{\Delta=-\frac{4}{7}} & = \frac{q^{\frac{3}{28}}}{(q;q^7)_\infty (q^2;q^7)_\infty^2 (q^5;q^7)_\infty^2 (q^6;q^7)_\infty} = q^{\frac{3}{28}} \left(1 + q + 3q^2 + 3q^3 + 6q^4 + \ldots \right) \\
\nonumber
\chi_{\Delta=-\frac{5}{7}} & = \frac{q^{-\frac{1}{28}}}{(q;q^7)_\infty^2 (q^3;q^7)_\infty (q^4;q^7)_\infty (q^6;q^7)_\infty^2} = q^{-\frac{1}{28}} \left(1 + 2q + 3q^2 + 5q^3 + 8q^4 + \ldots \right)
\end{align}
\begin{table}
\centering
\begin{tabular}{|c|c|c|c|c|c|}
\hline
$\Delta$ & $\mathcal{I}_2$ & $\mathcal{I}_3$ & $(b_1,b_2,b_3)_{\mathcal{W}_3}$ & $(b_1,b_2,b_3,b_4)_{\mathcal{W}_4}$ & asymptotics \\
\hline
$0$ & $19/28$ & $0$ & $(0,-15,15)$ & $(0,-20,40,0)$ & $\emptyset$ \\
$-3/7$ & $7/28$ & $+2 \cdot 7^{-3/2}$ & $(0,-6,12)$ & $(0,-8,24,-24)$ & $\ydiagram{1,1}$ \\
$-3/7$ & $7/28$ & $-2 \cdot 7^{-3/2}$ & $(0,-6,0)$ & $(0,-8,8,0)$ & $\ydiagram{1}$ \\
$-4/7$ & $3/28$ & $0$ & $(0,-3,3)$ & $(0,-4,8,-8)$ & $\ydiagram{2}$ \\
$-5/7$ & $-1/28$ & $0$ & $(0,0,0)$ & $(0,0,0,0)$ & $\ydiagram{2,1}$ \\
\hline
\end{tabular}
\caption{List of primaries in $(\mathcal{W}_3,\mathcal{W}_4)$ minimal models as well as the corresponding coefficients of the differential operators.}
\label{w3w4primaries}
\end{table}
The associated differential operators of third and fourth degree have parameters listed in Table \ref{w3w4primaries}.

\subsubsection{Ground state primary $\Delta = -\frac{5}{7}$ ($\mathcal{W}_3$ point of view)}

\paragraph{Level $1$}
From \eqref{w3w4chars} we expect to have two states and indeed Bethe equations at level $1$ imply\footnote{Remember that if we use the non-orbifolded Bethe equations, we have $7$ roots $x_1, \ldots, x_7$ such that $X_1 = x_j^7$ and all of these have the same coefficient $\beta_1 = \ldots = \beta_7$.}
\begin{equation}
\beta_1 = \pm 3i
\end{equation}
and
\begin{equation}
X_1 = x_1^7 = -\frac{5}{2} \beta_1.
\end{equation}
The corresponding charges are
\begin{equation}
\label{w3w4gschargeslvl1}
(\mathcal{I}_2,\mathcal{I}_3,\mathcal{I}_4,\mathcal{I}_5,\mathcal{I}_6,\mathcal{I}_7,\mathcal{I}_8) = \left( \frac{27}{28}, \pm \frac{2i}{\sqrt{7}},0,0,\frac{4229}{28224},0,\frac{596573}{9219840}\right)
\end{equation}
and this agrees with the result obtained by using the commutation relations \cite{Prochazka:2014gqa,Prochazka:2023zdb}.

\paragraph{Level $2$}
At level $2$ we expect three states. For the first one, we have
\begin{equation}
\left\{X_1,X_2\right\} = \left\{ \frac{3\sqrt{39}i}{16}, -\frac{3\sqrt{39}i}{16} \right\}, \qquad \left\{\beta_1,\beta_2\right\} = \left\{ \frac{3\sqrt{39}i}{2}, -\frac{3\sqrt{39}i}{2} \right\}
\end{equation}
and
\begin{equation}
\label{w3w4gschargeslvl2a}
(\mathcal{I}_2,\mathcal{I}_3,\mathcal{I}_4,\mathcal{I}_5,\mathcal{I}_6,\mathcal{I}_7,\mathcal{I}_8) = \left( \frac{55}{28}, 0, 0, 0, -\frac{25135}{28224}, 0, \frac{2958351}{3073280} \right).
\end{equation}
The second and the third are complex conjugates,
\begin{equation}
\left\{X_1,X_2\right\} = \left\{ \pm 11.394i, \mp 71.093i \right\}, \qquad \left\{\beta_1,\beta_2\right\} = \left\{ \pm 7.266i, \pm 2.684i \right\}
\end{equation}
solving fourth order equations
\begin{align}
0 & = X^4 + 5184 X^2 + 656100 \\
0 & = 4\beta^4 + 240\beta^2 + 1521.
\end{align}
The corresponding charges are
\begin{equation}
\label{w3w4gschargeslvl2b}
(\mathcal{I}_2,\mathcal{I}_3,\mathcal{I}_4,\mathcal{I}_5,\mathcal{I}_6,\mathcal{I}_7,\mathcal{I}_8) = \left( \frac{55}{28}, \pm 2i\sqrt{\frac{11}{7}}, 0, 0, \frac{85745}{28224}, 0, \frac{16239311}{3073280} \right).
\end{equation}
Here we see an example of non-vanishing odd charge, $\mathcal{I}_3$, that distinguishes between the two states. The $\mathcal{I}_3$ charge differs simply by a sign because the representation with $\Delta=-\frac{5}{7}$ that we are considering now is self-conjugate.

\paragraph{Level $3$}
Finally let us have a look at level $3$. From the character formulas, we see that there are five states. The first charge conjugate pair of solutions is
\begin{equation}
\left\{X_1,X_2,X_3\right\} = \left\{ \mp 112.49, \pm 23.988, \pm 0.78627 \right\}, \qquad \left\{\beta_1,\beta_2,\beta_3\right\} = \left\{\pm 6.2098, \pm 4.3326, \mp 1.3091\right\}
\end{equation}
with charges
\begin{multline}
\label{w3w4gschargeslvl3a}
(\mathcal{I}_2,\mathcal{I}_3,\mathcal{I}_4,\mathcal{I}_5,\mathcal{I}_6,\mathcal{I}_7,\mathcal{I}_8) = \\
= \left( \frac{83}{28}, \pm 2.3266, 0, 0, \frac{310061-5040 \sqrt{4089}}{28224}, 0, \frac{128191711-2140320 \sqrt{4089}}{3073280} \right).
\end{multline}
The charges $\mathcal{I}_3$ are roots of the equation
\begin{equation}
49\mathcal{I}_3^4 + 1260 \mathcal{I}_3^2 - 8256 = 0.
\end{equation}
The second charge conjugate pair of solutions is
\begin{align}
\nonumber
\left\{X_1,X_2,X_3\right\} & = \left\{ -59.161 \mp 42.023i, \mp 126.30i, +59.161 \mp 42.023i \right\}, \\
\left\{\beta_1,\beta_2,\beta_3\right\} & = \left\{-6.169 \pm 19.97i, \mp 17.81i, +6.169 \pm 19.97i \right\}
\end{align}
with charges
\begin{multline}
\label{w3w4gschargeslvl3b}
(\mathcal{I}_2,\mathcal{I}_3,\mathcal{I}_4,\mathcal{I}_5,\mathcal{I}_6,\mathcal{I}_7,\mathcal{I}_8) = \\
= \left( \frac{83}{28}, \pm 5.5792i, 0, 0, \frac{310061+5040\sqrt{4089}}{28224}, 0, \frac{128191711+2140320\sqrt{4089}}{3073280} \right).
\end{multline}
The charges $\mathcal{I}_3$ are roots of the same equation as above, and the even charges differ by choice of a sign of square root, i.e. Galois group of the equation connects all four eigenstates at this level.

We expect one more state at level $3$. The correct eigenvalues of $\mathcal{I}_n$ are reproduced by
\begin{equation}
\left\{X_1,X_2,X_3\right\} = \left\{ 0,0,0 \right\}, \qquad \left\{\beta_1,\beta_2,\beta_3\right\} = \left\{ 0,0,0\right\}
\end{equation}
and the values of first $\mathcal{I}_n$ are
\begin{equation}
\label{w3w4gschargeslvl3c}
(\mathcal{I}_2,\mathcal{I}_3,\mathcal{I}_4,\mathcal{I}_5,\mathcal{I}_6,\mathcal{I}_7,\mathcal{I}_8) = \left( \frac{83}{28}, 0, 0, 0, -\frac{164419}{28224}, 0, \frac{34574271}{3073280} \right)
\end{equation}
which we independently verified by the explicit action of $\mathcal{W}_3$ algebra. We see that this solution is singular, but we need to include it in order to reproduce all $5$ states that we have at level $3$.

It is interesting to note that if we calculate the action of $\mathcal{I}_3 \sim W_{3,0}$ in the corresponding Verma module, the action is not diagonalizable because there is one non-trivial Jordan block of size $2$ corresponding to doubly degenerate generalized eigenvalue zero. Nevertheless, the zero eigenvector lies in the maximal proper submodule, so the quotient from Verma module to the irreducible module is well-defined and the action of $W_{3,0}$ on the irreducible quotient is diagonalizable \footnote{We would like to thank to Shigenori Nakatsuka for a useful discussion on this point.}.

\subsubsection{Ground state primary $\Delta = -\frac{5}{7}$ ($\mathcal{W}_4$ point of view)}
We can compare the previous solution from the point of view of $\mathcal{W}_3$ Bethe ansatz to the corresponding solution of $\mathcal{W}_4$ Bethe ansatz equations. Remember that we are only focusing on $\mathbbm{Z}_{K+N} = \mathbbm{Z}_7$-invariant solutions.

\paragraph{Level $1$}
On level $1$, the equation for $\beta_1$ admits two solutions,
\begin{equation}
\beta_1 = \pm 4i.
\end{equation}
This determines
\begin{equation}
c_{42}^1 = -4 \mp 4i, \quad c_{43}^1 = 16 \mp 4i.
\end{equation}
Finally the second non-linear Bethe equation has solution
\begin{equation}
X_1 \equiv x_1^7 = -\frac{40}{3}.
\end{equation}
Plugging this into expressions for $\mathcal{I}_n$ charges in $\mathcal{W}_4$ reproduces \eqref{w3w4gschargeslvl1}.

\paragraph{Level $2$}
There are three physical states at level $2$. The first solution is real and has
\begin{alignat}{2}
\nonumber
\left\{ X_1, X_2 \right\} & = \left\{ -82.715, 4.715 \right\} & \qquad \left\{ \beta_1, \beta_2 \right\} & = \left\{ 0, 0 \right\} \\
\left\{ c_{42}^1, c_{42}^2 \right\} & = \left\{ 24.980, -24.980 \right\} & \qquad \left\{ c_{43}^1, c_{43}^2 \right\} & = \left\{ 113.430, -61.430 \right\}
\end{alignat}
It corresponds to \eqref{w3w4gschargeslvl2a}. The other two solutions form a complex conjugate pair,
\begin{align}
\nonumber
\left\{ X_1, X_2 \right\} & = \left\{ -74.434, 47.768 \right\} \\
\nonumber
\left\{ \beta_1, \beta_2 \right\} & = \left\{ \mp 3.499i, \pm 16.766i \right\} \\
\left\{ c_{42}^1, c_{42}^2 \right\} & = \left\{ -64.162 \pm 3.499i, -9.171 \mp 16.766i \right\} \\
\nonumber
\left\{ c_{43}^1, c_{43}^2 \right\} & = \left\{ 72.850 \pm 3.499i, 52.483 \mp 16.766i \right\}
\end{align}
and reproduce the charges given in \eqref{w3w4gschargeslvl2b}.

\paragraph{Level $3$}
Finally at level $3$ we have five states. The solutions corresponding to \eqref{w3w4gschargeslvl3a} are
\begin{align}
\nonumber
\left\{ X_1, X_2, X_3 \right\} & = \left\{ 2.387, 50.037+100.950i, 50.037-100.950i \right\} \\
\nonumber
\left\{ \beta_1, \beta_2, \beta_3 \right\} & = \left\{ -6.445, 9.378-13.965i, 9.378+13.965i \right\} \\
\left\{ c_{42}^1, c_{42}^2, c_{42}^3 \right\} & = \left\{ -6.034, -24.713+56.765i, -24.713-56.765i \right\} \\
\nonumber
\left\{ c_{43}^1, c_{43}^2, c_{43}^3 \right\} & = \left\{ 18.719, 66.059-57.701i, 66.059+57.701i \right\}
\end{align}
and
\begin{align}
\nonumber
\left\{ X_1, X_2, X_3 \right\} & = \left\{ 2.387, 50.037+100.950i, 50.037-100.950i \right\} \\
\nonumber
\left\{ \beta_1, \beta_2, \beta_3 \right\} & = \left\{ 6.445, -9.378+13.965i, -9.378-13.965i \right\} \\
\left\{ c_{42}^1, c_{42}^2, c_{42}^3 \right\} & = \left\{ -18.923, -5.957+28.836i, -5.957-28.836i \right\} \\
\nonumber
\left\{ c_{43}^1, c_{43}^2, c_{43}^3 \right\} & = \left\{ 5.830, 84.815-85.631i, 84.815+85.631i \right\}.
\end{align}
We see that $X_j$ for these solutions are the same, $\beta_j$ differ by a sign and the remaining coefficients $c_{42}^j$ and $c_{43}^j$ are related by a more complicated transformation. The next two solutions correspond to \eqref{w3w4gschargeslvl3b}. They are given by
\begin{align}
\nonumber
\left\{ X_1, X_2, X_3 \right\} & = \left\{ -44.087, 222.480 \pm 597.963i, 222.480 \mp 597.963i \right\} \\
\nonumber
\left\{ \beta_1, \beta_2, \beta_3 \right\} & = \left\{ \pm 10.284i, 4.305 \pm 9.619i, -4.305 \pm 9.619i \right\} \\
\left\{ c_{42}^1, c_{42}^2, c_{42}^3 \right\} & = \left\{ -2.363 \mp 10.284i, -34.843 \mp 14.508i, -26.234 \mp 4.730i \right\} \\
\nonumber
\left\{ c_{43}^1, c_{43}^2, c_{43}^3 \right\} & = \left\{ -46.582 \mp 10.284i, 110.706 \mp 262.135i, 119.315 \pm 242.896i \right\}.
\end{align}
Finally, the last state corresponds to the solution
\begin{align}
\nonumber
\left\{ X_1, X_2, X_3 \right\} & = \left\{ -6.735, -529.935, 110.004 \right\} \\
\nonumber
\left\{ \beta_1, \beta_2, \beta_3 \right\} & = \left\{ 0, 0, 0 \right\} \\
\left\{ c_{42}^1, c_{42}^2, c_{42}^3 \right\} & = \left\{ -21.690, 44.405, -22.715 \right\} \\
\nonumber
\left\{ c_{43}^1, c_{43}^2, c_{43}^3 \right\} & = \left\{ -45.221, 195.559, -30.338 \right\}
\end{align}
and the charges are as in \eqref{w3w4gschargeslvl3c}. This solution had all $X_j = 0$ in $\mathcal{W}_3$ Bethe equations, but from the point of view of $\mathcal{W}_4$ Bethe equations this degeneracy is lifted, even though we still have $\beta_j = 0$.

\subsubsection{Primary $\Delta = -\frac{3}{7}$ ($\mathcal{W}_3$ point of view)}
As a final example, we consider one of the two primaries with
\begin{equation}
\Delta = -\frac{3}{7}.
\end{equation}
The reason we choose this primary is that unlike primaries in Virasoro minimal models or other primaries in $(\mathcal{W}_3,\mathcal{W}_4)$ model, this primary is not self-conjugate. Therefore, the charge conjugation symmetry is not a symmetry of this representation and we find more non-vanishing odd $\mathcal{I}_n$ charges than in other examples we considered so far.

\paragraph{Level $0$}
We choose $W_{3,0}$ charge of the primary to be
\begin{equation}
W_{3,0} \ket{-\frac{3}{7}} = \frac{2}{7\sqrt{21}} \ket{-\frac{3}{7}}.
\end{equation}
The corresponding eigenvalues of $\mathcal{I}_n$ are
\begin{equation}
(\mathcal{I}_2,\mathcal{I}_3,\mathcal{I}_4,\mathcal{I}_5,\mathcal{I}_6,\mathcal{I}_7,\mathcal{I}_8) = \left( \frac{1}{4}, \frac{2}{7\sqrt{7}}, 0, 0, -\frac{697}{197568}, 0, \frac{313}{439040} \right).
\end{equation}

\paragraph{Level $1$}
At level $1$ we have one state and the corresponding solution of Bethe equations is
\begin{equation}
X_1 = 9, \qquad \beta_1 = -3.
\end{equation}
The eigenvalues of $\mathcal{I}_n$ are
\begin{equation}
\label{w3w4exchargeslvl1}
(\mathcal{I}_2,\mathcal{I}_3,\mathcal{I}_4,\mathcal{I}_5,\mathcal{I}_6,\mathcal{I}_7,\mathcal{I}_8) = \left( \frac{5}{4}, -\frac{12}{7 \sqrt{7}}, 0, 0, -\frac{18085}{197568}, 0, -\frac{41527}{439040} \right).
\end{equation}
Apart from this solution corresponding to a unique level $1$ state in the irreducible representation, we also see another solution of Bethe equations corresponding to a state in the Verma module which is removed when we go to the simple quotient. The corresponding solution is
\begin{equation}
X_1 = 0, \qquad \beta_1 = 3
\end{equation}
with charges
\begin{equation}
(\mathcal{I}_2,\mathcal{I}_3,\mathcal{I}_4,\mathcal{I}_5,\mathcal{I}_6,\mathcal{I}_7,\mathcal{I}_8) = \left( \frac{5}{4}, \frac{16}{7 \sqrt{7}}, 0, -\frac{16}{7 \sqrt{7}}, -\frac{179365}{197568}, 0, \frac{496073}{439040} \right).
\end{equation}
Since we are solving $\mathcal{W}_3$ Bethe equations, the $\mathcal{I}_4$ and $\mathcal{I}_7$ charges are automatically zero. The charges $\mathcal{I}_5, \mathcal{I}_9$ on the other hand should vanish for states in for $\mathcal{W}_4$ model (i.e. in the irreducible module), but typically do not vanish for the solutions corresponding to Verma module states.

\paragraph{Level $2$}
The character calculation shows us that we should expect two solutions at level $2$. They are
\begin{equation}
\left\{X_1,X_2\right\} = \left\{ 5.26 \mp 14.8i, -18.76 \pm 73.26i \right\}, \qquad \left\{\beta_1,\beta_2\right\} = \left\{ 0.22 \mp 6.55i, 1.27 \mp 1.79i \right\}.
\end{equation}
The charges are
\begin{multline}
\label{w3w4exchargeslvl2}
(\mathcal{I}_2,\mathcal{I}_3,\mathcal{I}_4,\mathcal{I}_5,\mathcal{I}_6,\mathcal{I}_7,\mathcal{I}_8) = \\
= \left( \frac{9}{4}, 0.486 \mp 2.104i, 0, 0, \frac{359327 \pm 15120 i \sqrt{31}}{197568}, 0, \frac{3559019 \pm 191520 i \sqrt{31}}{1317120}  \right).
\end{multline}
The $\mathcal{I}_3$ charges are related by a fourth-order equation
\begin{equation}
117649 \mathcal{I}_3^4+986468 \mathcal{I}_3^2+2560000 = 0.
\end{equation}
Two of its roots give us charges of two states we are studying while the other two solutions correspond to two states in the conjugate representation.

\paragraph{Level $3$}
Finally at level $3$ we expect three states. The first two solutions are
\begin{align}
\nonumber
\left\{X_1,X_2,X_3\right\} & = \left\{ -59.3858 \mp -23.7372i, -21.3028 \mp 166.6815i, 70.3056 \mp 27.2091i \right\} \\
\left\{\beta_1,\beta_2,\beta_3\right\} & = \left\{ -6.437 \pm 13.802i, 3.259 \mp 7.723i, 3.736 \pm 14.648i \right\}.
\end{align}
with charges
\begin{equation}
\label{w3w4exchargeslvl3a}
(\mathcal{I}_2,\mathcal{I}_3,\mathcal{I}_4,\mathcal{I}_5,\mathcal{I}_6,\mathcal{I}_7,\mathcal{I}_8) = \left( \frac{13}{4}, 0.249 \pm 5.22i, 0, 0, 18.9 \mp 0.475i, 0, 68.3 \mp 2.39i \right).
\end{equation}
The irrational charges satisfy
\begin{align}
\label{jnchargesdelta37l3}
\nonumber
0 & = -10241440000+29411075568 \mathcal{I}_3^2 + 2182624248 \mathcal{I}_3^4 + 40353607 \mathcal{I}_3^6 \\
\nonumber
0 & = 8558030406036357541 + 1868075948123751744 \mathcal{I}_6 \\
& - 268440760779706368 \mathcal{I}_6^2 + 7711694390034432 \mathcal{I}_6^3 \\
\nonumber
0 & = -1508018278425929430777 + 439158357852658510080 \mathcal{I}_8 \\
\nonumber
& - 11879956659062374400 \mathcal{I}_8^2 + 84627647627264000 \mathcal{I}_8^3.
\end{align}
The third solution of Bethe equations is
\begin{equation}
\left\{X_1,X_2,X_3\right\} = \left\{ -5.0300, 0.0781, -19.332 \right\}, \qquad \left\{\beta_1,\beta_2,\beta_3\right\} = \left\{ -15.87, -1.34, 19.10 \right\}.
\end{equation}
with charges
\begin{equation}
\label{w3w4exchargeslvl3b}
(\mathcal{I}_2,\mathcal{I}_3,\mathcal{I}_4,\mathcal{I}_5,\mathcal{I}_6,\mathcal{I}_7,\mathcal{I}_8) = \left( \frac{13}{4}, 0.5828, 0, 0, -3.089, 0, 3.817 \right).
\end{equation}
Their charges again satisfy the same equations \eqref{jnchargesdelta37l3} which show that all the three states are connected by the action of the Galois group that goes beyond simple complex conjugation. The odd charge $\mathcal{I}_3$ solves bi-cubic equation while the charges $\mathcal{I}_6$ and $\mathcal{I}_8$ solve simple cubic equation. These facts reflect the $\mathbbm{Z}_2$ conjugation symmetry in $\mathcal{W}_3$ algebra (which in this case maps the state to an associated state in charge conjugate representation).

\subsubsection{Primary $\Delta = -\frac{3}{7}$ ($\mathcal{W}_4$ point of view)}
Let us consider $\Delta = -\frac{3}{7}$ primary from point of view of $\mathcal{W}_4$ Bethe equations.

\paragraph{Level $1$}
At level $1$, we have solution
\begin{equation}
\left\{ X_1, \beta_1, c_{42}^1, c_{43}^1 \right\} = \left\{ 0, -4, 8, 20 \right\}
\end{equation}
and the corresponding charges agree with \eqref{w3w4exchargeslvl1}. In particular, we have a non-trivial $\mathcal{I}_3$ charge at the first excited level.

\paragraph{Level $2$}
At level $2$ we expect two states and there is indeed a complex conjugate pair,
\begin{align}
\nonumber
\left\{ X_1, X_2 \right\} & = \left\{ 65.469 \pm 24.801i, -109.802 \mp 63.775i \right\} \\
\nonumber
\left\{ \beta_1, \beta_2 \right\} & = \left\{ -0.105 \mp 14.497i, 2.105 \pm 3.362i \right\} \\
\left\{ c_{42}^1, c_{42}^2 \right\} & = \left\{ -9.761 \pm 15.926i, -46.496 \mp 0.495i \right\} \\
\nonumber
\left\{ c_{43}^1, c_{43}^2 \right\} & = \left\{ 10.848 \mp 6.394i, 109.409 \pm 13.234i \right\}
\end{align}
These correctly reproduce the charges \eqref{w3w4exchargeslvl2}.

\paragraph{Level $3$}
Finally, at level $3$ we have a pair of conjugate solutions
\begin{align}
\nonumber
\left\{ X_1, X_2, X_3 \right\} & = \left\{ 176.197 \mp 583.057i, -62.905 \pm 6.780i, 180.176 \pm 630.060i \right\} \\
\nonumber
\left\{ \beta_1, \beta_2, \beta_3 \right\} & = \left\{ -5.696 \pm 10.474i, -0.685 \pm 7.858i, 7.125 \pm 9.303i \right\} \\
\left\{ c_{42}^1, c_{42}^2, c_{42}^3 \right\} & = \left\{ -13.595 \mp 10.722i, -5.262 \mp 2.469i, -25.467 \mp 13.826i \right\} \\
\nonumber
\left\{ c_{43}^1, c_{43}^2, c_{43}^3 \right\} & = \left\{ 152.322 \pm 230.767i, -105.638 \pm 17.770i, 140.153 \mp 276.791i \right\}
\end{align}
reproducing the charges of \eqref{w3w4exchargeslvl3a} and a real solution
\begin{align}
\nonumber
\left\{ X_1, X_2, X_3 \right\} & = \left\{ -401.002, 165.148, -10.412 \right\} \\
\nonumber
\left\{ \beta_1, \beta_2, \beta_3 \right\} & = \left\{ -1.552, 3.581, 0.484 \right\} \\
\left\{ c_{42}^1, c_{42}^2, c_{42}^3 \right\} & = \left\{ 35.844, -32.634, -1.856 \right\} \\
\nonumber
\left\{ c_{43}^1, c_{43}^2, c_{43}^3 \right\} & = \left\{ 164.915, -19.035, -8.258 \right\}
\end{align}
which exactly reproduces \eqref{w3w4exchargeslvl3b}.

\subsection{Summary}
Even though we considered for concreteness only two minimal models and focused on lower energy levels, we saw that the structure and properties of the solutions are rather rich.

\begin{enumerate}
\item We studied Bethe equations in $(A_{K-1},A_{N-1})$ Argyres-Douglas VOAs. These admit two sets of Bethe equations -- equations corresponding to $\mathcal{W}_K$ algebra and equations corresponding to $\mathcal{W}_N$ algebra. For physical states, i.e. states in the irreducible module, we found an exact agreement between the eigenvalues of higher Hamiltonians, even though we do not know an explicit map identifying the solutions in the two duality frames.
\item There are other solutions that for generic values of parameters correspond to states in the associated Verma module that are quotiented out when going to the irreducible module, i.e. correspond to null states. The number of these solutions is related to whatever system of Bethe equations we are solving (either $\mathcal{W}_K$ or $\mathcal{W}_N$). The general properties of systems of algebraic equations such as those that we are studying imply that the number of solutions (if we correctly include also the singular ones) does not change as we vary the parameters, so after specialization of parameters we find spurious solutions corresponding to the null states.
\item The duality exchanging $K$ and $N$ does not preserve the number of states in the corresponding Verma module so the number of solutions is different between the two frames and we can only hope to identify the states in the irreducible module.
\item Some solutions corresponding to physical states are singular, i.e. some or all of the Bethe roots vanish (which is violating the assumptions that we used while deriving the Bethe equations) or the corresponding $\beta$ parameters are infinite, but nevertheless we reproduce the correct charges corresponding to the physical states.
\item We have no systematic way of eliminating the solutions corresponding to null states, but for Argyres-Douglas VOAs we have additional constraints coming from the vanishing of a subset of conserved charges (i.e. for physical states, both $\mathcal{I}_n$ with $n = jN+1$ and $n = jK+1$ should vanish while for Verma module states only one of these families is guaranteed to vanish). Requiring that all of these charges vanish eliminates the null states and gives a useful practical way of selecting the physical states.
\item Feigin-Frenkel duality should identify the solutions of the Bethe equations for the same $\mathcal{W}_N$ but different parameter $K$. Since for $K$ integer the dual $K^\prime$ is not an integer, we have to use the projected Bethe equations. We expect the duality in this situation to work also on the level of corresponding Verma modules, but again we do not have any explicit formula that would identify the solutions on both sides of the duality.
\item At levels where there is more than one solution, we saw that all of these were algebraically connected by the action of the Galois group. This is a reflection of the familiar property of quantum states in integrable models, the fact that as we vary parameters, all the states at given energy level are connected to one another \cite{Prochazka:2023zdb}\footnote{This is visually illustrated by an applet written by Miroslav Vel'k and available at \texttt{www.fzu.cz/\textasciitilde prochazkat/ilwsolver3}. It solves the ILW Bethe equations as a function of the twist parameter $q$ and one can study the monodromy of the solutions as $q$ is varied.}. In the related context, this important property is used in TBA literature to study excited states \cite{Dorey:1996re}.
\item In most of this paper we assume that the $\mathcal{I}_n$ are simultaneously diagonalizable and for generic values of the parameters this is expected to be the case. We however encountered one interesting example (level $3$, $\Delta=-\frac{5}{7}$ in $(\mathcal{W}_3,\mathcal{W}_4)$ minimal model) where $\mathcal{I}_3$ was not diagonalizable in the Verma module, although it became diagonalizable when we passed to the simple quotient. It would be interesting to see if such non-diagonalizability can happen also in the irreducible quotients.
\end{enumerate}

\section{Large $K$ and $N$ limit}
\label{seclargelimit}

Let us have a look what happens with the eigenvalues of the $\mathcal{I}_n$ of the highest weight state as $K$ and $N$ go to infinity. We see from \eqref{primarytricharges} that if the Young diagram $\mu$ parametrizing the highest weight is kept fixed as $K$ and $N$ go to infinity, for $n$ even the leading behavior is independent of $\mu$\footnote{For $n$ odd the corresponding coefficient of $(\lambda_1\lambda_2\lambda_3)^{n/2}$ vanishes for primaries with finite $\mu$ as well.}. We can therefore restrict to $\mu = \emptyset$ and we find
\begin{align}
\nonumber
\mathcal{I}_2 & = \frac{(K-1)(N-1)(KN+K+N)}{24(K+N)} \\
\nonumber
& \to \frac{K^2 N^2}{24(K+N)} = -\frac{\lambda_1 \lambda_2 \lambda_3}{24} \\
\nonumber
\mathcal{I}_4 & = \frac{(K-1)(K-3)(N-1)(N-3)(KN+K+N)(KN+3K+3N)}{1920(K+N)^2} \\
\nonumber
& \to \frac{K^4 N^4}{1920 (K+N)^2} = \frac{\lambda_1^2 \lambda_2^2 \lambda_3^2}{1920} \\
\nonumber
\mathcal{I}_6 & = \frac{(K-1)(K-5)(N-1)(N-5) (KN+K+N) (KN+5K+5N) }{580608(K+N)^3} \times \\
\nonumber
& \times \left(13K^2N^2-61K^2-61KN+139K-61N^2+139N\right) \\
& \to \frac{13K^6N^6}{580608(K+N)^3} = -\frac{13\lambda_1^3 \lambda_2^3 \lambda_3^3}{580608} \\
\nonumber
\mathcal{I}_8 & = \frac{(K-1)(K-7)(N-1)(N-7)(KN+K+N)(KN+7K+7N)}{199065600 (K+N)^4} \times \\
\nonumber
& \times \Big(281K^4N^4 - 4210K^4N^2 - 4210K^3N^3 - 4210K^2N^4 + 12888K^3N^2 +12888K^2N^3 \\
\nonumber
& + 12569K^4 + 25138K^3N +37707K^2N^2 + 25138KN^3 + 12569N^4 - 69576K^3 \\
\nonumber
& - 139152K^2N - 139152KN^2 - 69576N^3 + 90727K^2 + 181454KN + 90727N^2\Big) \\
\nonumber
& \to \frac{281\lambda_1^4\lambda_2^4\lambda_3^4}{199065600}
\end{align}
(the odd charges vanish for $\mu = \emptyset$ primary). This can be calculated both on VOA side, where we need to know the explicit expressions for the charges (and we use the fact that the $u_j^{(pl)}$ charges of this state are zero), or from the ODE side. The result is manifestly triality invariant, because the translationally invariant vacuum state $\mu = \emptyset$ preserves the triality symmetry. The leading order behavior of the even charges $\mathcal{I}_{2n}$ with $(\lambda_1\lambda_2\lambda_3)^n$ factored out is
\begin{multline}
\Big\{ -\frac{1}{24}, \frac{1}{1920}, -\frac{13}{580608}, \frac{281}{199065600}, -\frac{163}{1513881600}, \\
\frac{20177107}{2191186722816000}, -\frac{37826207}{44497945755648000}, \ldots \Big\}.
\end{multline}

Another interesting primary state whose large $N$ and $K$ limit we can study is the lowest dimension primary, i.e. the ground state on the cylinder. Since the corresponding Young diagram $\mu$ has roughly $\frac{KN}{2}$ boxes, which grows as $N, K \to \infty$, the leading order behavior of the charges in large $N$ and $K$ limit is different from the vacuum state. We find
\begin{align}
\nonumber
\mathcal{I}_2 & = -\frac{(K-1)(N-1)}{24 (K+N)} \to \frac{1}{24} \lambda_2 \\
\nonumber
\mathcal{I}_4 & = -\frac{(K-1)(K-3)(N-1)(N-3)(2K+2N-1)}{1920(K+N)^2} \\
& \to \frac{1}{960} \lambda_1 \lambda_2 \lambda_3 \\
\nonumber
\mathcal{I}_6 & = -\frac{(K-1)(K-5)(N-1)(N-5)}{580608(K+N)^3} \times \\
\nonumber
& \times \Big(8K^3N^2 + 48K^3N - 152K^3 + 8K^2N^3 + 96K^2N^2 - 474K^2N + 478K^2 \\
\nonumber
& + 48KN^3 - 474KN^2 + 969KN - 309K - 152N^3 + 478N^2 - 309N + 61 \Big) \\
& \to -\frac{1}{72576} (\lambda_1 \lambda_2 \lambda_3)^2 \\
\mathcal{I}_8 & = -\frac{(K-1)(K-7)(N-1)(N-7)}{199065600(K+N)^4} \times \\
\nonumber
& \Big(96K^5N^4 + 768K^5N^3 - 1584K^5N^2 - 18048K^5N + 36048K^5 + 96K^4N^5 \\
\nonumber
& + 1536K^4N^4 - 3616K^4N^3 - 57964K^4N^2 + 235168K^4N - 238100K^4 + 768K^3N^5 \\
\nonumber
& - 3616K^3N^4 - 79832K^3N^3 + 525836K^3N^2 - 963976K^3N + 480980K^3 - 1584K^2N^5 \\
\nonumber
& - 57964K^2N^4 + 525836K^2N^3 - 1452033K^2N^2 + 1456108K^2N - 337523K^2 \\
\nonumber
& - 18048KN^5 + 235168KN^4 - 963976KN^3 + 1456108KN^2 - 679256KN + 104884K \\
\nonumber
& + 36048N^5 - 238100N^4 + 480980N^3 - 337523N^2 + 104884N - 12569\Big) \\
& \to \frac{1}{2073600} (\lambda_1 \lambda_2 \lambda_3)^3
\end{align}
In general, the leading coefficient of $\mathcal{I}_{2n}$ after we factor out $(\lambda_1 \lambda_2 \lambda_3)^{n-1}$ is
\begin{equation}
\Big\{ 0, \frac{1}{960}, -\frac{1}{72576}, \frac{1}{2073600}, -\frac{1}{35481600}, \frac{691}{285310771200}, \ldots \Big\}.
\end{equation}
We recognize $691$ as a numerator of $12$th Bernoulli number and in fact this sequence is of the form
\begin{equation}
\mathcal{I}_n \sim (-1)^{n/2-1} \frac{(n-1)B_{n-2}B_n}{2n(n-2)!} (\lambda_1 \lambda_2 \lambda_3)^{n-1} + \ldots.
\end{equation}
We can compare this to asymptotic expansion of the MacMahon function
\begin{equation}
\mathbb{M}(q) = \prod_{n=1}^\infty \frac{1}{(1-q^n)^n}
\end{equation}
which has asymptotic expansion
\begin{equation}
\log \mathbb{M}(q = e^{i\hbar}) \sim -\frac{\zeta(3)}{\hbar^2} + \frac{1}{12} \log(-i\hbar) + \zeta^\prime(-1) + \sum_{g=2}^\infty F_g \hbar^{2g-2}
\end{equation}
with
\begin{equation}
F_g = (-1)^{g-1} \frac{B_{2g-2}B_{2g}}{2g(2g-2)(2g-2)!}, \qquad g \geq 2
\end{equation}
as is very well known from the topological strings literature \cite{Grassi:2022zuk,Marino:2005sj,zagier2006appendix}. We have
\begin{align}
\mathcal{I}_n & \sim (-1)^{n/2-1} \frac{(n-1)B_{n-2}B_n}{2n(n-2)!} (\lambda_1 \lambda_2 \lambda_3)^{n-1} + \ldots \\
& \sim {n-1 \choose 2} F_{n/2} \, (\lambda_1 \lambda_2 \lambda_3)^{n-1} + \ldots
\end{align}
so that
\begin{equation}
\log \mathbb{M}(e^{-\epsilon}) \sim \frac{\zeta(3)}{\epsilon^2} + \frac{1}{12} \log\epsilon + \zeta^\prime(-1) + \frac{1}{\lambda_1 \lambda_2 \lambda_3} \sum_{g=2}^\infty \frac{(-1)^{g-1} \mathcal{I}_{2g}}{(g-1)(2g-1)} \left(\frac{\epsilon}{\lambda_1 \lambda_2 \lambda_3}\right)^{2g-2}.
\end{equation}
We see that the logarithm of MacMahon function is an asymptotic generating function of the charges of the ground state, i.e. that we can extract from it the ground state eigenvaluesvalues of $\mathcal{I}_n$ (the higher Casimir energies).

\subsection{$N=2$ and large $K$}
\label{secschrliouville}
Instead of considering both $N$ and $K$ large, we can also study the case of finite rank $N$ and large $K$. For concreteness, let us first focus on $N=2$ and $K \to \infty$, i.e. the large central charge limit of the Virasoro integrals of motion. We start with the differential operator
\begin{equation}
\hbar^2 \partial_x^2 + x^K - 1
\end{equation}
representing the Virasoro ground state. In the $K \to \infty$ limit this potential is singular. We therefore introduce a new variable
\begin{equation}
e^X = x^{K+2} \qquad \text{or} \qquad x = e^{\frac{X}{K+2}} \quad \text{or} \quad X = (K+2) \log x.
\end{equation}
We could have introduced another power of $x$ such as $x^K$ but these give the same large $K$ limit. Furthermore, since around $x = \infty$ there are $K+2$ Stokes sectors rotated by Symanzik symmetry, choosing $x^{K+2}$ seems to be more appropriate. Changing the coordinate and rescaling the wave function to remove the first derivative term, the differential operator is transformed to
\begin{equation}
\tilde{\hbar}^2 \partial_X^2 + e^X - e^{\frac{2X}{K+2}} - \frac{\tilde{\hbar}^2}{4(K+2)^2}
\end{equation}
with
\begin{equation}
\tilde{\hbar} \equiv (K+2)\hbar.
\end{equation}
So far we only changed the coordinate so locally the new differential operator is equivalent to the original one (globally this is no longer obvious as the coordinate $X$ gives an infinite sheeted covering of the complex $x$-plane).

As a next step, we take $K \to \infty$ limit keeping $\tilde{\hbar}$ fixed. We find a simple equation
\begin{equation}
\label{schrliouville}
\left(\hbar^2 \partial_X^2 + e^X - 1\right)\psi(X) = 0
\end{equation}
where from now on we write $\hbar$ instead of $\tilde{\hbar}$. This is a Schr\"odinger equation in the background Liouville field. Just as before, we can study the small $\hbar$ asymptotics of solutions of this equation using the WKB ansatz. Writing the wave function in the form
\begin{equation}
\psi(X) = \exp \left[ \frac{1}{\hbar} \int^X Y(X^\prime)dX^\prime \right],
\end{equation}
we see that $Y(X)$ satisfies the Riccati equation
\begin{equation}
\hbar Y^\prime(X) + Y^2(X) + e^X - 1.
\end{equation}
We look for an asymptotic expansion of $Y(X)$ of the form
\begin{equation}
Y(X) = \sum_{n=0}^\infty \hbar^n Y_n(X).
\end{equation}
The leading term gives the Liouville WKB curve
\begin{equation}
y^2 + e^X = 1
\end{equation}
where we identify $y$ with $Y_0$. As before, the WKB ansatz generates an infinite collection of $1$-forms living on the WKB curve, the first few being
\begin{align}
\label{n2largekasymptotics}
\nonumber
Y_1(X)dX & = \frac{e^X dX}{4y^2} \\
\nonumber
Y_2(X)dX & = \left( -\frac{1}{8} \frac{e^X}{y^2} - \frac{5}{32} \frac{e^{2X}}{y^4} \right) \frac{dX}{y} \\
Y_3(X)dX & = \left( \frac{1}{16} \frac{e^X}{y^2} + \frac{9}{32} \frac{e^{2X}}{y^4} + \frac{15}{64} \frac{e^{3X}}{y^6} \right) \frac{dX}{y^2} \\
\nonumber
Y_4(X)dX & = \left( -\frac{1}{32} \frac{e^X}{y^2} - \frac{47}{128} \frac{e^{2X}}{y^4} - \frac{221}{256} \frac{e^{3X}}{y^6} - \frac{1105}{2048} \frac{e^{4X}}{y^8} \right) \frac{dX}{y^3}.
\end{align}
Just as before, the WKB curve projects to three-punctured sphere where the period integrals are easy to evaluate. In fact, introducing the coordinate
\begin{equation}
t = e^X
\end{equation}
so that
\begin{equation}
y^2 + t = 1, \qquad y = \sqrt{1-t}, \qquad X = \log t,
\end{equation}
we can express the $1$-forms as living on the $t$-plane,
\begin{align}
\nonumber
Y_1(X) dX & = \frac{dt}{4(1-t)} \\
\nonumber
Y_2(X) dX & = -\frac{5}{32} \frac{tdt}{(1-t)^{5/2}} - \frac{1}{8} \frac{dt}{(1-t)^{3/2}} \\
Y_3(X) dX & = \frac{t^2 dt}{64 (1-t)^4} + \frac{5t dt}{32(1-t)^4} + \frac{dt}{16 (1-t)^4} \\
\nonumber
Y_4(X) dX & = -\frac{25t^3 dt}{2048(1-t)^{11/2}} - \frac{57t^2 dt}{256(1-t)^{11/2}} - \frac{35t dt}{128(1-t)^{11/2}} - \frac{dt}{32(1-t)^{11/2}}
\end{align}
etc. The $Y_{2n-1}(X) dX$ are single-valued so their contour integrals give vanishing contribution as is expected for Virasoro algebra. The even $1$-forms $Y_{2n}(X)dX$ have square-root type branch points at $t = 1$ and $t = \infty$.

The integrals along figure eight contours around these two points vanish and so does the integral along the Pochhammer contour, because one of the three singular points is in this case regular and the contour integral around the other two points can be contracted to a point. In order to get a non-trivial contour integral, we consider instead an open Hankel contour $\mathcal{H}$ that starts at $t=0$ (which is a regular point of $Y_n$), encircles $t=1$ counter-clockwise and returns back to $t=0$. Because we are going once around square root type branch point at $t=1$, when we return to $t=0$, we have the opposite sign of $Y_n$, i.e. the contour is not closed. Nevertheless, this prescription gives up to an overall factor periods which nicely agree with large $K$ limit of integrals over closed contour. We need to evaluate the integrals of the form
\begin{equation}
\int_{\mathcal{H}} t^j (1-t)^{-\frac{3n-1}{2}} dt = 2B\left(j+1,\frac{3(1-n)}{2}\right).
\end{equation}
Using this formula, we find
\begin{equation}
\int_{\mathcal{H}} Y_{2n}(t)dt = \left\{ \frac{1}{12}, -\frac{1}{1440}, \frac{1}{20160}, -\frac{1}{107520}, \frac{1}{304128}, -\frac{691}{369008640}, \ldots \right\}
\end{equation}
and more generally
\begin{equation}
\int_{\mathcal{H}} Y_{2n}(t)dt = \frac{B_{2n}}{2^{2n-1}n(2n-1)}.
\end{equation}
The large $K$ limits of ground state expectation values of $\mathcal{I}_{2n}$ (rescaled by a suitable power of $K$) are instead
\begin{equation}
\lim_{K \to \infty} K^{1-n} \mathcal{I}_{2n} = \left\{-\frac{1}{24}, \frac{1}{960}, -\frac{1}{8064}, \frac{1}{30720}, -\frac{1}{67584}, \frac{691}{67092480}, \ldots \right\}.
\end{equation}
Comparing these, we see that
\begin{equation}
\label{largevirasoromap}
\lim_{K \to \infty} K^{1-n} \mathcal{I}_{2n} = -\frac{2n-1}{2} \int_{\mathcal{H}} Y_{2n}(t)dt.
\end{equation}
We were only comparing eigenvalues evaluated on one state and using it to fix the relative normalization between $\mathcal{I}_{2n}$ and the contour integral (so in particular we could always fix the normalization to get equality between the two calculations). Note however that the relative normalization in \eqref{largevirasoromap} is a very simple linear function, in particular much simpler than the eigenvalues themselves (which involve the Bernoulli numbers).

\subsection{$N$ fixed and large $K$}
We can repeat the discussion of the previous section in the case of $N>2$ (Boussinesq operator for $N=3$). The starting point is the ground state operator for $\mathcal{W}_N$ algebra
\begin{equation}
\hbar^N \partial_x^N + x^K - 1
\end{equation}
and the new variable is
\begin{equation}
e^X = x^{K+N} \qquad \text{or} \qquad x = e^{\frac{X}{K+N}} \qquad \text{or} \qquad X = (K+N) \log X.
\end{equation}
The rescaled Planck constant is now
\begin{equation}
\tilde{\hbar} = (K+N) \hbar
\end{equation}
and the differential operator in the $K \to \infty$ limit keeping $\tilde{\hbar}$ fixed becomes simply
\begin{equation}
\label{largeNoper}
\tilde{\hbar}^N \partial_X^N + e^X - 1,
\end{equation}
i.e. we still have Liouville potential but the kinetic term is generalized to $N$-th power of the momentum. In the following, we will write $\hbar$ instead of $\tilde{\hbar}$. The first few $1$-forms that we have to integrate are
\begin{align}
\nonumber
Y_1(X)dX & = -\frac{N-1}{2} \frac{Y_0^\prime(X)}{Y_0(X)} \\
\nonumber
Y_2(X)dX & = \frac{N^2-1}{12} \frac{Y_0^{\prime\prime}(X)}{Y_0(X)^2} -\frac{N^2-1}{8} \frac{Y_0^\prime(X)^2}{Y_0(X)^3} \\
\nonumber
Y_3(X)dX & = - \frac{N^2-1}{4} \frac{Y_0^\prime(X)^3}{Y_0(X)^5} + \frac{N^2-1}{4} \frac{Y_0^\prime(X) Y_0^{\prime\prime}(X)}{Y_0(X)^4} - \frac{N^2-1}{24} \frac{Y_0^{\prime\prime\prime}(X)}{Y_0(X)^3} \\
Y_4(X)dX & = \frac{(N^2-1)(7N^2-127)}{128} \frac{Y_0^\prime(X)^4}{Y_0(X)^7} -\frac{(N^2-1)(7N^2-127)}{96} \frac{Y_0^\prime(X)^2 Y_0^{\prime\prime}(X)}{Y_0(X)^6} \\
\nonumber
& +\frac{(N^2-1)(N^2-19)}{72} \frac{Y_0^{\prime\prime\prime}(X) Y_0^\prime(X)}{Y_0(X)^5} +\frac{(N^2-1)(N^2-17)}{96} \frac{Y_0^{\prime\prime}(X)^2}{Y_0(X)^5} \\
\nonumber
& -\frac{(N^2-1)(N^2-19)}{720} \frac{Y_0^{(4)}(X)}{Y_0(X)^4}.
\end{align}
Transforming these to $t$-coordinate (i.e. writing them as $1$-forms on the mirror curve), we find
\begin{align}
\nonumber
Y_1(X)dX & = \frac{N-1}{2N} \frac{dt}{1-t} \\
\nonumber
Y_2(X)dX & = \left( -\frac{N^2-1}{12N} -\frac{N^2-1}{24N^2} t \right) \frac{dt}{(1-t)^{2+\frac{1}{N}}} \\
\nonumber
Y_3(X)dX & = \left( \frac{(N-1) (N+1) t^2}{24 N^3}+\frac{(N-1) (N+1) (N+3) t}{24 N^2}+\frac{(N-1) (N+1)}{24 N} \right) \frac{dt}{(1-t)^{3+\frac{2}{N}}} \\
Y_4(X)dX & = \left( \frac{(N^2-1)(N^2-19)}{720 N} + \frac{(N^2-1) (8N^3+21N^2-152N-369)t}{1440N^2} \right. \\
\nonumber
& \left. +\frac{(N^2-1)(2N^4+12N^3-11N^2-228N-483)t^2}{1440N^3} + \frac{(N^2-1)(9N^2-161)t^3}{1920N^4} \right) \frac{dt}{(1-t)^{4+\frac{3}{N}}}.
\end{align}
In general, these are of the form
\begin{equation}
Y_n(X)dX = \frac{P_n(t)}{(1-t)^{\frac{Nn+n-1}{N}}} dt
\end{equation}
where $P_n(t)$ is a polynomial of degree $n-1$ in $t$. We integrate these again along the Hankel contour $\mathcal{H}$ as described in the previous section. The integrals are of the form
\begin{equation}
\int_{\mathcal{H}} \frac{t^j}{(1-t)^{\frac{Nn+n-1}{N}}} dt = \left( 1 - e^{-\frac{2\pi i (n-1)}{N}} \right) B\left(j+1,-\frac{(N+1)(n-1)}{N}\right)
\end{equation}
In order to reproduce large $K$ limits of $\mathcal{I}_{2n}$, we normalize the integrals as
\begin{equation}
\lim_{K \to \infty} \frac{\mathcal{I}_{2n}}{K^{n-1}} \leftrightarrow \frac{1-2n}{1 - e^{-\frac{2\pi i (2n-1)}{N}}} \mathcal{P}_{2n} \equiv \frac{1-2n}{1 - e^{-\frac{2\pi i (2n-1)}{N}}} \int_{\mathcal{H}} Y_{2n}(X) dX.
\end{equation}
For small $N$, the first of these expectation values are:
\begin{center}
\begin{tabular}{|c|c|c|c|c|c|c|}
\hline
$N$ & $\mathcal{I}_2$ & $K^{-1} \mathcal{I}_4$ & $K^{-2} \mathcal{I}_6$ & $K^{-3} \mathcal{I}_8$ & $K^{-4} \mathcal{I}_{10}$ & $K^{-5} \mathcal{I}_{12}$ \\
\hline
$2$ & $-1/24$ & $1/960$ & $-1/8064$ & $1/30720$ & $-1/67584$ & $691/67092480$ \\
$3$ & $-1/12$ & $0$ & $1/2268$ & $-1/6480$ & $0$ & $691/7960680$ \\
$4$ & $-1/8$ & $-1/320$ & $1/1152$ & $17/30720$ & $-1/2048$ & $-691/1064960$ \\
$5$ & $-1/6$ & $-1/120$ & $0$ & $1/400$ & $7/3300$ & $-691/163800$ \\
$6$ & $-5/24$ & $-1/64$ & $-265/72576$ & $671/165888$ & $31/2048$ & $69093781/3260694528$ \\
$7$ & $-1/4$ & $-1/40$ & $-1/84$ & $0$ & $13/308$ & $691/3528$ \\
\hline
\end{tabular}
\end{center}
Just like in the limit of both $N$ and $K$ large, we recognize the Bernoulli numbers. Factoring these out, we find that the sequence of eigenvalues
\begin{equation}
a_n \equiv \frac{n}{B_{2n}} \lim_{K \to \infty} \frac{\mathcal{I}_{2n}}{K^{n-1}}
\end{equation}
for each fixed $N$ satisfies a linear recurrence relation of order $N-1$ which allows for very fast evaluation of these quantities. For example $N=2$ we have a linear relation
\begin{equation}
a_{n+1} = \frac{1}{4} a_{n}
\end{equation}
from which we can easily calculate
\begin{equation}
\lim_{K \to \infty} \frac{\mathcal{I}_{2n}}{K^{n-1}} = -\frac{B_{2n}}{n 2^{2n}}
\end{equation}
Analogously, for $N=3$ we have relation
\begin{equation}
a_{n+2} = \frac{1}{3} a_{n+1} - \frac{1}{9}
\end{equation}
with initial condition
\begin{equation}
a_1 = -\frac{1}{2}, \qquad a_2 = 0.
\end{equation}
This can be solved and we find
\begin{equation}
a_n = \left(\sqrt{3}\right)^{1-2n} (-1)^n \cos\left(\frac{2n-1}{6}\pi\right)
\end{equation}
so that for $N=3$
\begin{equation}
\lim_{K \to \infty} \frac{\mathcal{I}_{2n}}{K^{n-1}} = -\frac{B_{2n}}{n} \left(\sqrt{3}\right)^{1-2n} (-1)^n \cos\left(\frac{2n-1}{6}\pi\right)
\end{equation}
and similarly for higher values of $N$.

\subsection{$K$ fixed and large $N$}
Let us finally sketch what happens if we instead send the order of the differential operator $N$ to infinity while keeping $K$ fixed. This corresponds to formally Fourier transforming the differential operator \eqref{largeNoper}. The result is
\begin{equation}
e^{-\hbar \partial_x} + x^K - 1.
\end{equation}
Acting on wave functions, this gives simply
\begin{equation}
\label{deltaeqnlargen}
\left(e^{-\hbar \partial_x} + x^K - 1\right) \psi(x) = \psi(x-\hbar) - \left( 1-x^N \right) \psi(x),
\end{equation}
i.e. we have a finite difference operator acting on $\psi(x)$. In the semiclassical limit $\hbar \to 0$ this finite difference operator reduces to a differential operator and we can apply the WKB analysis analogously to the case of ordinary differential operators. The WKB curve is
\begin{equation}
e^{-y} + x^K = 1.
\end{equation}
The advantage of considering the difference operator is that one can formally find solutions in terms of gamma functions. Let us show this on the example of $N=2$, i.e. Schr\"odinger equation with Liouville potential. In this case the equation is solvable in both duality frames. In the Schr\"odinger frame, there are two linearly independent solutions of the differential equation \eqref{schrliouville} written in terms of Bessel functions,
\begin{equation}
J_{\pm 2 \hbar^{-1}}\left(2\hbar^{-1} e^{X/2} \right)
\end{equation}
(here we assume that $\hbar$ is generic, i.e. that $2\hbar^{-1}$ is not an integer in which case the two solutions would become linearly dependent). The solution with the upper choice of the sign,
\begin{equation}
\label{schrliousol}
\psi(X) = J_{2 \hbar^{-1}}\left(2\hbar^{-1} e^{X/2} \right)
\end{equation}
has asymptotic expansion for $\hbar \to 0$ compatible with the analysis of Section \ref{secschrliouville}. Fourier transforming the wave function
\begin{equation}
\hat{\psi}(Y) = \int_{-\infty}^{+\infty} e^{-\hbar^{-1}XY} J_{2\hbar^{-1}}\left(2\hbar^{-1} e^{X/2}\right) dX,
\end{equation}
changing the variable to
\begin{equation}
w = 2\hbar^{-1} e^{X/2} \qquad \text{and} \qquad X = 2\log \left(\frac{\hbar w}{2}\right)
\end{equation}
and using the formula \texttt{[DLMF 10.22.43]}
\begin{equation}
\int_0^\infty x^\mu J_\nu(t)dt = 2^\mu \frac{\Gamma\left(\frac{\nu+\mu+1}{2}\right)}{\Gamma\left(\frac{\nu-\mu+1}{2}\right)}
\end{equation}
for the resulting Mellin integral, we arrive at the expression
\begin{equation}
\hat{\psi}(Y) = \hbar^{-2\hbar^{-1}Y} \frac{\Gamma\left(\frac{1-Y}{\hbar}\right)}{\Gamma\left(\frac{1+Y}{\hbar}+1\right)}.
\end{equation}
It is easy to check explicitly that it satisfies \eqref{deltaeqnlargen} with $K=2$ and the variable $x$ replaced by $Y$. The inverse Fourier transformation gives a representation of the solution \eqref{schrliousol} as a contour integral \texttt{[DLMF 10.9.22]}
\begin{equation}
\psi(X) = \frac{1}{2\pi i} \int_{1-i\infty}^{1+i\infty} e^{\hbar^{-1} XY} \hbar^{-2\hbar^{-1} Y} \frac{\Gamma\left(\frac{1-Y}{\hbar}\right)}{\Gamma\left(\frac{1+Y}{\hbar}+1\right)} dY = \frac{1}{2\pi i} \int_{1-i\infty}^{1+i\infty} e^{\hbar^{-1} XY} \hat{\psi}(Y) dY
\end{equation}
and the integration contour should lie to the left of the sequence of poles of $\hat{\psi}(Y)$ that are at
\begin{equation}
Y = 1+n\hbar, \quad n=0,1,2,\ldots.
\end{equation}

\section{Discussion and open questions}
In this article we discussed some aspects of ODE/IM correspondence applied to $\mathcal{W}_N$ family of algebras. The main new results are the following:

\begin{itemize}
\item We explicitly constructed local integrals of motion in $\mathcal{W}_N$ or $\mathcal{W}_\infty$ up to current density of dimension $12$. These charges are essentially unique: the local charges are uniquely determined (up to normalization) by the requirement that they commute with the zero mode of the unique non-trivial dimension $3$ current. The currents themselves can be uniquely determined (up to normalization) by requiring them to be quasi-primary. There is a choice of normalization that makes them triality invariant.

\item Via ODE/IM correspondence, we can associate simultaneous eigenvectors of these local charges with ordinary differential operators. The general form of these is given in \cite{Feigin:2007mr,Masoero:2018rel} and we determined explicit Bethe equations for Virasoro algebra, $\mathcal{W}_3$ algebra and $\mathcal{W}_4$ algebra.

\item According to ODE/IM correspondence, the corresponding eigenvalues can be determined from the spectral or the monodromy data of these differential operators. In our case we focused on the formal WKB periods. We showed how introducing correctly the Planck parameter $\hbar$ allows to calculate all WKB periods systematically, i.e. we can easily generate explicit expressions for the eigenvalues of quantum Hamiltonians in terms of solutions of Bethe ansatz equations at any level (this is in contrast with many examples in the literature where only first one or two integrals are given or where the authors specialize to states at first levels \cite{Bazhanov:2003ni,Ashok:2024zmw,Ito:2024kza,Ashok:2024ygp}).

\item The description of the eigenvalues in terms of Bethe roots has the following nice factorization property: the usual calculation using commutation relations requires us to calculate for every Virasoro level and every quantum Hamiltonian the corresponding matrix elements and find their eigenvalues. If we approach the same problem from the point of view of quantum integrability, we can discuss both things separately: first we associate to every quantum eigenstate the corresponding solution of Bethe equations. Every such solution compactly encodes the eigenvalues of the whole infinite family of quantum Hamiltonians. The second step is for each quantum Hamiltonian to find the expression for its eigenvalues in terms of Bethe roots. The resulting formulas are universal symmetric polynomials in Bethe roots and apply uniformly to all quantum eigenstates.

\item We tried as much as possible to work with the whole $\mathcal{W}_\infty$ family, making the symmetries such as the triality symmetry manifest. In particular, by focusing on differential operators of the form \eqref{hwoper} and using the Argyres-Douglas parametrization (see for example \cite{Fioravanti:2004cz} for related discussion), we kept the duality $K \leftrightarrow N$ at the level of the quantum curve manifest. For the other choices of coordinates (such as the one used in \cite{Litvinov:2013zda}) this duality is not manifest. The full triality symmetry is only manifest at the level of the associated classical WKB curve $\mathcal{W}$ and especially its projection, the mirror curve $\mathcal{M}$.

\item We considered the mirror curve $\mathcal{M}$ as the most relevant geometric object that controls the WKB calculations. The triality symmetry is realized as permutation of punctures, the Calabi-Yau condition \eqref{winfcentralcharge} is realized by the fact that the total abelian monodromy around all three punctures is tautologically trivial. The basic structure function of $\mathcal{W}_\infty$ which captures the box as a basic building block of the representation theory of the algebra \cite{Prochazka:2015deb} is reflected in the intersection form in the homology of $\mathcal{W}$.

We discussed the geometry of the WKB curves $\mathcal{W}$ and the mirror curve $\mathcal{M}$ in detail, in particular we explained the procedure to put a natural hyperbolic metric on these curves and used this to find geodesic representative classes for the homology classes. We used the basis of homology that respects the symmetries of the problem instead using the usual symplectic basis.

\item We tested Bethe equations as well as the universal expressions for higher charges on examples of $(A_1,A_2)$ and $(A_2,A_3)$ Argyres-Douglas VOAs at lower levels. We found exact match between explicit calculations using commutation relations on one hand and the results of WKB calculations on the other hand. We illustrated on examples the difficulties that one faces when selecting the right solutions of BAE.
\end{itemize}

Let us list some of possible unsolved questions and future directions:

\begin{itemize}

\item \textbf{Bethe ansatz equations for $\mathcal{W}_N$ and affine Gaudin} \\
Even though the oper description works for any $\mathcal{W}_N$, we see that the Bethe equations themselves in the form that we derived them for $\mathcal{W}_3$ and $\mathcal{W}_4$ do not seem to show any simple structure as we increase the rank $N$. It would be very nice to find a system of equations that would be analogous to nested Bethe ansatz equations such as those for $\mathfrak{gl}(N)$ Gaudin model or Heisenberg XXX spin chain \cite{Kulish:1983rd,Jurco:1989mg,Slavnov:2019hdn}. One issue is that not even the parameters parametrizing the descendants are of the same type, i.e. for $\mathcal{W}_3$ oper the parameters $x_j$ parametrize positions of the regular points on $\mathbbm{C}P^1$, while the parameters such as $\beta_j$ in \eqref{w3bethe1} and \eqref{w3bethe2} are coefficients of these poles. It would be useful to have a parametrization where the different sets of parameters are on the same footing. One possible way would be to link the present discussion to the affine Gaudin models as considered in \cite{Feigin:2007mr,Gaiotto:2020dhf,masoero2023q}. These integrable models should also shed light on the integrability properties of more general Grassmannian VOAs such as those considered in \cite{Eberhardt:2019xmf,Eberhardt:2020zgt}.

\item \textbf{Link to Litvinov's Bethe equation and ILW} \\
Bethe equations of BLZ type are not the only Bethe equations we can write down for $\mathcal{W}_N$ family. There are also generalizations of Heisenberg XXX-type equations with third degree magnon interaction \cite{Nekrasov:2009rc,Litvinov:2013zda,Kozlowski:2016too,Bonelli:2014iza,Galakhov:2022uyu,Prochazka:2023zdb} that diagonalize the Hamiltonians of the quantum ILW hierarchy. These equations have several advantages over those that we considered here. First of all, they work uniformly and take the same form for all $\mathcal{W}_N$ algebras (and even for other quotients of $\mathcal{W}_{1+\infty}$ such as the parafermion algebra). Second, they include an additional twist parameter $q$ which appears naturally if the ILW Hamiltonians are constructed following the algebraic Bethe ansatz procedure starting from the instanton $\mathcal{R}$-matrix \cite{Maulik:2012wi,Smirnov:2013hh,Litvinov:2013zda,Zhu:2015nha,Prochazka:2019dvu,Prochazka:2023zdb}. This twist parameter is also very convenient for numerical studies of solutions of the equations \cite{Prochazka:2023zdb}. In the related context of quantum toroidal algebra, the expressions for the corresponding eigenvalues of Hamiltonians as functions of Bethe roots were found in \cite{feigin2017finite,feigin2017finite2} and these equations were extensively tested in \cite{Prochazka:2023zdb}. The ILW Hamiltonians are in general non-local and one can only extract local Hamiltonians in the singular limit as $q \to 1$. It would be interesting to understand a link between these two approaches to quantum integrability of $\mathcal{W}_\infty$. In particular, a regularized version of the generating function of ILW eigenvalues in $q \to 1$ limit would give a natural generating function of local $\mathcal{W}_\infty$ conserved charges.

\item \textbf{Non-perturbative completion, TBA equations} \\
We focused on eigenvalues of local quantum Hamiltonians and found that these are encoded in the asymptotic WKB expansion of the associated differential equation. But all these calculations were formal. The WKB ansatz for the wave function is typically not convergent and we cannot evaluate them for complex values of the spectral parameter (Planck constant). It would be quite useful to have access to non-perturbatively completed generating functions of quantities we are interested in. In particular \cite{Ito:2017ypt,Ito:2018eon,Ito:2025pfo} define quantum periods as well-defined analytic functions of $\hbar$ whose asymptotic expansion encodes the conserved quantities. These non-perturbative completions can be defined using Wronskians of solutions \cite{Ito:2021boh,Ito:2021sjo} of the relevant differential equation or even more interestingly as solution of integral equations of TBA type \cite{Zamolodchikov:1989cf,Gaiotto:2009hg,Gaiotto:2014bza}. The structure of these equations reflects the T-system or Y-system which in turn is closely related to the fusion ring of representations of the corresponding algebra \cite{Kuniba:2010ir}. It would be nice to understand the form of the corresponding TBA-like system of equations for any $\mathcal{W}$-algebra of our class and in particular see if the geometrical structures studied here translate directly to such systems of equations.

\item \textbf{Modularity and holomorphic anomaly equations, mirror symmetry} \\
Another interesting aspect is the relation to modularity \cite{Iles:2014gra,Maloney:2018hdg,Downing:2021mfw}. The partition functions of minimal models of $\mathcal{W}_N$ are modular invariant and the question is what happens if one refines these by inserting higher Hamiltonians into the partition functions. The local densities of conserved currents usually transform in a simple way under such transformation, but the contour integral of the current which results in the conserved charge breaks the modular symmetry by a preferred choice of an integration cycle. An argument of Dijkgraaf \cite{Dijkgraaf:1996iy,Maloney:2018hdg} shows that as long as we insert just one conserved charge in our partition function, the modularity is preserved, essentially due to the fact that we can use time independence of the quantity to average the charge over the dual cycle. If we try to repeat this argument with two or more insertions, we encounter the problem of contact terms. Even if by construction the zero modes of conserved currents commute, this is not true for the currents themselves -- they typically have short distance singularities. Dijkgraaf's analysis shows that as a consequence of these short distance singularities, there are certain modular anomalies. These are in turn controlled by the structure of the symmetry algebra itself. It would be interesting to understand how this is generalized to the whole $\mathcal{W}_\infty$ family. For recent discussion of Lee-Yang $(A_1,A_2)$ model see \cite{Downing:2023lnp}.

The modular invariance connects in a non-trivial way the low energy and high energy properties of CFT. In particular, the properties of highly excited states are correlated with the properties near the ground state of the theory and having access to local conserved quantities gives us refined observables that we can use as probes. Calculating these for typical high energy states using the usual commutation relations seems to be quite difficult, but in contrast the Bethe equations for highly excited states can simplify and often give a geometric description of a typical highly excited state (for example in the form of the geometry of the corresponding limit shape \cite{okounkov2016limit}). This is quite parallel to the mirror symmetry in topological strings: the low energy and highly quantum expansion around $q \to 0$ corresponds to box counting picture \cite{Okounkov:2003sp} and can be identified with calculations of characters of $\mathcal{W}_\infty$ as power series \cite{feigin2012quantum,Prochazka:2015deb}. On the other hand, the mirror description starts with a geometric object such as the mirror curve corresponding to $q \to 1$ limit of the partition function and which captures the shape of a typical highly excited state. The corrections around the geometric limit $q \to 1$ can be calculated using similar tools as we employed here, by first promoting the relevant geometric curve to a quantum curve and applying WKB techniques to it \cite{Grassi:2014zfa,Marino:2015nla,Grassi:2019coc,Grassi:2022zuk}.



\end{itemize}

\section*{Acknowledgments}
We would like to thank to Federico Ambrosino, Takashi Aoki, Tomoyuki Arakawa, Christopher Beem, Vinicius Bernardes, Giulio Bonelli, Federico Bonetti, Jean-Emile Bourgine, Max Downing, Lorenz Eberhardt, Davide Fioravanti, Pavlo Gavrylenko, Alba Grassi, Tamara Grava, Andreas Hohl, Katsushi Ito, Konstantin Jakob, Saebyeok Jeong, Taro Kimura, Reinier Kramer, Shota Komatsu, Oleg Lisovyy, Marcos Mari\~no, Davide Masoero, Shigenori Nakatsuka, Stefano Negro, Andrew Neitzke, Fabrizio Nieri, Go Noshida, Ulisses Portugal, Paolo Rossi, Martin Schnabl, Alessandro Tanzini and Yegor Zenkevich for useful discussions. TP would like to thank the organizers and participants of the workshop Supersymmetric Quantum Field Theory and Mathematics in Pollica, the workshop BPS Dynamics and Quantum Mathematics at GGI Firenze and the programme Supersymmetric Quantum Field Theories, Vertex Operator Algebras, and Geometry at SCGP for many stimulating discussions. TP would also like to thank for the hospitality to CERN Theory group where part of this work was done. The research leading to these results has received support from the European Structural and Investment Funds and the Czech Ministry of Education, Youth and Sports (project No. \texttt{FORTE-CZ.02.01.01/00/22\_008/0004632}).

\appendix

\section{Summation over $\mathbbm{Z}_n$ orbits}
\label{apporbitsums}
The basic identity for summation over $\mathbbm{Z}_n$ orbits is the identity between rational functions
\begin{equation}
\sum_{j=0}^{n-1} \frac{1}{x-e^{\frac{2\pi i j}{n}} y} = \frac{nx^{n-1}}{x^n-y^n} = \partial_x \log (x^n-y^n)
\end{equation}
which one can derive by comparing singularities as well as asymptotic behavior at infinity. The identities for higher order poles can be derived easily by differentiating this basic one, i.e. we have
\begin{equation}
\sum_{j=0}^{n-1} \frac{1}{\left(x-e^{\frac{2\pi i j}{n}}y\right)^2} = \frac{nx^{n-2}(x^n+(n-1)y^n)}{\left(x^n-y^n\right)^2}
\end{equation}
and
\begin{equation}
\sum_{j=0}^{n-1} \frac{1}{\left(x-e^{\frac{2\pi i j}{n}}y\right)^3} = \frac{nx^{n-3} \left(2x^{2n} + (n-1)(n+4) x^n y^n + (n-1)(n-2) y^{2n}\right)}{2 \left(x^n-y^n\right)^3}.
\end{equation}
We also need a specialization where $x^n=y^n$ but $x \neq y$. In this case, we need to subtract the $j=0$ contribution before taking the limit and find
\begin{align}
\nonumber
\sum_{j=1}^{n-1} \frac{1}{1-e^{\frac{2\pi i j}{n}}} & = \frac{n-1}{2} \\
\sum_{j=1}^{n-1} \frac{1}{\left(1-e^{\frac{2\pi i j}{n}}\right)^2} & = -\frac{(n-1)(n-5)}{12} \\
\nonumber
\sum_{j=1}^{n-1} \frac{1}{\left(1-e^{\frac{2\pi i j}{n}}\right)^3} & = -\frac{(n-1)(n-3)}{8}.
\end{align}

\section{Uniformization}
\label{appuniform}

\subsection{Construction of the uniformizing map}

Although not necessary for the evaluation of the period integrals, it is useful to put a metric on $\mathcal{W}$ (and on $\mathcal{M}$) \cite{caratheodory1954theory,siegel1969topics,gray2008linear}. This in particular allows us to represent the homology classes in terms of canonical representatives, the geodesics of shortest length in each homology class. As a secondary result, we also find a triangulation of $\mathcal{W}$ which is useful for visualizations.

One of the basic results in the theory of compact Riemann surfaces is the uniformization theorem which states that in each equivalence class of metrics under the action of Weyl transformations, there is up to an overall scale a unique metric of constant scalar curvature. For genus zero the curvature is positive, for genus $1$ it is zero and for all higher genera we have a metric of constant negative curvature . Each choice of the metric determines uniquely a compatible complex structure and Weyl-equivalent metrics lead to the same complex structure (this is essentially due to the fact that the complex structure corresponds to $\frac{\pi}{2}$ rotation in the tangent space and Weyl transformations preserve the angles). The opposite implication is true as well, a choice of complex structure determines the metric up to a choice of the Weyl factor. This is our starting point in order to find the constant curvature metric: given a complex coordinate $z$, the constant curvature metric should be of the form
\begin{equation}
\label{unifmetricansatz}
ds^2 = e^{2\Phi(z,\bar{z})} dz d{\bar{z}} = e^{2\Phi(x,y)}(dx^2+dy^2).
\end{equation}
We would like to find the Weyl factor $\Phi$ such that the metric has curvature equal to $-1$ (since the genus of $\mathcal{W}$ is greater than one except for the case of Lee-Yang model $(K,N)=(2,3)$). It is easy to calculate from the metric \eqref{unifmetricansatz} the Christoffel symbols and the components of the curvature tensors. The components of the metric and of the inverse metric are
\begin{equation}
g_{z\bar{z}} = g_{\bar{z}z} = \frac{1}{2}e^{2\Phi} \qquad \text{and} \qquad g^{z\bar{z}} = g^{\bar{z}z} = 2e^{-2\Phi},
\end{equation}
the non-vanishing Christoffel symbols are
\begin{equation}
\Gamma_{zz}^z = g^{z\bar{z}} \partial_z g_{z\bar{z}} = 2\partial_z \Phi \qquad \text{and} \qquad \Gamma_{\bar{z}\bar{z}}^{\bar{z}} = 2 \partial_{\bar{z}} \Phi,
\end{equation}
the Riemann tensor has non-vanishing components
\begin{equation}
{{R_{z\bar{z}}}^z}_z = -2\partial_z \partial_{\bar{z}} \Phi \qquad \text{and} \qquad R_{z\bar{z}z\bar{z}} = e^{2\Phi} \partial_z \partial_{\bar{z}} \Phi
\end{equation}
and the Ricci tensor and scalar curvature are
\begin{equation}
R_{z\bar{z}} = R_{\bar{z}z} = -2\partial_z \partial_{\bar{z}} \Phi \qquad \text{and} \qquad R = -8e^{-2\Phi} \partial_z \partial_{\bar{z}} \Phi.
\end{equation}
In general, the differential equation for $\Phi$ that we need to solve to get the metric of scalar curvature $R$ is
\begin{equation}
\label{liouvilleequation}
\partial_z \partial_{\bar{z}} \Phi = -\frac{R}{8} e^{2\Phi}
\end{equation}
which is the well-known Liouville equation. For this reason, we sometimes call $\Phi$ the \emph{Liouville field}. Constant shifts of $\Phi$ correspond to overall rescalings of the metric and of $R$. It is not difficult to find solutions of this equation, for example the round two-sphere of unit curvature radius (and the scalar curvature $R=2$) has
\begin{equation}
e^{2\Phi} = \frac{4}{(1+|z|^2)^2}
\end{equation}
The difficulty of finding $\Phi$ lies in finding a solution on a higher genus surface with the correct periodicity conditions along all the cycles of $\mathcal{W}$.

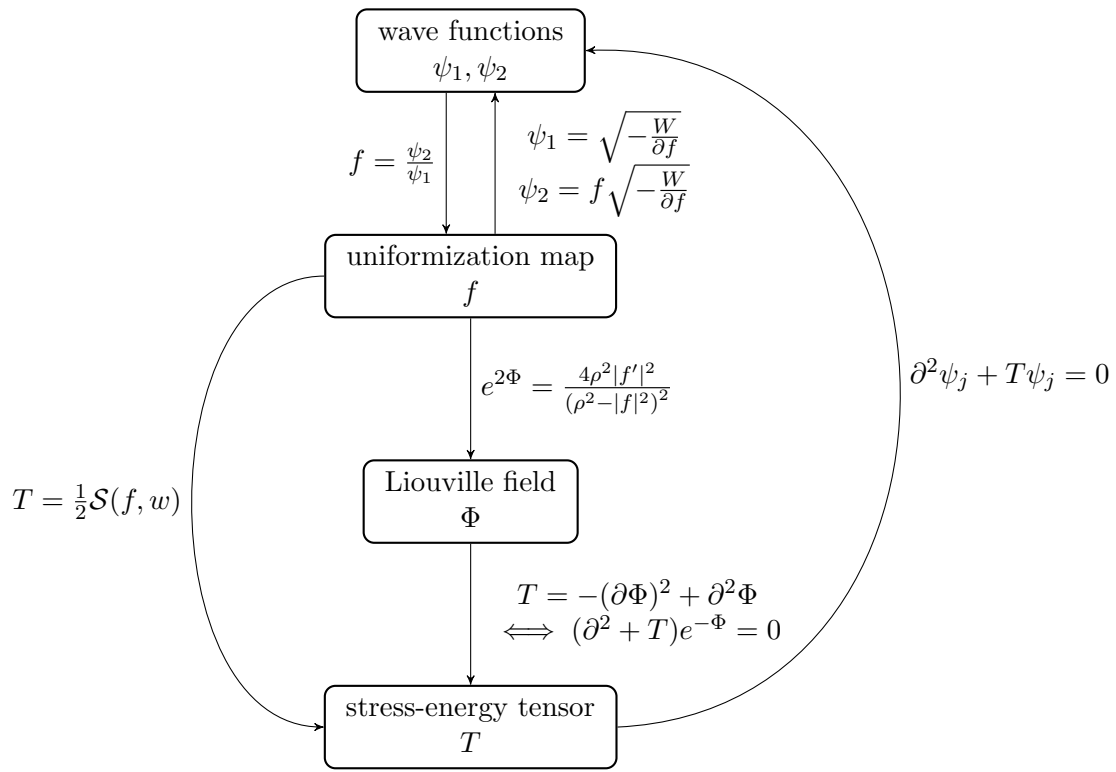
\begin{figure}
\centering
\begin{tikzpicture}[->,>=stealth']

\node[state] (wavefn) {\begin{tabular}{c} wave functions \\ $\psi_1,\psi_2$ \end{tabular}};
\node[state,below of=wavefn,node distance=3cm,anchor=center] (uni) {\begin{tabular}{c} uniformization map \\ $f$ \end{tabular}};
\node[state,below of=uni,node distance=3cm,anchor=center] (liou) {\begin{tabular}{c} Liouville field \\ $\Phi$ \end{tabular}};
\node[state,below of=liou,node distance=3cm,anchor=center] (stresst) {\begin{tabular}{c} stress-energy tensor \\ $T$ \end{tabular}};

\path (wavefn.240) edge node[anchor=east,left]{$f = \frac{\psi_2}{\psi_1}$} (uni.120);
\path (uni.60) edge node[anchor=west,right]{$\begin{array}{c} \psi_1 = \sqrt{-\frac{W}{\partial f}} \\ \psi_2 = f\sqrt{-\frac{W}{\partial f}} \end{array}$} (wavefn.300);
\path (uni) edge node[anchor=west,right]{$e^{2\Phi} = \frac{4\rho^2 |f^\prime|^2}{\left(\rho^2-|f|^2\right)^2}$} (liou);
\path (liou) edge node[anchor=west,right]{$\begin{array}{c} T = -(\partial \Phi)^2 + \partial^2 \Phi \\ \iff (\partial^2+T)e^{-\Phi}=0 \end{array}$} (stresst);
\path (stresst) edge[bend right=90,looseness=1.5] node[anchor=west,right]{$\partial^2 \psi_j + T \psi_j = 0$} (wavefn);
\path (uni) edge[bend right=90,looseness=1] node[anchor=east,left]{$T = \frac{1}{2}\mathcal{S}(f,w)$} (stresst);
\end{tikzpicture}
\caption{Uniformization procedure that we use to find the constant curvature metric on $\mathcal{W}$ and $\mathcal{M}$.}
\label{unifscheme}
\end{figure}

\paragraph{Uniformizing map}
We are going to solve \eqref{liouvilleequation} by finding a multiple-valued holomorphic map from $\mathcal{M}$ to the covering surface of $\mathcal{W}$ which is the hyperbolic plane (in our situation represented by the Poincar\'e disk model). Using this map, we can pull back the hyperbolic metric from the Poincar\'e disk to $\mathcal{W}$, to the $x$-plane, $y$-plane or all the way to $\mathcal{M}$. Let us for future convenience rescale the hyperbolic disk $\mathcal{D}$ so that the coordinate range is $|z|<\rho$ and the metric is
\begin{equation}
\label{hyperbolicdiskmetric}
ds^2_{\mathcal{D}} = \frac{4\rho^2 dz d\bar{z}}{(\rho^2-|z|^2)^2}.
\end{equation}
The corresponding Liouville field is
\begin{equation}
\Phi_{\mathcal{D}} = \frac{1}{2} \log \left( \frac{4\rho^2}{(\rho^2-|z|^2)^2} \right)
\end{equation}
and we can verify that it solves the Liouville equation \eqref{liouvilleequation} with $R=-2$ which is the correct scalar curvature for hyperbolic space of the curvature radius $1$. The parameter $\rho$ in \eqref{hyperbolicdiskmetric} does not change the scalar curvature but only rescales the coordinate region representing the hyperbolic disk $\mathcal{D}$ in the $z$-plane. Having this flexibility will be convenient in the following, in particular, with suitable $\rho$ the uniformization map will have Taylor expansion with rational coefficients.

Given a holomorphic map to $\mathcal{D}$ (represented by a multi-valued function $z(w)$), we can pull back the metric on $\mathcal{D}$ to $w$-plane and get the metric
\begin{equation}
ds^2 = \frac{4\rho^2 dw d\bar{w}}{(\rho^2-|z(w)|^2)^2} \left| \frac{dz}{dw} \right|^2.
\end{equation}
The corresponding Liouville field in the $w$-plane is
\begin{equation}
\label{unifpullbackmetric}
\Phi^{(w)} = \frac{1}{2} \log \left( \frac{4\rho^2}{(\rho^2-|z(w)|^2)^2} \left| \frac{dz}{dw} \right|^2 \right) = \log \left( \frac{2\rho|dz/dw|}{\rho^2-|z(w)|^2} \right).
\end{equation}
This works for any holomorphic function $z(w)$ in any region where it is single-valued so locally we can produce infinitely many solutions of the Liouville equation as pull-backs with respect to holomorphic maps\footnote{In the context of the field theory, this is just the B\"{a}cklund transformation between the classical Liouville equation and the wave equation in two dimensions whose solutions are holomorphic and anti-holomorphic functions. Given any holomorphic solution of the wave function, the B\"{a}cklund transformation produces a local solution of the Liouville equation.}.

\paragraph{Stress-energy tensor}
The Liouville field $\Phi$ is a real function and in particular neither holomorphic nor anti-holomorphic. We can extract from it a holomorphic quantity\footnote{Ultimately we will be interested in studying $\Phi$ on $\mathcal{M}$ so we could use the $t$-coordinate, but for the current discussion let us just consider an abstract surface with holomorphic coordinate $w$.}
\begin{equation}
\label{uniftdef}
T(w) = -(\partial_w \Phi(w,\bar{w}))^2 + \partial_w^2 \Phi(w,\bar{w}).
\end{equation}
Using the Liouville equation \eqref{liouvilleequation} we see that it is indeed holomorphic,
\begin{equation}
\partial_{\bar{w}} T = -2 \partial_w \partial_{\bar{w}} \Phi \partial_w \Phi + \partial_w^2 \partial_{\bar{w}} \Phi = \frac{R}{4} e^{2\Phi} \partial_w \Phi - \frac{R}{8} \partial_w e^{2\Phi} = 0.
\end{equation}
For Liouville field on the sphere or on the hyperbolic disk $\mathcal{D}$ we have $T = 0$.

\paragraph{Conformal transformations}
Let us now have a look how $\Phi$ and $T$ transform under holomorphic maps. Given a holomorphic map $z(w)$ between two Riemann surfaces, the invariance of the metric (i.e. the fact that we can use $z(w)$ to pull back the metric)
\begin{equation}
e^{2\Phi^{(w)}(w)} dw d\bar{w} = e^{2\Phi^{(z)}(z)} dz d\bar{z} = e^{2\Phi^{(z)}(z(w))} \left| \frac{\partial z}{\partial w} \right|^2 dw d\bar{w}
\end{equation}
implies that the transformation of $\Phi$ is
\begin{equation}
2\Phi^{(w)} = 2 \Phi^{(z)} + \log \left(\frac{\partial z}{\partial w}\right) + \log \left(\frac{\partial \bar{z}}{\partial \bar{w}}\right).
\end{equation}
Therefore the corresponding holomorphic quantity $T$ transforms as
\begin{equation}
\label{unifttransf}
T^{(w)} = -(\partial_w \Phi^{(w)})^2 + \partial_w^2 \Phi^{(w)} = \left( \frac{\partial z}{\partial w} \right)^2 T^{(z)} + \frac{1}{2} \mathcal{S}(z,w)
\end{equation}
where
\begin{equation}
\mathcal{S}(z,w) \equiv \frac{\frac{\partial^3 z}{\partial w^3}}{\frac{\partial z}{\partial w}} - \frac{3}{2} \frac{\left(\frac{\partial^2 z}{\partial w^2}\right)^2}{\left(\frac{\partial z}{\partial w}\right)^2}
\end{equation}
is the Schwarzian derivative. This is the classical transformation property of the stress-energy tensor (sometimes called the projective connection). The homogeneous part is that of quadratic differential, but there is an additional inhomogeneous contribution controlled by the Schwarzian derivative.

\paragraph{Schr\"odinger equation}
Given a classical stress-energy tensor transforming under conformal transformations as \eqref{unifttransf}, it is natural to consider the associated differential operator
\begin{equation}
\label{unifschr}
\partial_z^2 + T^{(z)}(z).
\end{equation}
Under holomorphic map $z(w)$ the differential operator transforms to a form which also involves the first order derivative. We can eliminate this first derivative term (and therefore preserve the form \eqref{unifschr}) by letting wave functions on which \eqref{unifschr} acts transform as weight $-1/2$-form, i.e.
\begin{equation}
\psi^{(w)}(w) = \left( \frac{\partial z}{\partial w} \right)^{-1/2} \psi^{(z)}(z(w)).
\end{equation}
and transforming $T(z)$ with Schwarzian derivative as in \eqref{unifttransf}. The image of the action of \eqref{unifschr} on $\psi(z)$ turns out to be a $3/2$-form. The wave function acquired this particular transformation property because we required that there is no first order derivative term generated in the transformation, i.e. that the conformal transformation preserves the form of the differential operator \eqref{unifschr}.

In terms of the Schr\"odinger operator \eqref{unifschr}, the stress-energy tensor \eqref{uniftdef} and the Liouville field are connected via relation
\begin{equation}
\label{unifphischrk}
(\partial_z^2 + T)e^{-\Phi} = 0.
\end{equation}

\paragraph{Liouville field from solutions of Schr\"odinger equation}
We have seen how to associate an auxiliary Schr\"odinger equation to a solution of Liouville equation. Now we want to invert this, i.e. reconstruct the Liouville field from the associated Schr\"odinger equation. The first step is to find two linearly independent wave functions in the kernel of \eqref{unifschr},
\begin{equation}
\label{unifschrpsi}
\partial_w^2 \psi_j(w) + T(w) \psi_j(w) = 0, \qquad j=1,2.
\end{equation}
The Wronskian of the two solutions is constant because the ODE does not have the first derivative term \eqref{unifschr}. The ratio of the two solutions
\begin{equation}
\label{unifsolrat}
f_{\psi}(w) \equiv \frac{\psi_2(w)}{\psi_1(w)}
\end{equation}
is holomorphic and satisfies
\begin{equation}
\partial_w f_\psi = \frac{\psi_1 \partial_w \psi_2 - \psi_2 \partial_w \psi_1}{\psi_1^2} = \frac{W}{\psi_1^2}.
\end{equation}
This relation can be solved for $\psi_1$ and $\psi_2$ and we find
\begin{equation}
\psi_1 = \sqrt{\frac{W}{\partial_w f_\psi}} \qquad \text{and} \qquad \psi_2 = f_{\psi} \sqrt{\frac{W}{\partial_w f_\psi}}.
\end{equation}
Under a linear transformation of solutions $\psi_1$ and $\psi_2$ the function $f_{\psi}$ undergoes a fractional linear transformation. It is easy to see that the Schwarzian derivative of $f_\psi$ gives back the stress-energy tensor (the Schr\"odinger potential),
\begin{equation}
\frac{1}{2} \mathcal{S}(f_\psi,w) = T(w).
\end{equation}
The relation \eqref{unifphischrk} and its complex conjugate together with the fact that $\psi_1$ and $\psi_2$ form a basis of solutions of \eqref{unifschrpsi} implies that $e^{-\Phi}$ can be expanded as a linear combination of products of $\psi_j$ and their complex conjugates,
\begin{equation}
\label{unifexpmphi}
e^{-\Phi} = \sum_{j,k=1}^2 \psi_j^*(\bar{w}) \Lambda_{jk} \psi_k(w) = \psi^\dagger \Lambda \psi.
\end{equation}
The reality of $\Phi$ requires that $\Lambda$ is a hermitian matrix. Plugging \eqref{unifexpmphi} in the Liouville equation \eqref{liouvilleequation} we find an additional constraint on $\Lambda$
\begin{equation}
\partial_w \partial_{\bar{w}} \Phi = \frac{(\psi^\dagger \Lambda \partial_w \psi)(\partial_{\bar{w}}\psi^\dagger \Lambda \psi)-(\psi^\dagger \Lambda \psi)(\partial_{\bar{w}}\psi^\dagger \Lambda \partial_w  \psi)}{\left(\psi^\dagger \Lambda \psi\right)^2} = \frac{1}{4\left(\psi^\dagger \Lambda \psi\right)^2}.
\end{equation}
Using the Fierz rearrangement for a $2 \times 2$ matrix $\Lambda$ and four commuting $2$-vectors $\alpha$, $\beta$, $\gamma$ and $\delta$
\begin{equation}
(\alpha \cdot \Lambda \cdot \beta)(\gamma \cdot \Lambda \cdot \delta) - (\alpha \cdot \Lambda \cdot \delta)(\gamma \cdot \Lambda \cdot \beta) = \det \Lambda \cdot (\alpha \cdot \epsilon \cdot \gamma) (\beta \cdot \epsilon \cdot \delta),
\end{equation}
the requirement on $\Lambda$ becomes
\begin{equation}
-\det \Lambda \cdot |W|^2 = \frac{1}{4},
\end{equation}
i.e. the determinant of $\Lambda$ is fixed to be
\begin{equation}
\det \Lambda = -\frac{1}{4|W|^2}.
\end{equation}
There are various consistent choices of $\Lambda$ satisfying this condition and the hermiticity condition. These choices are related to choices of a basis of solutions $\psi_1$ and $\psi_2$ and correspond to different $SL(2,\mathbbm{C})$ images of the hyperbolic disk in the complex plane. The choice that we will make is
\begin{equation}
\Lambda = \frac{1}{2|W|} \begin{pmatrix} \rho & 0 \\ 0 & -\rho^{-1} \end{pmatrix}.
\end{equation}
Given this choice, the Liouville field becomes
\begin{equation}
e^{2\Phi} = \frac{4|W|^2}{\left(\rho|\psi_1|^2-\rho^{-1}|\psi_2|^2\right)^2} = \frac{4\rho^2|\partial_w f_\psi|^2}{\left(\rho^2-|f_\psi|^2\right)^2}.
\end{equation}
Comparing this with \eqref{unifpullbackmetric}, we see that we can identify $f_\psi$ with the uniformizing map mapping the $w$-plane into Poincar\'e disk of radius $\rho$ in the $z$-plane, such that the pull-back of the hyperbolic metric with respect to $f_\psi(w)$ gives a solution of Liouville equation in $w$-plane we were looking for.

\paragraph{Branch points}
The last element we have to look at before solving the uniformization problem that we are interested in is the behavior of $T(z)$ and $\Phi(z)$ under branched covering maps. Until this point we were assuming that the coordinate transformations $z(w)$ were locally holomorphic and one-to-one (i.e. conformal). Consider now what happens for a transformation of the form $w = z^n$ which is what in our case happens around the branch points $t=0$, $t=1$ and $t=\infty$. We have
\begin{align}
\label{uniftpoles}
T^{(w)}(w) & = \left(\frac{\partial z}{\partial w}\right)^2 T^{(z)}(z(w)) + \frac{1}{2} \mathcal{S}(z,w) \\
& = \frac{1}{w^2} \left[ \frac{w^{\frac{2}{n}}}{n^2} T^{(z)}(z(w)) + \frac{n^2-1}{4n^2} \right].
\end{align}
Assuming that in the $z$-coordinate $T^{(z)}$ vanishes, i.e. it comes from a constant curvature metric, we see that the branched covering transformation $w = z^n$ produces a double pole at the branch point $w = 0$ with its coefficient related to the deficit angle $2\pi\left(1-\frac{1}{n}\right) \equiv 2\pi(1-\alpha)$ that the pulled-back metric has around $w = 0$. For $n \pm 1$ this double pole disappears as it has to because in this case the transformation is locally holomorphic.

To summarize, the stress-energy tensor $T^{(w)}(w)$ is a meromorphic object in the $w$-plane transforming as \eqref{unifttransf} under changes of coordinates and with double points at the positions of the branch points.

\paragraph{Uniformization of the three-punctured sphere}
Let us now turn to our main example, the one of three-punctured sphere. We choose three punctures to be at $w=0$, $w=1$ and $w=\infty$ and the associated angles $2\pi\alpha$, $2\pi\beta$ and $2\pi\gamma$. The stress-energy tensor $T(w)$ should have double poles as in \eqref{uniftpoles} at $w=0$ and $w=1$ and is otherwise holomorphic in the finite part of the $w$-plane, so has the form
\begin{equation}
\label{uniftexp}
T(w) = \frac{1-\alpha^2}{4} \frac{1}{w^2} + \frac{1-\beta^2}{4} \frac{1}{(w-1)^2} + \frac{\rho}{w} + \frac{\sigma}{w-1} + reg.
\end{equation}
We should still impose the condition of having a double pole at infinity with the correct coefficient. Remembering that $T(w)$ transforms as in \eqref{unifttransf} under changes of coordinates, i.e.
\begin{equation}
T^{(1/w)}(1/w) = w^4 T^{(w)}(w),
\end{equation}
the requirement of vanishing third order pole at $w=\infty$ is $\rho + \sigma = 0$ while having the correct coefficient of the double pole imposes
\begin{equation}
\frac{1-\gamma^2}{4} = \frac{4\sigma-\alpha^2-\beta^2+2}{4}.
\end{equation}
These two equations completely determine the undetermined parameters $\rho$ and $\sigma$ and simultaneously fix the regular terms \eqref{uniftexp} to zero, so the stress-energy tensor in $w$-plane is
\begin{equation}
T(w) = \frac{1-\alpha^2}{4} \frac{1}{w^2} + \frac{1-\beta^2}{4} \frac{1}{(w-1)^2} + \frac{\alpha^2+\beta^2-\gamma^2-1}{4w(w-1)}.
\end{equation}
We see that $T(w)$ (living on $\mathcal{M}$) is completely determined the branch points and the ramification data.

It remains to solve the corresponding second order ordinary differential equation \eqref{unifschr}. In our situation this differential equation is a Fuchsian ODE with three regular singular points and as such its solutions can be written in terms of Gauss hypergeometric functions. The Frobenius indices around $w=0$ are $\frac{1\pm\alpha}{2}$ and solutions themselves can be chosen as
\begin{align}
\nonumber
\psi_1 & = w^{\frac{1-\alpha}{2}} (w-1)^{\frac{1-\beta}{2}} {}_2 F_1 \left( \frac{1-\alpha-\beta+\gamma}{2}, \frac{1-\alpha-\beta-\gamma}{2}; 1-\alpha; w \right) \\
\psi_2 & = w^{\frac{1+\alpha}{2}} (w-1)^{\frac{1-\beta}{2}} {}_2 F_1 \left( \frac{1+\alpha-\beta+\gamma}{2}, \frac{1+\alpha-\beta-\gamma}{2}; 1+\alpha; w \right).
\end{align}
The ratio of these solutions \eqref{unifsolrat} gives us the uniformization map that we were looking for, the Schwarz triangle function
\begin{equation}
\label{unifmap}
z = s_{\alpha,\beta,\gamma}(w) \equiv w^{\alpha} \frac{{}_2 F_1 \left( \frac{1+\alpha-\beta+\gamma}{2}, \frac{1+\alpha-\beta-\gamma}{2}; 1+\alpha; w \right)}{{}_2 F_1 \left( \frac{1-\alpha-\beta+\gamma}{2}, \frac{1-\alpha-\beta-\gamma}{2}; 1-\alpha; w \right)}.
\end{equation}
This is the main result of this section, an explicit expression for the multi-valued map from the three-punctured sphere to Poincar\'e disk which uniformizes the three-punctured sphere. In the following section we will study properties of this map.

\subsection{Properties of the uniformizing map}

\paragraph{Schwarz triangle}
The map \eqref{unifmap} is not a single-valued meromorphic function due to prefactor $w^\alpha$ and also due to the fact that the standard hypergeometric function defined by power series expansion has a branch cut in the interval $(1,\infty)$. If we choose the principal branch of \eqref{unifmap}, i.e. $s_{\alpha,\beta,\gamma}(w)$ single valued and holomorphic in $\mathbbm{C}$ with cuts at $(-\infty,0)$ and $(1,\infty)$, the image of the upper and lower half $w$-plane are two hyperbolic triangles in the $z$-space. It is obvious that the point $w=0$ is mapped to the origin of the $z$-plane
\begin{equation}
w = 0 \mapsto z = 0,
\end{equation}
which we label $A$. Using Gauss's summation theorem (i.e. formula for ${}_2 F_1$ at $w=1$), we can find the image of $w=1$,
\begin{equation}
w = 1 \mapsto z = \frac{\Gamma(1+\alpha)\Gamma\left(\frac{1-\alpha+\beta-\gamma}{2}\right)\Gamma\left(\frac{1-\alpha+\beta+\gamma}{2}\right)}{\Gamma(1-\alpha)\Gamma\left(\frac{1+\alpha+\beta-\gamma}{2}\right)\Gamma\left(\frac{1+\alpha+\beta+\gamma}{2}\right)},
\end{equation}
the vertex $B$ of the two hyperbolic triangles. Finally, due to multi-valued nature of $s_{\alpha,\beta,\gamma}$, the image of $w=\infty$ depends on whether we approach it from the upper- or lower-half $w$-plane. We have
\begin{equation}
w = \pm i \infty \mapsto z = e^{\pm i \pi \alpha} \frac{\Gamma(1+\alpha)\Gamma\left(\frac{1-\alpha-\beta+\gamma}{2}\right)\Gamma\left(\frac{1-\alpha+\beta+\gamma}{2}\right)}{\Gamma(1-\alpha)\Gamma\left(\frac{1+\alpha-\beta+\gamma}{2}\right)\Gamma\left(\frac{1+\alpha+\beta+\gamma}{2}\right)}
\end{equation}
and we will label these two points $C$ and $\bar{C}$ (with $C$ being in the upper half $z$-plane).

The interval $(0,1)$ is mapped to the segment of the real axis in $z$-plane connecting $A$ to $B$. Similarly, the negative real axis $(-\infty,0)$ is mapped to straight lines of slope $e^{\pm i\pi\alpha}$ connecting the point $A$ to $C$ and $\bar{C}$ (there is a branch cut in the definition of $s_{\alpha,\beta,\gamma}(t)$ along the negative real axis and approaching it from above gives $AC$ while approaching it from below gives $A\bar{C}$). The third side, connecting points $B$ and $C$ in $z$-plane is not a straight line but a circular segment. We can parametrize it as a M\"obius transformation
\begin{equation}
z \mapsto \frac{\alpha a e^{i\pi(\alpha+\gamma)}z - \alpha \gamma b e^{i\pi\alpha}}{c e^{i\pi\gamma}z - \gamma d}
\end{equation}
of the interval
\begin{equation}
\left( 0, \frac{\Gamma(1+\gamma) \Gamma\left( \frac{1-\alpha+\beta-\gamma}{2} \right) \Gamma\left( \frac{1+\alpha+\beta-\gamma}{2} \right)}{\Gamma(1-\gamma) \Gamma\left( \frac{1-\alpha+\beta+\gamma}{2} \right) \Gamma\left( \frac{1+\alpha+\beta+\gamma}{2} \right)} \right) \subset \mathbbm{R}
\end{equation}
and where the coefficients $a, b, c$ and $d$ define a $SL(2,\mathbbm{R})$ matrix
\begin{equation}
\label{unifdefabcd}
\begin{pmatrix} a & b \\ c & d \end{pmatrix} = \begin{pmatrix} \frac{\Gamma(1-\gamma)\Gamma(\alpha)}{\Gamma\left(\frac{1+\alpha-\beta-\gamma}{2}\right)\Gamma\left(\frac{1+\alpha+\beta-\gamma}{2}\right)} & \frac{\Gamma(\gamma)\Gamma(\alpha)}{\Gamma\left(\frac{1+\alpha-\beta+\gamma}{2}\right)\Gamma\left(\frac{1+\alpha+\beta+\gamma}{2}\right)} \\ \frac{\Gamma(1-\gamma)\Gamma(1-\alpha)}{\Gamma\left(\frac{1-\alpha-\beta-\gamma}{2}\right)\Gamma\left(\frac{1-\alpha+\beta-\gamma}{2}\right)} & \frac{\Gamma(\gamma)\Gamma(1-\alpha)}{\Gamma\left(\frac{1-\alpha-\beta+\gamma}{2}\right)\Gamma\left(\frac{1-\alpha+\beta+\gamma}{2}\right)} \end{pmatrix}.
\end{equation}

\paragraph{Monodromy properties and symmetries}
As already discussed, the Schwarz triangle map $s_{\alpha,\beta,\gamma}(w)$ is not a single valued function, i.e. the analytic continuation around branch points $w=0, w=1$ and $w=\infty$ gives a M\"obius transformation of $s_{\alpha,\beta,\gamma}(w)$ corresponding to a symmetry transformation of the hyperbolic disk.

\begin{figure}[!t]
\centering
\includegraphics[scale=0.40]{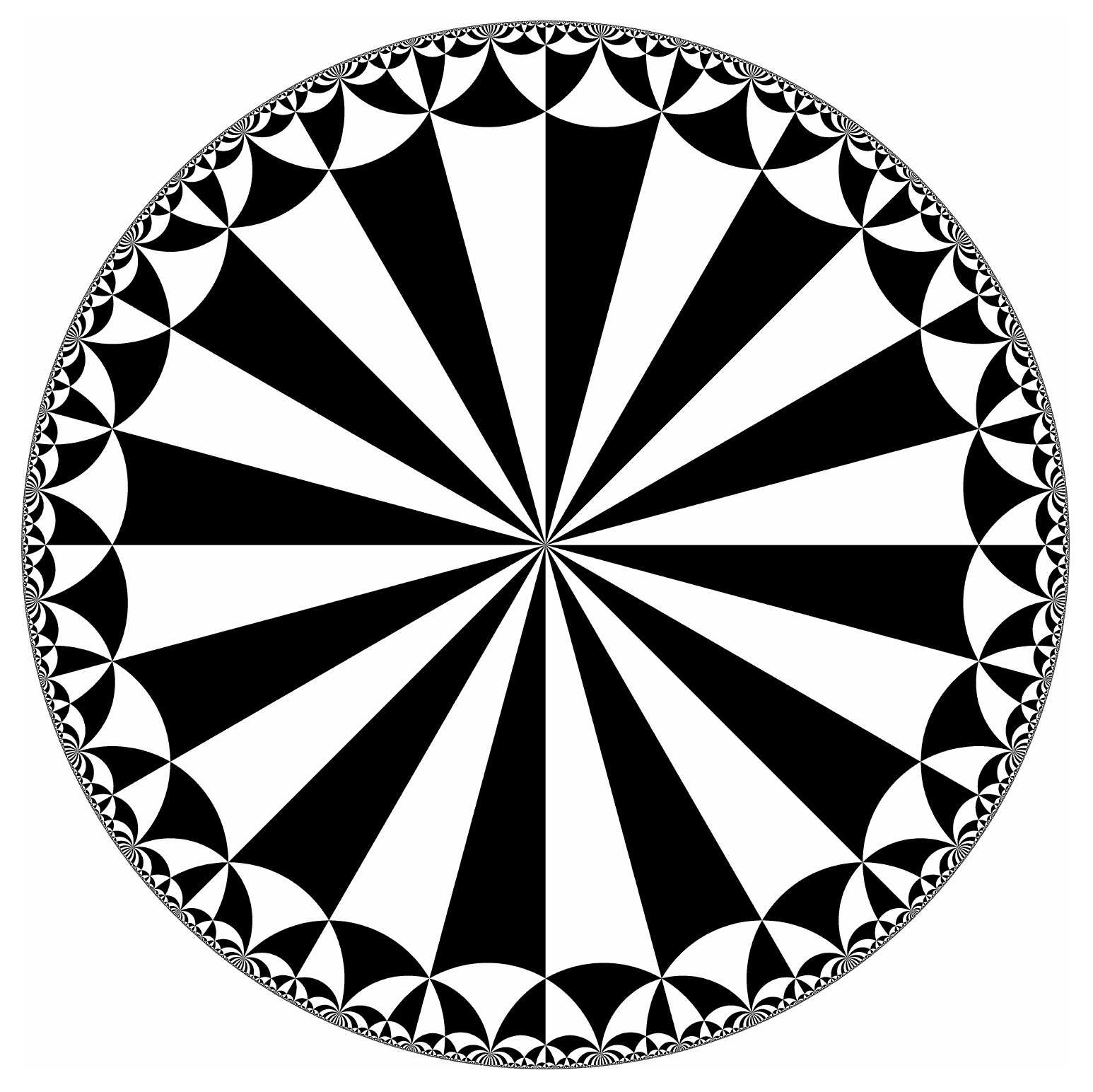}
\caption{Tiling of the hyperbolic disk as an image of analytic continuation of the Schwarz triangle function $s_{\alpha,\beta,\gamma}(w)$, in this case plotted with the choice $\alpha = \frac{1}{12}$, $\beta = \frac{1}{3}$ and $\gamma = \frac{1}{4}$ corresponding to minimal model of $\mathcal{W}_3$ or $\mathcal{W}_4$ algebra with central charge $c=-114/7$. We see that the basic white triangle has angle $\frac{\pi}{12}$ at the center $A$ of the hyperbolic disk, angle $\frac{\pi}{3}$ at vertex $B$ along positive real axis and angle $\frac{\pi}{4}$ at vertex $C$ in correspondence with the parameters of the algebra.}
\label{figbwtiling}
\end{figure}

Let us first discuss the basic symmetries generated by the analytic continuation of $s_{\alpha,\beta,\gamma}(z)$. These are generated by the reflections with respect to the sides of the hyperbolic triangle $ABC$. We have the reflection with respect to the side $AB$,
\begin{equation}
\Refl_{AB}: z \mapsto \bar{z},
\end{equation}
the reflection with respect to side $AC$,
\begin{equation}
\Refl_{AC}: z \mapsto e^{2\pi i \alpha} \bar{z}
\end{equation}
and finally the reflection with respect to the circular side $BC$
\begin{equation}
\label{reflbc}
\Refl_{BC}: z \mapsto \frac{e^{i\pi(\alpha-\gamma)}(1-ad(1-e^{2\pi i\gamma}))\bar{z}+\alpha ab e^{-i\pi\gamma} (1-e^{2\pi i \gamma})}{-\alpha^{-1}cd e^{-i\pi\gamma} (1-e^{2\pi i \gamma})\bar{z}+e^{-i\pi(\alpha+\gamma)}(1+bc(1-e^{2\pi i \gamma}))}
\end{equation}
where $a, b, c$ and $d$ are as in \eqref{unifdefabcd}. We can use white and black colors for the image of the upper half $w$-plane and lower half $w$-plane. The other triangles of the tiling can be obtained by a sequence of reflections. Since a reflection changes the orientation, all triangles can be colored write or black depending on whether they are obtained by an even or an odd number of reflections starting from the fundamental white triangle $ABC$. For the values of parameters $\alpha, \beta$ and $\gamma$ that we are going to use, the image of the upper half $w$-plane will always produce a white triangle while the image of the lower half $w$-plane produces a black triangle. These triangles will never partially overlap (this would not be the case for instance if one of the parameters were irrational, in our case the non-overlapping triangles these are sometimes called M\"obius triangles) and produce tiling such as the one in Figure \ref{figbwtiling}.

Compositions of pairs of reflections give rotations which preserve the tiling. The simplest one is
\begin{equation}
\Rot_\alpha = \Refl_b \circ \Refl_c: z \mapsto e^{2\pi i \alpha} z
\end{equation}
and corresponds to rotation around vertex $A$ by an angle $2\pi\alpha$. Similarly (and cyclically) we have for the other two rotations
\begin{align}
\Rot_\beta & = \Refl_c \circ \Refl_a: z \mapsto \frac{e^{-i\pi\alpha}(e^{i\pi\gamma}-2iad\sin(\pi\gamma)) z+2i\alpha ab\sin(\pi\gamma)}{-2i\alpha^{-1} cd\sin(\pi\gamma)z+e^{i\pi\alpha}(e^{i\pi\gamma}+2ibc\sin(\pi\gamma))} \\
\Rot_\gamma & = \Refl_a \circ \Refl_b: z \mapsto \frac{(e^{-i\pi\gamma}+2iad\sin(\pi\gamma)) z -2i\alpha ab e^{i\pi\alpha}\sin(\pi\gamma)}{2i\alpha^{-1}cde^{-\pi i \alpha} \sin(\pi\gamma) z + (e^{-i\pi\gamma}-2ibc\sin(\pi\gamma))}.
\end{align}
These rotations can be written as a fractional linear transformation represented by $SL(2,\mathbbm{C})$ matrices
\begin{align}
\nonumber
\Rot_\alpha & \leftrightarrow \begin{pmatrix} e^{i\pi\alpha} & 0 \\ 0 & e^{i\pi\alpha} \end{pmatrix} \\
\Rot_\beta & \leftrightarrow \begin{pmatrix} e^{-i\pi\alpha+i\pi\gamma}(1-ad(1-e^{-2i\pi\gamma})) & ab\alpha e^{i\pi\gamma}(1-e^{-2i\pi\gamma}) \\ -cd\alpha^{-1} e^{i\pi\gamma}(1-e^{-2i\pi\gamma}) & e^{i\pi\alpha+i\pi\gamma}(1+bc(1-e^{-2i\pi\gamma})) \end{pmatrix} \\
\nonumber
\Rot_\gamma & \leftrightarrow \begin{pmatrix} e^{-i\pi\gamma}(1-ad(1-e^{2i\pi\gamma})) & ab\alpha e^{i\pi\alpha-i\pi\gamma}(1-e^{2i\pi\gamma}) \\ -cd\alpha^{-1} e^{-i\pi\alpha-i\pi\gamma}(1-e^{2i\pi\gamma}) & e^{-i\pi\gamma}(1+bc(1-e^{2i\pi\gamma})) \end{pmatrix}
\end{align}
and from their definition they satisfy the identity
\begin{equation}
\label{uniftotalmonotriv}
\Rot_\alpha \circ \Rot_\beta \circ \Rot_\gamma = \mathbbm{1}.
\end{equation}
It is straightforward to check that the subgroup of M\"obius transformations of the complex plane $SL(2,\mathbbm{C})$ that preserve the disk of the radius $\rho$ centered at the origin of the $z$-plane are transformations of the form
\begin{equation}
\label{hyperrotation}
z \mapsto \rho \frac{\mu z + \nu \rho}{\bar{\nu} z + \bar{\mu} \rho} = \frac{\mu z + \nu \rho}{\bar{\nu} \rho^{-1} z + \bar{\mu}}
\end{equation}
with
\begin{equation}
|\mu|^2-|\nu|^2=1
\end{equation}
i.e. these can be represented by matrices in $SU(1,1)$. One can check explicitly that all three rotations $\Rot_{\alpha}, \Rot_{\beta}$ and $\Rot_{\gamma}$ are of this form with
\begin{equation}
\label{unifdiskradius}
\rho = \alpha \sqrt{\frac{ab}{cd}}
\end{equation}
which determines the coordinate radius of the Poincar\'e disk (this should not be confused with the associated curvature radius). We also see that the basic reflections $\Refl_{AB}, \Refl_{AC}$ and $\Refl_{BC}$ preserve the disk of radius $\rho$ centered at the origin because the reflection $\Refl_{AB}$ certainly does and the other two are compositions of $\Refl_{AB}$ with the basic rotations. Therefore, starting from the basic white triangle $ABC$ and applying to it the reflections we obtain a tiling of hyperbolic disk of radius \eqref{unifdiskradius}, see Figure \ref{figbwtiling}.

Having discussed the symmetries of the tiling, we can write how these are related to analytic continuation of $s_{\alpha,\beta,\gamma}(w)$ around the branch points. This is quite simple around $w = 0$:
\begin{equation}
s_{\alpha,\beta,\gamma}\left(e^{2\pi i} w \right) = e^{2\pi i \alpha} s_{\alpha,\beta,\gamma}(w) = \Rot_{\alpha}(s_{\alpha,\beta,\gamma}(w)),
\end{equation}
i.e. the analytic continuation of $s_{\alpha,\beta,\gamma}$ around $w=0$ counter-clockwise corresponds to rotation of the corresponding white triangle around $z=0$ by angle $2\pi\alpha$ which is exactly what we expect, see Figure \ref{figbwtiling}. Similarly, the analytic continuation around $w=1$ counter-clockwise corresponds to rotation around the vertex $B$,
\begin{equation}
s_{\alpha,\beta,\gamma}\left(e^{2\pi i} (w-1) + 1 \right) = \Rot_{\beta}(s_{\alpha,\beta,\gamma}(w)),
\end{equation}
The analytic continuation around $w=\infty$ can be obtained from these two using the identity \eqref{uniftotalmonotriv} which expresses the fact that product of three monodromies around all three branch points is trivial. In other words, continuing $s_{\alpha,\beta,\gamma}(t)$ along a large circle at infinity amounts to a product of continuation around $w=0$ and $w=1$, but this large circle when viewed from a coordinate chart around $t=\infty$ is clockwise instead of being counter-clockwise and this is exactly the content of \eqref{uniftotalmonotriv}.

\subsection{Fundamental polygon}

\paragraph{Tiling from geodesic reflections}
In order to generate the tiling of the hyperbolic disk as in Figure \ref{figbwtiling}, we use the following simple facts from hyperbolic geometry. First of all, the geodesics are either straight lines through the center of the hyperbolic disk or circular arcs intersecting orthogonally the (ideal) boundary of the hyperbolic disk. As such, they can be parametrized by pairs of points at which they intersect the boundary (the ideal circle at infinity),
\begin{equation}
\label{geolabelq}
z_1 = \rho e^{i\varphi_1} \qquad \text{and} \qquad z_2 = \rho e^{i\varphi_2}.
\end{equation}
The coordinate of the center of the Euclidean circle which represents the geodesic is then
\begin{equation}
z = \frac{2 z_1 z_2}{z_1 + z_2}
\end{equation}
while the Euclidean radius of this circle is
\begin{equation}
r = i\rho \frac{z_1-z_2}{z_1+z_2}.
\end{equation}
Every geodesic determines uniquely a reflection of the hyperbolic disk that preserves the geodesic pointwise. The transformation takes the form analogous to \eqref{hyperrotation}, but with $z$ being replaced by $\bar{z}$,
\begin{equation}
\label{hyperreflection}
z \mapsto \frac{\mu \bar{z} + \nu \rho}{\bar{\nu} \rho^{-1} \bar{z} + \bar{\mu}}
\end{equation}
with
\begin{equation}
|\mu|^2 - |\nu|^2 = 1 \qquad \text{and} \qquad \nu = -\bar{\nu}
\end{equation}
(the second condition is required by imposing the involutivity of the reflection). The parameters $\mu$ and $\nu$ can be written in terms of the parameters $(z_1,z_2)$ labeling the geodesic as
\begin{equation}
\mu = -\frac{2z_1 z_2}{\rho(z_1-z_2)}, \qquad \bar{\mu} = \frac{2\rho}{z_1-z_2}, \qquad \nu = \frac{z_1+z_2}{z_1-z_2} = -\bar{\nu}.
\end{equation}
These obviously satisfy $|\mu|^2-|\nu|^2=1$. Exchanging $z_1 \leftrightarrow z_2$ flips sign of both $\mu$ and $\nu$ simultaneously which does not affect \eqref{hyperreflection}. The inverse transformation is
\begin{equation}
z_1 = \frac{\rho \mu}{1-\nu}, \qquad z_2 = -\frac{\rho \mu}{1+\nu}.
\end{equation}
The action of the reflection \eqref{hyperreflection} on an arbitrary point $z$ is
\begin{equation}
z \mapsto \frac{2z_1 z_2 \bar{z} - (z_1+z_2)\rho^2}{(z_1+z_2)\bar{z}-2\rho^2}
\end{equation}
while for point on the boundary $z = \rho q$, $|q|=1$ this simplifies to
\begin{equation}
\label{georeflection}
z \mapsto z^\prime = \frac{2z_1z_2-(z_1+z_2)z}{(z_1+z_2)-2z}.
\end{equation}
We can start with the three pairs of points corresponding to three geodesic sides of the basic triangle $ABC$ and apply these reflections to generate all the geodesics in Figure \ref{figbwtiling}. The geodesic on which points $A$ and $B$ lie has $z$-parameters
\begin{equation}
(z_1,z_2) = (\rho,-\rho),
\end{equation}
the geodesic corresponding to side $AC$ has
\begin{equation}
(z_1,z_2) = \left(\rho e^{\pi i \alpha},-\rho e^{\pi i \alpha}\right).
\end{equation}
Finally, the most interesting and non-trivial are the points of intersection of side $BC$ with the boundary of the hyperbolic disk. These are
\begin{equation}
(z_1,z_2) = \rho \left(e^{i\pi\alpha}\frac{\alpha a e^{i\pi\gamma}+\rho c}{\alpha a + \rho c e^{i\pi\gamma}}, e^{i\pi\alpha}\frac{\alpha a e^{i\pi\gamma}-\rho c}{-\alpha a + \rho c e^{i\pi\gamma}}\right).
\end{equation}
Using these three geodesics as an initial seed and iterating the map \eqref{georeflection}, we can produce an infinite number of geodesics that cut the hyperbolic disk into tiling as sketched in Figure \ref{figbwtiling}.

\paragraph{Fundamental polygon}
Recall that the map $\pi: \mathcal{W} \to \mathcal{M}$ was $KN$-sheeted covering branched at $t = 0$, $t = 1$ and $t = \infty$. Therefore, the white and black half-spheres in $\mathcal{M}$ should be covered by corresponding $KN$ white and black regions in $\mathcal{W}$. Furthermore, since the ramification index of $\pi: \mathcal{W} \to \mathcal{M}$ corresponding to $t = \infty$ was $KN$, locally the map $t(z)$ as in \eqref{automorphict} around $z=0$ behaves as $z^{-KN}$. This means that we can identify the fundamental domain in $\mathcal{D}$ covering $\mathcal{W}$ exactly once under the map $(x(z),y(z))$ with $KN$ white and black triangles around the center of the hyperbolic disk $\mathcal{D}$, i.e. with the polygon made of $KN$ white and $KN$ black triangles with vertex at the origin of $\mathcal{D}$. This polygon is shown in Figure \ref{fighypergeo}. Another way that we can view this is that there are $2KN$ geodesic arcs on $\mathcal{W}$ corresponding to image of sides $BC$ and $B\bar{C}$ as well as their rotated images. Cutting out these $2KN$ arcs from $\mathcal{W}$ leaves us with a simply connected domain homeomorphic to a disk and the map $(x(z),y(z))$ restricted to the fundamental polygon of Figure \ref{fighypergeo} provides a bi-holomorphic map between this open set in $\mathcal{W}$ and the fundamental polygon.

If we worked with the symplectic basis of $H_1(\mathcal{W},\mathbbm{Z})$, we would obtain another fundamental polygon by cutting out $2g = (K-1)(N-1)$ closed curves from $\mathcal{W}$, and so the polygon would have $2(K-1)(N-1)$ edges. These edges and their identifications would be arranged so that the edges come in quadruples in order $A_i$, $B_i$, $A_i^{-1}$, $B_i^{-1}$, each of these topologically producing one handle. The representation of $\mathcal{W}$ as a polygon with identified edges that we are using here respects the $\mathbbm{Z}_K \times \mathbbm{Z}_N$ group of automorphisms of $\mathcal{W}$, but the corresponding polygon has $2KN$ edges\footnote{For $K$ or $N$ equal to $2$, i.e. Virasoro case, the angle between two edges is $\pi$ so they look like a single edge of double length.}. In order to understand the geodesics representing the homology classes, we should first understand the pairwise identifications of the edges of the fundamental polygon.

Recall that we identify the parameters $\alpha$, $\beta$ and $\gamma$ as in \eqref{alphabetagammank}. We will label with triangles sequentially in the \emph{clockwise} direction as in Figure \ref{fighypergeo}. The reason why we choose to label the triangles clockwise instead of counter-clockwise is that our choice of map \eqref{uniformizationmap} involves the parameter $t^{-1}$ of the hypergeometric function and as a result the upper half $t$-plane is mapped to the hyperbolic triangle in the \emph{lower half} of $\mathcal{D}$, i.e. a black triangle in Figure \ref{figbwtiling}. The basic (black) triangle with side $AB$ along the positive real axis and third vertex in the lower half plane is labeled $T_0$ (and $0$ in Figure \ref{fighypergeo}). We call the next black triangle in the clockwise direction $T_1$ (and $1$ in the Figure) etc., i.e. $T_j$ is obtained from $T_0$ by rotation by angle $-\frac{2\pi j}{KN}$ around the center of the hyperbolic disk $\mathcal{D}$. The white triangles are instead labeled in such a way that $\bar{T}_j$ is the white triangle that is reached from the black triangle $T_j$ if we cross the boundary of the fundamental polygon, i.e. $\bar{T}_0$ is the white triangle that we reach from $T_0$ by crossing the side $B\bar{C}$. By rotation symmetry these triangles also satisfy the condition that rotation by $-\frac{2\pi k}{KN}$ around the center of $\mathcal{D}$ maps $\bar{T}_j$ to $\bar{T}_{j+k}$.

We need to figure out the relative position of $T_j$ and $\bar{T}_j$ triangles. Let $\bar{T}_n$ be the triangle that is sharing sides with $T_0$ and $T_1$. We want to determine the value of $n$. The neighboring triangles of $T_0$ are therefore $\bar{T}_{n-1}$ in the counter-clockwise direction and $\bar{T}_n$ in the clockwise direction. By rotation symmetry the neighboring triangles of $\bar{T}_0$ are $T_{-n}$ in the counter-clockwise direction and $T_{1-n}$ in the clockwise direction. Let us proceed in few steps:
\begin{enumerate}
\item Let us start in the black triangle $T_0$ near the vertex $B$. We want to encircle the vertex $B$ in the counter-clockwise direction (as viewed in $z$-coordinate). First we cross the side $B\bar{C}$ and we appear in the white triangle $\bar{T}_0$. We continue moving counter-clockwise around vertex $B$ and we reach the black triangle $T_{1-n}$. Up to now we moved around vertex $B$ by angle $\frac{2\pi}{N}$ and reached $T_0 \to \bar{T}_0 \to T_{1-n}$. After $N$ steps of this rotation we should be back where we started so we have a condition
\begin{equation}
N(n-1) = 0 \mod KN.
\end{equation}
\item We can do the analogous analysis around vertex $C$. Starting in $\bar{T}_0$ near the vertex $\bar{C}$ and going counter-clockwise, we first reach $T_0$ and then $\bar{T}_n$. We did the $\frac{2\pi}{K}$ of the full rotation around $z = \bar{C}$ and reached $\bar{T}_0 \to T_0 \to \bar{T}_n$. Continuing $K-1$ steps like this, we reach $\bar{T}_{Kn}$ and this should be identified with $\bar{T}_0$. We thus find a second condition
\begin{equation}
Kn = 0 \mod KN.
\end{equation}
\item To solve these two conditions for $n$, we use the fact that $K$ and $N$ are coprime. The Euclid's algorithm in this case implies that there exist two integers $\tilde{K}$ and $\tilde{N}$ such that
\begin{equation}
\label{euclidknab}
\tilde{K}N+\tilde{N}K = 1 \mod KN.
\end{equation}
Furthermore, $\tilde{K}$ is uniquely determined modulo $K$ and $\tilde{N}$ is uniquely determined modulo $N$. Given such pair $(\tilde{K},\tilde{N})$, we find that the value of $n$ is
\begin{equation}
n = N\tilde{K} = -K\tilde{N}+1 \mod KN.
\end{equation}
As an illustration, for the case $(N,K)=(3,4)$ the solution to \eqref{euclidknab} is $(\tilde{K},\tilde{N})=(3,1)$ as one can easily check. Therefore, the white triangle between $T_0$ and $T_1$ is $\bar{T}_9$ which agrees with Figure \ref{fighypergeo}.
\end{enumerate}

To summarize the previous discussion, for each given coprime $N$ and $K$ there exist integers $\tilde{K}$ (unique modulo $K$) and $\tilde{N}$ (unique modulo $N$) such that \eqref{euclidknab} is satisfied and in terms of these two numbers, starting from the triangle $T_0$ and crossing images of sides $AB$, $BC$ and $AC$ we arrive at triangles
\begin{equation}
T_0 \stackrel{AB}{\longrightarrow} \bar{T}_{-K\tilde{N}}, \qquad T_0 \stackrel{BC}{\longrightarrow} \bar{T}_0, \qquad T_0 \stackrel{\bar{C}A}{\longrightarrow} \bar{T}_{N\tilde{K}}.
\end{equation}

The previous discussion allows us also to determine the action of the group of automorphisms $\mathbbm{Z}_K \times \mathbbm{Z}_N$ on the fundamental polygon. The rotation in the $x$-plane by $\frac{2\pi}{K}$, i.e. the transformation
\begin{equation}
x \mapsto e^{\frac{2\pi i}{K}} x
\end{equation}
corresponds to one full counter-clockwise rotation around $t = 0$,
\begin{equation}
t \mapsto e^{2\pi i} t.
\end{equation}
In the fundamental polygon this translates to going counter-clockwise around vertex $\bar{C}$. This sends $T_0 \to T_{N\tilde{K}}$, i.e. the rotation around $x$-axis by $\frac{2\pi}{K}$ in the positive direction corresponds to rotation
\begin{equation}
\label{xrotationtoz}
z \mapsto e^{-\frac{2\pi i n}{KN}} z = e^{-\frac{2\pi i \tilde{K}}{K}} z.
\end{equation}
Analogously, the rotation in the $y$-plane
\begin{equation}
y \mapsto e^{\frac{2\pi i}{N}} y
\end{equation}
corresponds to a full counter-clockwise rotation around $t = 1$. Therefore we are doing a counter-clockwise rotation around $z = B$ by $\frac{2\pi}{N}$, i.e. rotation that maps $T_0 \to T_{K\tilde{N}}$. This corresponds to rotation
\begin{equation}
\label{yrotationtoz}
z \mapsto e^{\frac{2\pi i (n-1)}{KN}} z = e^{-\frac{2\pi i \tilde{N}}{N}} z.
\end{equation}
The integers $\tilde{K}$ and $\tilde{N}$ provide us with a map between the rotation symmetry $\mathbbm{Z}_{KN}$ of the fundamental polygon around the center and the symmetry group $\mathbbm{Z}_K \times \mathbbm{Z}_N$ of the Catalan curve. Mathematically, for $K$ and $N$ coprime these groups are isomorphic and $(\tilde{K},\tilde{N})$ are two integers that provide an explicit isomorphism.

\subsection{Geodesics}
We want to describe how do various geodesics representing the sides of the fundamental polygon and the canonical representatives of homology classes look like in $\mathcal{D}$. In Poincar\'e disk model geodesics are segments of Euclidean circles orthogonal to the boundary of the hyperbolic disk. The boundary of the hyperbolic disk, i.e. a circle with radius $\rho$ \eqref{unifdiskradius}, is the ideal circle. The group of orientation-preserving isometries of hyperbolic plane is $PSU(1,1)$, but sometimes it is useful to consider the larger group of M\"obius transformation $PSL(2,\mathbbm{C})$ (fractional linear transformations). This group in general acts non-trivially on the ideal circle, but keeping track of it allows to write formulas that apply both to Poincar\'e disk as well as upper half plane model of the hyperbolic space.

One such quantity of interest is the geodesic distance. Any two points $z_1$ and $z_2$ in the hyperbolic space (this applies to both upper half plane or disk model) determine a unique geodesic. This geodesic approaches the ideal circle at two points, $z_0$ and $z_3$. Let us choose these such that the points appear along the geodesic in the order $z_0, z_1, z_2$ and $z_3$. The geodesic distance between the points $z_1$ and $z_2$ can then be written as
\begin{equation}
\label{geodesicdistance}
d(z_1,z_2) = \log \frac{(z_2-z_0)(z_1-z_3)}{(z_1-z_0)(z_2-z_3)}.
\end{equation}
If we exchange $z_0$ and $z_3$, the logarithm would just exchange the sign, so with the absolute value we would get a formula that does not depend on the order of the points. Notice that up to a logarithm, the geodesic distance of the points is just the cross-ratio of the four points. If we make a $PSL(2,\mathbbm{C})$ transformations of all four points, their cross-ratio is invariant. Therefore this formula applies both to upper half plane model as well as the Poincar\'e disk model because these two models are related by a M\"obius transformation. The covariance of geodesic length under $PSL(2,\mathbbm{C})$ also provides a simple way of proving the formula, because we can always use a M\"obius transformation to go to a frame where the ideal circle is the real axis and the geodesic we are interested in is a line parallel to imaginary axis. In this configuration the geodesic distance is very easy to evaluate and agrees with the formula \eqref{geodesicdistance}.

A similar argument allows us to describe the relative position of two geodesics. Notice that unlike in the Euclidean or spherical case, we can have two geodesics that either intersect at one finite point in the hyperbolic plane, or have finite minimal distance between themselves (so called ultra-parallel geodesics) or as a limiting case asymptotically approach each other as we approach the ideal circle. Let us label the first geodesic by its intersection points $(z_1,z_2)$ with the ideal circle and analogously $(\tilde{z}_1,\tilde{z}_2)$ for the second geodesic. This is the parametrization used already in \eqref{geolabelq}. We can now form the following combination of cross-ratios of these four points on the ideal circle:
\begin{equation}
\Delta = \frac{(z_1-\tilde{z}_1)(z_1-\tilde{z}_2)(z_2-\tilde{z}_1)(z_2-\tilde{z}_2)}{(z_1-z_2)^2(\tilde{z}_1-\tilde{z}_2)^2}.
\end{equation}
This expression is invariant under $PSL(2,\mathbbm{C})$ acting on all four points and also under exchanges $z_1 \leftrightarrow z_2$, $\tilde{z}_1 \leftrightarrow \tilde{z}_2$ as well as the interchange of pairs $(z_1,z_2) \leftrightarrow (\tilde{z}_1,\tilde{z}_2)$. Since all $4$ points lie on the ideal circle, this combination is also real (as can be seen by mapping any circle to the real axis). For ultra-parallel geodesics we have $\Delta > 0$ and the distance $d$ between these satisfies
\begin{equation}
\Delta = \frac{1}{4} \sinh^2 d, \qquad d = \log \left( \sqrt{4\Delta} + \sqrt{1+4\Delta} \right).
\end{equation}
For intersecting geodesics, we have instead $\Delta \in \left(-\frac{1}{4},0\right)$ and the angle $\alpha$ between them at the intersection is
\begin{equation}
\Delta = -\frac{1}{4} \sin^2 \alpha.
\end{equation}
Note that formally under analytic continuation we can think of the distance as being the imaginary angle and vice versa. This is one of the dualities of the hyperbolic geometry.

\subsubsection{Sides of polygon and homology}
\begin{figure}
\centering
\includegraphics[scale=0.6]{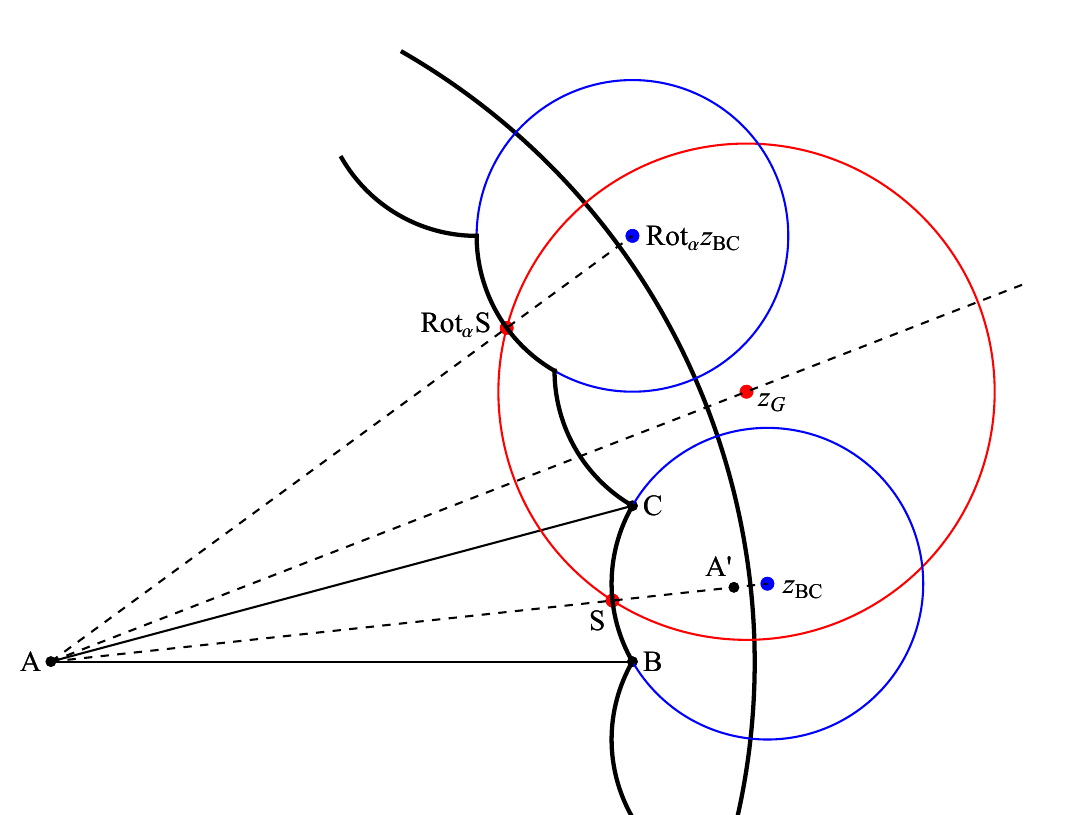}
\caption{Construction of the circles representing the sides of the polygon and the geodesics generating $H_1(\mathcal{W},\mathbbm{Z})$.}
\label{figgeoconstr}
\end{figure}

\paragraph{Side of the polygon}
We want to determine the circle representing the shortest geodesic whose rotations generate the first homology group $H_1(\mathcal{W},\mathbbm{Z})$. This is illustrated in Figures \ref{fighypergeo} and \ref{figgeoconstr}. First of all, the side $BC$ of the fundamental black triangle lies on an Euclidean circle. The center of this circle is such that the reflection \eqref{reflbc} maps this point to infinity, i.e.
\begin{equation}
z_{BC} = -\frac{e^{i\pi\alpha}\alpha(1-ad(1-e^{2\pi i\gamma}))}{cd(1-e^{2\pi i \gamma})}.
\end{equation}
The Euclidean radius of this circle is
\begin{equation}
r_{BC} = \frac{\alpha}{2cd\sin(\pi\gamma)}
\end{equation}
where we used the fact that given a geodesic reflection of the form \eqref{hyperreflection} such as \eqref{reflbc}, the radius of the corresponding Euclidean circle is
\begin{equation}
r = \frac{\rho}{|\nu|}.
\end{equation}

\paragraph{Image of $A$ under reflection in $BC$}
Another important point is the pole of $t(z)$ nearest to the origin, i.e. the image $A^\prime$ of the center of the polygon $A$ under the reflection with respect to the side $BC$, \eqref{reflbc}. The absolute value of coordinate of this point determines the radius of convergence of the automorphic function $t(z)$, \eqref{automorphict}. We have
\begin{equation}
A^\prime = \frac{\alpha ab e^{i\pi\alpha} (1-e^{2\pi i \gamma})}{1+bc(1-e^{2\pi i \gamma})}.
\end{equation}
Since this point is a reflection of $A$ with respect to a circle, the points $A$, $A^\prime$ and $z_{BC}$ are on the same Euclidean line and also
\begin{equation}
|A-z_{BC}||A^\prime-z_{BC}| = r_{BC}^2.
\end{equation}
Because the circle with the center $z_{BC}$ and radius $r_{BC}$ is orthogonal to the ideal circle of radius $\rho$, we also have the relation
\begin{equation}
\rho^2 + r_{BC}^2 = |z_{BC}|^2.
\end{equation}

\paragraph{Point $S$}
The straight line connecting $A$ to $A^\prime$ intersects the side $BC$ at point $S$ orthogonally. The point $S$ has the following characterizations:
\begin{enumerate}
\item it is the point of intersection of the fundamental cycle with the side $BC$
\item it is also the foot of an altitude (perpendicular foot) of the fundamental triangle $ABC$
\item it is the intersection of $BC$ with the line connecting $A$ to $z_{BC}$ (because $A$, $A^\prime$ and $z_{BC}$ lie on the same line)
\item it is the midpoint of the geodesic connecting $A$ and $A^\prime$, with side $BC$ being its perpendicular bisector
\end{enumerate}
Its coordinate is
\begin{equation}
S = \frac{z_{BC}}{|z_{BC}|} \left( |z_{BC}| - r_{BC} \right).
\end{equation}

\paragraph{Fundamental class}
The geodesic representing the fundamental cycle $\mathcal{C}$ is orthogonal to the ideal circle (as is every geodesic) and passes through the point $S$ as well as its rotated image under the rotation by angle $\frac{2\pi}{KN}$ around the center of the disk $A$, see Figures \ref{fighypergeo} and \ref{figgeoconstr}. This determines the center of the Euclidean circle representing the geodesic to be at
\begin{equation}
z_G = \frac{e^{i\pi\alpha}z_{BC}}{\cos\pi\alpha}
\end{equation}
and its radius to be
\begin{equation}
r_G^2 = \frac{|z_{BC}|^2}{\cos^2\pi\alpha} - \rho^2.
\end{equation}

\subsection{Special cases and limits}

\begin{figure}[!t]
\centering
\includegraphics[scale=0.55]{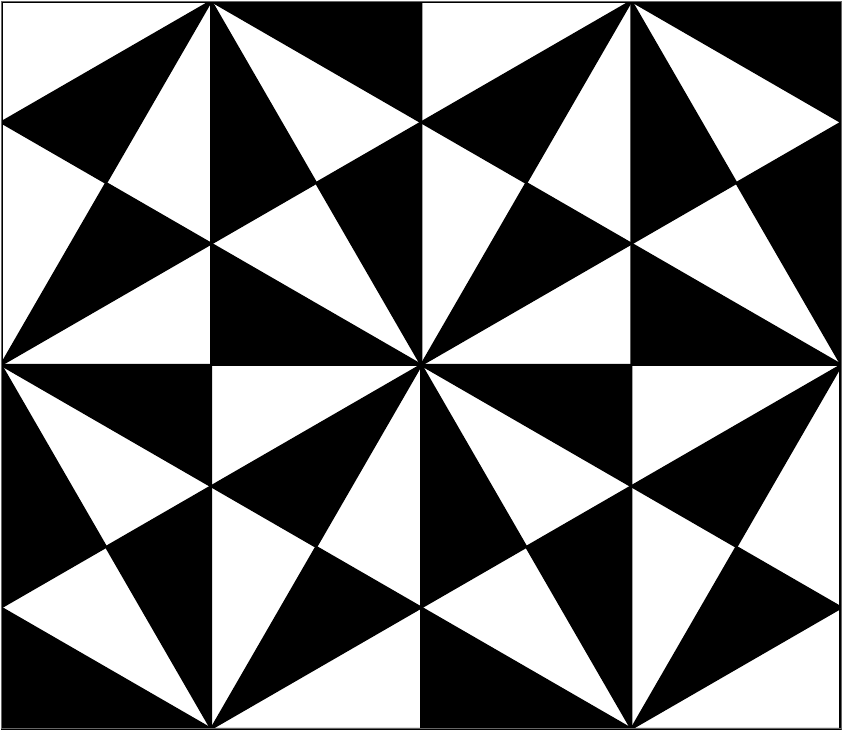}
\caption{Tiling corresponding to parameters $(N,K)=(2,3)$ of Lee-Yang model. In this case the genus of $\mathcal{W}$ is one so the universal covering space is simply the flat complex plane and this is tiled by Euclidean triangles of angles $\frac{\pi}{6}, \frac{\pi}{2}$ and $\frac{\pi}{3}$.}
\label{leeyangtiling}
\end{figure}

\subsubsection{Lee-Yang model}
\label{leeyangelliptic}
Lee-Yang model with $(N,K)=(2,3)$ and $c=-\frac{22}{5}$ is special because in this case the genus of $\mathcal{W}$ is one, i.e. it is a complex torus. Therefore, many properties of the hyperbolic surfaces do not apply to this flat case, but formulas like \eqref{automorphicx} or \eqref{automorphicy} still apply. In fact, it is well-known that the points on a curve
\begin{equation}
\label{catalan23}
y^2 + x^3 = 1
\end{equation}
can be parametrized by Weierstrass elliptic functions. In our situation, the parameters $\alpha, \beta$ and $\gamma$ take values $\frac{1}{6}, \frac{1}{2}$ and $\frac{1}{3}$ which sum up to $1$, so the Schwarz triangle is Euclidean (i.e. sum of the internal angles is $\pi$), see Figure \ref{leeyangtiling}. The corresponding tiling of the $z$-plane can be characterized by modular parameter $\tau = e^{\frac{\pi i}{3}}$ (this is one of the special values where there is $\mathbbm{Z}_3$ symmetry). Weierstrass elliptic function $\wp(z)$ satisfies the differential equation
\begin{equation}
\wp^{\prime 2}(z) = 4\wp^3(z) - g_2 \wp(z) - g_3
\end{equation}
where $g_2$ and $g_3$ characterize the shape (the complex or conformal structure) of the torus. For $\tau = e^{\frac{\pi i}{3}}$ we have $g_2 = 0$ and we can choose $g_3$ by simply rescaling the $z$-coordinate. The value that is convenient for us is $g_3 = 4$. The differential equation becomes
\begin{equation}
\wp^3(z) + \left( -\frac{i}{2} \wp^{\prime}(z) \right)^2 = 1
\end{equation}
so we can parametrize the solutions of \eqref{catalan23} by
\begin{align}
x(z) & = \wp(z) = \frac{1}{z^2} + \frac{z^4}{7} + \frac{z^{10}}{637} + \frac{z^{16}}{84721} + \frac{3z^{22}}{38548055} + \mathcal{O}\left(z^{28}\right) \\
y(z) & = -\frac{i}{2} \wp^{\prime}(z) = \frac{i}{z^3} - \frac{2iz^3}{7} - \frac{5iz^9}{637} - \frac{8iz^{15}}{84721} - \frac{33iz^{21}}{38548055} + \mathcal{O}\left(z^{27}\right)
\end{align}
which upon specialization of parameters agree with more general automorphic functions \eqref{automorphicx} and \eqref{automorphicy}. The third power of $x(z)$ is $t(z)$,
\begin{equation}
t(z) = \wp^3(z) = \frac{1}{z^6} + \frac{3}{7} + \frac{6z^6}{91} + \frac{4z^{12}}{931} + \frac{33z^{18}}{289835} + \frac{36z^{24}}{18384457}  + \mathcal{O}\left(z^{30}\right)
\end{equation}
which agrees with the expansion \eqref{automorphict} if we specialize to $(N,K)=(2,3)$. In the flat case $(N,K)=(2,3)$ the canonical bundle of $\mathcal{W}$ is trivial so we can have a relation between $x(z)$ and $y(z)$ such as
\begin{equation}
\frac{dx(z)}{dz} + 2i y(z) = 0
\end{equation}
which together with the algebraic equation \eqref{catalan23} provides a very convenient way of finding a series expansions of $x(z)$ and $y(z)$.

The tiling corresponding to Lee-Yang model is sketched in Figure \ref{leeyangtiling}. The fundamental domain is the hexagon around the central vertex that has 6 white and 6 black triangles. The torus with $\tau = e^{\frac{\pi i}{3}}$ is usually represented by a rhombus and we can identify such rhombi in Figure \ref{leeyangtiling} as well (these are still made of 6 white and 6 black triangles). Since the genus $1$ Riemann surface is flat, it has continuous group of isometries so in particular the geodesics representing homology classes in $H_1(\mathcal{W},\mathbbm{Z})$ are not unique unlike in the hyperbolic case.

\subsubsection{Large $K$ and $N$ limit}

\begin{figure}
\centering
\includegraphics[scale=0.60]{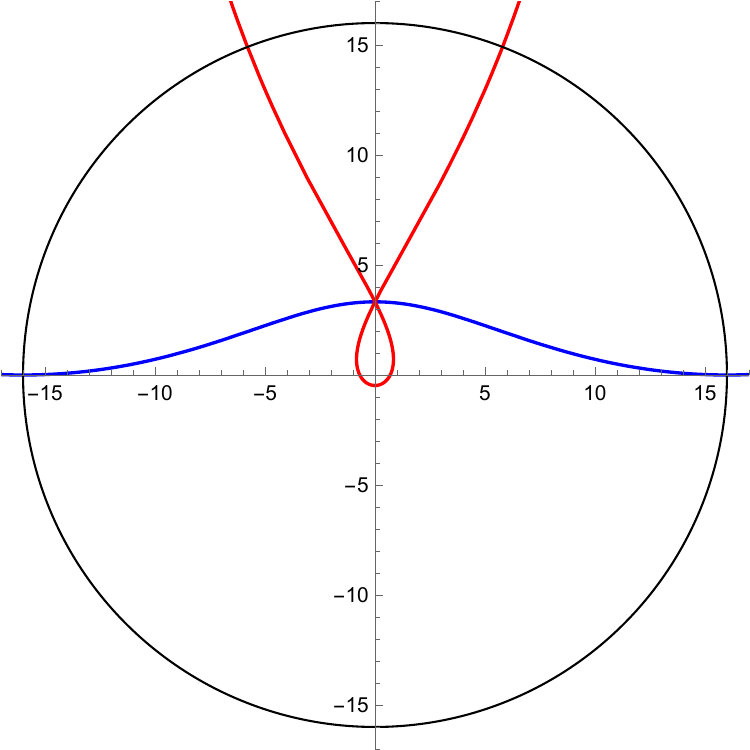}
\caption{The image of the geodesic arcs parametrizing the curve $\mathcal{C}$ (red) as well as the sides of the fundamental polygon (blue curve). The complete red curve intersects the pure imaginary axis at $6$ points (counted with multiplicity) while the blue curve intersects it only at two points.}
\label{figlargencoords}
\end{figure}

\paragraph{Limit of geodesics}

Let us consider the limit where $K$ and $N$ go to infinity such that their ratio goes to a constant. The quantities associated to the geodesics have the following asymptotic behavior:
\begin{align}
\nonumber
\rho & = 1 + \frac{4 \log 2}{KN} + \mathcal{O}\left(\frac{1}{N^3}\right) \\
\nonumber
z_{BC} & = 1 + \frac{i \pi + 8 \log 2}{2KN} + \mathcal{O}\left(\frac{1}{N^3}\right) \\
r_{BC} & = \frac{\pi}{2KN} + \mathcal{O}\left(\frac{1}{N^3}\right) \\
\nonumber
z_G & = 1 + \frac{3 \pi i + 8 \log 2}{2KN} + \mathcal{O}\left(\frac{1}{N^3}\right) \\
\nonumber
r_G & = \frac{\sqrt{5}\pi}{2KN} + \mathcal{O}\left(\frac{1}{N^3}\right).
\end{align}
We see that that as $N$ and $K$ go to infinity, the circle representing the geodesic generator of $H_1(\mathcal{W},\mathbbm{Z})$ shrinks to zero size. This is simply due to the fact that the fundamental black or white triangles shrink as well because the opening angle at the center of the hyperbolic disk is $\frac{\pi}{KN}$. For this reason it is convenient to introduce a new coordinate $s = z^{KN}$ which covers the stable limit of the fundamental triangle. We have
\begin{equation}
\rho^{KN} \to 16,
\end{equation}
i.e. the radius of the ideal circle in $s$-coordinate asymptotes $16$. Under the map $z \mapsto s = z^{KN}$ all the images of $BC$ sides under the group of automorphisms of $\mathcal{W}$ collapse to a single curve parametrized by
\begin{equation}
z = z_S + r_S e^{i\varphi} \qquad \to \qquad s = 16i \exp \left[ \frac{\pi}{2} e^{i\varphi} \right]
\end{equation}
while the circles that parametrize $\mathcal{C}$ map to
\begin{equation}
\label{geocirccoords}
z = z_G + r_G e^{i\varphi} \qquad \to \qquad s = -16i \exp \left[ \frac{\sqrt{5}\pi}{2} e^{i\varphi} \right].
\end{equation}
The first curve intersects the pure imaginary axis when $e^{i\varphi} = \pm 1$, i.e. at two points
\begin{equation}
s = 16i e^{\frac{\pi}{2}} \qquad \text{or} \qquad s = 16i e^{-\frac{\pi}{2}}.
\end{equation}
The second curve intersects the pure imaginary axis for
\begin{equation}
\frac{\sqrt{5}\pi}{2} \sin \varphi = n\pi, \qquad n \in \mathbbm{N}.
\end{equation}
This happens for
\begin{equation}
\varphi \in \left\{ 0, \pi, \pm \arcsin \frac{2}{\sqrt{5}}, \pi \pm \arcsin \frac{2}{\sqrt{5}}\right\} \qquad \text{(modulo $2\pi$)}
\end{equation}
and the corresponding four intersection points (two simple and two double) are
\begin{equation}
s = -16i e^{\pm \frac{\sqrt{5}\pi}{2}} \qquad \text{and} \qquad s = 16i e^{\pm \frac{\pi}{2}}.
\end{equation}
The part of the geodesic arc corresponding to $\mathcal{C}$ and passing through three triangles from the point $S$ to $e^{\frac{2\pi i}{KN}}S$ (this covers half of the fundamental cycle $\mathcal{C}$ on $\mathcal{W}$) corresponds to the small node in the Figure \ref{figlargencoords} and is parametrized by
\begin{equation}
\label{uniflargengeo}
s = -16i \exp \left[ -\frac{\sqrt{5}\pi}{2} e^{i\varphi} \right], \qquad \varphi \in \left[ -\arcsin \left( \frac{2}{\sqrt{5}} \right), +\arcsin \left( \frac{2}{\sqrt{5}} \right) \right].
\end{equation}

\paragraph{Limit of the uniformization map}

\begin{figure}
\centering
\includegraphics[scale=0.60]{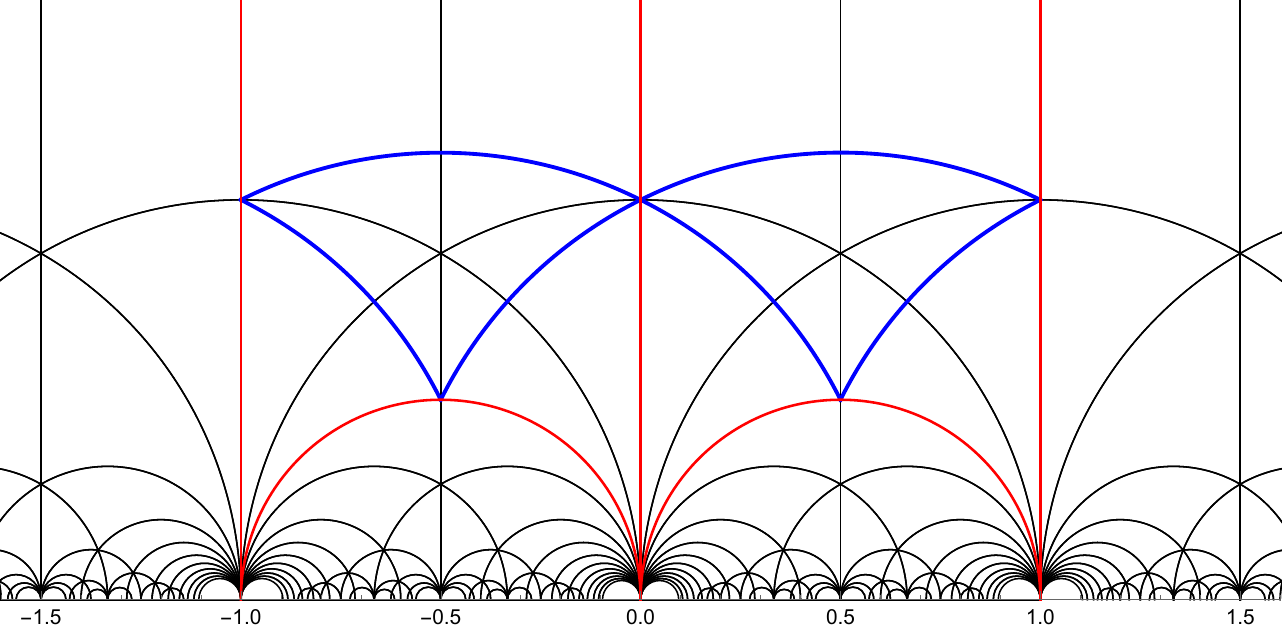}
\caption{Limiting large $K$ and $N$ picture showing the sides of the fundamental polygon (red) and $6$ segments of the generating geodesic in the upper half plane model (in $\tau$ coordinate). All the curves shown in this figure are euclidean circles with center on the real axis, i.e. geodesics.}
\label{figgeouhp}
\end{figure}

\begin{figure}
\centering
\includegraphics[scale=0.55]{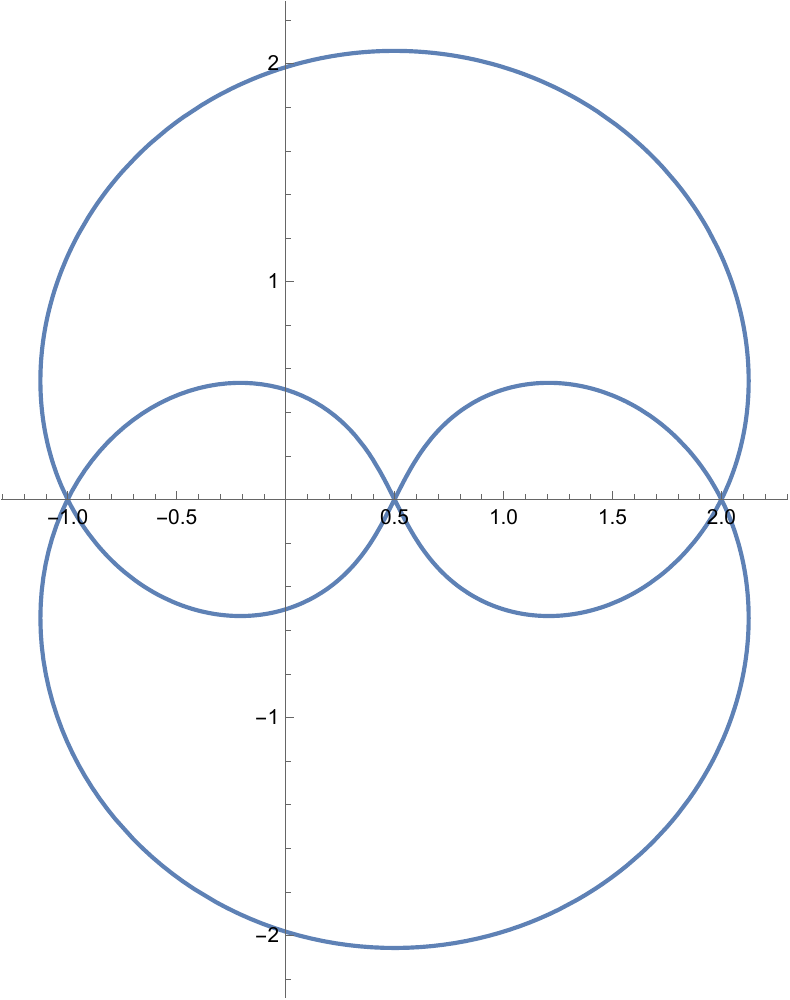}
\caption{Parametrization of the Pochhammer contour using modular $\lambda$-function in the large $K$ and $N$ limit.}
\label{figpochlargekn}
\end{figure}

Let us now have a look at the behavior of \eqref{unifmap} as $K$ and $N$ go to infinity with their ratio going to a constant. Plugging in the explicit values \eqref{alphabetagammank}, we see that the leading behavior of the ratio of hypergeometric functions is
\begin{equation}
z \to w^{\frac{1}{KN}} \left[ 1 + \frac{1}{KN} \frac{2\partial^{(1)} {}_2 F_1\left(\frac{1}{2},\frac{1}{2};1;w\right) + 2 \partial^{(3)} {}_2 F_1\left(\frac{1}{2},\frac{1}{2};1;w\right)}{{}_2 F_1\left(\frac{1}{2},\frac{1}{2};1;w\right)} + \mathcal{O}\left(\frac{1}{N^3}\right) \right]
\end{equation}
where $\partial^{(1)}$ denotes the derivative with respect to the first parameter of the hypergeometric function, i.e. $a$ in the usual notation. The hypergeometric function that appears is the complete elliptic integral of the first kind\footnote{Here we write $K$ as a function of the elliptic parameter $m$ which in terms of the usual elliptic modulus $k$ is $m=k^2$. In Mathematica, the corresponding function is $\mathtt{EllipticK}[m]$.},
\begin{equation}
K(m) = \int_0^1 \frac{dt}{\sqrt{(1-t^2)(1-mt^2)}} = \frac{\pi}{2} {}_2 F_1\left(\frac{1}{2},\frac{1}{2};1;m\right).
\end{equation}
To evaluate its derivatives, we can differentiate the Euler's transformation
\begin{equation}
{}_2 F_1(a,b;c;w) = (1-w)^{c-a-b} {}_2 F_1(c-a,c-b;c;w)
\end{equation}
with respect to $a$ and specialize the parameters to $a = \frac{1}{2} = b$, $c = 1$ to find
\begin{equation}
\partial^{(1)} {}_2 F_1 \left( \frac{1}{2}, \frac{1}{2}; 1; w \right) = -\frac{1}{2} \log(1-w) \, {}_2 F_1 \left( \frac{1}{2}, \frac{1}{2}; 1; w \right).
\end{equation}
Analogously, starting from the identity
\begin{multline}
{}_2 F_1 (a,b;c;z) = \frac{\Gamma(c)\Gamma(c-a-b)}{\Gamma(c-a)\Gamma(c-b)} {}_2 F_1(a,b;a+b+1-c;1-z) \\
+ \frac{\Gamma(c)\Gamma(a+b-c)}{\Gamma(a)\Gamma(b)} (1-z)^{c-a-b} {}_2 F_1(c-a,c-b;c-a-b+1;1-z)
\end{multline}
and differentiating it with respect to $a$ and expanding around $c=1$ we find
\begin{multline}
\partial^{(3)} {}_2 F_1 \left( \frac{1}{2}, \frac{1}{2}; 1; w \right) = \\
= -\frac{\pi}{2} {}_2 F_1 \left( \frac{1}{2}, \frac{1}{2}; 1; 1-w \right) + \frac{1}{2} {}_2 F_1 \left( \frac{1}{2}, \frac{1}{2}; 1; w \right) \left[ \log(16(1-w)) - \log(w) \right].
\end{multline}
Using these, we find that the uniformization map behaves as
\begin{equation}
z \to w^{\frac{1}{KN}} \left[ 1 + \frac{1}{KN} \left( - \log \left( \frac{w}{16} \right) -\pi \frac{{}_2 F_1\left(\frac{1}{2},\frac{1}{2};1;1-w\right)}{{}_2 F_1\left(\frac{1}{2},\frac{1}{2};1;w\right)} \right) + \mathcal{O}\left(\frac{1}{N^3}\right) \right]
\end{equation}
In terms of variable $s = z^{KN}$ this simplifies to
\begin{equation}
s(w) \to 16 \exp \left[ -\pi \frac{K(1-w)}{K(w)} \right].
\end{equation}
In other words,
\begin{equation}
-\frac{1}{\pi} \log \frac{s}{16} = \frac{K(1-w)}{K(w)} \equiv -i\tau.
\end{equation}
It is well-known that if we put the right-hand side equal to $-i\tau$ with $\tau$ being the modular parameter, the inverse of this relation is given by the modular $\lambda$-function, i.e.
\begin{equation}
t^{-1} = w = m = k^2 = \lambda(\tau)
\end{equation}
and the $KN$-th power of the uniformization map is given simply by
\begin{equation}
s = 16 e^{i\pi\tau}.
\end{equation}
This can be compared to the large $K$ and $N$ limit of \eqref{automorphict}. We have
\begin{align}
\label{uniflambdatos}
\nonumber
\lambda(\tau) & = 16e^{i\pi\tau} - 128e^{2i\pi\tau} + 704e^{3i\pi\tau} - 3072e^{4i\pi\tau} + 11488e^{5i\pi\tau} - 38400e^{6i\pi\tau} + \ldots \\
& = s - \frac{1}{2} s^2 + \frac{11}{64} s^3 - \frac{3}{64} s^4 + \frac{359}{32768} s^5 - \frac{75}{32768} s^6 + \ldots
\end{align}
while the large $K$ and $N$ limit of \eqref{automorphict} is
\begin{align}
t & = w^{-1} = s^{-1} \left[ 1 + \frac{1}{2} s + \frac{5}{64} s^2 - \frac{31}{32768} s^4 + \frac{27}{2097152} s^6 + \ldots \right].
\end{align}
Taylor expanding $w$, we reproduce \eqref{uniflambdatos}. In terms of the modular parameter $\tau$, the curve \eqref{uniflargengeo} parametrizing half of the geodesic is given by the equation
\begin{equation}
\tau = -\frac{1}{2} + i\pi \frac{\sqrt{5}}{2} e^{i\varphi}, \qquad \varphi \in \left[ -\arcsin \left( \frac{2}{\sqrt{5}} \right), +\arcsin \left( \frac{2}{\sqrt{5}} \right) \right].
\end{equation}
Using these explicit parametrizations, we can plot the geodesics. Figure \ref{figgeouhp} shows how the sides of the fundamental polygon and the geodesics look like in $\tau$ coordinate system corresponding to the upper half plane model. Figure \ref{figpochlargekn} shows how Pochhammer contour in the $t$-plane looks like if we parametrize it using modular $\lambda$ function in the large $K$ and $N$ limit.

\section{Integrals of motion and their eigenvalues for Virasoro and $\mathcal{W}_3$}
\label{appiom}

\subsection{Period integrals for Virasoro algebra}
\label{appvirasoroperiods}
The first few local densities (normalized but up to total derivatives, i.e. not chosen to be quasi-primary) for Virasoro algebra are
\begin{align}
\nonumber
\mathcal{J}_2 & = T \\
\nonumber
\mathcal{J}_4 & = -\frac{1}{2} (TT) \\
\nonumber
\mathcal{J}_6 & = \frac{1}{2} \left[ (T(TT)) - \frac{c+2}{12}(\partial T \partial T) \right] \\
\nonumber
\mathcal{J}_8 & = -\frac{5}{8} \left[ (T(T(TT))) + \frac{1}{15} \left(c^2+14 c-21\right) (\partial T(\partial T T)) + \frac{1}{30} \left(c^2+19 c+19\right) (\partial^2 T (TT)) \right] \\
\nonumber
\mathcal{J}_{10} & = \frac{7}{8} \Big[ (T(T(T(TT)))) +\frac{1}{6} \left(c^2+24 c+59\right) (\partial T(\partial T(TT))) \\
\nonumber
& +\frac{1}{18} \left(c^2+29 c+129\right) (\partial^2 T(T(TT))) +\frac{1}{504} \left(10 c^3+287 c^2+1448 c-3842\right) (\partial^2 T(\partial T \partial T))\Big] \\
\nonumber
\mathcal{J}_{12} & = -\frac{21}{16} \Big[ (T(T(T(T(TT))))) \\
& +\frac{140 c^4+6444 c^3+85129 c^2+349106 c-198800}{3528} (\partial T(\partial T(T(TT)))) \\
\nonumber
& +\frac{-35 c^4-1506 c^3-16538 c^2-30678 c+180040}{2646} (\partial T(\partial T(\partial T \partial T))) \\
\nonumber
& +\frac{140 c^4+6444 c^3+85129 c^2+354986 c-81200}{14112} (\partial^2 T(T(T(TT)))) \\
\nonumber
& +\frac{-420 c^4-18492 c^3-216943 c^2-575518 c+1766912}{7056} (\partial^2 T(\partial T(\partial T T))) \\
\nonumber
& +\frac{-140 c^4-6444 c^3-83953 c^2-309122 c+418712}{14112} (\partial^2 T(\partial^2 T(TT))) \\
\nonumber
& +\frac{420 c^4+18940 c^3+228535 c^2+517726 c-2969568}{508032} (\partial^2 T(\partial^2 T \partial^2 T)) \Big].
\end{align}
Longer list is in the attached Mathematica notebook. The following is the list of the eigenvalues of the first few higher Hamiltonians (normalized periods) in Virasoro algebra. The variables $p_j$ denote the power sum polynomials of the \emph{orbifolded} Bethe roots $X_j$.
\begin{align}
\mathcal{I}_2 & = -\frac{K-1}{24(K+2)} + \frac{\ell(\ell+1)}{2(K+2)} + M \\
\nonumber
\mathcal{I}_4 & = -\frac{1}{2} M^2 + \frac{K+1}{8(K+2)} M - \frac{\ell (\ell+1)}{2(K+2)} M - \frac{(\ell-1)\ell(\ell+1)(\ell+2)}{8(K+2)^2} \\
& + \frac{(K-3)(K-1)(2K+3)}{1920(K+2)^2} + \frac{\ell(\ell+1)(K-3)}{16(K+2)^2} + \frac{K}{2(K+2)} p_1 \\
\nonumber
\mathcal{I}_6 & = \frac{1}{2} M^3 - \frac{(5K+7)}{16(K+2)} M^2 + \frac{3\ell(\ell+1)}{4(K+2)} M^2 - \frac{5(K-1)\ell(\ell+1)}{16(K+2)^2} M \\
\nonumber
& -\frac{(K+1)(2K^2-21K-41)}{384(K+2)^2} M + \frac{3(\ell-1)\ell(\ell+1)(\ell+2)}{8(K+2)^2} M - \frac{3K(K+3)}{2(K+2)(K+5)} p_1 M \\
& -\frac{(K-5)(2K^2-9K+13) \ell(\ell+1)}{768(K+2)^3} - \frac{3K(K+3)\ell(\ell+1)}{4(K+5)(K+2)^2} p_1 \\
\nonumber
& -\frac{5(K-5)(\ell-1)\ell(\ell+1)(\ell+2)}{64(K+2)^3} + \frac{(\ell-2)(\ell-1)\ell(\ell+1)(\ell+2)(\ell+3)}{16(K+2)^3} \\
\nonumber
& +\frac{K(K+3)(2K+3)}{16(K+2)^2} p_1 + \frac{3K^2}{4(K+2)^2(K+5)} p_2 - \frac{(K-5)(K-1)(24K^3+22K^2-117K-139)}{193536(K+2)^3} \\
\nonumber
\mathcal{I}_8 & = -\frac{5}{8} M^4 - \frac{5\ell(\ell+1)}{4(K+2)} M^3 + \frac{5(7K+11)}{48(K+2)} M^3 \\
\nonumber
& -\frac{15(\ell-1)\ell(\ell+1)(\ell+2)}{16(K+2)^2} M^2 + \frac{(14K^3-231K^2-826K-661)}{768(K+2)^2} M^2 \\
\nonumber
& +\frac{5\ell(\ell+1)(7K-1)}{32(K+2)^2} M^2 + \frac{15K(K+3)}{4(K+2)(K+7)} p_1 M^2 \\
\nonumber
& +\frac{(K+1)(24K^4-310K^3+805K^2+6010K+6071)}{27648(K+2)^3} M \\
\nonumber
& -\frac{5(\ell-2)(\ell-1)\ell(\ell+1)(\ell+2)(\ell+3)}{16(K+2)^3} M + \frac{35(K-3)(\ell-1)\ell(\ell+1)(\ell+2)}{64(K+2)^3} M \\
\nonumber
& +\frac{7(2K^3-33K^2+2K-43)\ell(\ell+1)}{768(K+2)^3} M - \frac{5K(K+3)(2K^2+15K+19)}{16(K+2)^2 (K+7)} p_1 M \\
\nonumber
& +\frac{15\ell(\ell+1)K(K+3)}{4(K+2)^2(K+7)} p_1 M - \frac{15K^2(2K+5)}{4(K+2)^2(K+7)(2K+7)} p_2 M \\
\nonumber
& -\frac{15K^2(K+3)^2}{8(K+2)^2(K+7)(2K+7)} p_1^2 -\frac{5K(K+3)(2K+1)\ell(\ell+1)}{32(K+2)^3} p_1 \\
\nonumber
& -\frac{(15K^2(2K+5))\ell(\ell+1)}{8(K+2)^3(K+7)(2K+7)} p_2 - \frac{5(\ell-3)(\ell-2)(\ell-1)\ell(\ell+1)(\ell+2)(\ell+3)(\ell+4)}{128(K+2)^4} \\
\nonumber
& +\frac{(K-7)(K-1)(432K^5-300K^4-5188K^3-1621K^2+14716K+12961)}{13271040(K+2)^4} \\
\nonumber
& +\frac{35(K-7)(\ell-2)(\ell-1)\ell(\ell+1)(\ell+2)(\ell+3)}{384(K+2)^4} \\
& +\frac{7(K-7)(2K^2-19K+69)(\ell-1)\ell(\ell+1)(\ell+2)}{3072(K+2)^4} \\
\nonumber
& +\frac{(K-7)(24K^4-118K^3+173K^2-290K+475)\ell(\ell+1)}{55296 (K+2)^4} \\
\nonumber
& +\frac{K(K+3)(2K+3)(12K^2+85K+113)}{768(K+2)^3} p_1 + \frac{15K(K+3)(\ell-1)\ell(\ell+1)(\ell+2)}{16(K+2)^3(K+7)} p_1 \\
\nonumber
& +\frac{5K^2(2K+5)(3K+5)}{32(K+2)^3(K+7)} p_2 + \frac{15K^3}{8(K+2)^3(K+7)(2K+7)} p_3
\end{align}

\subsection{Local conserved quantities for $\mathcal{W}_3$}
\label{appw3im}
Here we summarize the conventions we use for $\mathcal{W}_3$-algebra and list first few conserved charges. Apart from the stress-energy tensor $T(z)$ satisfying the usual Virasoro OPE, we have another primary field $W$ of conformal dimension $3$,
\begin{equation}
T(z) W(w) \sim \frac{3W(w)}{(z-w)^2} + \frac{\partial W(w)}{z-w} + reg.
\end{equation}
The associativity conditions determine the OPE of $W$ with itself up to an overall normalization,
\begin{align}
\nonumber
W(z) W(w) & \sim C_{33}^0 \Bigg[ \frac{1}{(z-w)^6} + \frac{\frac{6}{c}T(w)}{(z-w)^4} + \frac{\frac{3}{c} \partial T(w)}{(z-w)^3} \\
& + \frac{1}{(z-w)^2} \left(\frac{96}{c(5c+22)} \Lambda(w) + \frac{9}{10c} \partial^2 T(w) \right) \\
\nonumber
& + \frac{1}{z-w} \left( \frac{48}{c(5c+22)} \partial \Lambda(w) + \frac{1}{5c} \partial^3 T(w) \right) \Bigg] + reg.
\end{align}
were $\Lambda$ is the quasi-primary composite
\begin{equation}
\Lambda(z) = (TT)(z) - \frac{3}{10} \partial^2 T(z).
\end{equation}
We follow the conventions where we normalize $W(z)$ as
\begin{equation}
W(z) = -U_3(z) + \ldots
\end{equation}
which implies
\begin{equation}
C_{33}^0 = \frac{c(5c+22)}{144}
\end{equation}
and agrees with the conventions used in \cite{Bazhanov:2001xm}. Following these conventions, the first few non-zero local currents are
\begin{align}
\nonumber
\mathcal{J}_2 & = T \\
\nonumber
\mathcal{J}_3 & = \sqrt{3} W \\
\nonumber
\mathcal{J}_5 & = -\sqrt{3} (TW) \\
\mathcal{J}_6 & = -\frac{1}{3} \left[ (T(TT)) + 9(WW) - \frac{c-10}{32} (\partial T \partial T) \right] \\
\nonumber
\mathcal{J}_8 & = \frac{1}{3} \Big[ (T(T(TT))) + 18(T(WW)) +\frac{1}{320} \left(5 c^2+64 c-852\right) (\partial T(\partial T T)) \\
\nonumber
& - \frac{3(c+30)}{8} (\partial W \partial W) +\frac{1}{640} \left(5 c^2+124 c-492\right) (\partial^2 T(TT)) \Big] \\
\nonumber
\mathcal{J}_9 & = \frac{5\sqrt{3}}{3} \int \Big[ (T(T(TW))) +\frac{1}{16} (c+26) (T(T\partial W)) + 3(W(WW)) \\
\nonumber
& -\frac{c+14}{32} (\partial T(\partial T W)) +\frac{1}{960} \left(-c^2-22 c+440\right) (\partial^3 T \partial W) \Big]
\end{align}
The eigenvalues of the corresponding charges in terms of Bethe roots are
\begin{align}
\mathcal{I}_2 & = \frac{b_1^2}{9(K+3)} - \frac{b_1}{3(K+3)} - \frac{b_2}{3(K+3)} - \frac{K-1}{12(K+3)} + M \\
\mathcal{I}_3 & = \frac{b_1 M}{3\sqrt{K+3}}+\frac{2b_1^3-9b_1^2-9b_1 b_2+9b_1+27b_2+27b_3}{81(K+3)^{3/2}}+\frac{2}{3\sqrt{K+3}} \sum_j \beta_j \\
\mathcal{I}_4 & = 0 \\
\nonumber
\mathcal{I}_5 & = \frac{2b_2}{9(K+3)^{3/2}} \sum_{j=1}^M \beta_j -\frac{2b_1^2}{27(K+3)^{3/2}} \sum_{j=1}^M \beta_j +\frac{2b_1}{9(K+3)^{3/2}} \sum_{j=1}^M \beta_j \\
\nonumber
& +\frac{2(K+2)}{9(K+3)^{3/2}} \sum_{j=1}^M \beta_j -\frac{6M}{9\sqrt{K+3}} \sum_{j=1}^M \beta_j -\frac{K}{3(K+3)^{3/2}} \sum_{j=1}^M X_j -\frac{b_1 M^2}{3\sqrt{K+3}} \\
\nonumber
& -\frac{5 b_1^3 M}{81(K+3)^{3/2}} + \frac{2b_1^2 M}{9(K+3)^{3/2}} + \frac{2b_1 b_2 M}{9(K+3)^{3/2}} + \frac{(K+1) b_1 M}{9(K+3)^{3/2}} - \frac{b_2 M}{3(K+3)^{3/2}} \\
& - \frac{b_3M}{3(K+3)^{3/2}} -\frac{2b_1^5}{729(K+3)^{5/2}} +\frac{5b_1^4}{243(K+3)^{5/2}} +\frac{5b_1^3 b_2}{243(K+3)^{5/2}} \\
\nonumber
& +\frac{2(K-4)b_1^3}{243(K+3)^{5/2}} -\frac{b_1^2 b_2}{9(K+3)^{5/2}} -\frac{(K+1) b_1^2}{27(K+3)^{5/2}} -\frac{b_1^2 b_3}{27(K+3)^{5/2}} +\frac{b_1 b_3}{9(K+3)^{5/2}} \\
\nonumber
& -\frac{b_1 b_2^2}{27(K+3)^{5/2}} +\frac{(K+2) b_1}{27(K+3)^{5/2}} -\frac{(K-2) b_1 b_2}{27(K+3)^{5/2}} +\frac{b_2^2}{9(K+3)^{5/2}} +\frac{(K+2) b_2}{9(K+3)^{5/2}} \\
\nonumber
& +\frac{b_2 b_3}{9(K+3)^{5/2}} +\frac{(K+2) b_3}{9(K+3)^{5/2}} \\
\nonumber
\mathcal{I}_6 & = -\frac{(K+1)(2K^2-13K-49)b_1}{432(K+3)^3} - \frac{(K+1)(2K^2-13K-49)b_2}{432(K+3)^3} \\
\nonumber
& +\frac{(2K^3-11K^2-2K+67)b_1^2}{1296(K+3)^3} - \frac{(5K-11)b_2 b_1^2}{162(K+3)^3} -\frac{7b_1^6}{6561(K+3)^3} +\frac{7b_1^5}{729(K+3)^3} \\
\nonumber
& +\frac{5(3K-11)b_1^4}{2916(K+3)^3} +\frac{7b_1^4 b_2}{729(K+3)^3}-\frac{5(K+1)b_1^3}{162(K+3)^3} - \frac{16 b_1^3 b_2}{243(K+3)^3} -\frac{4 b_1^3 b_3}{243(K+3)^3} \\
\nonumber
& +\frac{2b_1^2 b_3}{27(K+3)^3} -\frac{2b_1^2 b_2^2}{81(K+3)^3} +\frac{b_1 b_2^2}{9(K+3)^3} +\frac{(5K+7)b_1 b_2}{54(K+3)^3} +\frac{2b_1 b_2 b_3}{27(K+3)^3} \\
\nonumber
& -\frac{2 b_1 b_3}{27(K+3)^3} +\frac{b_2^3}{81(K+3)^3} +\frac{(5K-1) b_2^2}{108(K+3)^3} -\frac{b_3^2}{9(K+3)^3} -\frac{2b_2 b_3}{9(K+3)^3} \\
& +\frac{(K-5)(K-1)(16K^3+34K^2-93K-167)}{36288(K+3)^3} \\
\nonumber
& +\frac{4b_1^2 b_2 M}{27(K+3)^2} -\frac{7b_1^4 M}{243(K+3)^2} +\frac{4b_1^3 M}{27(K+3)^2} +\frac{(5K+1)b_1^2 M}{54(K+3)^2} -\frac{4b_1 b_2 M}{9(K+3)^2} \\
\nonumber
& -\frac{2b_1 b_3 M}{9(K+3)^2} -\frac{(5K+11) b_1 M}{18(K+3)^2} -\frac{b_2^2 M}{9(K+3)^2} -\frac{(5K+11) b_2 M}{18(K+3)^2} \\
\nonumber
& +\frac{(K+1) \left(2 K^2-13 K-49\right) M}{144 (K+3)^2} + \frac{b_1 M^2}{3(K+3)} +\frac{b_2 M^2}{3(K+3)} -\frac{2 b_1^2 M^2}{9(K+3)} \\
\nonumber
& +\frac{5K M^2}{12(K+3)} +\frac{11M^2}{12(K+3)} -\frac{M^3}{3} -\frac{2K(K+4) b_1}{9(K+3)^2(K+5)} \sum_{j=1}^M X_j -\frac{2K}{9(K+3)^2} \sum_{j=1}^M \beta_j X_j \\
\nonumber
& +\Bigg(-\frac{8 b_1^3}{243(K+3)^2} +\frac{4b_1^2}{27(K+3)^2} +\frac{4b_1 b_2}{27(K+3)^2} -\frac{4b_1}{27(K+3)^2} \\
\nonumber
& -\frac{4b_2}{9(K+3)^2} -\frac{4b_3}{9(K+3)^2}\Bigg) \sum_{j=1}^M \beta_j -\frac{4b_1 M}{9(K+3)} \sum_{j=1}^M \beta_j -\frac{4}{9(K+3)} \left(\sum_{j=1}^M \beta_j\right)^2.
\end{align}

\section{Transformation of third order differential operator}

\label{appw3comparison}
In this section we will transform the third order differential operator considered by Masoero and Raimondo \cite{Masoero:2019wqf} to the form that we considered in Section \ref{secw3}. The starting point of \cite{Masoero:2019wqf} is the operator
\begin{equation}
\label{mroper}
\partial_z^3 \psi - W_1 \partial_z \psi + W_2 \psi = 0
\end{equation}
with
\begin{align}
W_1 & = \frac{\bar{r}^1}{z^2} + \sum_{j=1}^M \left( \frac{3}{(z-w_j)^2} + \frac{a_{11}^{(j)}}{z(z-w_j)} \right) \\
W_2 & = \frac{\bar{r}^2}{z^3} + \frac{1}{z^2} + \lambda z^k + \sum_{j=1}^M \left( \frac{3}{(z-w_j)^3} + \frac{a_{21}^{(j)}}{z(z-w_j)^2} + \frac{a_{22}^{(j)}}{z^2(z-w_j)} \right)
\end{align}
which was also used in \cite{Ashok:2024zmw}. Let us first focus on the primary (i.e. for now we ignore the descendants -- the sums over $j$ are zero). Let us put
\begin{equation}
\label{transmrtoad}
x = A z^\alpha, \qquad z = A^{-1/\alpha} x^{1/\alpha}.
\end{equation}
This change of coordinates generates a second derivative term, but we can eliminate it by multiplicatively transforming the wave function (i.e. by pre-composing the differential operator by a scalar multiplication such as $z^{1-\alpha}$ and post-composing by an analogous factor). We find an operator
\begin{align}
\partial_x^3 + \frac{\alpha^2-1-\bar{r}_1}{\alpha^2 x^2} \partial_x + \left(\frac{\lambda A^{-\frac{k+3}{\alpha}} x^{\frac{k+3-3\alpha}{\alpha}}}{\alpha^3}+\frac{A^{-1/\alpha} x^{\frac{1-3\alpha}{\alpha}}}{\alpha^3}-\frac{\alpha(\alpha^2-1)-(\alpha-1)\bar{r}_1-\bar{r}_2}{\alpha^3 x^3}\right).
\end{align}
After the transformation, the term proportional to spectral parameter $\lambda$ should be $x$-independent (this is one of the characteristic features of the frame which we consider in the main part of the article). Therefore we need to require
\begin{equation}
\alpha = \frac{k+3}{3}.
\end{equation}
The powers of $x$ that appear in the potential term with no derivatives are $0, -3$ and
\begin{equation}
-\frac{3(k+2)}{k+3}
\end{equation}
which (comparing to powers $0, -3$ and $K$ in \eqref{opern3}) leads to symmetric identification
\begin{equation}
K = -\frac{3(k+2)}{k+3} \qquad \text{or} \qquad k = -\frac{3(K+2)}{K+3}.
\end{equation}
These equations are also compatible with the expressions for the central charge in terms of $k$, see \cite{Bazhanov:2001xm}. Fixing the normalization of $x^K$ term by choosing
\begin{equation}
A = \left(K+3\right)^{\frac{3}{K+3}},
\end{equation}
we finally find an operator of the form \eqref{opern3} if we identify
\begin{align}
\nonumber
b_1 & = 0 \\
\nonumber
b_2 & = -(K+2)(K+4) - (K+3)^2 \bar{r}_1 \\
b_3 & = (K+2)(K+4)-(K+2)(K+3)^2\bar{r}_1+(K+3)^3\bar{r}_2 \\
\nonumber
u & = -\left(K+3\right)^{\frac{3K}{K+3}} \lambda.
\end{align}
We can now consider the contribution of the descendants. Since the full transformation \eqref{transmrtoad} is
\begin{equation}
x =  \left(K+3\right)^{\frac{3}{K+3}} z^{\frac{1}{K+3}},
\end{equation}
every singular point $z = w_j$ in \eqref{mroper} gives rise to $K+3$ singularities at $x = x_{j,\ell}$
\begin{equation}
x_{j,\ell} = e^{\frac{2\pi i \ell}{K+3}} \left(K+3\right)^{\frac{3}{K+3}} w_j^{\frac{1}{K+3}}, \qquad \ell=0,1,\ldots,K+2.
\end{equation}
The descendant part of the transformed differential operator is
\begin{multline}
\sum_{j=1}^M \left( -\frac{3(K+3)^2 x^{2K+4}}{\left(x^{K+3}-X_j\right)^2} - \frac{(K+3)^2 x^{K+1} a_{11}^{(j)}}{x^{K+3}-X_j} \right) \partial_x + \sum_{j=1}^M \left( -\frac{(K+2) (K+3)^2 x^K a_{11}^{(j)}}{x^{K+3}-X_j} \right. \\
\left. + \frac{(K+3)^3 x^{2K+3} a_{21}^{(j)}}{\left(x^{K+3}-X_j\right)^2} + \frac{(K+3)^3 x^K a_{22}^{(j)}}{x^{K+3}-X_j} + \frac{3(K+3)^2 x^{2K+3} \left(x^{K+3}+KX_j+2X_j\right)}{\left(x^{K+3}-X_j\right)^3} \right)
\end{multline}
where $X_j = x_{j,\ell}^{K+3}$. Next we can use the orbit sums from Appendix \ref{apporbitsums}. We find that the descendant contribution can be written as
\begin{multline}
\sum_{j,\ell} \left( -\frac{3}{\left(x-x_{j,\ell}\right)^2} - \frac{3(K+2)+(K+3)a_{11}^{(j)}}{x(x-x_{j,\ell})} \right) \partial_x + \sum_{j,\ell} \left( \frac{3}{\left(x-x_{j,\ell}\right)^2} + \frac{2(K+3)a_{21}^{(j)}+3(K+2)}{2x\left(x-x_{j,\ell}\right)^2} \right. \\
\left. + \frac{-2(K+2)(K+3)\left(a_{11}^{(j)}-a_{21}^{(j)}\right)+2(K+3)^2a_{22}^{(j)}+3(K+2)}{2x^2(x-x_{j,\ell})} \right).
\end{multline}
Comparing this to \eqref{opern3}, we find the further identifications
\begin{align}
\nonumber
c_{21}^j & = -3(K+2)-(K+3) a_{11}^{(j)} \\
c_{31}^j & = (K+3) a_{21}^{(j)} + \frac{3}{2}(K+2) \\
\nonumber
c_{32}^j & = \frac{-2(K+2)(K+3)\left(a_{11}^{(j)}-a_{21}^{(j)}\right) + 2(K+3)^2 a_{22}^{(j)}+3(K+2)}{2}.
\end{align}

\paragraph{Comparison of Bethe equations}
Let us write the Bethe equations for \eqref{mroper}. We can derive them following the same steps as in \cite{Masoero:2019wqf} or in Section \ref{secw3}. If we expand the potentials as
\begin{align}
\nonumber
W_1 & = \sum_{m=0}^\infty q_{1m}^{(j)} (z-w_j)^{m-2} \\
W_2 & = \sum_{m=0}^\infty q_{2m}^{(j)} (z-w_j)^{m-3}
\end{align}
with $q_{10}^{(j)} = 3 = q_{20}^{(j)}$ as in \cite{Masoero:2019wqf}, the equation has trivial monodromy around $z = w_j$ if and only if
\begin{align}
\label{mrfirsteqn}
q_{12}^{(j)} & = \frac{1}{3} \left( q_{11}^{(j)2} - q_{11}^{(j)} q_{21}^{(j)} + q_{21}^{(j)2} \right) \\
\label{mrsecondqn}
q_{22}^{(j)} & = -\frac{1}{3} q_{11}^{(j)} \left( q_{11}^{(j)} - 2q_{21}^{(j)} \right)
\end{align}
and
\begin{equation}
\label{mrthirdeqn}
q_{14}^{(j)}+q_{24}^{(j)} = q_{11}^{(j)}q_{23}^{(j)} + \frac{1}{3} q_{13}^{(j)} (2q_{11}^{(j)}-q_{21}^{(j)}) + \frac{1}{27} q_{11}^{(j)} (q_{11}^{(j)}+q_{21}^{(j)}) (q_{11}^{(j)}-2q_{21}^{(j)}) (2q_{11}^{(j)}-q_{21}^{(j)})
\end{equation}
The explicit expressions for the expansion coefficients $q_{kl}^{(j)}$ are given in \cite{Masoero:2019wqf} below equation (2.18). The parameter $\lambda$ appears only linearly in $q_{23}^{(j)}$ and $q_{24}^{(j)}$. Requiring absence of logarithmic terms for any value of $\lambda$ in \eqref{mrthirdeqn} gives two conditions: the terms linear in $\lambda$ impose
\begin{equation}
\frac{k}{w_j} = q_{11}^{(j)} \qquad \text{or} \qquad a_{11}^{(j)} = k.
\end{equation}
Using this in \eqref{mrsecondqn}, we have \footnote{Here we slightly differ with respect to equation (2.21) in \cite{Masoero:2019wqf} but we are consistent with equations stated in \cite{Ashok:2024zmw,Ashok:2024ygp}.}
\begin{equation}
a_{22}^{(j)} = \frac{2k+3}{3} a_{21}^{(j)} - \frac{k^2}{3}.
\end{equation}
Following \cite{Masoero:2019wqf}, we label the remaining parameter $a_j \equiv a_{21}^{(j)}$. The equation \eqref{mrfirsteqn} reduces to a quadratic equation for $a_j$,
\begin{equation}
\label{mrbethe1}
a_j^2- k a_j + k^2 + 3k -3\bar{r}_1 = \sum_{\ell \neq j} \left( \frac{9w_j^2}{(w_j-w_\ell)^2} + \frac{3k w_j}{w_j-w_\ell} \right)
\end{equation}
which is equation (1.2a) of \cite{Masoero:2019wqf} and is a quadratic equation for the coefficients $a_j$ if we know the positions of the singularities $w_j$. Finally, we have the $\lambda$-independent part of the equation \eqref{mrthirdeqn}:
\begin{multline}
\label{mrbethe2}
A a_{21}^{(j)} +B-9 (k+2) w_j = \sum_{\ell \neq j} \Bigg[ 18(a_j+a_\ell-k) \frac{w_j^3}{(w_j-w_\ell)^3} \\
+ 3\left(3ka_j+5ka_\ell+6a_\ell-4k^2-3k\right) \frac{w_j^2}{(w_j-w_\ell)^2} + (2k+3)\left(ka_j+3k a_\ell+6a_\ell-2k^2-3k\right) \frac{w_j}{w_j-w_\ell} \Bigg]
\end{multline}
with
\begin{equation}
A = 2(k+1)(k+3)^2-2(k+3)\bar{r}_1
\end{equation}
and
\begin{equation}
B = -k(k+1)(k+3)^2+(k+3)(k+9)\bar{r}_1-9(k+3)\bar{r}_2.
\end{equation}
The coefficients $A$ and $B$ agree with the equations given in \cite{Ashok:2024zmw} if we replace $r_j$ by $\bar{r}_j$. The equations \eqref{mrbethe1} and \eqref{mrbethe2} are Bethe equations for $\mathcal{W}_3$ in coordinates used in \eqref{mroper}.

\bibliography{winfode}

\end{document}